\documentclass[journal]{IEEEtran}

\newtheorem{prop}{Proposition}

\usepackage{enumitem}
\usepackage{caption}
\newtheorem{lemma}{Lemma}
\newtheorem{remark}{Remark}
\usepackage{ifpdf}

\usepackage{cite}

\ifCLASSINFOpdf
   \usepackage[pdftex]{graphicx}
\else
   \usepackage[dvips]{graphicx}
\fi

\usepackage{amsmath}
\usepackage{amssymb}

\usepackage{array}
\usepackage{multirow}

\ifCLASSOPTIONcompsoc
  \usepackage[caption=false,font=normalsize,labelfont=sf,textfont=sf]{subfig}
\else
  \usepackage[caption=false,font=footnotesize]{subfig}
\fi

\usepackage{stfloats}

\usepackage{url}

\usepackage{color}

\begin{document}
\setlength{\abovedisplayskip}{3pt}
\setlength{\belowdisplayskip}{3pt}
\title{Extremely Sparse Co-Prime Sensing: Low Latency Estimation is not a Dream but a Reality}
\author{Usham~V.~Dias\\ Department of Electrical Engineering, Indian Institute of Technology Delhi, India\\
\vspace{8mm}

}

\maketitle
\begin{abstract}
Co-prime sensing is a sub-Nyquist technique for signal acquisition. Several modifications to the prototype co-prime array have been proposed in the literature. Researchers have also demonstrated low latency estimation. This paper describes the functioning of a Family of Adjustable Pivot Co-Prime Arrays. However, the main focus is to introduce the reader to a concept called Extremely Sparse Co-Prime Arrays and Samplers. Adjustable pivot co-prime arrays are a special case of the extremely sparse arrays. The closed-form expressions for the weight function and the correlogram bias window are derived. Low latency estimation is demonstrated using the extremely sparse co-prime scheme. Furthermore, a multidimensional and a hybrid extremely sparse co-prime array is proposed as a straightforward extension of the 1D-theory. Finally, a generalized extremely sparse structure is developed with several design parameters. Most existing structures may be viewed as a special case of the generalized scheme. 
\end{abstract}
%
%
%
%
\IEEEpeerreviewmaketitle
%
\section{Introduction}
\label{sec:intro}
Spatial or temporal sampling of a signal is the interface between the analog and digital domains. Nyquist theorem forms the basis for accurate sampling (acquisition) of the analog signal. The initial work in this area is attributed to Shannon, Nyquist, Whittaker, and Kotel’nikov~\cite{1.10,1.11,1.13,1.14}. The evolution of Nyquist and sub-Nyquist strategies is described in~\cite{1.9}.
Compressive sensing is a sub-Nyquist technique that uses the sparsity property of a signal for acquisition~\cite{1.5,1.6,1.7}. The three main design components in compressive sensing include projection matrix, sparsifying basis, and reconstruction algorithm. However, hardware implementation is a major challenge. Literature surveys on compressive sensing have been conducted in~\cite{survey_CSS1,survey_CSS2,survey_CSS3,survey_CSS4,survey_CSS5,survey_CSS6,survey_CSS7,survey_CSS8,survey_CSS9,survey_CSS10,survey_CSS11,survey_CSS12,survey_CSS13,survey_CSS14,survey_CSS15}. They consider several aspects including applications. In the field of communication, wide-band spectrum sensing for cognitive radios  is an important research topic~\cite{10.3,1.1}. 
Modulated wide-band converters (MWC) can be used to acquire signals that cannot be handled by commercial analog-to-digital converters~\cite{1.21}. Xampling is a technique to implement MWC with available analog devices~\cite{1.24,1.25}. It processes signals in the Union of sub-spaces, reduces input signal bandwidth, and employ non-linear algorithms. 

However, the focus in this paper is on sub-Nyquist co-prime sensing schemes. Minimum redundancy array~\cite{1.22} and minimum hole array~\cite{MHA} are some of the early works reported on sparse arrays. Subsequently, the nested~\cite{4.52} and co-prime arrays~\cite{4.7} were proposed with analytical expressions. These structures can estimate the second order statistics like autocorrelation and cross-correlation. The nested array has several modifications. Super nested array is proposed in~\cite{Nested1}. A compressed symmetric nested array is proposed in~\cite{Nested2}. A new nested array structure is described in ~\cite{Nested3}. Generalized nested array is proposed in~\cite{Nested5}. Nested array with displaced subarray (NADiS) is proposed in~\cite{Nested6}. A widened nested array is described in~\cite{Nested4}.

The co-prime array also has several modifications. The prototype array cannot estimate the autocorrelation at all difference values within a co-prime period. Hence, one of the sub-arrays of the prototype array is extended by an additional co-prime period. This concept is referred to as extended or conventional co-prime array~\cite{4.62}. Redundant elements in the extended co-prime array are done away with in the thinned co-prime array~\cite{20.1}. Another structure called co-prime array with compressed inter-element spacing (CACIS) has an additional parameter $p$ known as the compression factor. The prototype co-prime and the basic nested arrays are special cases of CACIS. In addition, co-prime array with displaced sub-arrays (CADiS) is described~\cite{4.44}. Co-prime arrays with multiple levels/n-tuple have been considered in~\cite{20.4,20.2,20.3,20.5,20.6,20.7} A semi-coprime array is proposed in~\cite{semiCA}. An attempt to unify the co-prime arrays is made in~\cite{CAMPs}.
Multidimensional co-prime arrays are considered in~\cite{multidim1, multidim2,multidim3, multidim4, rect1}. Other structures include L-shaped co-prime arrays~\cite{ShapeL1, ShapeL2, ShapeL3, ShapeL4, ShapeL5,ShapeL6,ShapeL7,ShapeL8, ShapeL9, ShapeL10}, V-shaped co-prime arrays~\cite{ShapeV1}, and circular co-prime arrays~\cite{ShapeCirc1, ShapeCirc2, ShapeCirc3, ShapeCirc4}.

MUSIC algorithm, DFT filter banks, Min processing, Multiplicative processing, etc. are some of the methods that have been used for co-prime based estimation~\cite{4.62,4.7,DFT_filter,MIN_defense,I2}. The initial paper had described several applications of co-prime sensing~\cite{4.7}.
Some of the applications of co-prime sensing are Direction-of-Arrival (DoA) estimation~\cite{4.45, F2, F3,F5}, beamforming~\cite{F1,F4, 16.1}, system identification~\cite{4.40}, power spectrum estimation~\cite{4.28,4.32}, speech~\cite{Speec, ShapeCirc1}, MIMO~\cite{multidim3}, \cite{mimo1, mimo2, mimo3,mimo4}, doppler~\cite{doppler1}, space-time~\cite{spacetime1}, cognitive radio~\cite{CR1}, cross-correlation, cross-spectrum, time-delay, range, velocity, and acceleration estimation~\cite{U_S_NCC2018,UVD_PHD}.

Low latency co-prime based modal analysis is discussed in~\cite{I1}. It uses ten snapshots for estimation. Frequency estimation~\cite{S1,U_S_1} is also demonstrated with few sapshots. A detailed study of low latency co-prime sensing is found in~\cite{UVD_PHD}. It describes the difference set theory for the prototype as well as the multiple period co-prime scheme. The closed-form expressions for the weight function, correlogram bias window and variance is derived. It was shown that consecutive integers of low value e.g. $M=7$, $N=6$ is a good choice. This choice of (M, N) reduces the side-lobe peak of the bias window with respect to the main-lobe peak. Therefore, it has the potential to reduce the spectral leakage. Furthermore, correlogram spectral estimation with few snapshots is demonstrated. Estimation was possible with just two snapshots but it may not be reliable. The estimate is shown to converge with around ten snapshots in (Fig.4.11~\cite{UVD_PHD}). In addition, correlogram min processor, correlogram multiplicative processor, and co-prime DFT filter banks with low latency is presented (Section 4.4~\cite{UVD_PHD}). It also considers multiple period co-prime correlogram with much better results. For example, $r=5$ periods could estimate the spectrum with just two snapshots with better resolution (Fig.4.21 in~\cite{UVD_PHD}). Besides autocorrelation and correlogram, it also discusses cross-correlation, cross-spectrum, time delay, range, velocity, and acceleration estimation. Another important contribution of that thesis is the proof that power spectrum can be estimated using the entire difference set (or combined set) including holes. It guarantees a valid power spectrum. However, the continuous set does not guarantee a valid power spectrum, and may require an external window function. Low latency correlogram estimation using extended co-prime array has been studied in~\cite{UVD_extended}. It notes that $M\approx\frac{N}{2}$ is a good choice. The CACIS structure has also been studied for low latency along similar lines in~\cite{UVD_CACIS}. However, it has not concluded on the choice of parameter and that remains an open question. Adjustable pivot co-prime array (APCA) is a modification of the prototype co-prime array. It allows the location of the pivot (or overlapping sub-array element) to be adjusted using a shifting parameter or pivot selection parameter $`s$'. APCA concept dates back to 2017 and demonstrates low latency estimation\footnote{APCA concept had three unsuccessful attempts at publication (two conferences and one letter). The last attempt at publication also included a family of adjustable pivot structures as in Fig.~\ref{fig:Extended_multi co-prime structures} and~\ref{fig:Truly_coprime_CACIS}. However, it will be shown that the APCA (and it's family) is only a special case of the extremely sparse co-prime sensing framework proposed in this paper.}. Another concept, referred as super-Nyquist co-prime sensing has also demonstrated low latency estimation~\cite{UVD_supernyquist}. It motivates the idea of extremely sparse co-prime arrays proposed in this paper. Recently, 2D estimation with twenty five snapshots has been presented in~\cite{rect1}. It proposes symmetry-imposed rectangular co-prime structure. 

Key contributions of this paper are mentioned below:
\begin{enumerate}
	\item Adjustable Pivot Co-Prime Arrays are presented in Section~\ref{sec:APCA} and low latency estimation is demonstrated.
	\item Extremely Sparse Co-Prime Array (ExSCA) is proposed in Section~\ref{EXTREMELY_proto}. The concept and the difference set is explained. The weight function and correlogram bias window expressions are developed. Low latency estimation is demonstrated.
	\item APCA is shown to be a special case of ExSCA. 
	\item Multidimensional ExSCA is described as an extension of the 1D-theory in Section~\ref{MultiExSCA}. A hybrid ExSCA is also proposed.
	\item A generalized ExSCA is proposed in Section~\ref{GenExSCA}.
\end{enumerate}
 
The reader is expected to have some background of the theory developed in~\cite{4.7} and~\cite{UVD_PHD}. The correlogram spectral estimation theory for Nyquist sampling is available in~\cite{7.1,7.17} and is developed for co-prime sensing in~\cite{UVD_PHD}. Note that the term arrays and samplers will be used interchangeably. The theory is valid for both spatial antenna arrays and temporal sampling. However, the simulations presented here will focus on temporal sampling. 
\section{Adjustable Pivot Co-Prime Arrays}
\label{sec:APCA}
It is well known that the first element of the two sub-arrays overlap for the prototype co-prime array. However, it can be designed to allow any other element to overlap. This idea is referred to as the Adjustable Pivot Co-Prime Arrays (APCA). Pivot is the overlapping element.

\subsection{Structure} 
\label{sec:structure}
\begin{figure}
\centering
\includegraphics[width=0.3\textwidth]{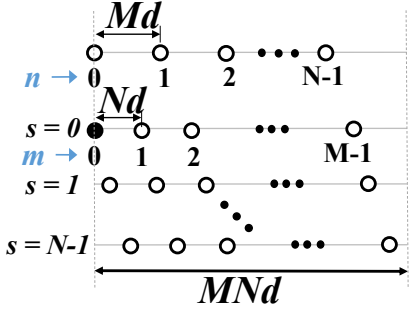}%
\caption{Adjustable pivot co-prime array structure.}
\label{fig:extreme_shifted_coprime_structure}
\end{figure}

The adjustable pivot co-prime array (APCA) structure is shown in Fig.~\ref{fig:extreme_shifted_coprime_structure}. The spatial (or temporal) positions of the elements of the two sub-arrays are given below:
\begin{eqnarray}
 \nonumber
  P1 &=& \{Mnd,\; 0\leq n \leq N-1\} \\ 
  P2 &=& \{(Nm+s)d,\; 0\leq m \leq M-1\}
\end{eqnarray} 
where $d$ is the Nyquist distance, $(M, N)$ is a co-prime pair, and $s$ is the pivot selection (or shifting) parameter. $s$ is an integer in the range $0\leq s\leq N-1$. As mentioned earlier, it determines the location of the overlapping array element. APCA structure for $(M, N)=(4, 3)$  is shown in Fig.~\ref{APCA_structure_M4N3}. The designer needs to select an appropriate value of $s$. In the remainder of this paper, the following is assumed without loss of generality: (a) $d=1$ for the ease of understanding and (b) the sub-array with inter-element spacing $Nd$ has a variable location, while the other sub-array has its first element at the origin. For every shift $s\in[0, N-1]$, one of the array elements in the shifted sub-array overlaps with exactly one element in the other sub-array. Therefore, APCA structure has the same number of elements as the prototype co-prime array, i.e. $M+N-1$ antennas or $M+N$ temporal samples. Thus, APCA does not incur additional costs. 
 \subsection{Difference Set Analysis}
\label{sec:Diff_set}
\begin{figure}
\centering
\subfloat[APCA structure]{\includegraphics[width=0.48\textwidth]{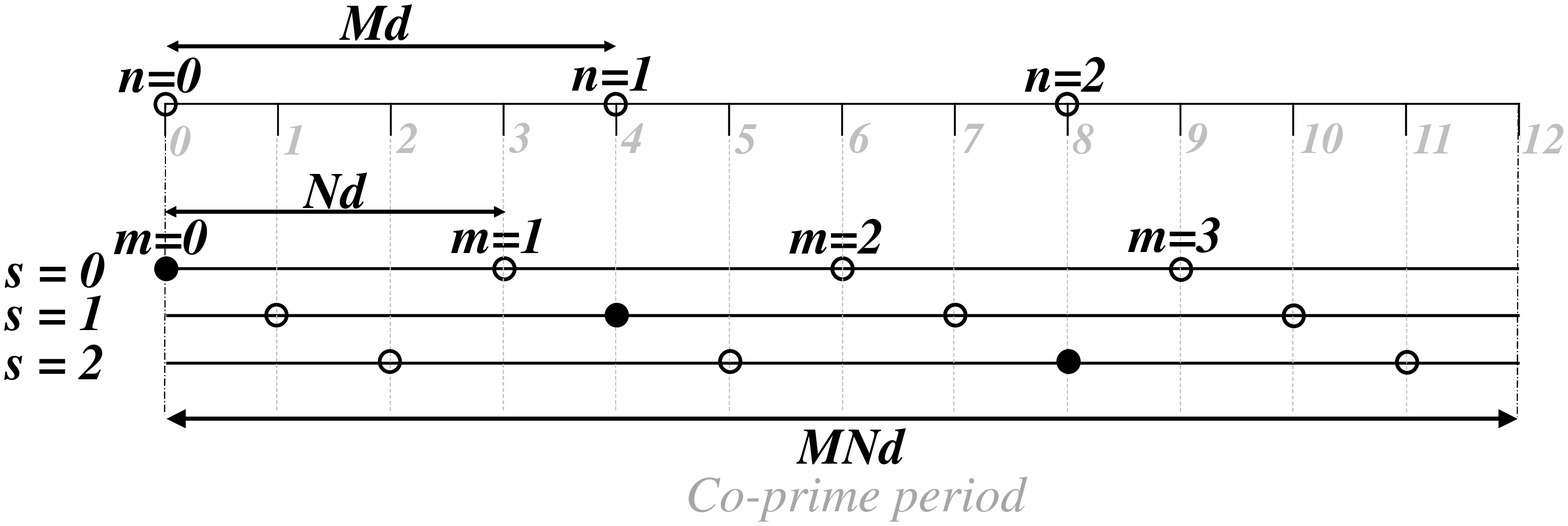}
\label{APCA_structure_M4N3}}
\hfil
\subfloat[Set $\mathcal{L}^+_{SN}\cup \mathcal{L}^-_{SN}$]{\includegraphics[width=0.24\textwidth]{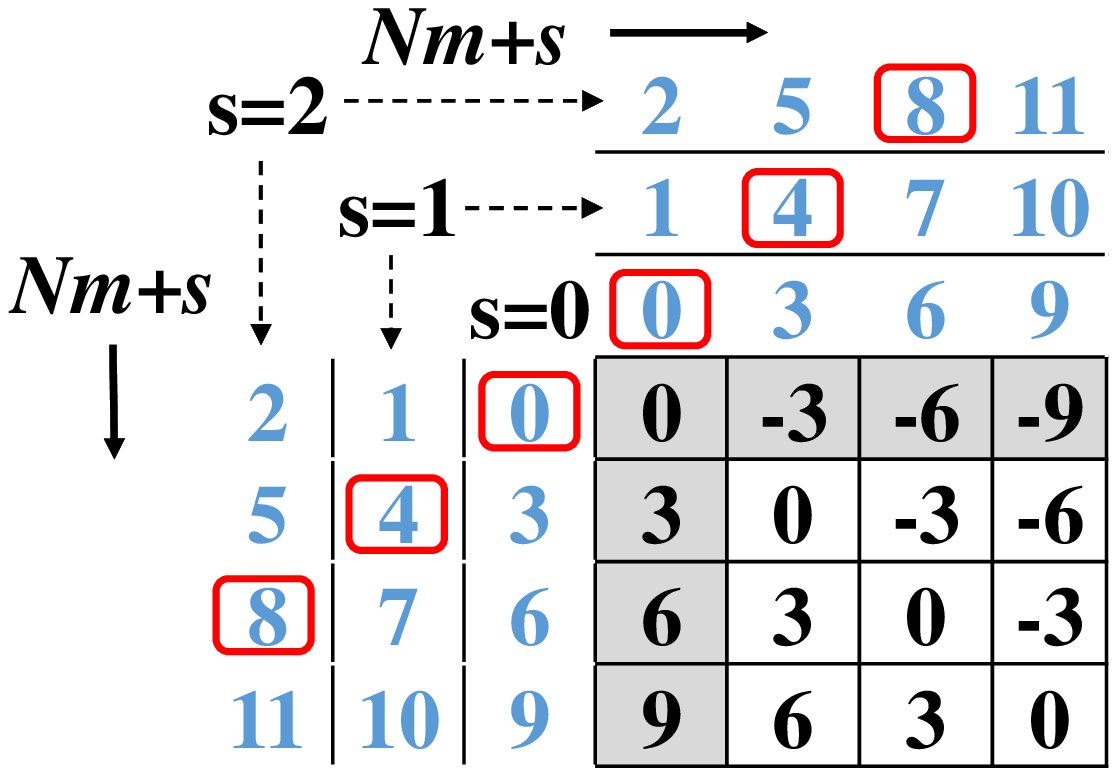}%
\label{extreme_shifted_coprime_self_diff_Nm}}
\hfil
\subfloat[Set $\mathcal{L}^+_{C}: s=0$]{\includegraphics[width=0.24\textwidth]{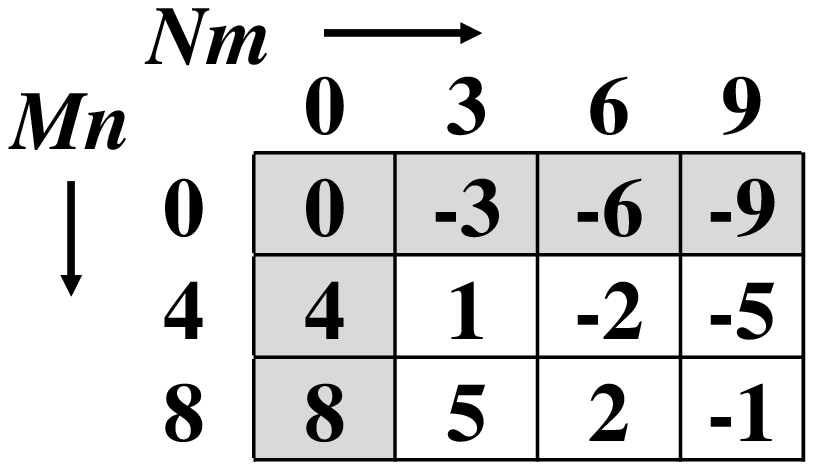}%
\label{extreme_shifted_coprime_cross_M4N3s0}}
\hfil
\subfloat[Set $\mathcal{L}^+_{C}: s=1$]{\includegraphics[width=0.24\textwidth]{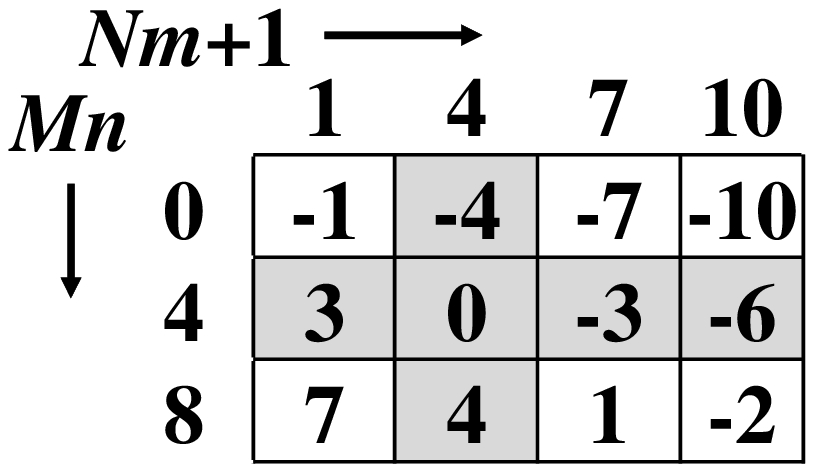}%
\label{extreme_shifted_coprime_cross_M4N3s1}}
\hfil
\subfloat[Set $\mathcal{L}^+_{C}: s=2$]{\includegraphics[width=0.24\textwidth]{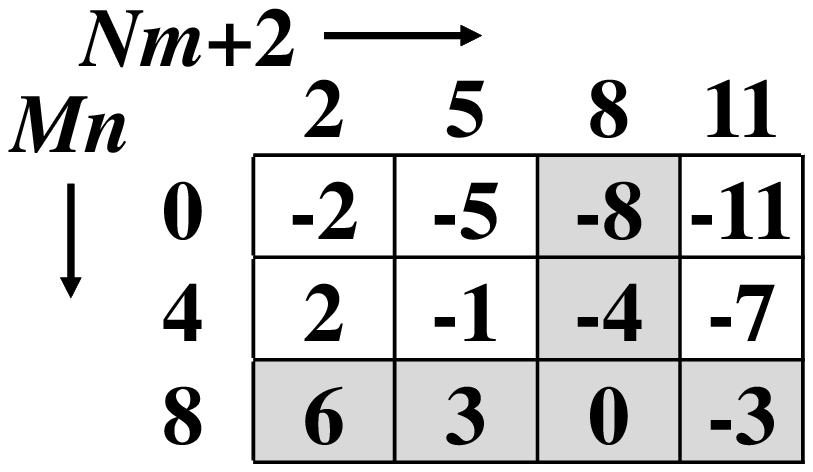}%
\label{extreme_shifted_coprime_cross_M4N3s2}}
\caption{APCA example with $(M,N)=(4,3)$.}
\label{fig:extreme_shifted_coprime_difference_set}
\end{figure}

The self differences for the adjustable pivot co-prime array are defined as in~\cite{U_S_1,UVD_PHD}. $\mathcal{L}^+_{SM}$ and $\mathcal{L}^+_{SN}$ denote the positive self difference sets for the sub-arrays with inter-element spacing $Md$ and $Nd$, respectively. Cross difference set for APCA structure is defined as:
\begin{equation}\label{eq:extreme_shifted_cross}
    \mathcal{L}^+_{C}=Mn-(Nm+s)
\end{equation}
where $n, s\in[0, N-1]$ and $m\in[0, M-1]$. Similar to the above sets; $\mathcal{L}^-_{SM}$, $\mathcal{L}^-_{SN}$, and $\mathcal{L}^-_{C}$ denote the negative difference sets. 
\begin{lemma}\label{lemma1}
The pivot location parameters $(n_p, m_p)$, as a function of the pivot selection parameter $s$, are given by:
$n_p=\left\{\frac{Nm+s}{M} \mid \frac{Nm+s}{M} \in \mathbb{Z}\right\}$~\text{or}~
$m_p=\left\{\frac{Mn-s}{N} \mid \frac{Mn-s}{N} \in \mathbb{Z}\right\}$
and the pivot location is $Mn_p$ or $\left(Nm_p+s\right)$.
\end{lemma}
\begin{IEEEproof}
Consider the cross difference set. It can be shown that zero belongs to the cross difference set. Let $l_c=Mn-(Nm+s)=0$, then $Nm=Mn-s$ with $n, s\in[0, N-1]$. Therefore, $-(N-1)\leq Nm \leq M(N-1)$ or $-N< Nm < MN$ which gives a valid range $0\leq m\leq M-1$.

When an element in the first sub-array overlaps with an element in the second sub-array, the cross difference between these element locations $(n_p,m_p)$ would be zero:
\begin{eqnarray}\label{eq:n_p}
Mn_p-(Nm_p+s)=0\implies n_p=\frac{Nm_p+s}{M}
\end{eqnarray}
For a given $s$, the value of $m$ that results in an integer $n$ gives the pivot element location, i.e. $(n_p,m_p)$. \eqref{eq:n_p} can be re-arranged as $m_p=\frac{Mn_p-s}{N}$. Let $P(m_p)=\frac{Nm_p+s}{M}\in \mathbb{Z}$ and $P(m)=\{\frac{Nm+s}{M}|m\neq m_p\}$, then $P(m_p)-P(m)=\frac{N(m_p-m)}{M}\notin \mathbb{Z}$,
since $|m_p-m|<M$. Therefore, $P(m)\notin \mathbb{Z}$ which proves the uniqueness of $(n_p, m_p)$ and hence Lemma~\ref{lemma1}. The solid circles in Fig.~\ref{APCA_structure_M4N3} and red boxes in Fig.~\ref{extreme_shifted_coprime_self_diff_Nm} represent the overlapping element locations. 
\end{IEEEproof}

$\mathcal{L}^+_{C}$ has difference values in the range $[-(N(M-1)+s),(M(N-1)-s)]$. This is obtained by substituting the ranges for $n$ and $m$ in~\eqref{eq:extreme_shifted_cross}. 
Similarly, $\mathcal{L}^-_{C}$ has difference values in the range $[-(M(N-1)-s), (N(M-1)+s)]$. Therefore, $\mathcal{L}_{C}=\mathcal{L}^+_{C}\cup \mathcal{L}^-_{C}$ will have its extreme values at $\pm \text{max}\{M(N-1)-s, N(M-1)+s\}$, and $\mathcal{L}=\mathcal{L}_{C}\cup \mathcal{L}_{S}$ will have values in the range $\pm \text{max}\{M(N-1)-s, N(M-1)+s, M(N-1)\}$. It can be shown that $\mathcal{L}^{+}_{C}$ and $\mathcal{L}^{-}_{C}$ have $MN$ unique difference values for all $s$ similar to the proof in~\cite{4.7} for the prototype array. 

Consider the APCA structure in Fig.~\ref{APCA_structure_M4N3}. With $s=0$, we obtain the prototype array with the pivot elements at the origin. For $s=1$, the pivot location parameter is $(n_p,m_p)=(1,1)$, and for $s=2$, $(n_p,m_p)=(2,2)$. The self difference matrix for the fixed sub-array (with inter-element spacing $Md$) is the same as was shown in~\cite{U_S_1,UVD_PHD}. The self difference matrix for the other sub-array for different values of $s$ is shown in Fig.~\ref{extreme_shifted_coprime_self_diff_Nm} with $(M, N)=(4, 3)$. Despite a shift in the location of the first element, the self difference values do not change. However, the values in the cross difference matrix change with the pivot selection parameter $s$. The cross difference matrix for different values of $s$ are shown in Fig.~\ref{extreme_shifted_coprime_cross_M4N3s0}-\ref{extreme_shifted_coprime_cross_M4N3s2}. For $s=0$ the self differences are a subset of the cross difference set $\mathcal{L}_{C}=\mathcal{L}^+_{C} \cup \mathcal{L}^-_{C}$. For $s\in [1, N-1]$, some of the self differences are not contained in the cross difference set. For $s=1$ shown in Fig.~\ref{extreme_shifted_coprime_cross_M4N3s1}, $\pm8$ and $\pm9$ are not contained in $\mathcal{L}_{C}$, while for $s=2$ (see Fig.~\ref{extreme_shifted_coprime_cross_M4N3s2}), $\pm9$ are not contained in $\mathcal{L}_{C}$.

\begin{figure*}
\centering
\includegraphics[width=0.24\textwidth]{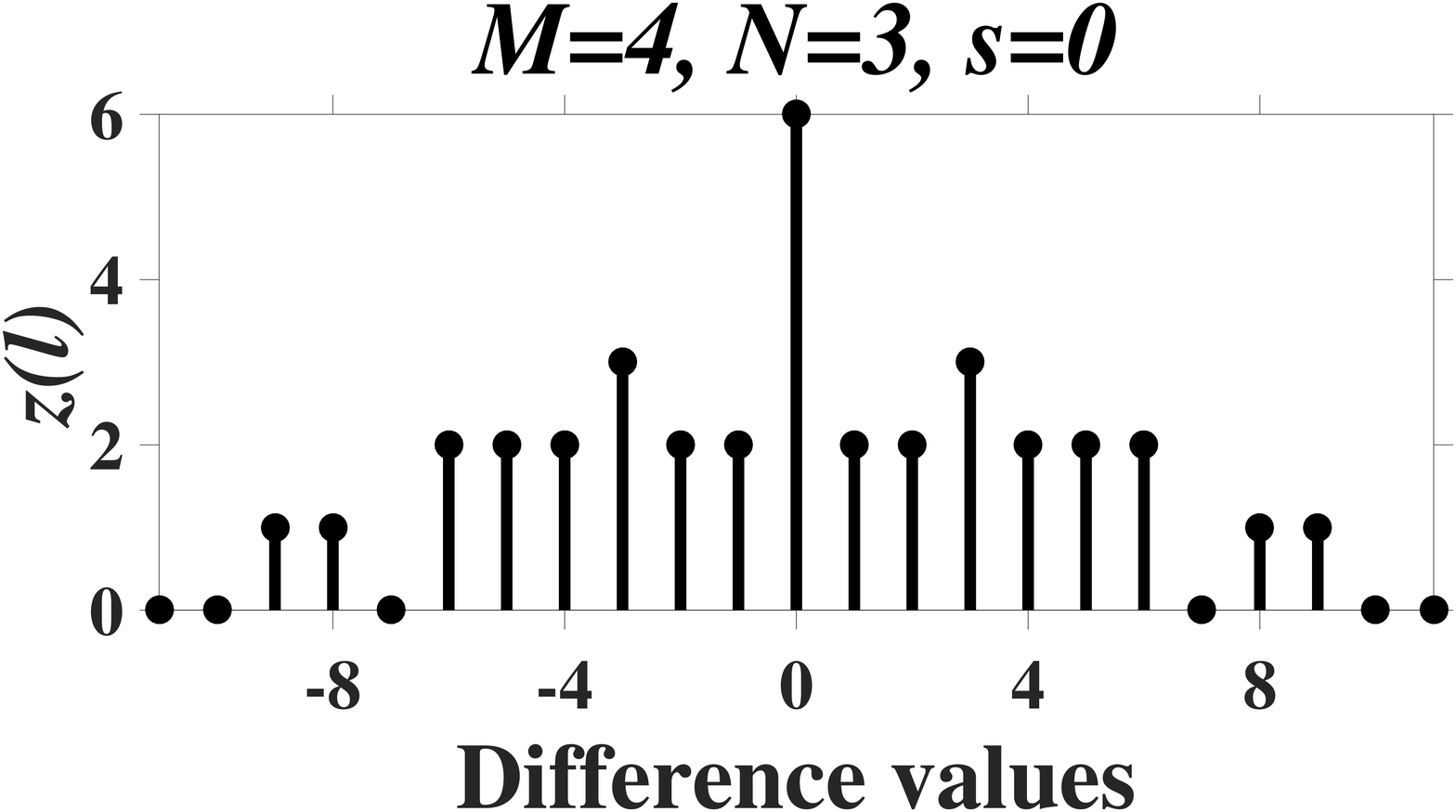}%
\includegraphics[width=0.24\textwidth]{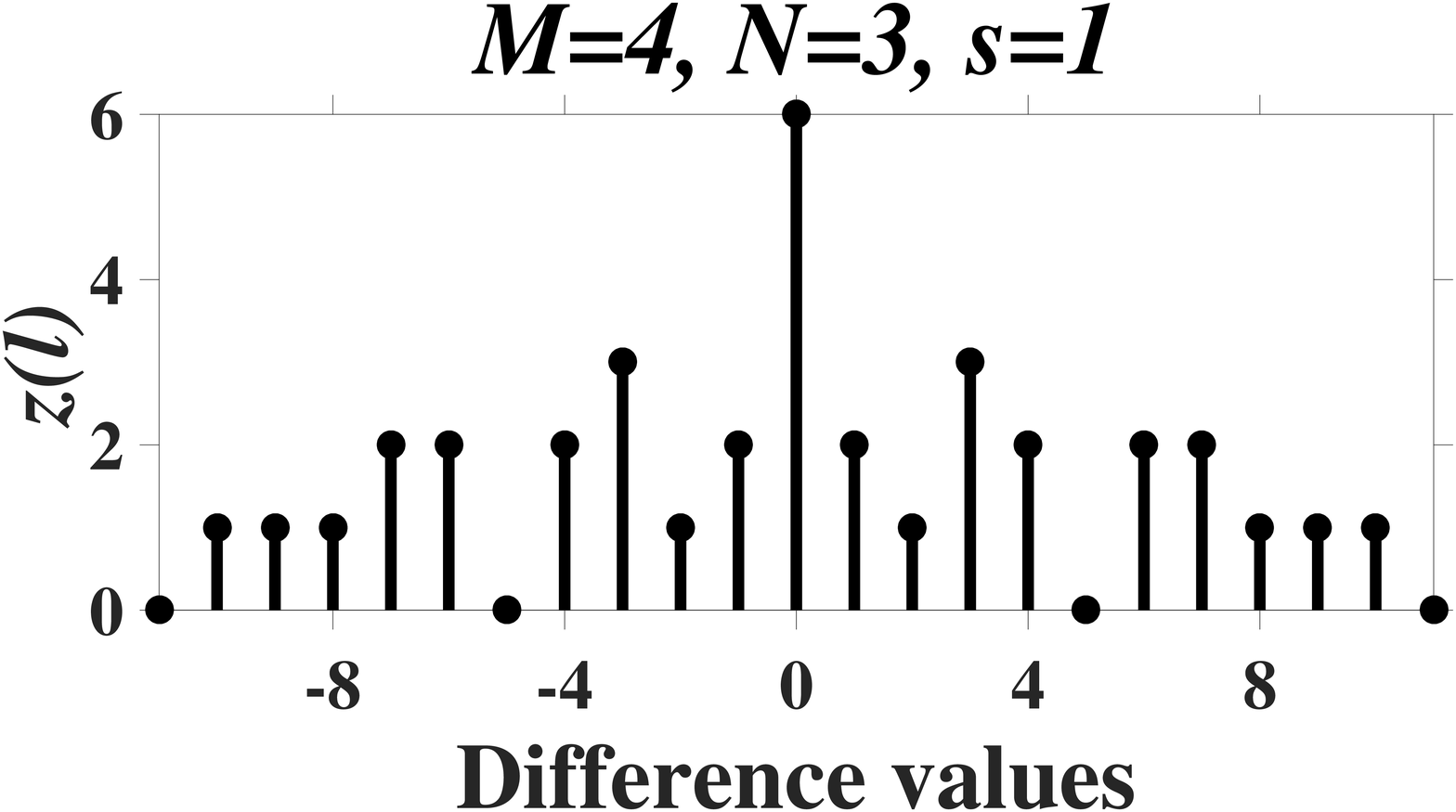}
\includegraphics[width=0.24\textwidth]{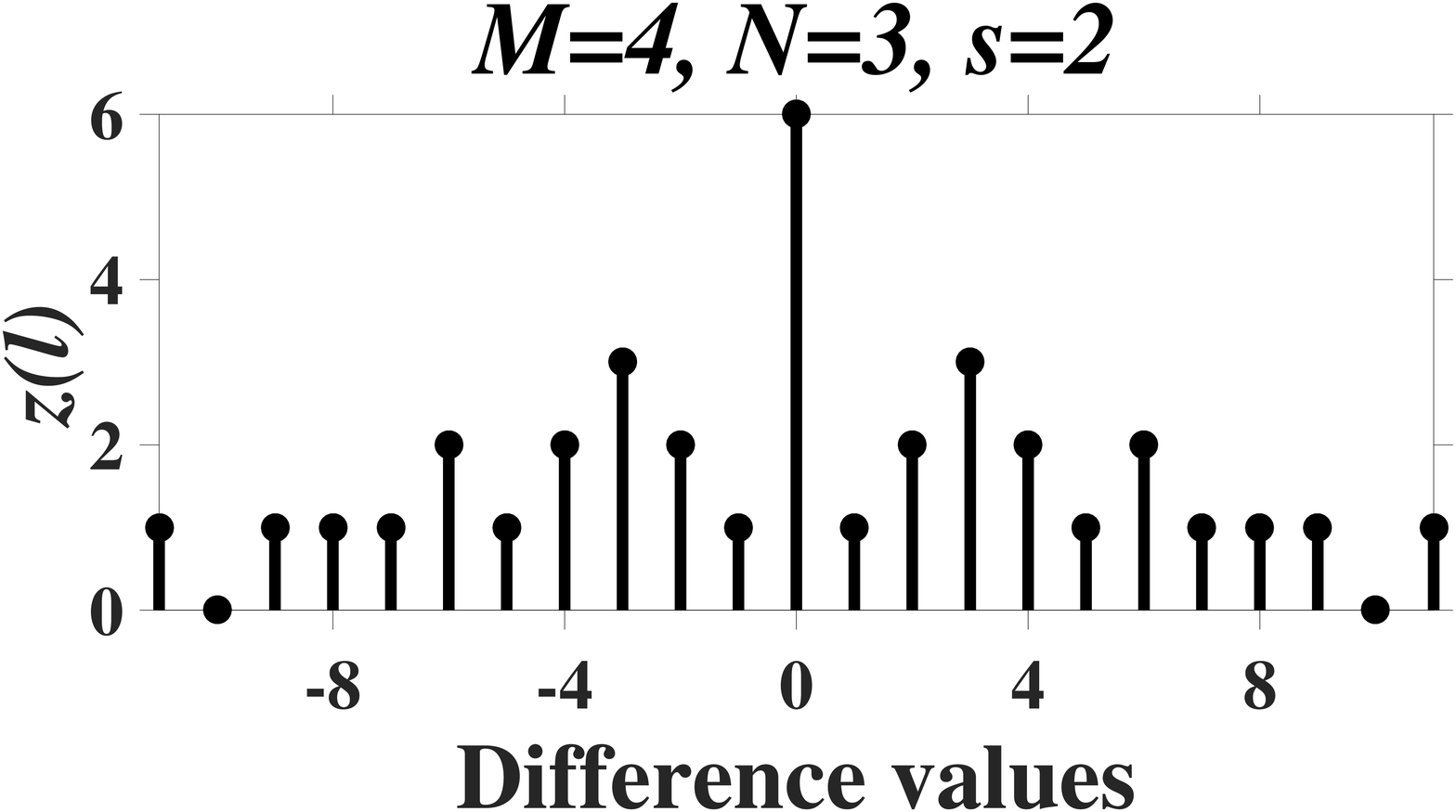}%
\includegraphics[width=0.25\textwidth]{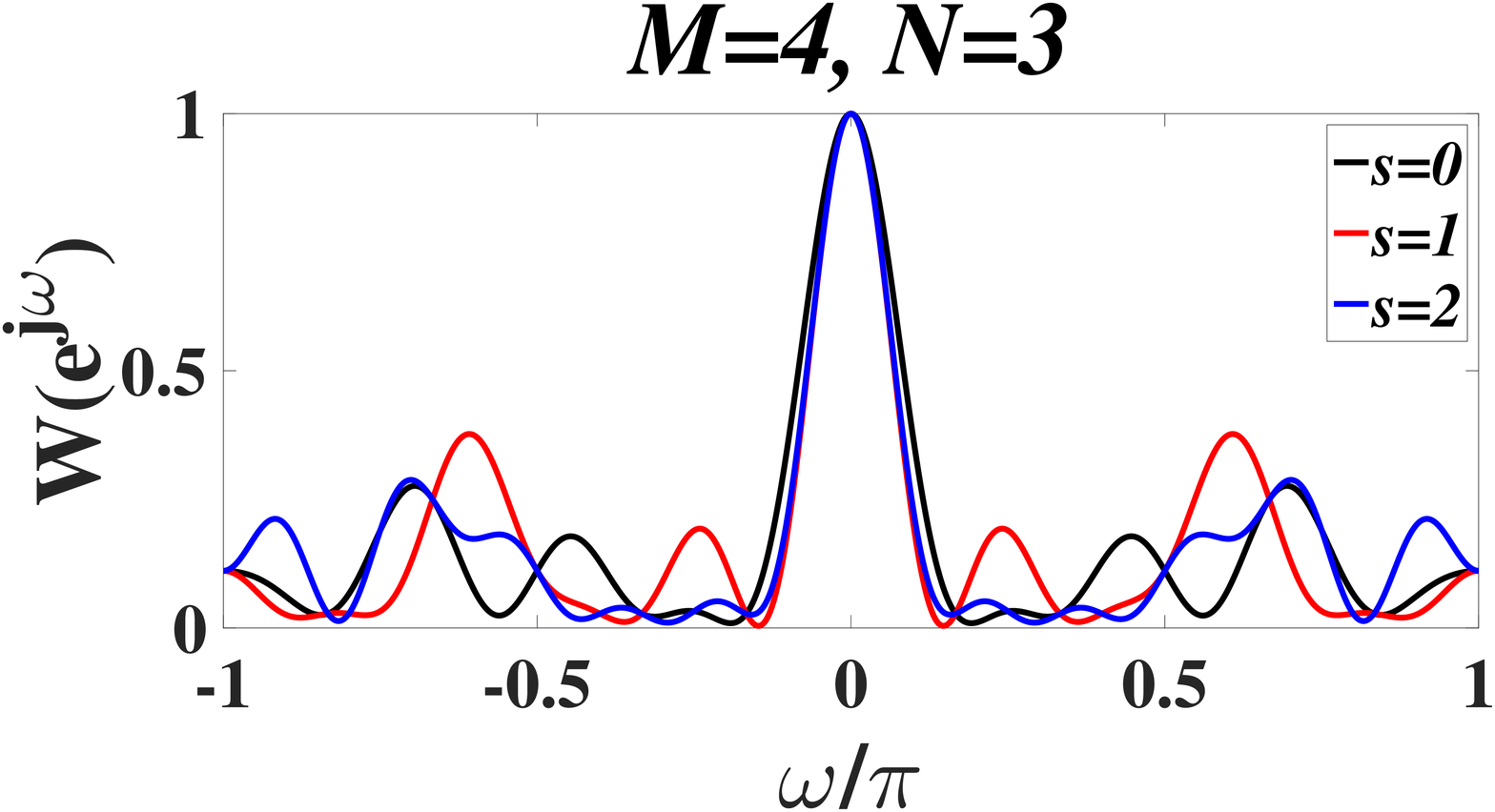}\\
\includegraphics[width=0.24\textwidth]{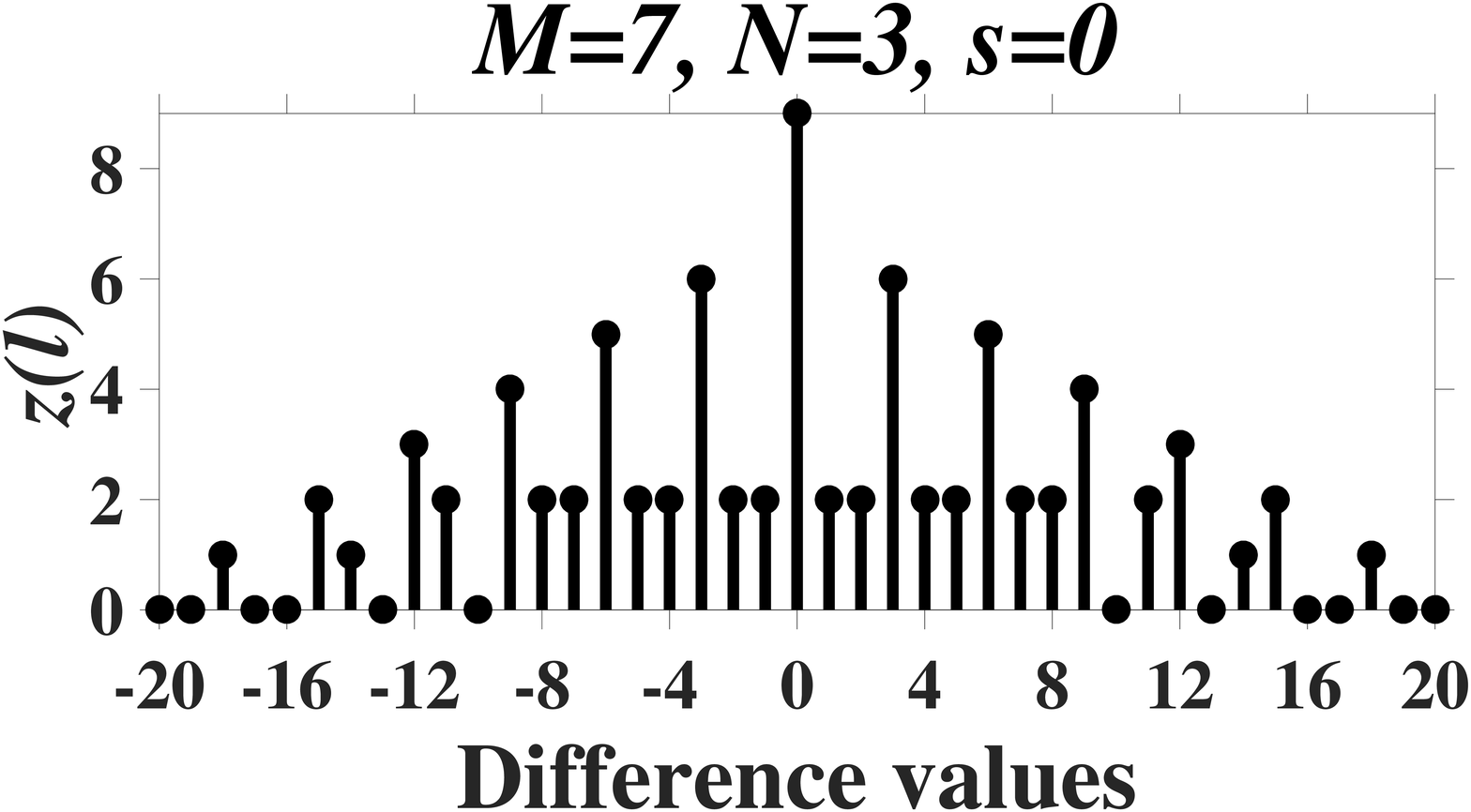}%
\includegraphics[width=0.24\textwidth]{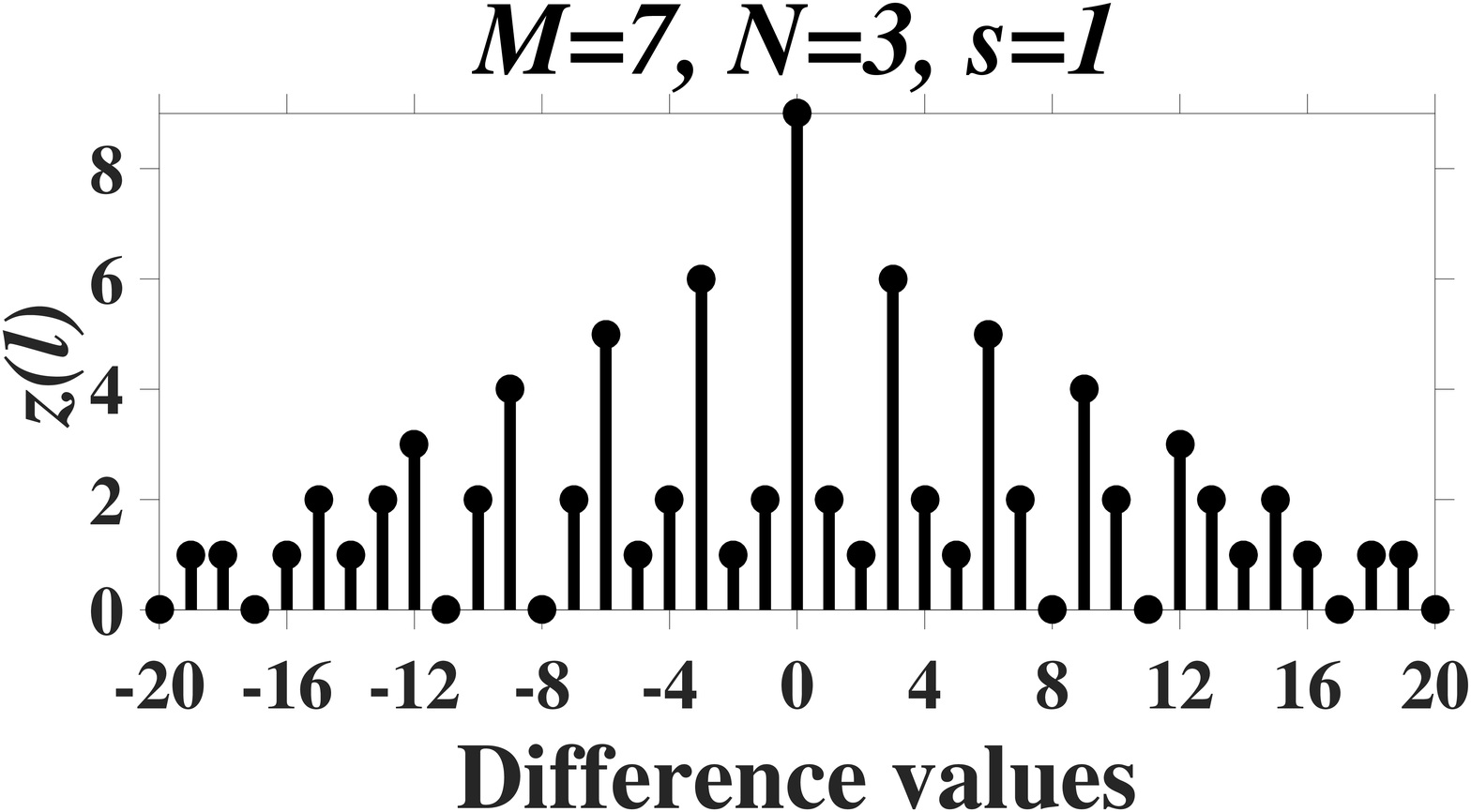}
\includegraphics[width=0.24\textwidth]{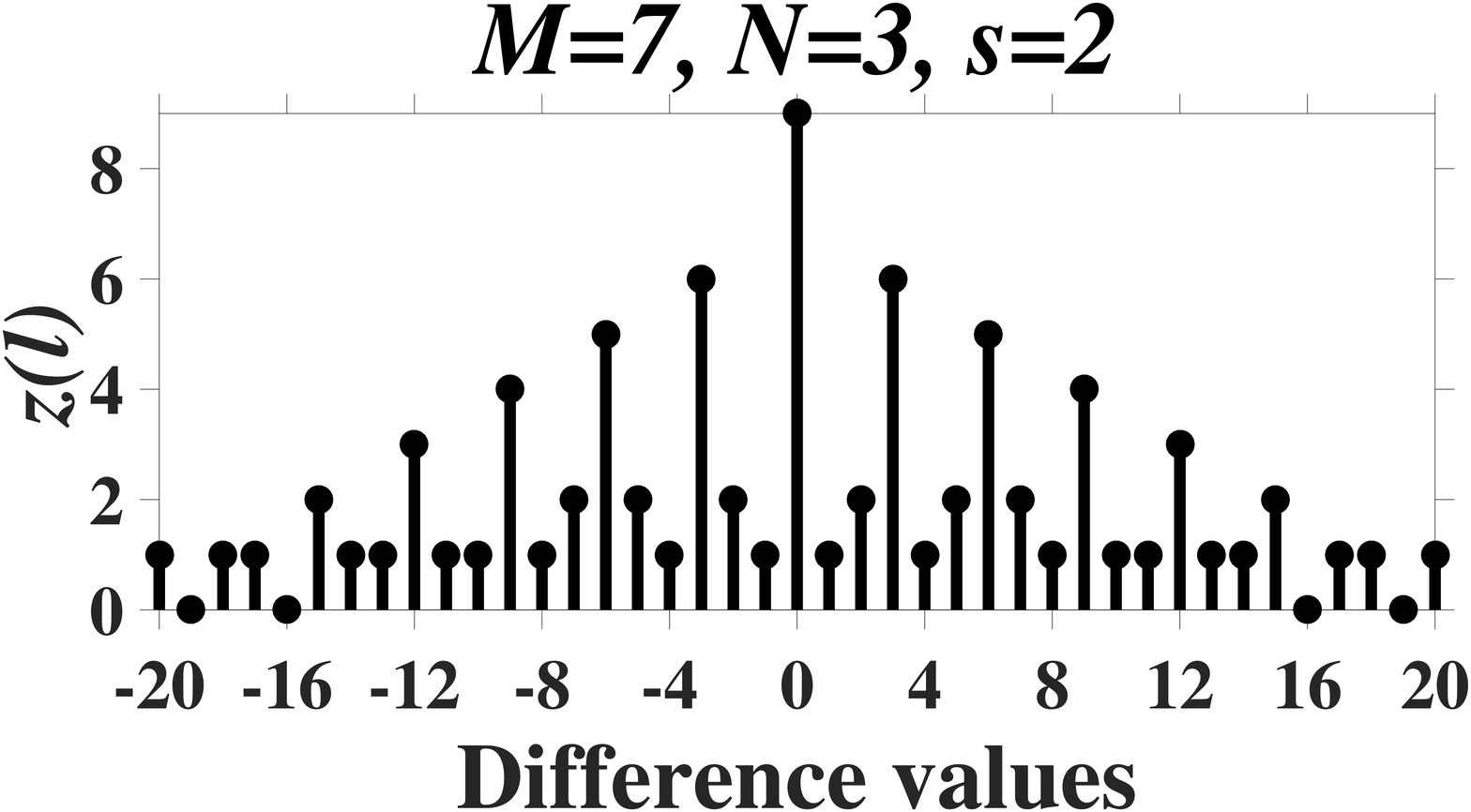}%
\includegraphics[width=0.25\textwidth]{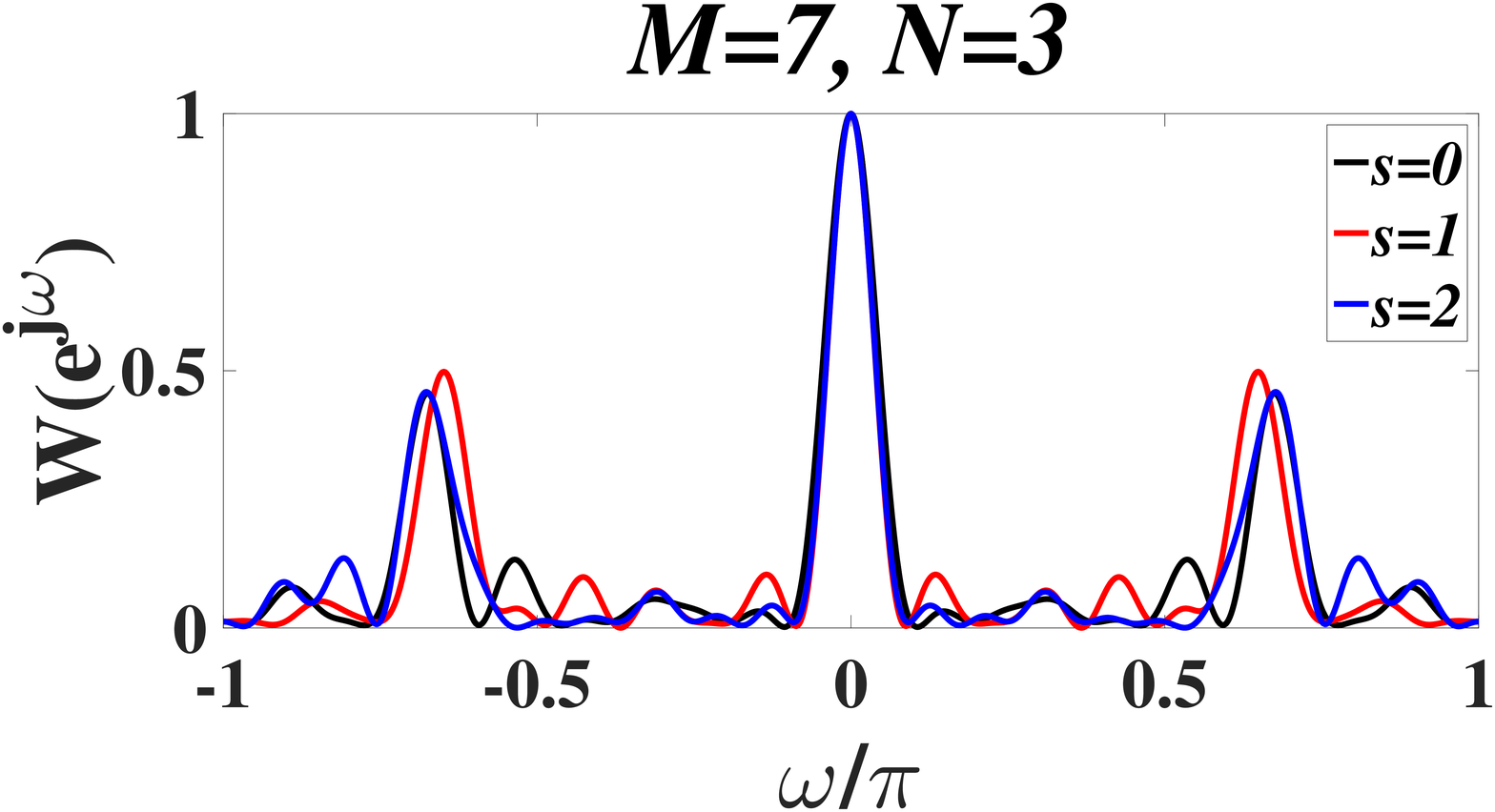}%
\caption{Weight function $z(l)$ and bias $W(e^{j\omega})$ of the correlogram estimate for different values of $s$: $(M,N)=(4,3)$ (first row), $(M,N)=(7,3)$ (second row).}
\label{fig:Truly_coprime_wts_bias_M4N3_M7N3}
\end{figure*}

\begin{figure*}[!t]
\centering
\includegraphics[width=0.24\textwidth]{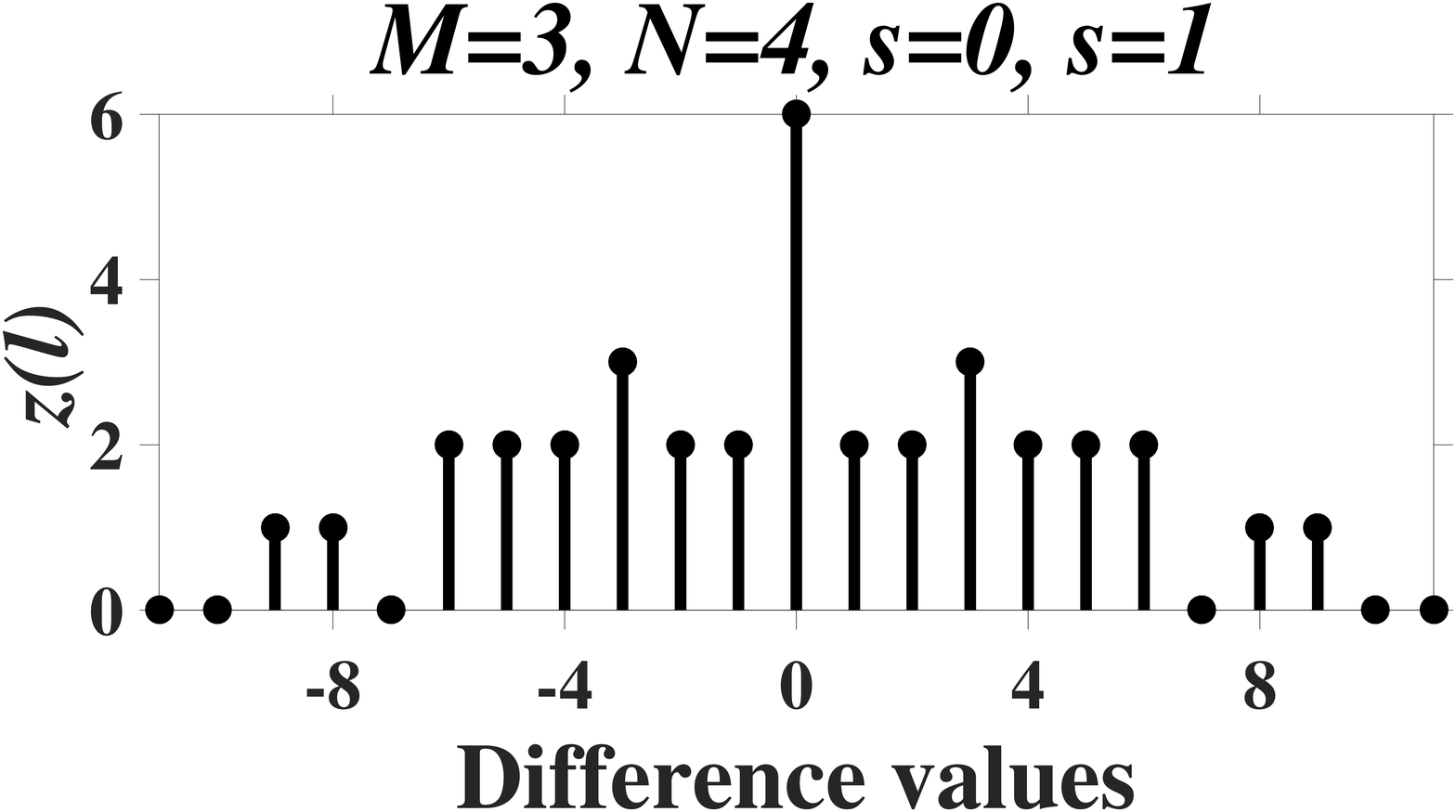}%
\includegraphics[width=0.24\textwidth]{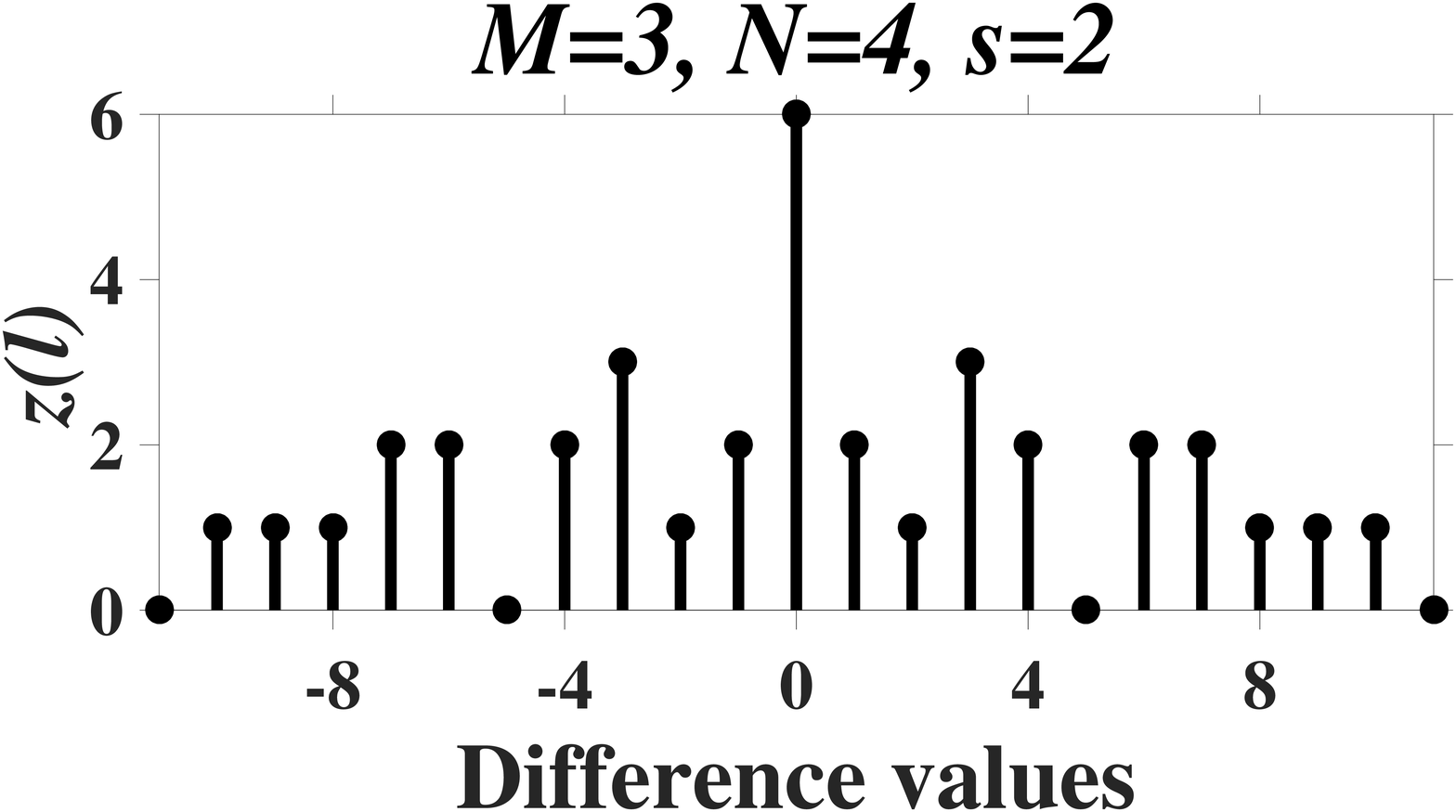}%
\includegraphics[width=0.24\textwidth]{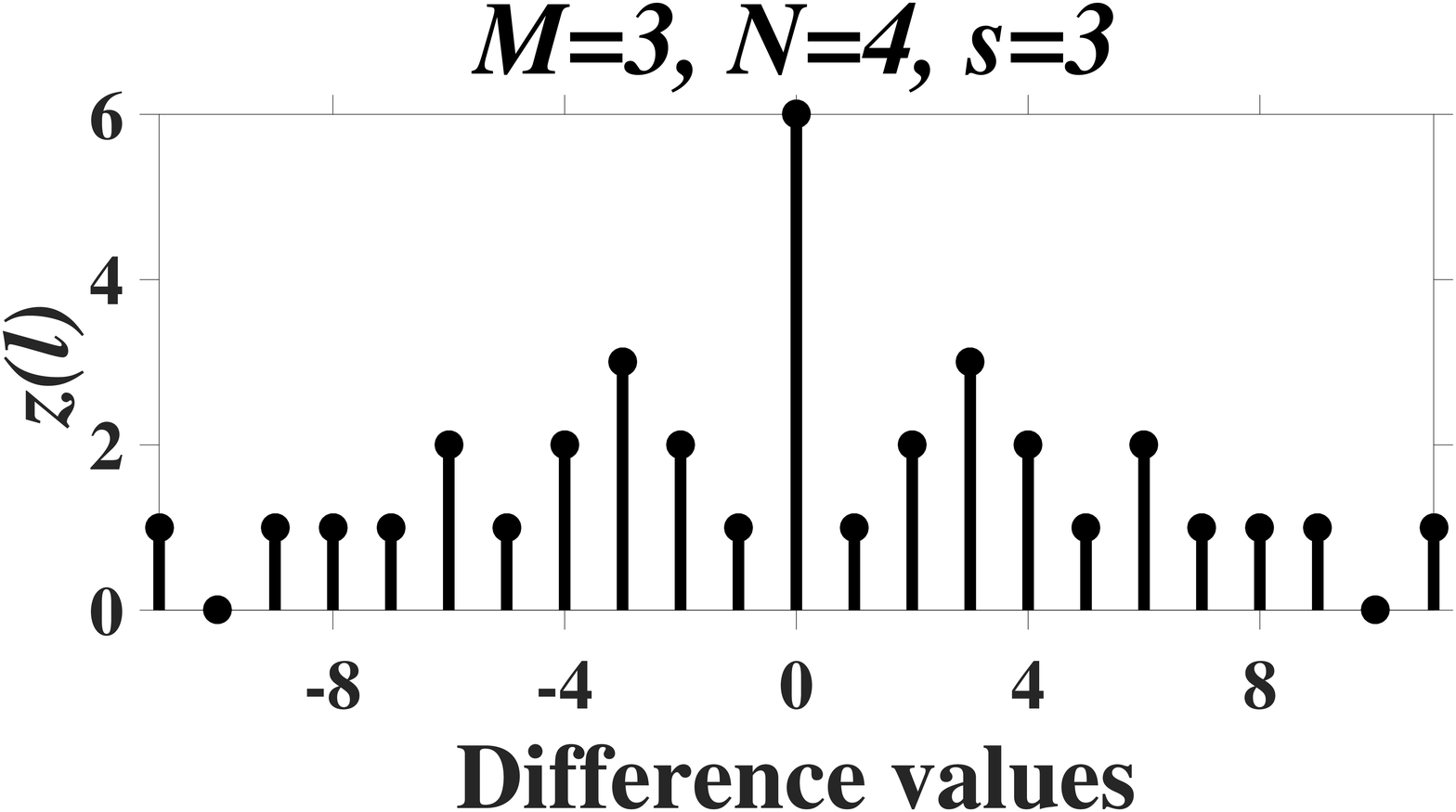}
\includegraphics[width=0.25\textwidth]{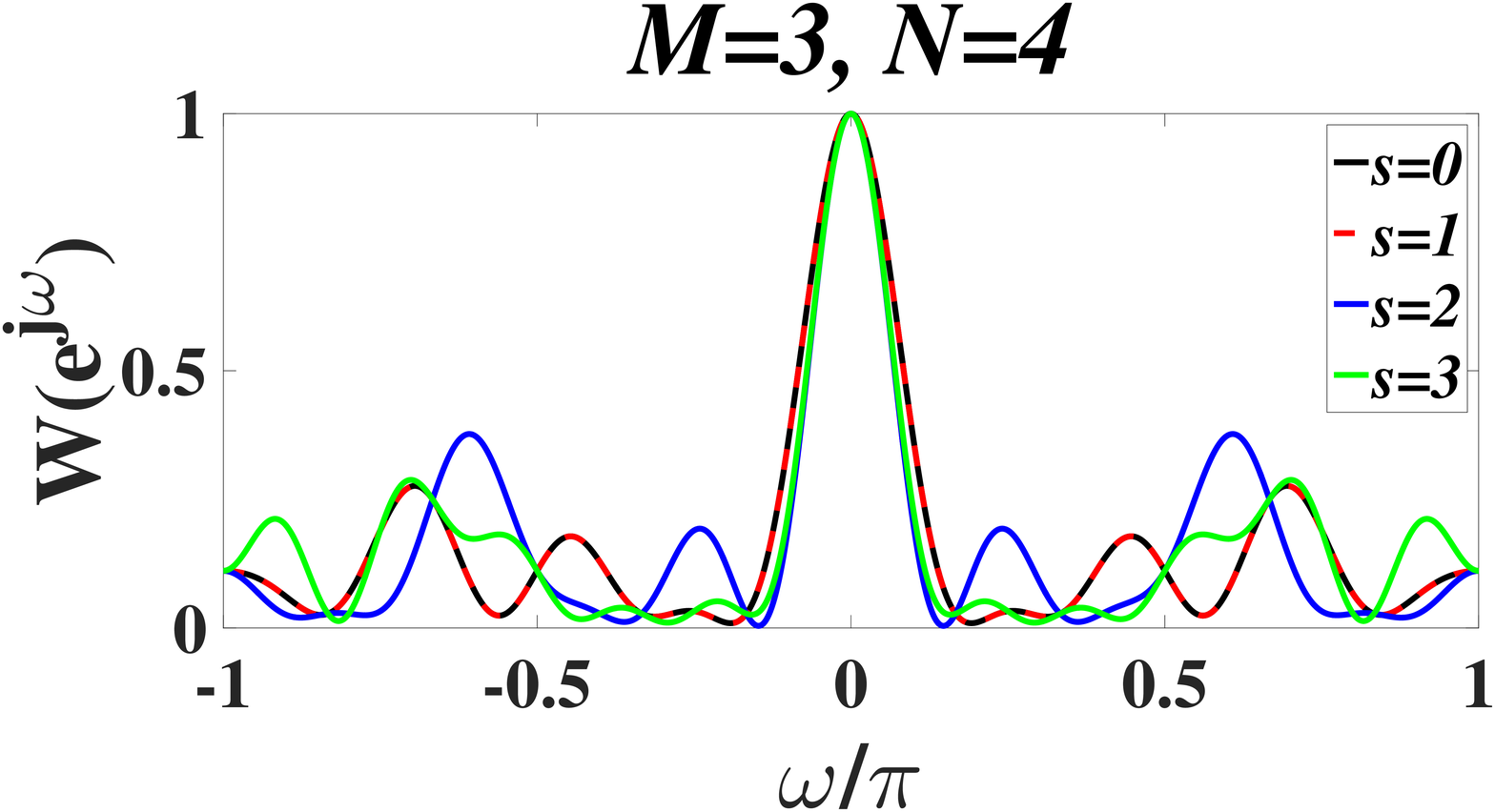}
\caption{Weight function $z(l)$ and bias $W(e^{j\omega})$ of the correlogram estimate for different values of $s$ with $(M,N)=(3,4)$.}
\label{fig:extreme_shifted_coprime_simultaion_M3N4}
\end{figure*}

The weight function, $z(l)$ where $l$ represents the difference value, is defined as the number of contributors available for autocorrelation estimation at each difference value. Note that $l$ is also referred to as \textit{lag}. Fig.~\ref{fig:Truly_coprime_wts_bias_M4N3_M7N3} depicts the weight function for the APCA structure for different values of $s$ and $(M, N)$. For $(M,N)=(4,3)$ and $s=\{0, 1, 2\}$, the number of unique difference values are $\{17, 19, 21\}$ and the continuous ranges are $\{[-6, 6], [-4, 4], [-9, 9]\}$, respectively. Among these $s=2$ gives the largest continuous range $[-9, 9]$ and highest number of unique difference values of $21$. For $(M,N)=(7,3)$ and $s=\{0,1,2\}$, the number of unique difference values are $\{29, 33, 37\}$ and continuous ranges are $\{$[-9, 9], [-7, 7], [-15, 15]$\}$. Again, $s=2$ gives larger number of unique difference values and continuous range. The weight function for $M<N$ is shown in Fig.~\ref{fig:extreme_shifted_coprime_simultaion_M3N4} with $(M,N)=(3,4)$. It is observed that a maximum of $21$ unique difference values and a continuous range of $[-9, 9]$ are obtained with $s=3$. Clearly, the proposed APCA scheme has a larger continuous range than the prototype array (refer Table~\ref{table_APCA} for additional examples). It is observed that holes (i.e. difference values for which $z(l)=0$) are present at large difference values which are relatively insignificant and can be suppressed by window functions during the estimation process. The APCA scheme can be employed even without interpolating the missing values. The entire difference set including holes guarantees a valid power spectral estimate as shown in~\cite{UVD_PHD}.
\begin{figure*}[!t]
\centering
\subfloat[One spectral peak.]{
\includegraphics[width=0.33\textwidth]{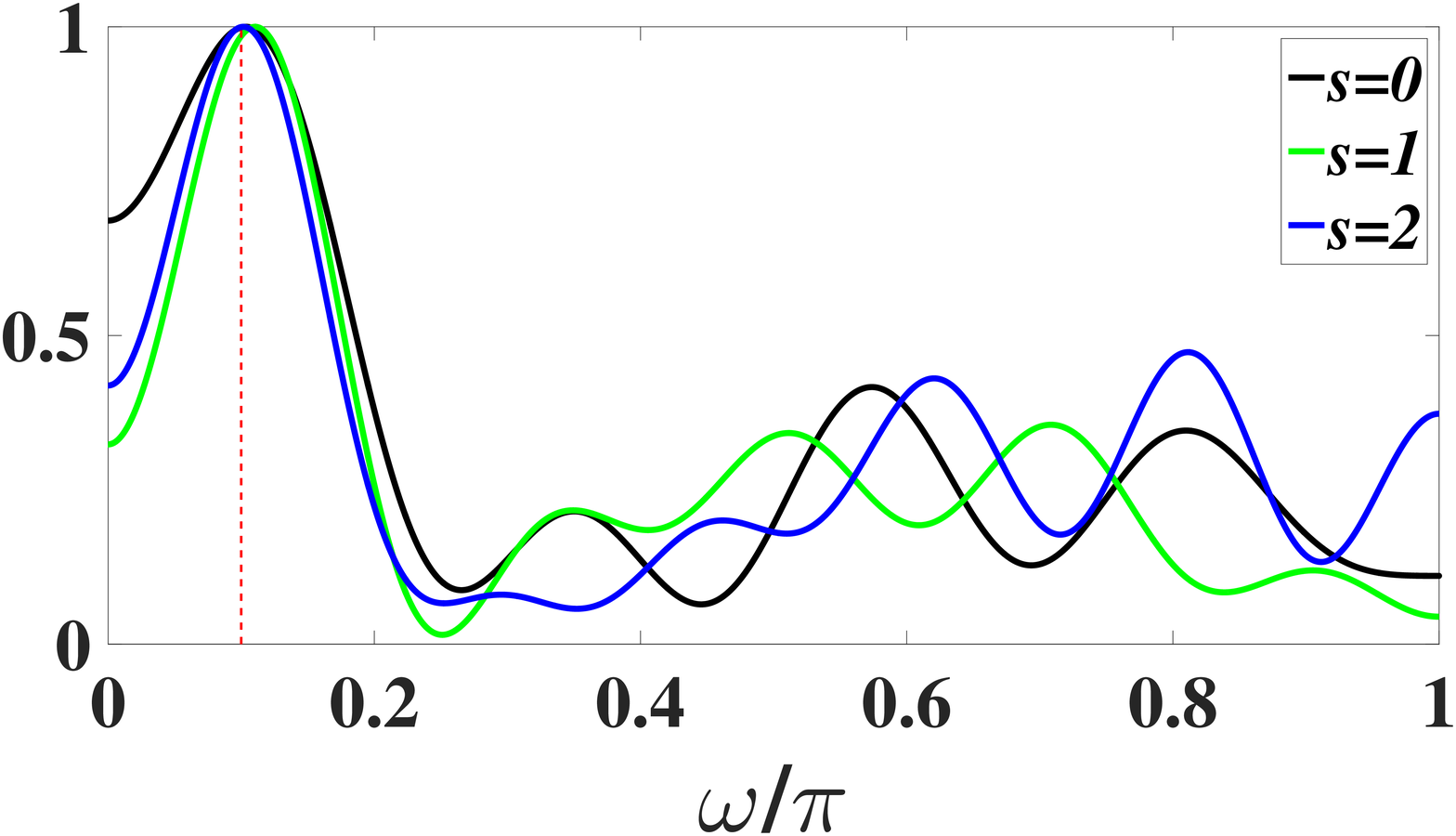}%
\label{Truly_coprime_sim_1peak_M4N3}}
\subfloat[Two spectral peaks.]{
\includegraphics[width=0.33\textwidth]{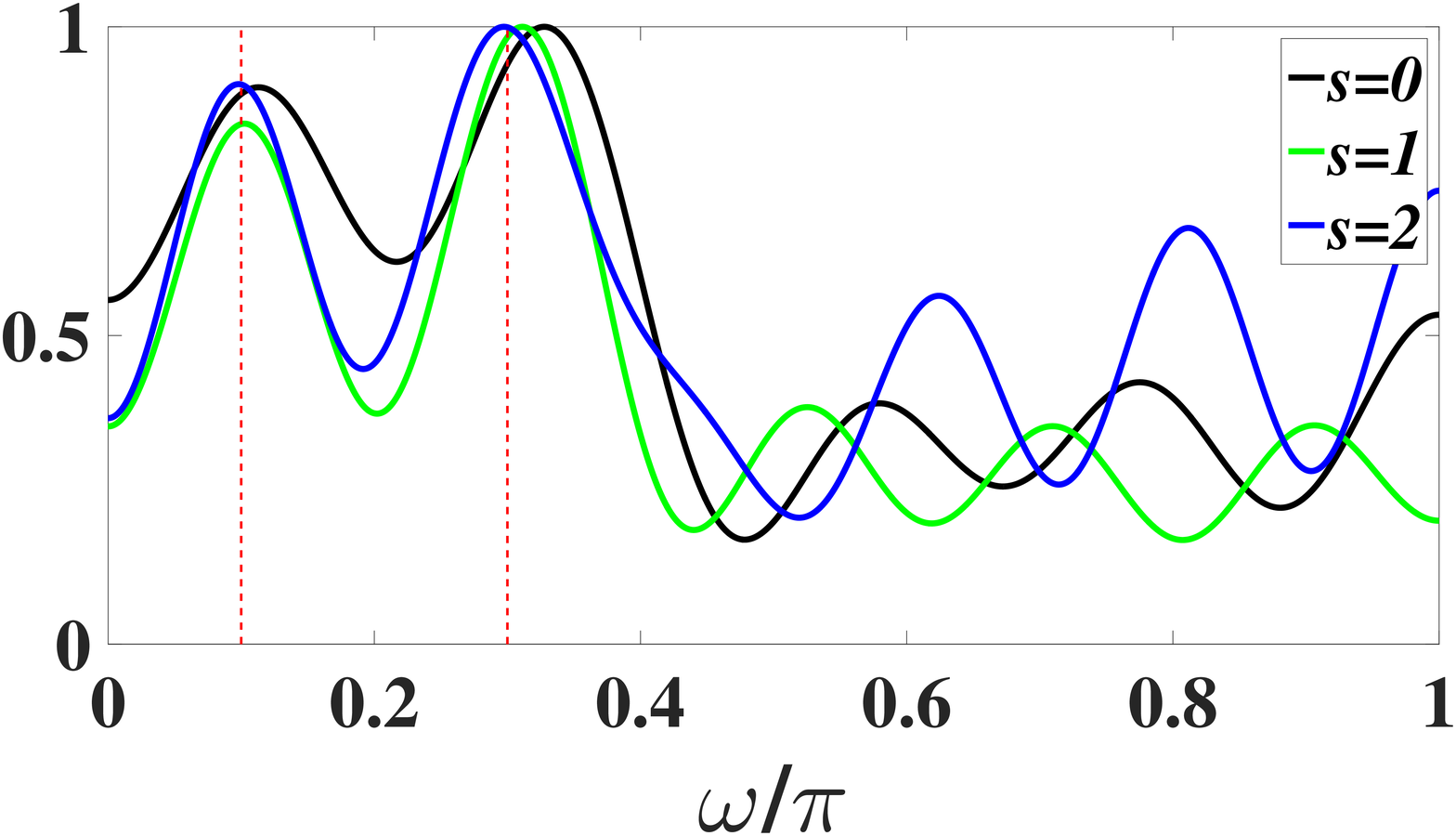}%
\label{Truly_coprime_sim_2peak_M4N3}}
\subfloat[Three spectral peaks.]{
\includegraphics[width=0.33\textwidth]{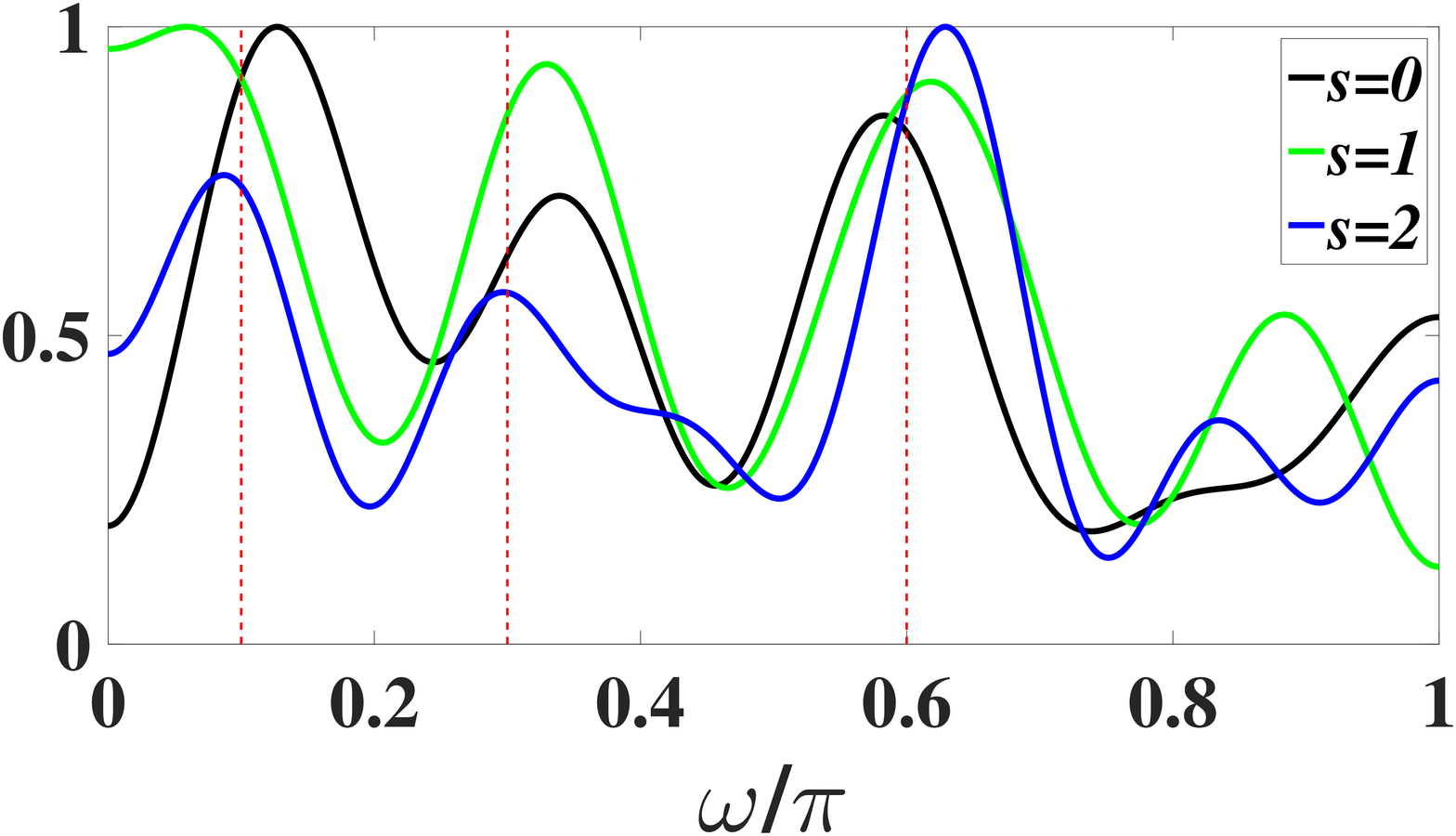}%
\label{Truly_coprime_sim_3peak_M4N3}}
\caption{Spectral estimation using APCA: $(M,N)=(4,3)$ and number of snapshots $=10$.}
\label{fig:Truly_coprime_sim_diff_peaks_M4N3}
\end{figure*}
\begin{table}
\caption{Comparison of the APCA and prototype array.}
\label{table_APCA}
\centering
\resizebox{0.5\textwidth}{!}{
\begin{tabular}{|l|c|c|c|l|}
\cline{1-5}
\multirow{2}{*}{\bf{(M, N})} & \multicolumn{2}{c}{\bf{Continuous Range}} &\multicolumn{2}{|c|}{\bf{Relative Amplitude}}\\  
\cline{2-5}
& \bf{Prototype} & \bf{APCA} & \bf{Prototype} & \multicolumn{1}{c|}{\bf{APCA}}\\
\hline
$(5, 9)$ &  $\pm 13$ & $\pm 27$\,$(s=8)$ & $0.5923$ & $0.6098$\,$(s=6)$\\
\hline
$(7, 10)$ & $\pm 16$ & $\pm 40$\,$(s=4)$ & $0.6435$ & $0.6668$\,$(s=4,6)$\\
\hline
$(9, 8)$ & $\pm 16$ & $\pm 41$\,$(s=6)$ & $0.6996$ & $0.7366$\,$(s=1)$\\
\hline
$(10, 9)$ & $\pm 18$ & $\pm 51$\,$(s=2)$ & $0.6993$ & $0.7335$\,$(s=1)$\\
\hline
$(29, 27)$ & $\pm 55$ & $\pm 418$\,$(s=13)$ & $0.6921$ & $0.7264$\,$(s=10)$\\
\hline
\end{tabular}
}
\end{table}

The correlogram spectral estimate, in a statistical sense, is given by the convolution of the bias window $W(e^{j\omega})$ (i.e. the Fourier transform of $z(l)$) and true spectrum. In Fig.~\ref{fig:Truly_coprime_wts_bias_M4N3_M7N3} and Fig.~\ref{fig:extreme_shifted_coprime_simultaion_M3N4}, $W(e^{j\omega})$ with $s=(N-1)$ results in side lobes with relatively low peaks close to the main lobe which prevent spectral leakage for closely spaced spectral peaks. However there may exist distant side lobes with relatively large peaks resulting in spurious spectral peaks away from the true peaks. Let relative amplitude ($R$) of the main lobe peak ($P_m$) with respect to the largest side lobe peak ($P_s$) be defined as $R=(P_m-P_s)/P_m$. This parameter was employed in the analysis reported in~\cite{UVD_PHD}. The relative amplitudes of the main lobe and largest side lobe for $(M, N)=(4, 3)$, $(7, 3)$, and $(3, 4)$ with $s\in [0, N-1]$ are $\{0.7237, 0.6229, 0.7121$\}$, $\{0.5440, 0.5018, 0.5410$\}$, and $\{0.7237, 0.7237, 0.6229, 0.7121\}$, respectively. Table~\ref{table_APCA} lists the maximum relative amplitude and corresponding pivot selection parameter. The APCA design involves optimal selection of three parameters ($M$, $N$, $s$) based on the application requirements. It is noted that $\sum_{l=0}^{MN-1}z(l)$ is the same for all shifts $s$, so the number of multiplications required for autocorrelation estimation is the same as for the prototype array~\cite{U_S_2, UVD_PHD,UVD_jitter_CAMP}.
\begin{figure*}[!t]
\centering
\subfloat[Extended adjustable pivot co-prime array $(M, N, s)$.]{
\includegraphics[width=0.38\textwidth]{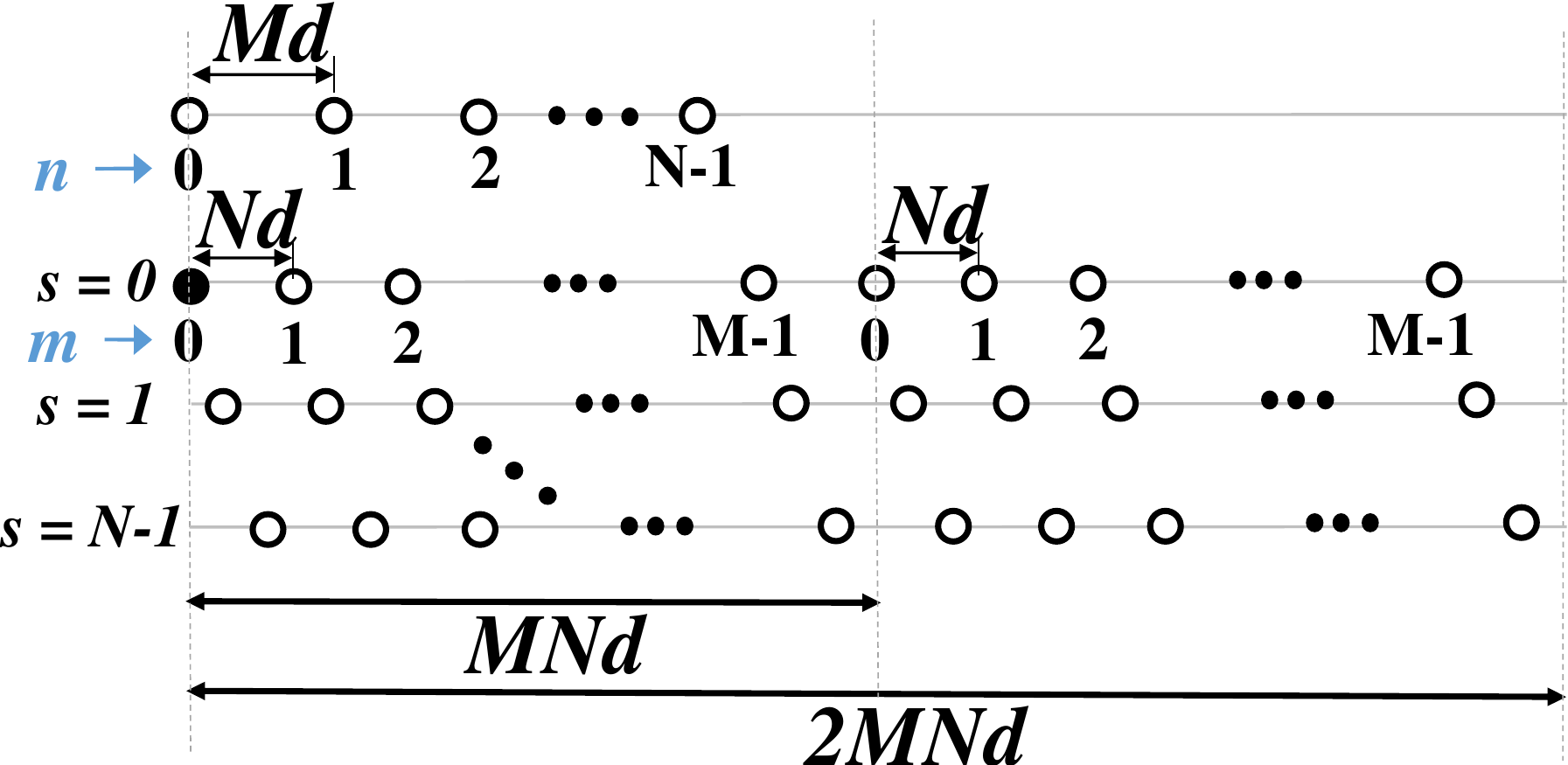}%
\label{Truly_coprime_extended}}
\hfil
\subfloat[Multi-level adjustable pivot co-prime arrays $(M, N, r, s)$.]{
\includegraphics[width=0.6\textwidth]{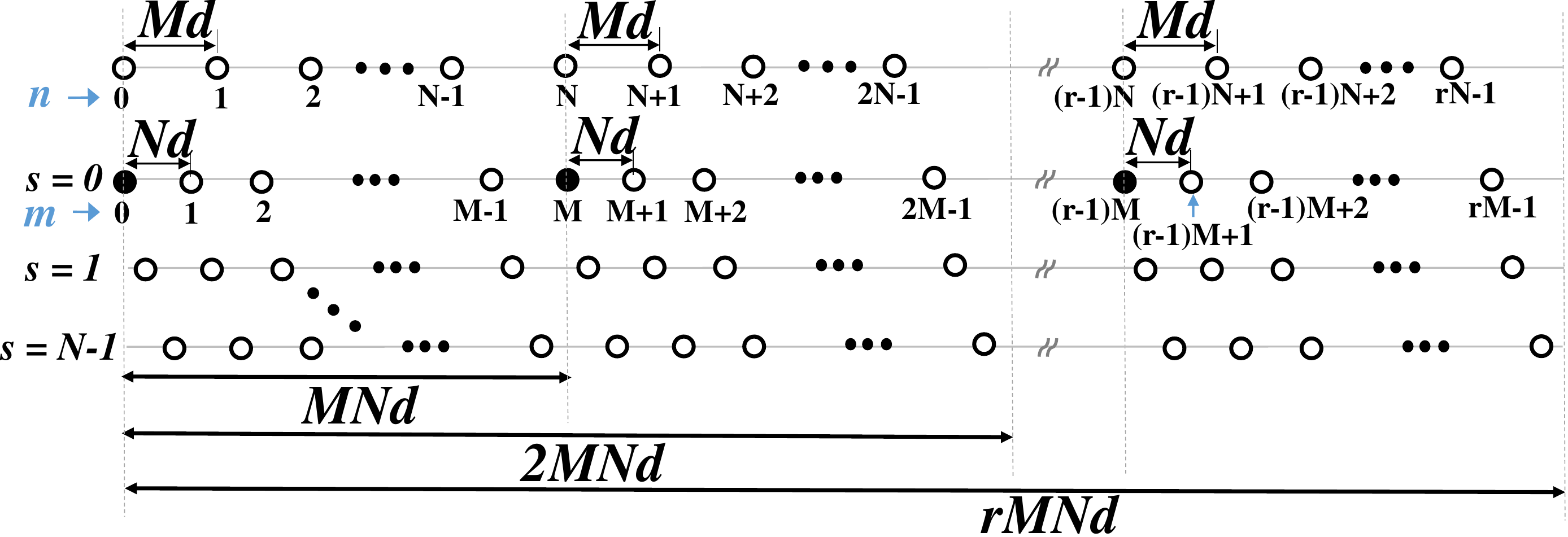}%
\label{Truly_coprime_multi}}
\caption{Extended and multi-level adjustable pivot co-prime structures.}
\label{fig:Extended_multi co-prime structures}
\end{figure*}
\begin{figure}\centering
\includegraphics[width=0.25\textwidth]{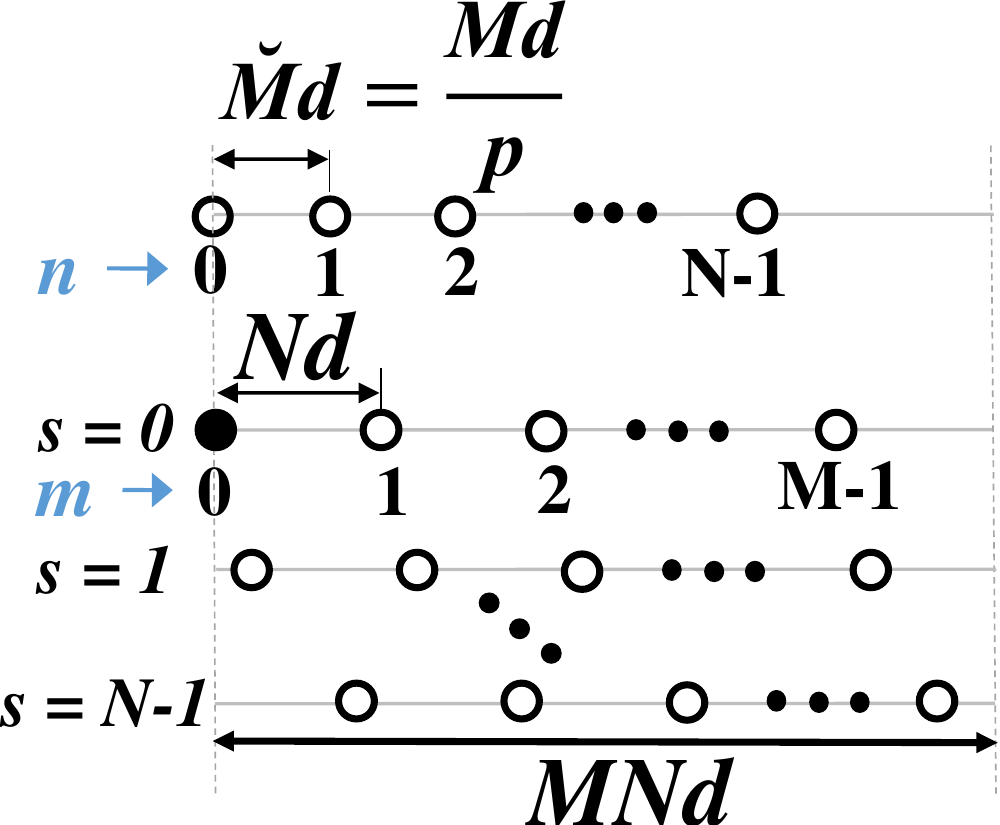}
\includegraphics[width=0.23\textwidth]{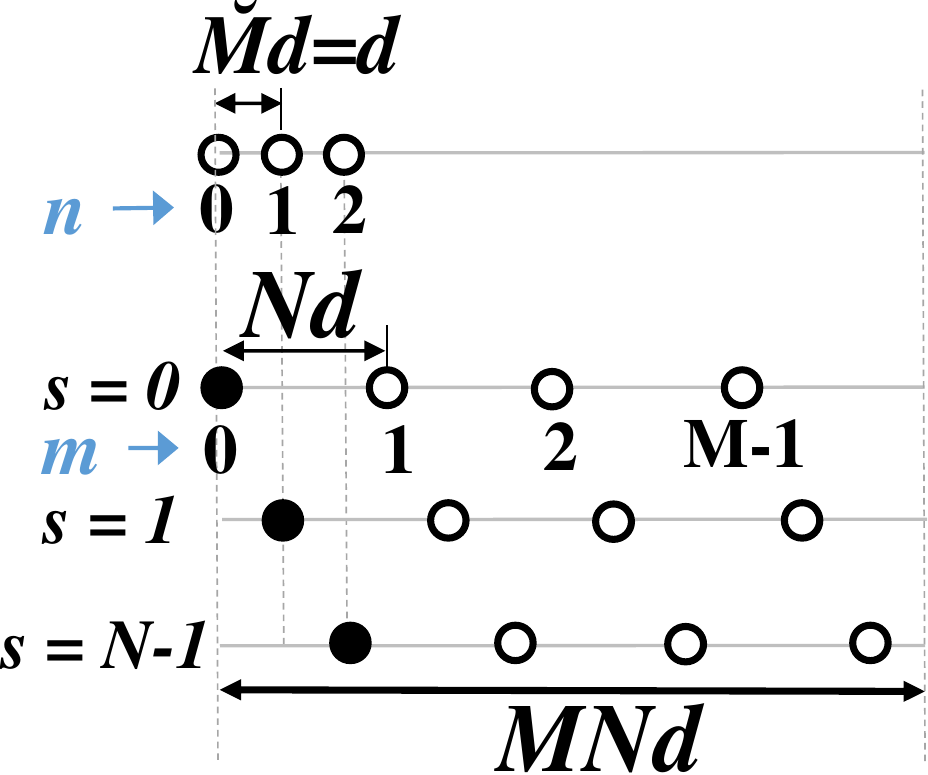}%
\caption{Adjustable pivot CACIS $(M, N, p, s)$: General (left), nested array (right).}
\label{fig:Truly_coprime_CACIS}
\end{figure}
\begin{table}
\caption{Average spectral peak location estimation error.}
\label{table_Error_peaks}
\centering
\resizebox{0.45\textwidth}{!}{
\begin{tabular}{|c|c|c|c|}
\cline{1-4}
    & {One peak} & {Two peaks} & {Three peaks}\\
\hline
$s=0$ & $0.0034$ &	$0.0203$ &	$0.0278$\\
\hline
$s=1$ & $0.0105$ &	$0.0063$ &	$0.0286$\\	
\hline
$s=2$ & $0.0014$ &	$0.0025$ &	$0.0150$\\
\hline
\end{tabular}
}
\end{table}
%
%
\subsection{Simulation Results}
\label{sec:simulation}
A noise-free signal representing a random process with frequency bands centered at $0.1\pi$, $0.3\pi$, and $0.6\pi$ is used for the simulations. Refer Section 4.2.3 in~\cite{UVD_PHD} for a detailed description of the signal model. Similar model is also considered in~\cite{4.32,U_S_1, UVD_extended, UVD_CACIS} 
The combined difference set, including holes, is employed for power spectrum estimation using the APCA design resulting.

To better understand the impact of the APCA scheme on spectral estimation and its frequency bias, three scenarios are shown in Fig.~\ref{fig:Truly_coprime_sim_diff_peaks_M4N3} with $(M,N)=(4,3)$. The estimate is 
averaged over $10$ snapshots (one co-prime period is referred to as a snapshot). 
For the case with single spectral peak in Fig.~\ref{Truly_coprime_sim_1peak_M4N3}, the peak location is estimated accurately for all values of $s$. With $s=2$, the spurious spectral peaks close to the true peak have low amplitudes. $s=1$ has the lowest overall spurious peak amplitude.
With two spectral peaks (Fig.~\ref{Truly_coprime_sim_2peak_M4N3}), $s=2$ estimates the spectral peak locations most accurately.
With three spectral peaks (Fig.~\ref{Truly_coprime_sim_3peak_M4N3}), $s=2$ estimates the spectral peak locations with reasonable accuracy. However, the spectral peak amplitudes vary significantly with the third peak having a larger amplitude due to spectral leakage. Table~\ref{table_Error_peaks} lists the average error in peak location for the spectrum estimation scenarios in Fig.~\ref{fig:Truly_coprime_sim_diff_peaks_M4N3}. It confirms that $s=2$ is a good choice for peak location estimation. It is important to note that this choice may not hold true for other signal models. Hence, it requires a more detailed analysis.
\subsection{Adjustable pivot co-prime array family}
\label{sec:implications_arrays}
The co-prime array with adjustable pivot can be used to redesign other arrays in the literature as shown in Fig.~\ref{fig:Extended_multi co-prime structures} and Fig.~\ref{fig:Truly_coprime_CACIS} along with the design parameters. The extended adjustable pivot co-prime array with the sub-array with inter-element spacing $Nd$ shifted by $s$ is shown in Fig.~\ref{Truly_coprime_extended}, where $0\leq s\leq N-1$. Fig.~\ref{Truly_coprime_multi} depicts the multiple period adjustable pivot co-prime array where $r$ is the number of periods. The adjustable pivot CACIS array (Fig.~\ref{fig:Truly_coprime_CACIS}) with $p=1$ results in the prototype adjustable pivot co-prime array and $p=M$ results in the adjustable pivot CACIS nested array. 

The adjustable pivot co-prime array described so far assumes $0\leq s\leq N-1$ so that the resulting difference values are within the co-prime range, i.e. $[-(MN-1), (MN-1)]$ and provide a fair comparison with the prototype array. If $s$ is allowed to take values outside this range other array configurations are possible. When $s>\breve{M}(N-1)$ with $M=p\breve{M}$, the structure is equivalent to the CADiS configuration. The author does not claim the superiority of a particular structure over the other. These structures need further investigation. However, the APCA structures will be shown to be a special case of the Extremely Sparse Co-Prime Array proposed in the next section. Motion based arrays have also been reported in~\cite{Dilation1,Dilation2,Dilation3}. 
\section{Extremely Sparse Co-Prime Arrays}
\label{EXTREMELY_proto}
The prototype co-prime array has two sub-arrays with an inter-element spacing of $Md$ and $Nd$. In this section, an array structure is proposed with antennas in the two sub-arrays at an inter-element spacing of $2Md$ and $2Nd$ apart. In general, inter-element spacing is $\mathcal{E}_xMd$ and $\mathcal{E}_xNd$ apart where $\mathcal{E}_x$ is the $\textit{Extremely Sparse}$ constant. This aspect would have consequences on the range and resolution of the reconstructed autocorrelation and hence, the spectral content. We will assume $\mathcal{E}_x =2$ throughout the paper and deviate from this assumption only in Section~\ref{GenExSCA}. For the case when $M=4$ and $N=3$, the co-prime period is $12d$. The same values of \textit{M} and \textit{N} for the extremely sparse co-prime array has a period of $24d$. The condition under which these advantages hold will be discussed later. This concept is derived from the super-Nyquist co-prime scheme~\cite{UVD_supernyquist} and APCA.

\subsection{Structure}
\begin{figure*}[!t]
	\centering
	\includegraphics[width=0.9\textwidth]{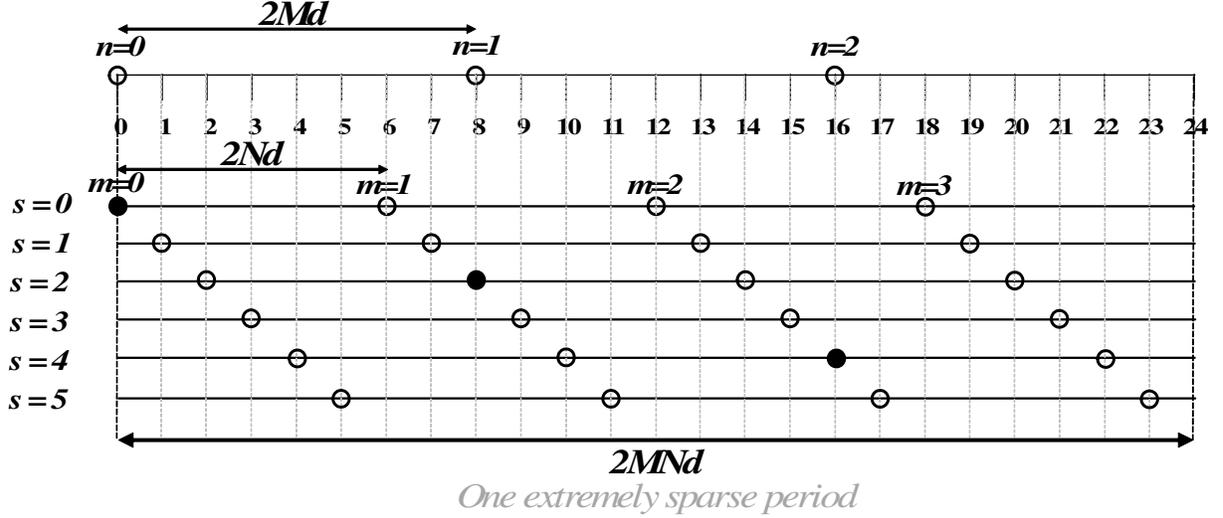}%
	\caption{Extremely sparse co-prime array structure.}
	\label{fig:extreme_proto_structure}
\end{figure*}

The structure for the extremely sparse co-prime array (ExSCA) is shown in Fig.~\ref{fig:extreme_proto_structure} for the case when $M=4$ and $N=3$. The array with inter-element spacing of $2Md$ is assumed to be fixed. The array with inter-element spacing $2Nd$ is kept variable with `$s$' representing the shift in the origin (similar to APCA). $`s$' lies in the range $0\leq s \leq 2N-1$. The ExSCA can reconstruct autocorrelation at Nyquist period `$d$'. It uses sparse/low rate samplers with inter-element spacing (or sampling period) of $2Md$ and $2Nd$. It works for only odd values of shift `$s$', i.e. 1, 3, 5. It requires $M+N$ antennas for its implementation. 

On the other hand for even values of `$s$', i.e. 0, 2, 4, the structure can reconstruct autocorrelation at a period of `$2d$'. It is similar to APCA with Nyquist distance of $2d$. It would require M+N-1 antennas for implementation. Refer Fig.~\ref{fig:extreme_proto_structure}. For example, with $s=0$ location $(Mn, Nm)=(0, 0)$ overlaps. For $s=2$, location $(8, 8)$ overlaps. For $s=4$, location $(16, 16)$ overlaps.

\begin{figure*}[!t]
	\centering
	\subfloat[Set $\mathcal{L}^+_{C}$]{\includegraphics[width=0.6\textwidth]{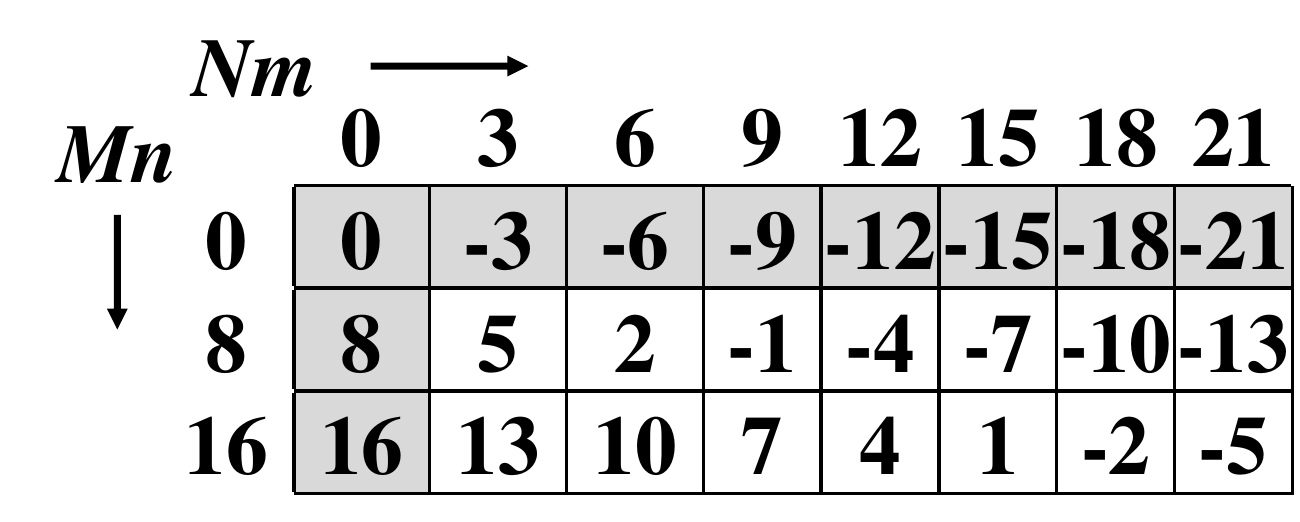}%
		\label{extreme_proto_compare_coprime_Lc+}}
	\hfil
	\subfloat[Set $\mathcal{L}$ with holes]{\includegraphics[width=0.99\textwidth]{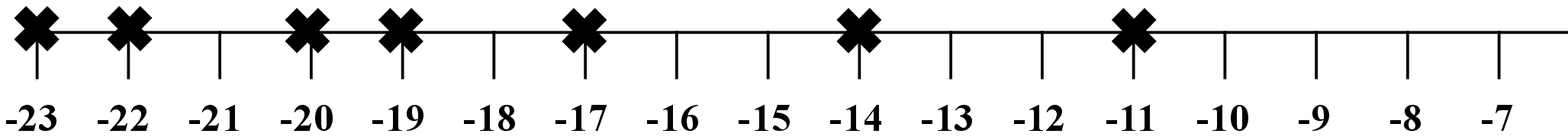}%
		\label{extreme_proto_compare_coprime_L}}
	\caption{Prototype co-prime difference set for $M=8$ and $N=3$.}
	\label{fig:extreme_proto_compare_coprime}
\end{figure*}

The ExSCA (\textit{s=odd}) has a period of $24d$ with $M+N=4+3=7$ antennas or samples. For the same range of $24d$, the prototype co-prime array requires $M=8$ and $N=3$ with $M+N-1=10$ antennas. The difference set for the prototype co-prime array with $M=8$ and $N=3$ is shown in Fig.~\ref{fig:extreme_proto_compare_coprime}, where `\textbf{x}' represents a missing difference value. Similarly, the extremely sparse array can generate a period of $84d$ with $M=7$, $N=6$ and 13 antennas, while the prototype co-prime array would require $M=12$, $N=7$ and 18 antennas. Hence, the proposed extremely sparse array can generate the same period with considerably lesser number of antennas. However, there are missing values and the ExSCA structure may not have a continuous range of difference values. But despite missing difference values, estimation is possible~\cite{UVD_PHD}.
\subsection{Difference Set}
This section develops the fundamentals of the difference set for the extremely sparse array which includes the range, degrees of freedom and its weight function.

\begin{figure}[!t]
	\centering
	\subfloat[Set $\mathcal{L}^+_{SM}\cup \mathcal{L}^-_{SM}$]{\includegraphics[width=0.18\textwidth]{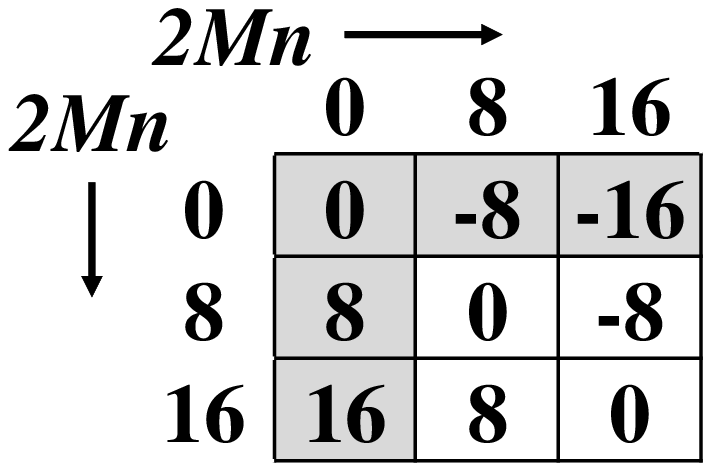}%
		\label{extreme_proto_self_diff_Mn}}
	\hfil
	\subfloat[Set $\mathcal{L}^+_{SN}\cup \mathcal{L}^-_{SN}$ for different values of shift `\textit{s}']{\includegraphics[width=0.3\textwidth]{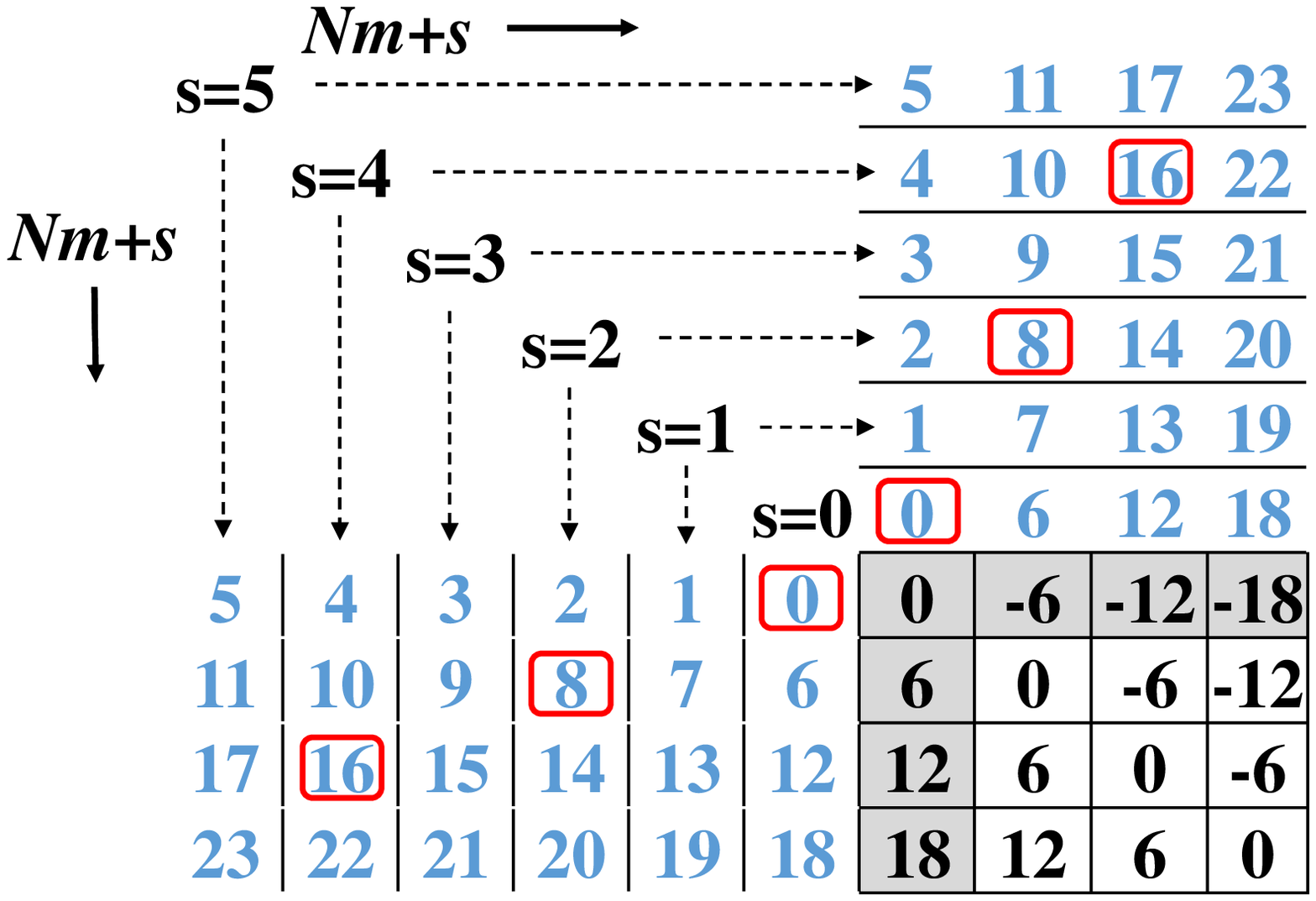}%
		\label{extreme_proto_self_diff_Nm}}
	\caption{Self difference set of an extremely sparse array: M=4, N=3.}
	\label{fig:extreme_proto_self_difference}
\end{figure}

\begin{figure}[!t]
	\centering
	\subfloat[$s=0$]{\includegraphics[width=0.24\textwidth]{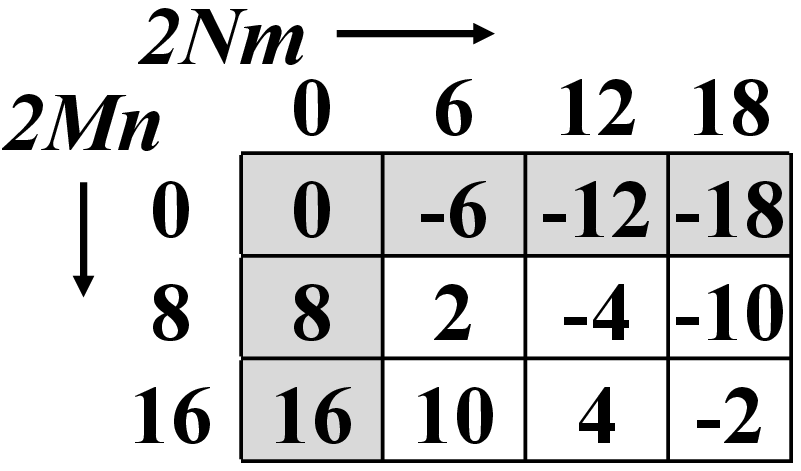}%
		\label{extreme_proto_M4N3s0}}
	\hfil
	\subfloat[$s=1$]{\includegraphics[width=0.24\textwidth]{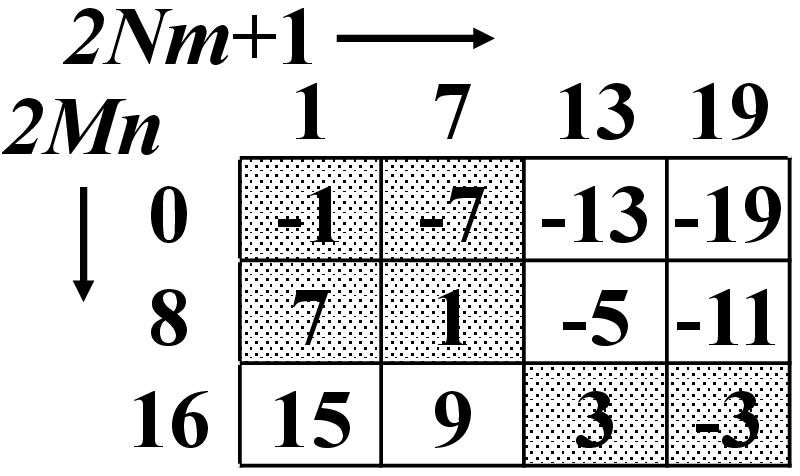}%
		\label{extreme_proto_M4N3s1}}
	\hfil
	\subfloat[$s=2$]{\includegraphics[width=0.24\textwidth]{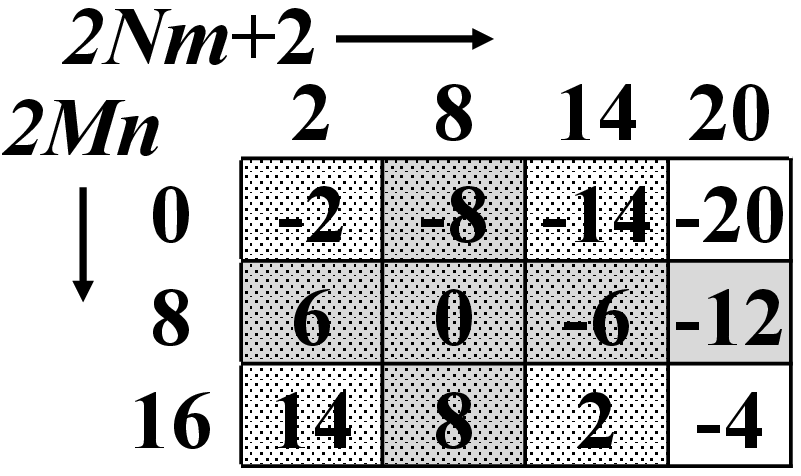}%
		\label{extreme_proto_M4N3s2}}
	\hfil
	\subfloat[$s=3$]{\includegraphics[width=0.24\textwidth]{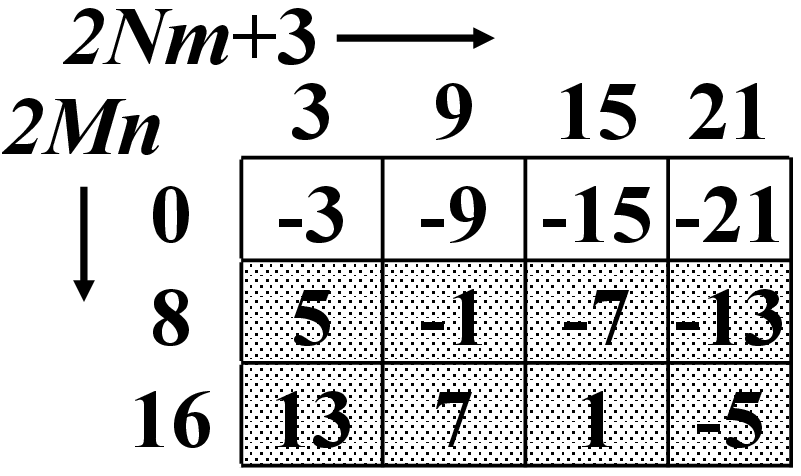}%
		\label{extreme_proto_M4N3s3}}
	\hfil
	\subfloat[$s=4$]{\includegraphics[width=0.24\textwidth]{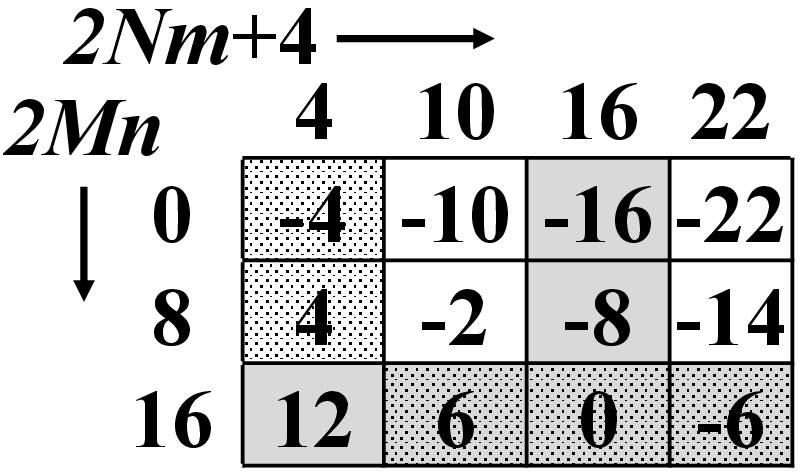}%
		\label{extreme_proto_M4N3s4}}
	\hfil
	\subfloat[$s=5$]{\includegraphics[width=0.24\textwidth]{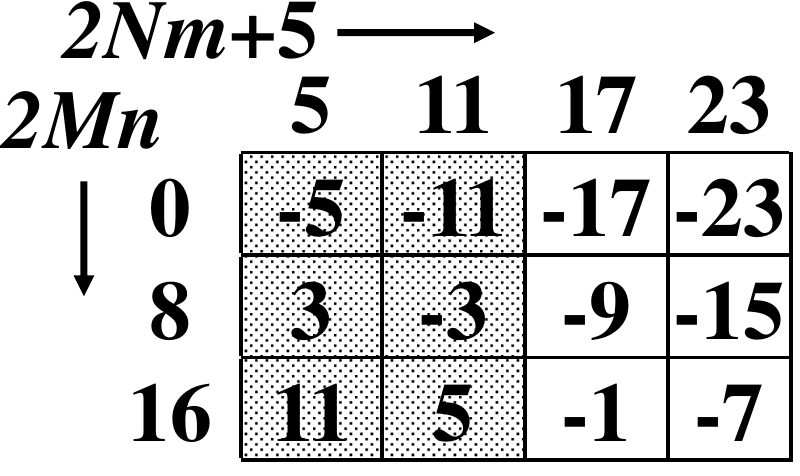}%
		\label{extreme_proto_M4N3s5}}
	\caption{Set $\mathcal{L}^+_{C}$ for different values of shift `\textit{s}'.}
	\label{fig:extreme_proto_difference_set}
\end{figure}

The definitions for the self and cross difference sets is given in~\eqref{eq:extreme_self} and~\eqref{eq:extreme_cross} respectively:
\begin{equation}\label{eq:extreme_self}
\mathcal{L}^+_{SM}=2Mn ~\text{and}~ \mathcal{L}^+_{SN}=2Nm
\end{equation}
where $n\in[0, N-1]$ and $m\in [0, M-1]$, with $\mathcal{L}^-_{SM}$ and $\mathcal{L}^-_{SN}$ representing the negative of the values in~\eqref{eq:extreme_self} and $\mathcal{L}_{S}=\mathcal{L}^+_{S}\cup\mathcal{L}^-_{S}=(\mathcal{L}^+_{SM}\cup\mathcal{L}^+_{SN})\cup (\mathcal{L}^-_{SM}\cup\mathcal{L}^-_{SN})$.
\begin{equation}\label{eq:extreme_cross}
\mathcal{L}^+_{C}=2Mn-(2Nm+s)
\end{equation}
and $\mathcal{L}^-_{C}$ represents the negative of the values in $\mathcal{L}^+_{C}$. For the case when $M=4$ and $N=3$, the set $\mathcal{L}^+_{SM}=\{0, 8, 16\}$, $\mathcal{L}^+_{SN}=\{0, 6, 12, 18\}$, $\mathcal{L}^+_{S}=\{0, 6, 8, 12, 16, 18\}$, $\mathcal{L}^-_{S}=\{0, -6, -8, -12, -16, -18\}$ and $\mathcal{L}_{S} = \{ -18, -16, -12, -8, -6, 0, 6, 8, 12, 16, 18 \}$ (Refer Fig.~\ref{fig:extreme_proto_self_difference}). The cross difference set $\mathcal{L}^+_{C}$ is shown in Fig.~\ref{fig:extreme_proto_difference_set} for different values of `$s$'. The shaded boxes represent the self differences while the dotted boxes represent the mirror pairs. When `$s$' is odd, the self differences do not form a subset of the cross differences while for the case when `$s$' is even some of the self differences form a subset of the cross differences. Specifically, when $s=0$ all the self differences are contained in the cross difference set and is same as the prototype co-prime array with a Nyquist period of $2d$.

For a shift of $s=\{0, 2, 4\}$ i.e. even, ExSCA is same as the APCA. The overlapping or pivot antenna is represented by an $(n,m)$ pair given by $\{(0,0), (1,1), (2,2)\}$ respectively. In general, it does not follow the pattern $(s,s)$. This aspect along with the range of an extremely sparse array is given in Proposition~\ref{exscaRange}.
\begin{prop}\label{exscaRange}
\begin{enumerate}
	\item The pivot or overlapping antenna location as a function of shifts `$s$', where $s$ is even, is given by:
	\begin{eqnarray}
	\nonumber   n=\left\{\frac{Nm+\frac{s}{2}}{M} \mid \frac{Nm+\frac{s}{2}}{M} \in \mathbb{Z}\right\}
	\end{eqnarray}
	which can also be written as:
	\begin{eqnarray}
	\nonumber   m=\left\{\frac{Mn-\frac{s}{2}}{N} \mid \frac{Mn-\frac{s}{2}}{N} \in \mathbb{Z}\right\}
	\end{eqnarray}
	\item Set $\mathcal{L}^+_{C}$ and $\mathcal{L}^-_{C}$ have difference values in the range $[-(2N(M-1)+s),(2M(N-1)-s)]$ and $[-(2M(N-1)-s), (2N(M-1)+s)]$ respectively.
	\item Set $\mathcal{L}_{C}$ has difference values in the range $[-R_{\mathcal{L}_{C}}, +R_{\mathcal{L}_{C}}]$, where 
	$R_{\mathcal{L}_{C}}=2N(M-1)+s$ if $(M+s)>N$,
	$R_{\mathcal{L}_{C}}=2M(N-1)-s$ if $N>(M+s)$.
	For the case when $M+s=N$, both the ranges are valid since $2N(M-1)+s=2M(N-1)-s$.
	\item  Set $\mathcal{L}=\mathcal{L}_{C}\cup \mathcal{L}_{S}$ has difference values in the range $[-R_{\mathcal{L}}, +R_{\mathcal{L}}]$, where $R_{\mathcal{L}}=max(R_{\mathcal{L}_{C}}, 2M(N-1),2N(M-1))$
\end{enumerate}
\end{prop}
\begin{IEEEproof}
\begin{enumerate}
	\item Let `s' be an even shift in the range $0<s<2N-1$. An antenna in the first array would overlap with the antenna in the second array for those values of $(n,m)$ such that the cross difference is zero.
	\begin{eqnarray}
	\nonumber        2Mn-(2Nm+s)&=&0\\
	\nonumber        2(Mn-Nm)&=&s\\
	\nonumber        Mn-Nm=\frac{s}{2}\\
	\nonumber        n=\frac{Nm+\frac{s}{2}}{M}
	\end{eqnarray}
	For a given value of `$s$', the value of `$m$' that generates an integer value `$n$' gives the location of the pivot antenna i.e. $(n_p,m_p)$. The above equation can be re-arranged to get $m$ as a function of $n$:
	\begin{equation}
	\nonumber m=\frac{Mn-\frac{s}{2}}{N}
	\end{equation}
	This proof is similar to the proof for Lemma 1.
	\item Given the cross difference set $\mathcal{L}^+_{C}=2Mn-(2Nm+s)$ where $n\in[0, N-1]$ and $m\in[0, M-1]$; the extreme values of set $\mathcal{L}^+_{C}$ can be obtained by substituting extreme values of $n$ and $m$. Substituting $m=0$ and $n=N-1$ gives a maximum difference value of $2M(N-1)-s$, while $n=0$ and $m=M-1$ gives a minimum difference value of $-(2N(M-1)+s)$. Thus proving the range for the set $\mathcal{L}^+_{C}$.
	
	Given a cross difference set $\mathcal{L}^-_{C}=(2Nm+s)-2Mn$ with $n\in[0, N-1]$ and $m\in[0, M-1]$. The extreme values of set $\mathcal{L}^-_{C}$ can be obtained by substituting extreme values of $n$ and $m$. Substituting $m=0$ and $n=N-1$ gives a minimum difference value of $-2M(N-1)+s$, while $n=0$ and $m=M-1$ gives a maximum difference value of $2N(M-1)+s$. Thus proving the range for the set $\mathcal{L}^-_{C}$.
	\item Since set $\mathcal{L}_{C}$ is the union of $\mathcal{L}^+_{C}$ and $\mathcal{L}^-_{C}$, the maximum value in set $\mathcal{L}_{C}$ is the $max(2M(N-1)-s, 2N(M-1)+s)$ while the minimum is the $min(-(2N(M-1)+s),-(2M(N-1)-s))$. Hence, set $\mathcal{L}_{C}$ has its range given by $[-R_{\mathcal{L}_{C}}, +R_{\mathcal{L}_{C}}]$ where $R_{\mathcal{L}_{C}}=max(2M(N-1)-s, 2N(M-1)+s)$. The condition under which $2N(M-1)+s$ is maximum is given by:
	\begin{eqnarray}
	\nonumber        2N(M-1)+s-(2M(N-1)-s)&>&0\\
	\nonumber        2(M-N+s)&>&0\\
	\nonumber        M+s&>&N\\
	\end{eqnarray}
	Similarly, the condition under which $2M(N-1)-s$ is maximum is given by:
	\begin{eqnarray}
	\nonumber        2M(N-1)-s-(2N(M-1)+s)&>&0\\
	\nonumber        2(N-M-s)&>&0\\
	\nonumber        N&>&M+s\\
	\end{eqnarray}
	It is straight forward to get $2M(N-1)-s=2N(M-1)+s$ when $M+s=N$, thus proving the claims.
	\item The maximum self difference value in the sets $\mathcal{L}^+_{SM}$ and $\mathcal{L}^+_{SN}$ is $2M(N-1)$ and $2N(M-1)$ respectively. The maximum value in the cross difference set $\mathcal{L}^+_{C}$ can be $2N(M-1)+s$ or $2M(N-1)-s$. Hence, the maximum value in the set $\mathcal{L}=\mathcal{L}_{C}\cup \mathcal{L}_{S}$ is given by the $max(R_{\mathcal{L}_{C}}, 2M(N-1), 2N(M-1))=max(2N(M-1)+s, 2M(N-1)-s, 2M(N-1), 2N(M-1))$. Thus proving the claim.
\end{enumerate}
\end{IEEEproof}%
The degrees of freedom in the self difference sets $\mathcal{L}^+_{SM}$, $\mathcal{L}^+_{SN}$, $\mathcal{L}^-_{SM}$, $\mathcal{L}^-_{SN}$, $\mathcal{L}^+_{S}$, $\mathcal{L}^-_{S}$ and $\mathcal{L}_{S}$ for an extremely sparse co-prime array is same as that of the prototype co-prime array. The cross difference set of an ExSCA as well as the prototype co-prime array ($\mathcal{L}^+_{C}$ and $\mathcal{L}^-_{C}$), have $MN$ unique values. It was proved for the prototype co-prime array.

The ExSCA does not have all the self differences within the cross difference set. In addition, all the elements in the set $\mathcal{L}^+_{C}-\mathcal{L}_{S}$ do not have a mirrored (negative) pair within the same set $\mathcal{L}^+_{C}$ (except for the case when `$s=0$'). 

For a given value of \textit{M}, \textit{N} and \textit{s}, $\mathcal{L}_{p}$ is defined as a set containing two elements with the same magnitude but different sign. For example, refer Fig.~\ref{extreme_proto_M4N3s0}, where value 2 is generated by $(n, m)$ = $(1, 1)$ and -2 by $(2, 3)$. Therefore, the set $\mathcal{L}_{p}=\{0, 2,-2,4,-4,10,-10\}$ for $s=0$. The elements `$l$' in the set $\mathcal{L}_{p}$ is given by:
\begin{eqnarray}\label{eq:extreme_Lc2}
\nonumber    \mathcal{L}_{p}&=&\left\{l=2Mn_1-(2Nm_1+s)\mid\right.\\ \nonumber&&M=\frac{N(m_1+m_2)+s}{(n_1+n_2)} ~\text{or}~ n_1=m_1=s=0 \\
&& \left . \forall m_1, m_2 \in[0, M-1], n_1, n_2 \in [0, N-1]\right\}
\end{eqnarray}
\textbf{Proof for expression~\eqref{eq:extreme_Lc2}:}
Let $(n_1, m_1)$ be an element in the set $\mathcal{L}^+_{C}$ such that, $2Mn_1-(2Nm_1+s)=x$. Let $(n_2, m_2)$ be another element in the set $\mathcal{L}^+_{C}$ such that, $2Mn_2-(2Nm_2+s)=-x$, then,
\begin{eqnarray}
\nonumber 2Mn_1-(2Nm_1+s)+2Mn_2-(2Nm_2+s)&=&0\\
\nonumber 2M(n_1+n_2)-2N(m_1+m_2)-2s&=&0\\
\nonumber M(n_1+n_2)=N(m_1+m_2)+s\\
\nonumber M=\frac{N(m_1+m_2)+s}{(n_1+n_2)}
\end{eqnarray}
This equation is not valid when the denominator is zero i.e. $(n_1+n_2)=0$. For the case when $n_1=n_2=0$, $x$ and $-x$ are expected to lie in the first row of the matrix $\mathcal{L}^+_{C}$ and in general would never hold true since $2Mn-(2Nm+s)=-(2Nm+s)$ when $n=0$, and is negative $\forall m>0$. Therefore the first row will not contain a mirror value. The only situation where $x$ and $-x$ lie in the first row is for the case when $x=0$, $s=0$, $n=0$ and $m=0$. Refer Fig.~\ref{fig:extreme_proto_difference_set}.

Similarly, a set containing an element without a mirror pair is denoted as $\mathcal{L}_{np}$.
As an example $\mathcal{L}_{np}=\{8, 16, -6, -12, -18\}$ for the case when $s=0$ (Fig.~\ref{extreme_proto_M4N3s0}) and $\mathcal{L}_{np}=\{9, 15, -5, -11, -13, -19\}$ for the case when $s=1$ (Fig.~\ref{extreme_proto_M4N3s1}). This set is not formally defined since it is of no consequence to the discussion that will follow. However, it important to note that when a difference value occurs in pair (except self-differences) it has two contributors for estimation. 

 Proposition~\ref{exscaProp2} describes the number of unique values in the ExSCA set, where `\#' denotes the cardinality of the set. Functions $f(n)$ and $f(m)$ are defined below.
\begin{eqnarray}\label{eq:f(x)}
\nonumber f(n)&=& \left\{ \begin{array}{cc}
n,& ~\textit{for}~ 0\leq n\leq \lfloor \frac{N-1}{2}\rfloor  \\
N-1-n,& ~\textit{for}~ \lfloor \frac{N-1}{2}\rfloor < n\leq N-1 
\end{array}\right.\\
\nonumber f(m)&=& \left\{ \begin{array}{cc}
m,& ~\textit{for}~ 0\leq m\leq \lfloor \frac{M-1}{2}\rfloor  \\
M-1-m,& ~\textit{for}~ \lfloor \frac{M-1}{2}\rfloor < m\leq M-1 
\end{array}\right.\\
\end{eqnarray}

\begin{prop}\label{exscaProp2}

\begin{enumerate}
	\item For odd values of `$s$', $\mathcal{L}_{S} \not\subset \mathcal{L}_{C}$. While some values of the self differences are contained in the cross difference set for even values of `$s$'. 
	\item The set $\mathcal{L}_{C}=\mathcal{L}^+_{C}\cup \mathcal{L}^-_{C}$ has $2MN-\#\mathcal{L}_{p}$ unique values (Note $l=0$ is included in $\mathcal{L}$).
	\item The set $\mathcal{L}=\mathcal{L}_{C}\cup \mathcal{L}_{S}$ has $2MN-\#\mathcal{L}_{p}+(2(M+N-1)-1)$ unique values for $s=odd$ and $2MN-\#\mathcal{L}_{p}+2(f(n_p)+f(m_p))$ for $s=even$.
\end{enumerate}
\end{prop}%

\begin{figure}[!t]
	\centering
	\includegraphics[width=0.45\textwidth]{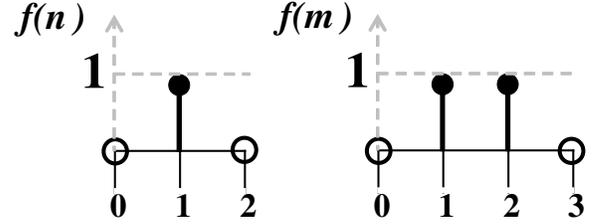}%
	\caption{Functions $f(n)$ and $f(m)$ when $M=4$ and $N=3$.}
	\label{fig:extreme_fn}
\end{figure}

\begin{IEEEproof}
\begin{enumerate}
	\item Let $l_c=2Mn-(2Nm+s)$ be an element in the cross difference set $\mathcal{L}^+_{C}$. For the case when $s$ is odd, $l_c=2Mn-s$ when $m=0$ and $l_c=-(2Nm+s)$ when $n=0$. The value $l_c$ is odd, for odd values of $s$ and cannot belong to the self difference set $\mathcal{L}_{S}$. Since the self difference set has even values i.e. $\pm 2Mn$ and $\pm 2Nm$. Refer Fig.~\ref{extreme_proto_M4N3s1},~\ref{extreme_proto_M4N3s3}, and~\ref{extreme_proto_M4N3s5}.
	
	For the case when $s$ is even, the location of the overlapping/pivot antenna was given by Proposition~\ref{exscaRange}-1. Let $n_s$ and $m_s$ denote this location for a given even value of $s$, where $n_s$ and $m_s$ denotes the row and the column respectively, of the matrix $\mathcal{L}^+_{C}$ which contains self difference values. Substituting $n_s$ for $n$ in the equation for $l_c$ gives:
	\begin{align}
	\nonumber   &2M\frac{(Nm_s+\frac{s}{2})}{M}-(2Nm+s)\\
	\nonumber   &2Nm_s+s-(2Nm+s)\\
	\nonumber   &2N(m_s-m) \in \mathcal{L}_{S}
	\end{align}
	This means that the row $n_s$ generates the self differences that belong to set $\mathcal{L}^+_{SM}\cup \mathcal{L}^-_{SM}$. A similar argument can be made when we substitute $m_s$ in place of $m$ in the equation for $l_c$:
	\begin{align}
	\nonumber   &2Mn-(2N\frac{(Mn_s-\frac{s}{2})}{N}+s)\\
	\nonumber   &2Mn-(2Mn_s)\\
	\nonumber   &2M(n-n_s) \in \mathcal{L}_{S}
	\end{align}
	This means that the column $m_s$ generates the self differences that belong to set $\mathcal{L}^+_{SN}\cup \mathcal{L}^-_{SN}$. Refer Fig.~\ref{extreme_proto_M4N3s0},~\ref{extreme_proto_M4N3s2}, and~\ref{extreme_proto_M4N3s4}.
	\item Set $\mathcal{L}^+_{C}$ and $\mathcal{L}^-_{C}$ have $MN$ unique values. If these two sets did not contain any common value, then the set $\mathcal{L}_{C}$ would have $2MN$ unique values. In case the set $\mathcal{L}^+_{C}$ and $\mathcal{L}^-_{C}$ have common values, it needs to be subtracted from $2MN$. The number of common values between set  $\mathcal{L}^+_{C}$ and $\mathcal{L}^-_{C}$ is given by the cardinality of the set $\mathcal{L}_{p}$. Thus proving the claims.
	\item For the case when $s=odd$, $\mathcal{L}_{S} \not\subset \mathcal{L}_{C}$ and hence the number of unique values in the set $\mathcal{L}$ is the sum of the number of unique values in $\mathcal{L}_{C}$ and $\mathcal{L}_{S}$ i.e. $(2MN-\#\mathcal{L}_{p})+(2(M+N-1)-1)$.
	
	For the case when $s=even$, the justification is developed with the help of the example given in Fig.~\ref{fig:extreme_proto_difference_set}. When $s=0$ (Fig.~\ref{extreme_proto_M4N3s0}), all the self differences are contained in the cross difference set. Therefore, the number of unique values in $\mathcal{L}$ is given by:
	\begin{align}
	2MN-\#\mathcal{L}_{p} + 2(0+0)
	\end{align}
	`0' implies no self-differences ($|l_s|$) are missing along the row $n_p$ and column $m_p$.
	When $s=2$ (Fig.~\ref{extreme_proto_M4N3s2}), self differences $\pm 16$ and $\pm 18$ from sets ($\mathcal{L}^+_{SM}\cup \mathcal{L}^-_{SM}$) and ($\mathcal{L}^+_{SN}\cup \mathcal{L}^-_{SN}$) are missing and needs to be added to the unique values of set $\mathcal{L}_{C}$. This is given by:
	\begin{align}
	2MN-\#\mathcal{L}_{p} + 2(1+1)
	\end{align}
	where $2$ is multiplied to take into account the positive as well as the negative value. `1' indicates missing ($|l_s|$) value along row and column. Similarly, when $s=4$ (Fig.~\ref{extreme_proto_M4N3s4}), the self differences $\pm 18$ from set ($\mathcal{L}^+_{SN}\cup \mathcal{L}^-_{SN}$) is missing while all the elements of the set ($\mathcal{L}^+_{SM}\cup \mathcal{L}^-_{SM}$) are present in the cross difference set. So, `1' missing ($|l_s|$) along column $m_p$ and `0' missing along row $n_p$. Therefore, the number of unique values in $\mathcal{L}$ is given by:
	\begin{align}
	2MN-\#\mathcal{L}_{p} + 2(0+1)
	\end{align}
	This variation in the number of self differences missing in the cross difference set is modelled as a function of $s$ and is given by $2(f(n_p)+f(m_p))$ where $(n_p, m_p)$ is the pivot location for shift $s$, $f(n_p)$ and $f(m_p)$ represent the missing self difference value in set $\mathcal{L}^+_{SM}$ and $\mathcal{L}^+_{SN}$ respectively. The functions $f(n_p)$ and $f(m_p)$ are defined in~\eqref{eq:f(x)} and have been shown in Fig.~\ref{fig:extreme_fn} for $M=4$ and $N=3$.
\end{enumerate}
\end{IEEEproof}

The next logical thing to do, is to analyse the weight function. This is presented as Proposition~\ref{exscaWtsProp} and is divided into two cases: odd values of shift (extremely sparse scheme) and even values of shift (up-sampled version of APCA). The proposition is followed by a brief explanation justifying the claims. Note that $n_p$ and $m_p$ represents the location of the overlapping antennas or the pivot, and is given by Proposition~\ref{exscaRange}-1.
%

\begin{prop}\label{exscaWtsProp}
\begin{enumerate}
	\item For the case when $s$ is odd:
	\begin{enumerate}
		\item $z(l)=(N-n), \{1 \leq n \leq N-1, l=\pm 2Mn\}$
		\item $z(l)=(M-m), \{1 \leq m \leq M-1, l=\pm 2Nm\}$
		\item $z(l)=M+N, \{l=0\}$
		\item $z(l)=2,  \{l \in \mathcal{L}_{p}-\{0\}\}$
		\item $z(l)=1,  \{l \in \mathcal{L}_{np}=\{\mathcal{L}_{C}-\mathcal{L}_{p}\}\}$
	\end{enumerate}
	\item For the case when $s$ is even:
	\begin{enumerate}
		\item $z(l)=(N-n), \{1 \leq n \leq N-1, l=\pm 2Mn\}$
\item $z(l)=(M-m), \{1 \leq m \leq M-1, l=\pm 2Nm\}$		\item $z(l)=M+N-1, \{l=0\}$
		\item $z(l)=2,  \{l \in \mathcal{L}_{p}-\mathcal{L}_{S}\}$
		\item $z(l)=1,  \{l \in \{\mathcal{L}_{C}-\mathcal{L}_{p}-\mathcal{L}_{S}\}\}$
	\end{enumerate}
\end{enumerate}
\end{prop}
For the case when $s$ is odd, the self differences are not contained in the cross difference set. Therefore the number of sample pairs that contribute to the autocorrelation estimation at the self difference values is given by the number of elements in the diagonal of the self difference matrix. For $l=0$, $z(l)$ is the sum of the individual principle diagonal elements in Fig.~\ref{extreme_proto_self_diff_Mn} and~\ref{extreme_proto_self_diff_Nm}, i.e. $M+N$. Note $s=\{1,3,5\}$. The elements in the set $\mathcal{L}_{p}$ have a mirror pair contained within it. Hence, every element in the set $\mathcal{L}_{p}$ has two pairs of $(n,m)$ that can generate the value $l$. The remaining values which neither belong to the self difference set nor $\mathcal{L}_{p}$; belong to $\mathcal{L}_{np}$. Only one pair $(n,m)$ produces the value in $\mathcal{L}_{np}$.

For the case when $s=0$ the weight function was described in~\cite{UVD_PHD}. In general, when $s$ is even, the number of contributors is similar to $s=odd$. However, there is a difference when $l=0$. Note that one of the contributors to the difference value zero is generated by the pivot location of the individual samplers and it is common. Therefore, it needs to be added once. Thus, justifying Proposition~\ref{exscaWtsProp}-2(c). This can be verified from Fig.~\ref{fig:extreme_proto_self_difference}, where $x(0)$, $x(8)$, and $x(16)$ is common to both the self difference matrices for $s=\{0, 2, 4\}$ respectively.
\begin{figure*}[!t]
	\centering
	\subfloat[$M=4$, $N=3$, $s=0$]{
		\includegraphics[width=0.45\textwidth]{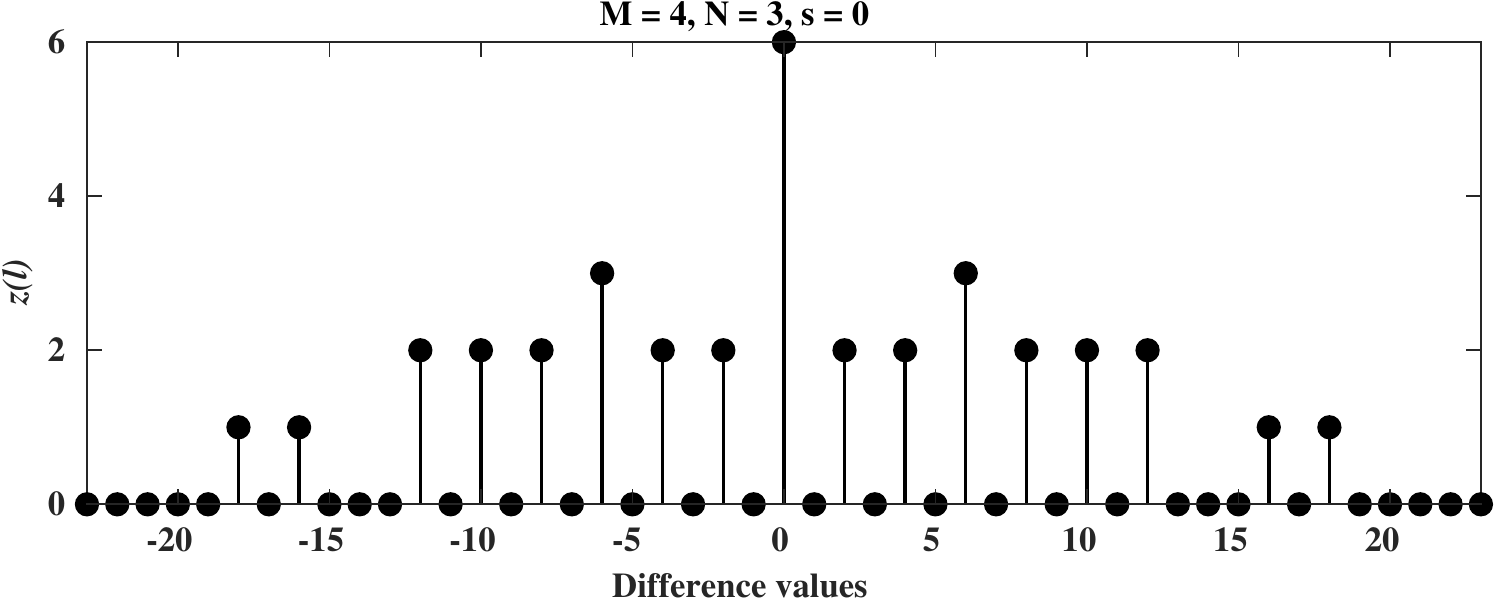}%
		\includegraphics[width=0.45\textwidth]{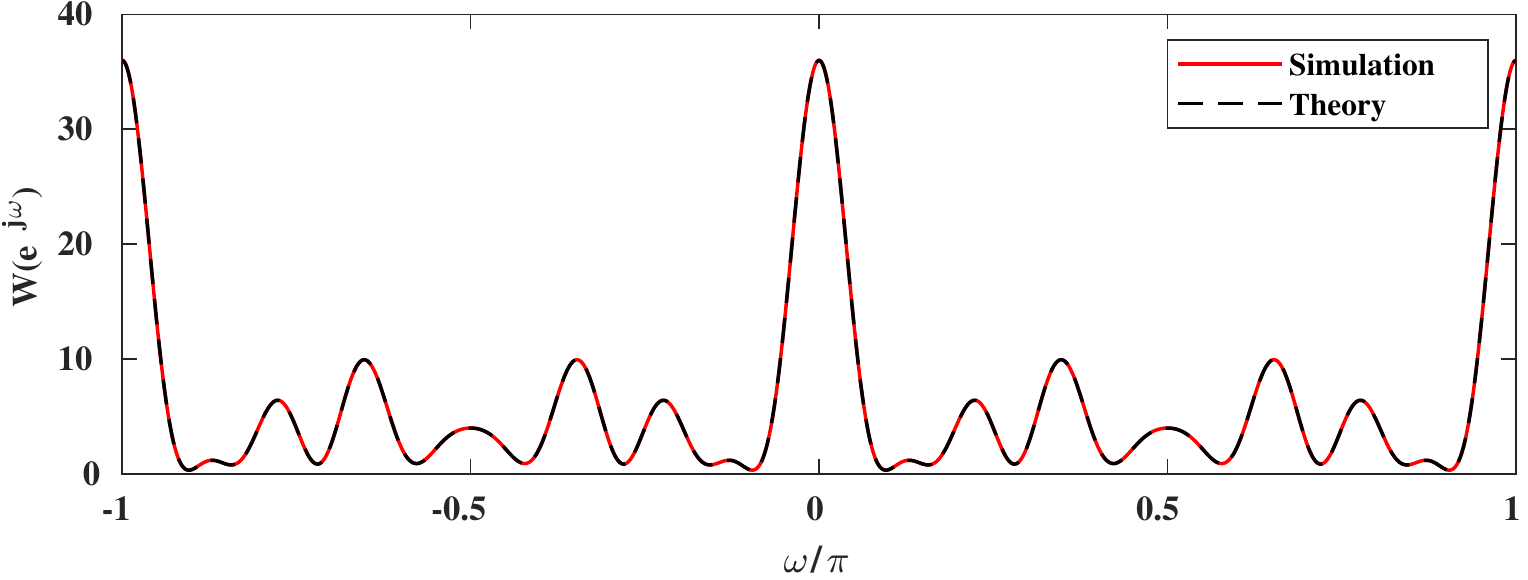}%
		\label{extreme_Ex2_wt_bias_M4N3s0}}
	\hfil
	\subfloat[$M=4$, $N=3$, $s=1$]{
		\includegraphics[width=0.45\textwidth]{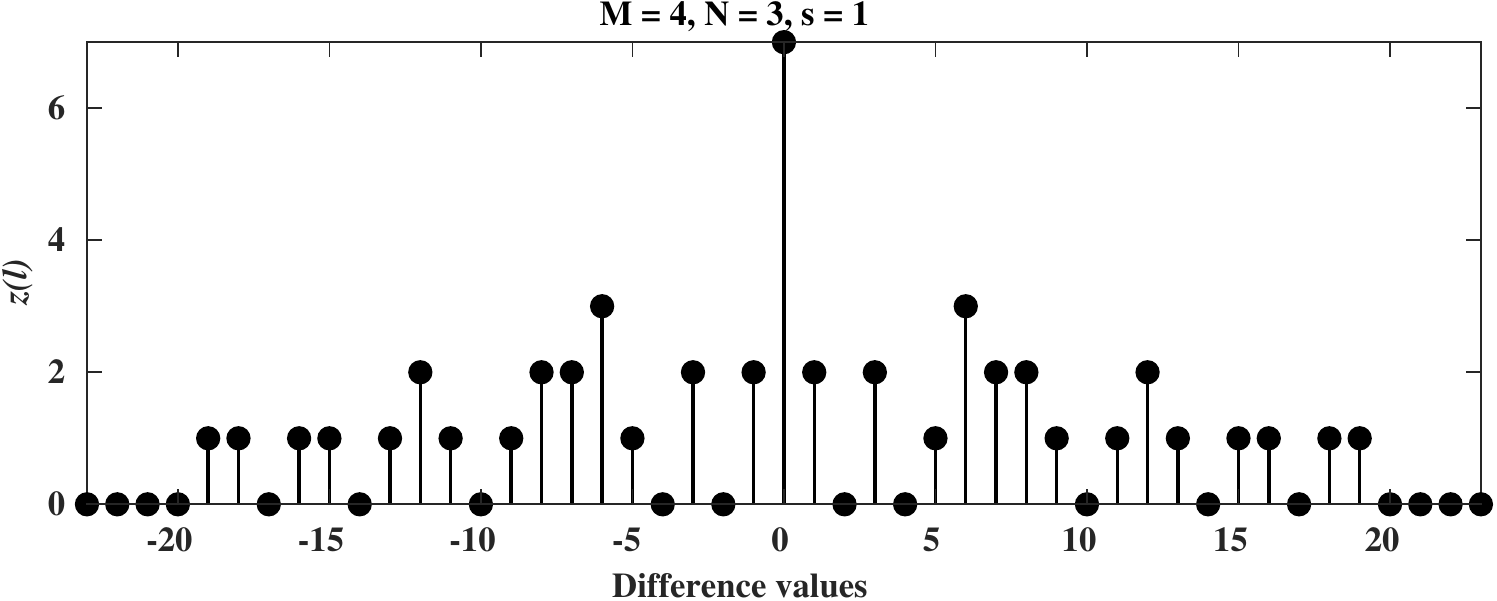}%
		\includegraphics[width=0.45\textwidth]{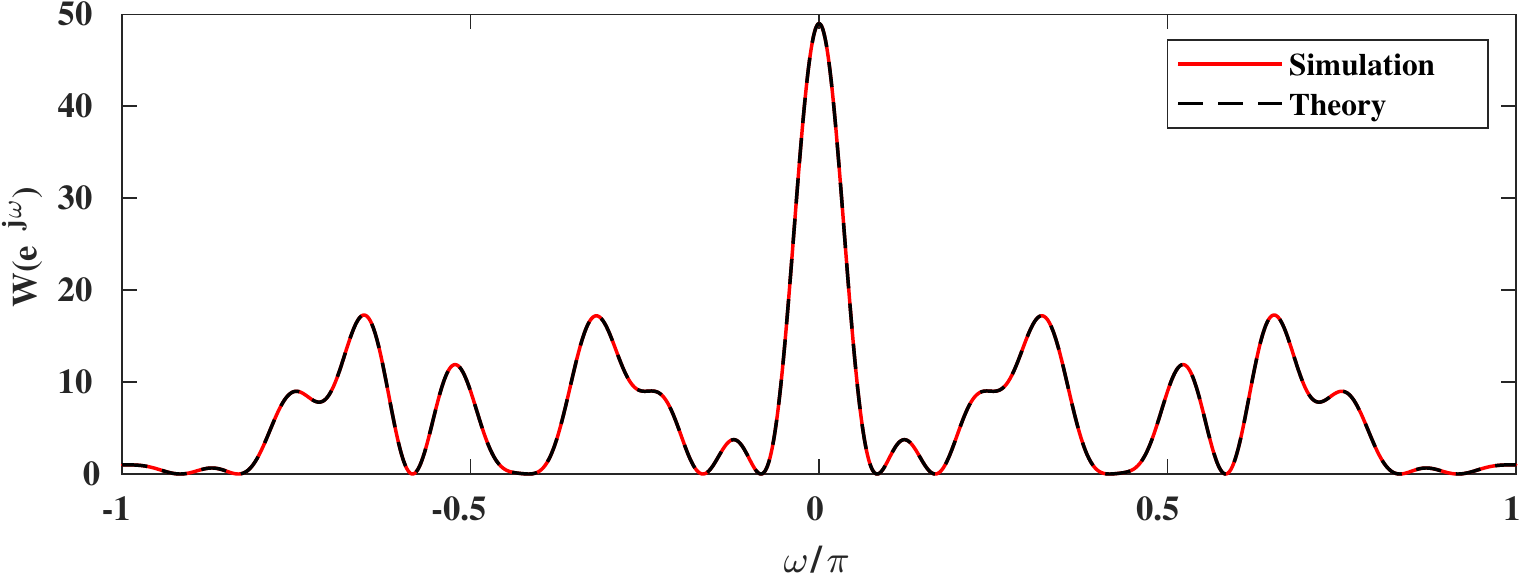}%
		\label{extreme_Ex2_wt_bias_M4N3s1}}
	\hfil
	\subfloat[$M=4$, $N=3$, $s=2$]{
		\includegraphics[width=0.45\textwidth]{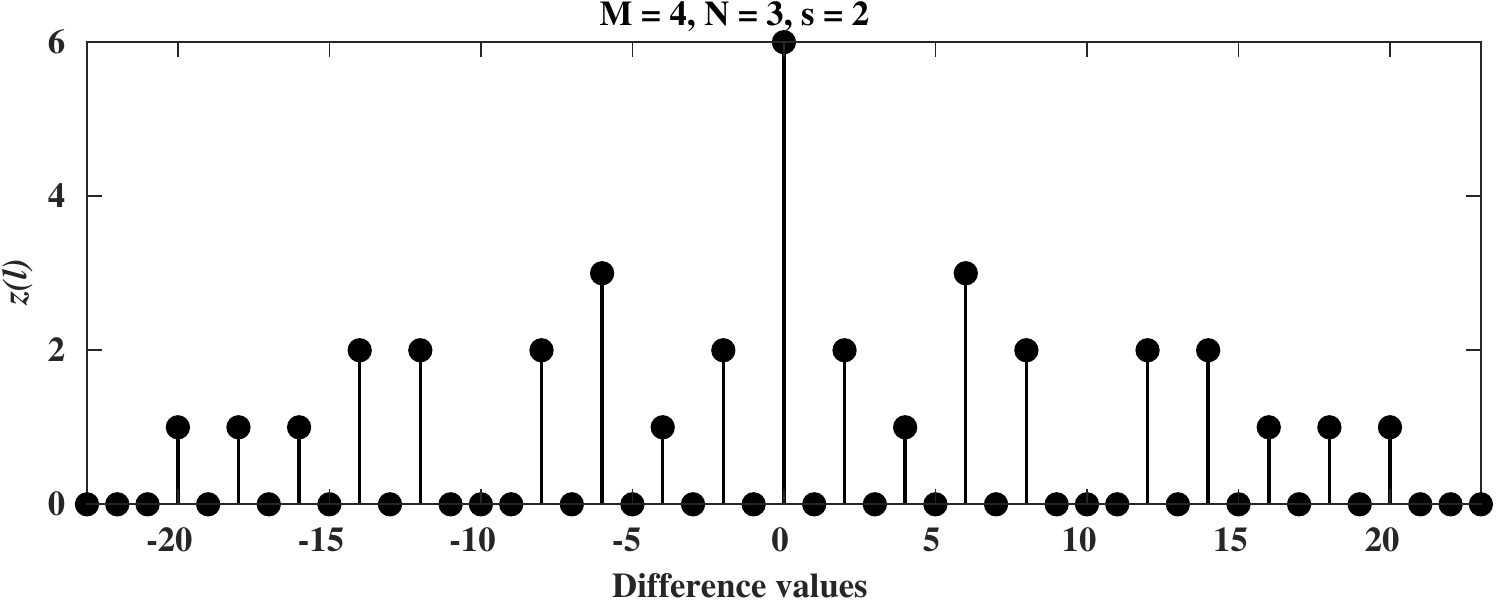}%
		\includegraphics[width=0.45\textwidth]{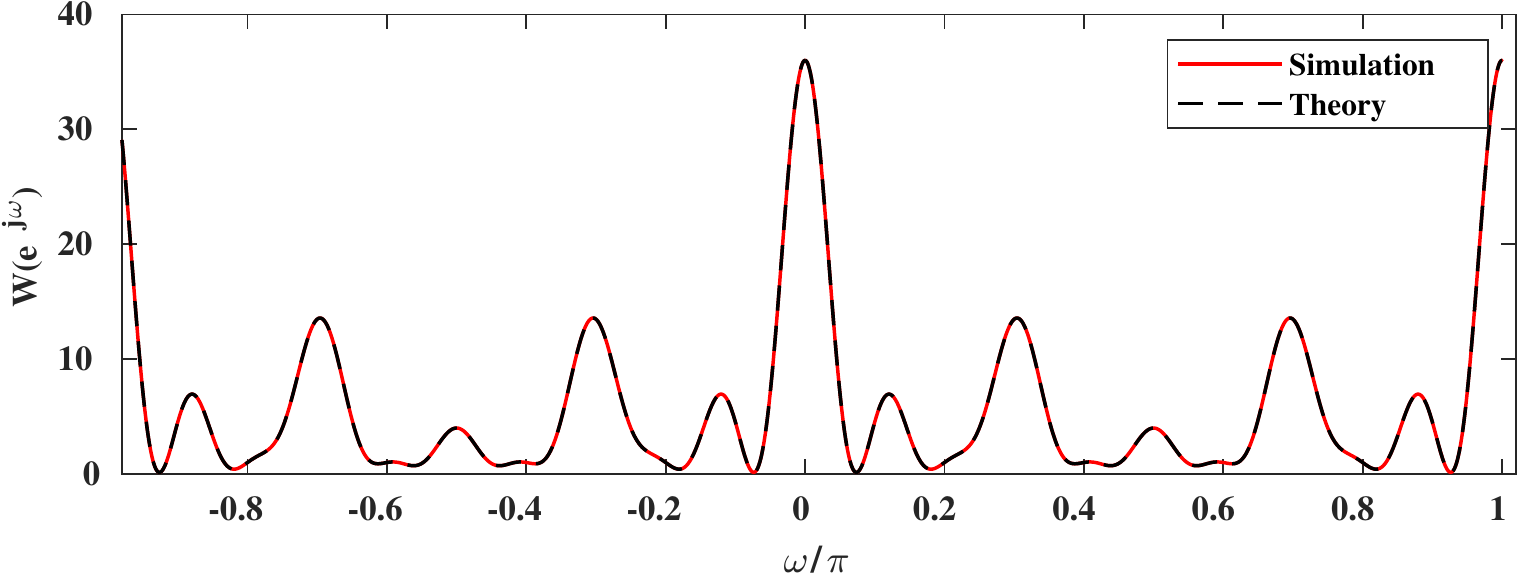}%
		\label{extreme_Ex2_wt_bias_M4N3s2}}
	\hfil		
	\subfloat[$M=4$, $N=3$, $s=3$]{
		\includegraphics[width=0.45\textwidth]{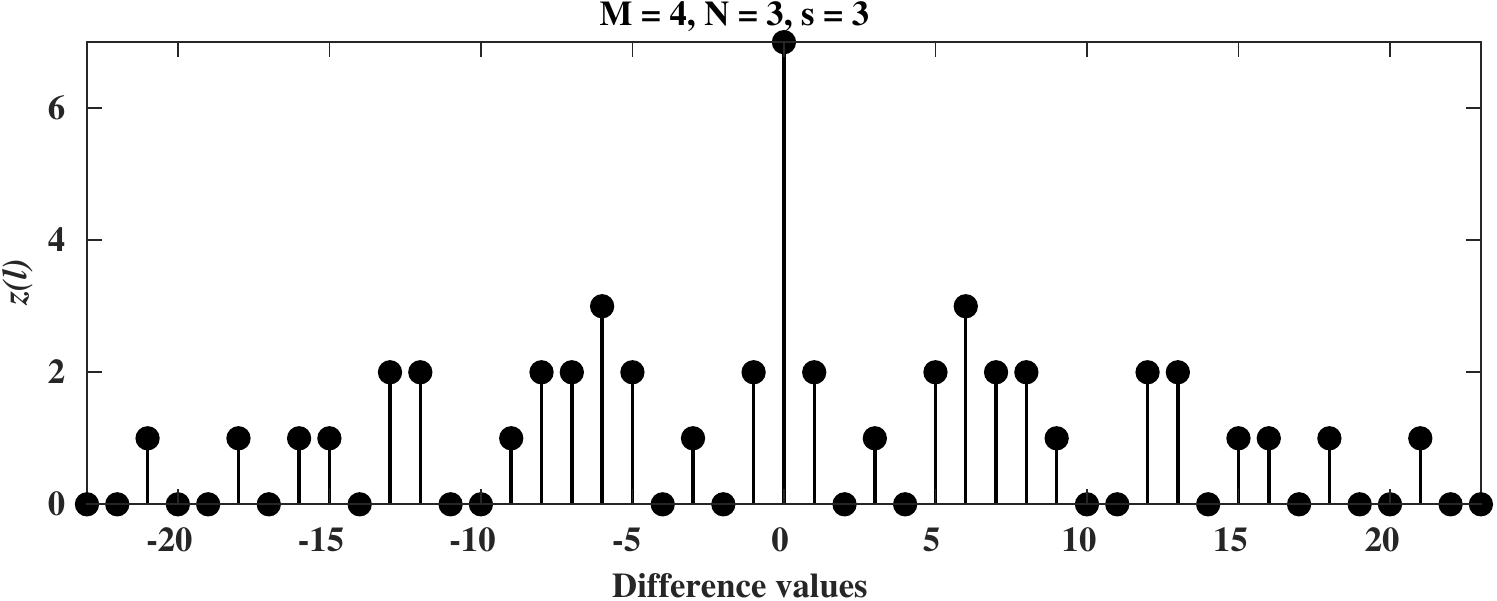}%
		\includegraphics[width=0.45\textwidth]{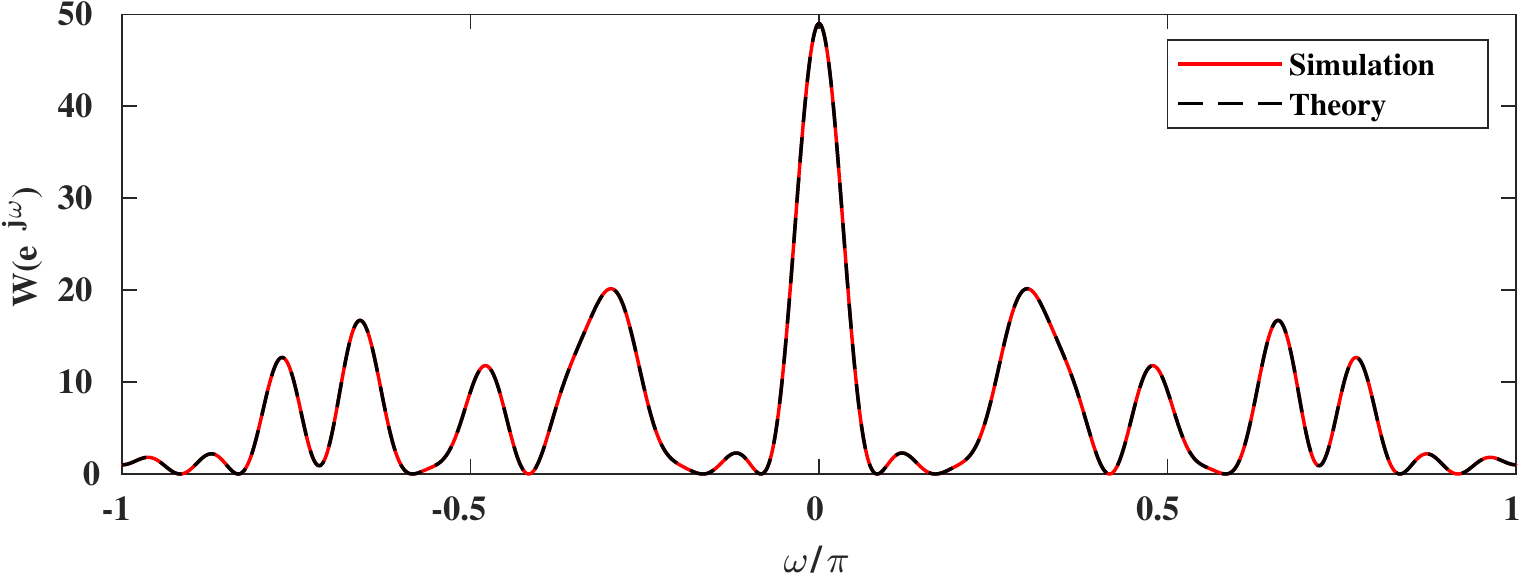}%
		\label{extreme_Ex2_wt_bias_M4N3s3}}
	\hfil
	\subfloat[$M=4$, $N=3$, $s=4$]{
		\includegraphics[width=0.45\textwidth]{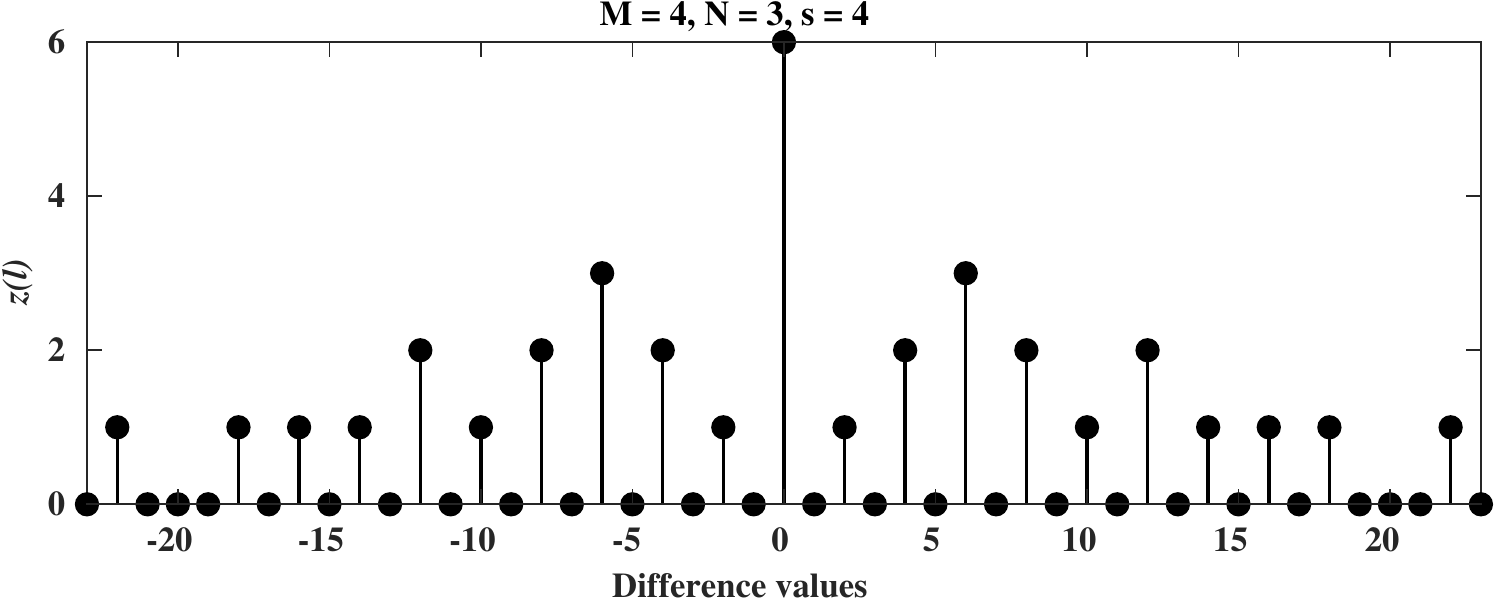}%
		\includegraphics[width=0.45\textwidth]{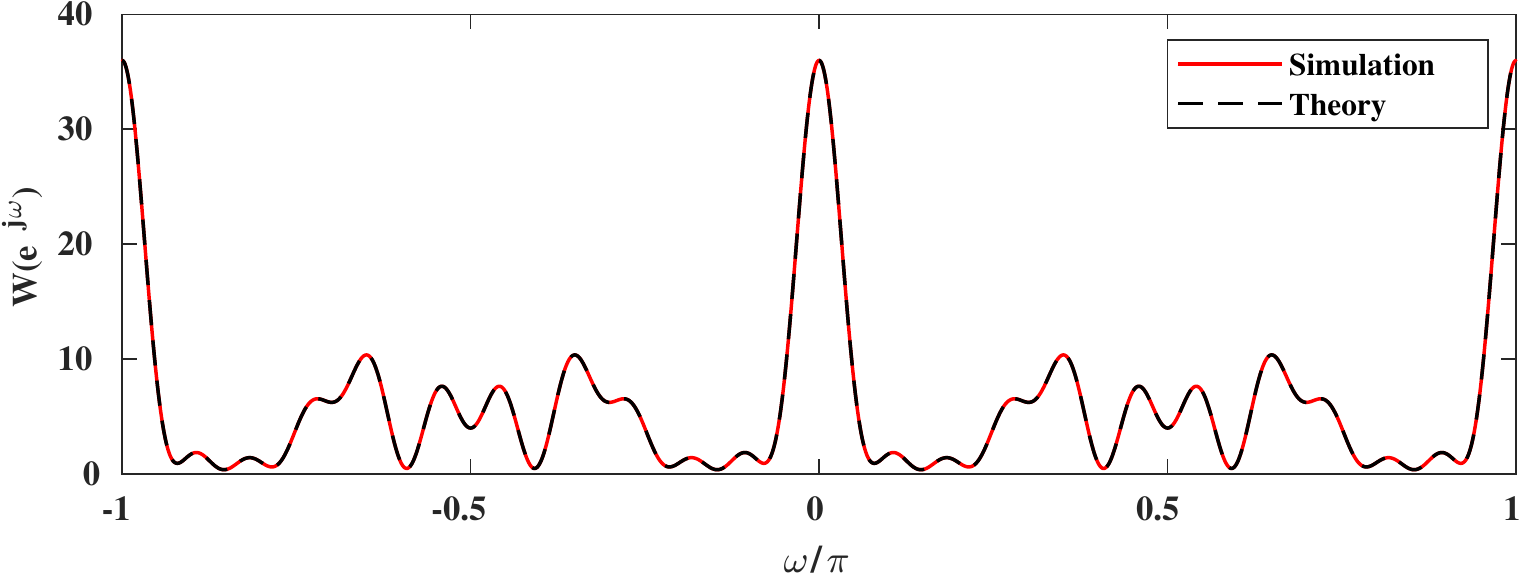}%
		\label{extreme_Ex2_wt_bias_M4N3s4}}
	\hfil
	\subfloat[$M=4$, $N=3$, $s=5$]{
		\includegraphics[width=0.45\textwidth]{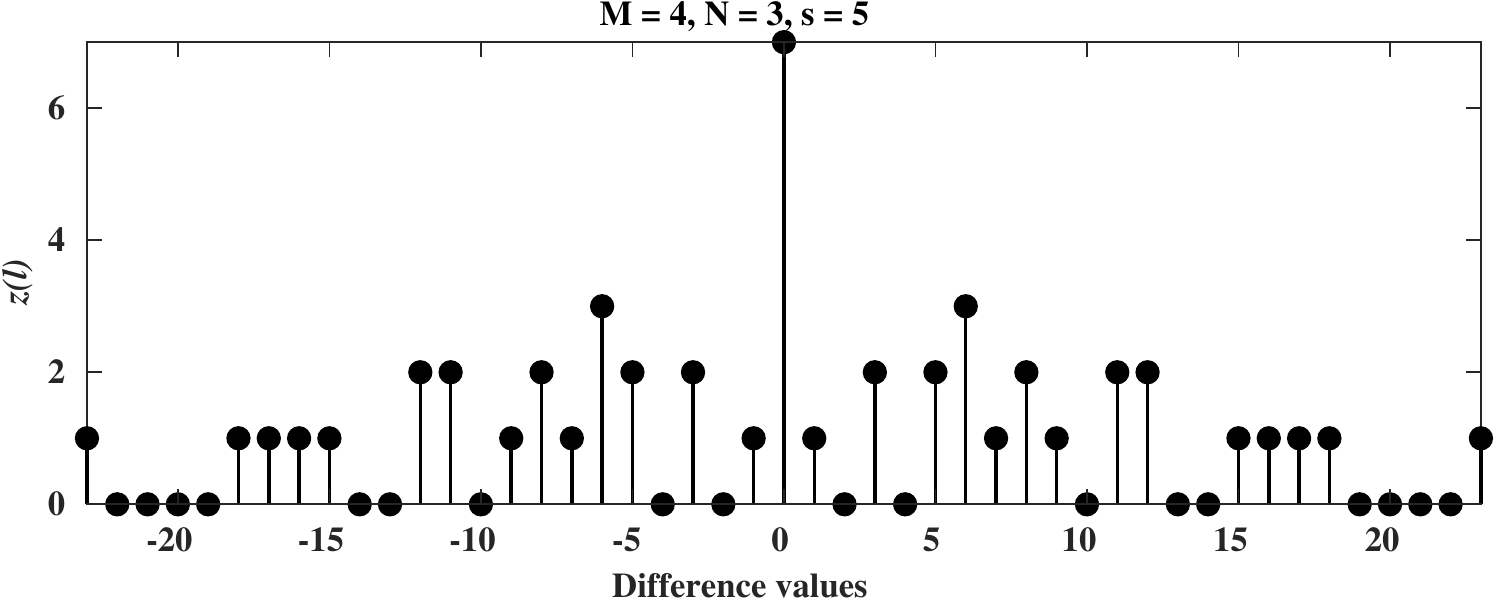}%
		\includegraphics[width=0.45\textwidth]{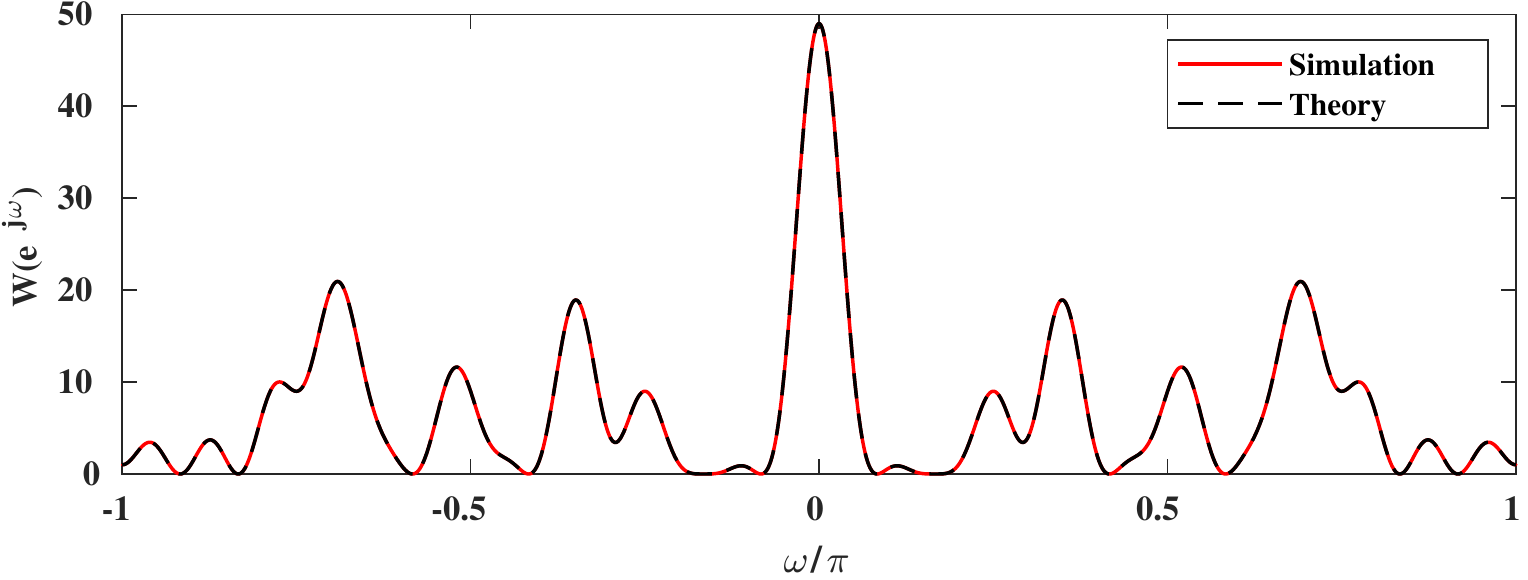}%
		\label{extreme_Ex2_wt_bias_M4N3s5}}
	\caption{Weight function and the associated bias of the correlogram estimate for Extremely sparse arrays with $M=4$, $N=3$.}
	\label{fig:extreme_Ex2_wt_bias_M4N3}
\end{figure*}
\subsection{Weight and Bias Analysis}
The weight function and the associated bias of the correlogram estimate for the ExSCA is shown in Fig.~\ref{fig:extreme_Ex2_wt_bias_M4N3} for $M=4$ and $N=3$. For even values of $s$, i.e. Figs.~\ref{extreme_Ex2_wt_bias_M4N3s0}, ~\ref{extreme_Ex2_wt_bias_M4N3s2} and ~\ref{extreme_Ex2_wt_bias_M4N3s4}, the weight function is an up-sampled version of the weight function of the APCA described in Fig.~\ref{fig:Truly_coprime_wts_bias_M4N3_M7N3}. Up-sampling a function in time-domain produces an image in the frequency domain, which is evident from the figures.
%
\begin{figure*}[!t]
	\centering
	\includegraphics[width=0.49\textwidth]{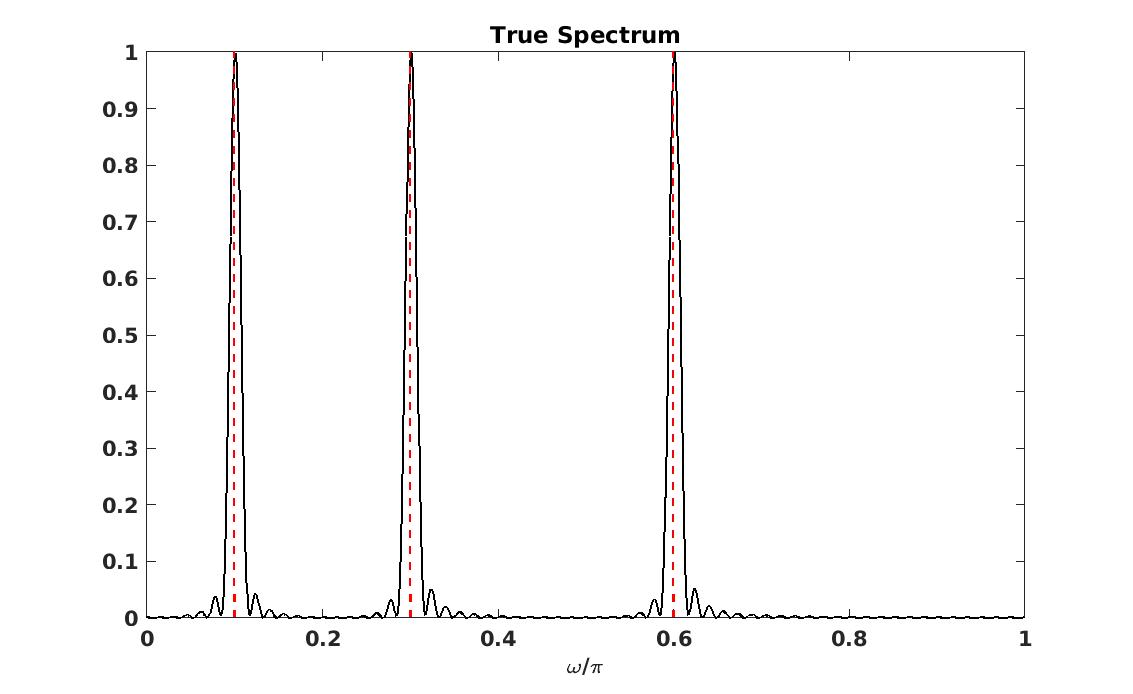}
	\includegraphics[width=0.49\textwidth]{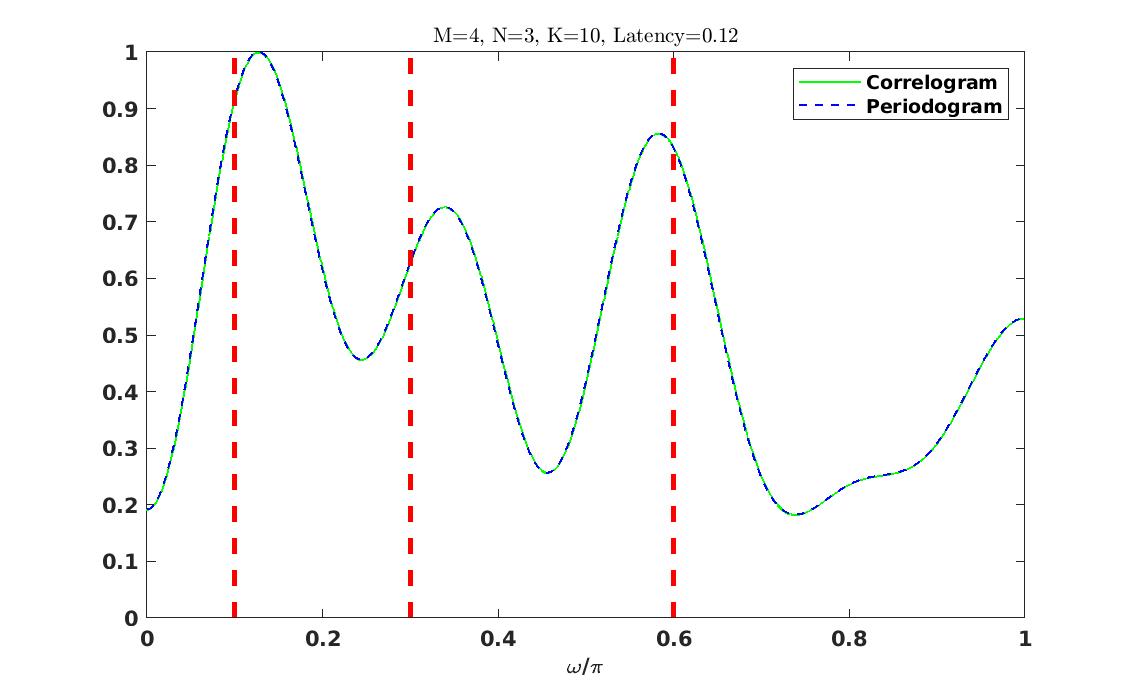}
	\caption{True spectrum (Nyquist) and prototype co-prime correlogram with $M=4$, $N=3$, and peaks at locations $[0.1, 0.3, 0.6]$.}
	\label{fig:extreme_Ex2_sim_true_spect}
\end{figure*}
\begin{figure*}[!t]
	\centering
		\includegraphics[width=0.49\textwidth]{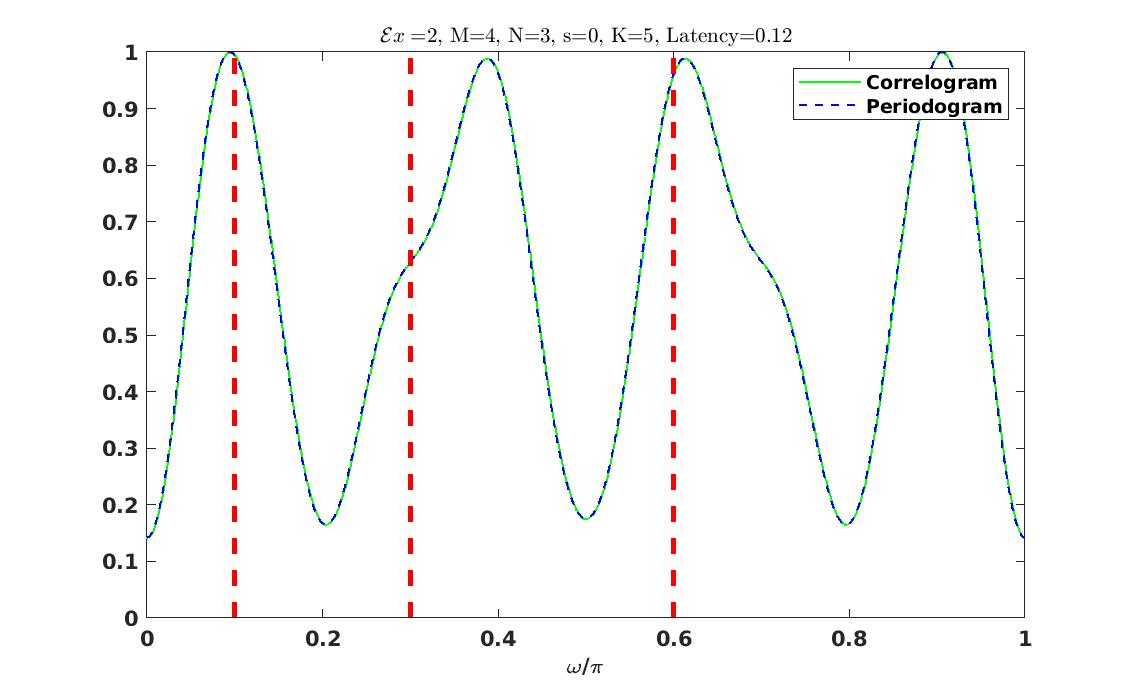}
		\includegraphics[width=0.49\textwidth]{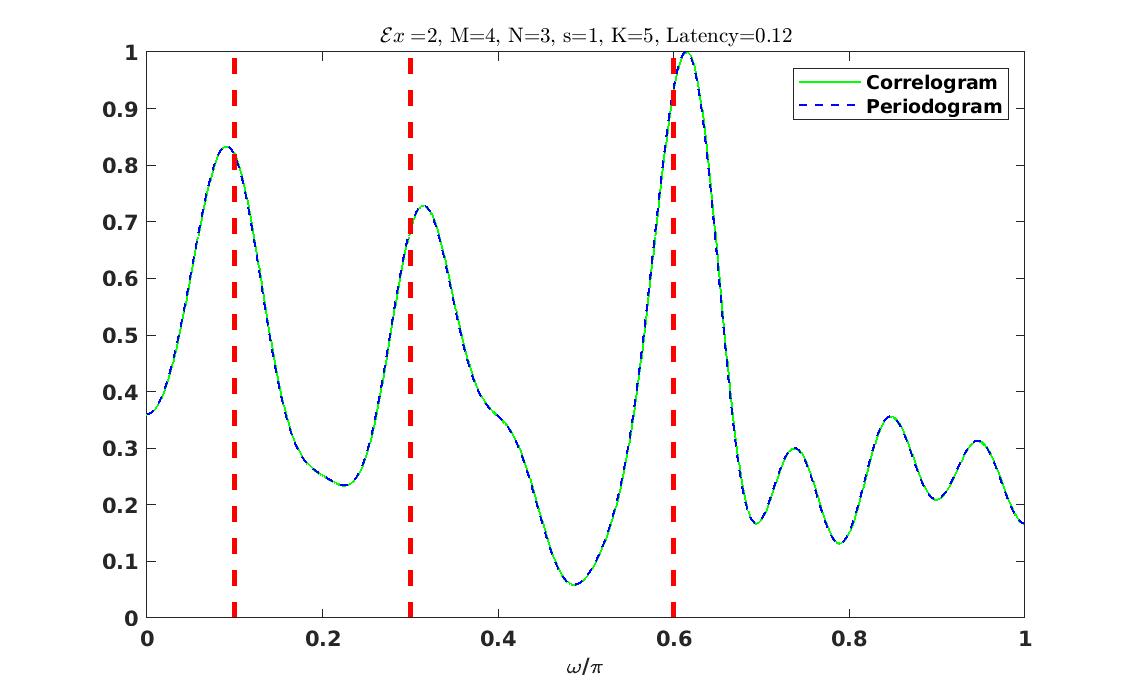}\\
		\includegraphics[width=0.5\textwidth]{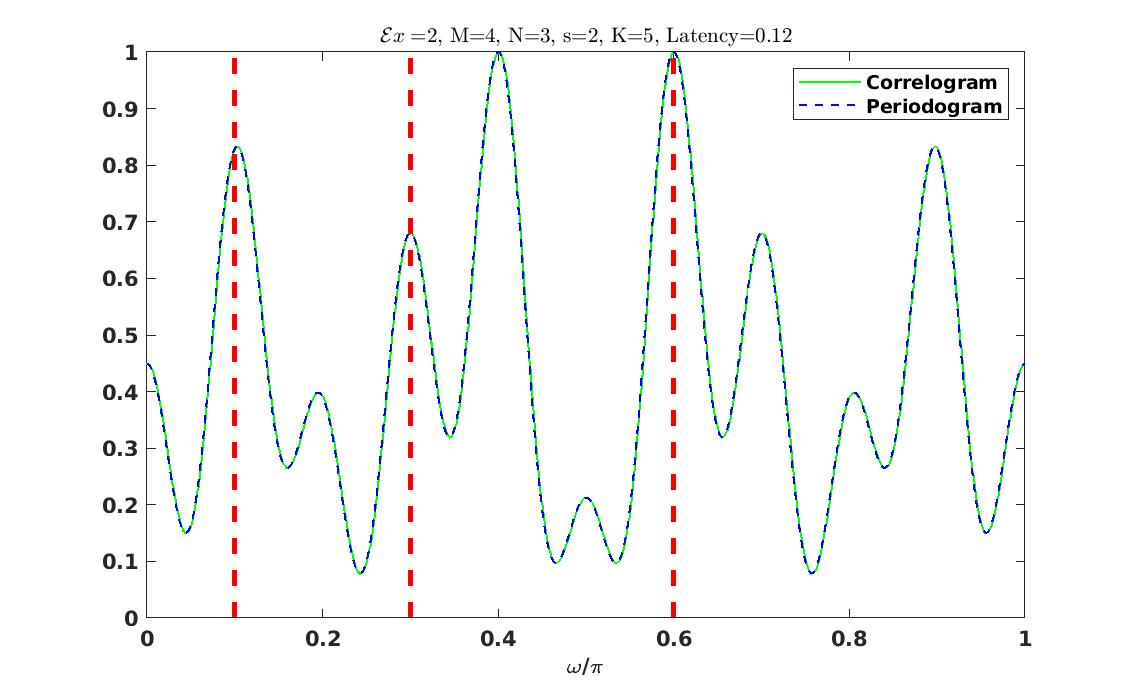}%
		\includegraphics[width=0.5\textwidth]{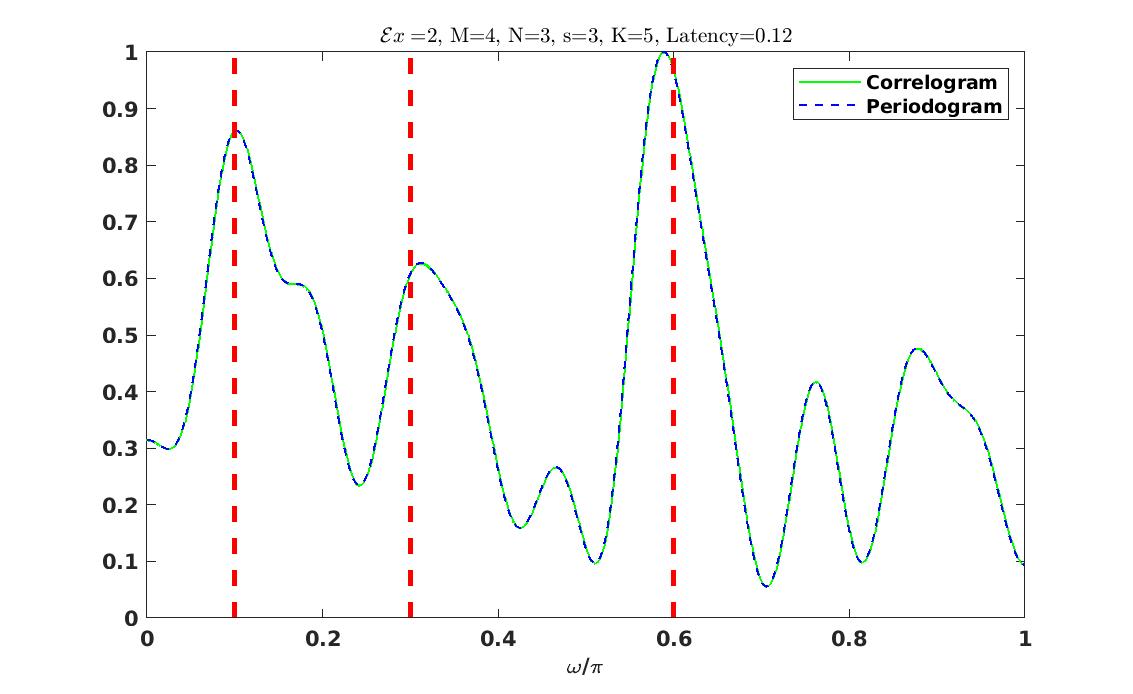}\\
		\includegraphics[width=0.5\textwidth]{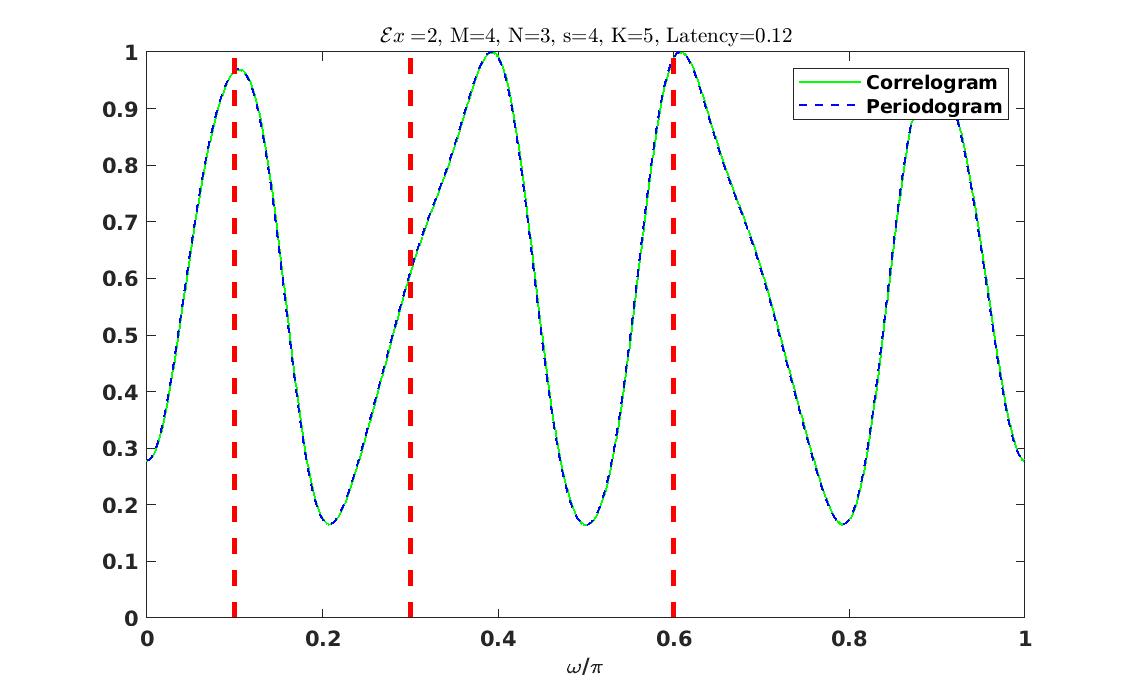}%
		\includegraphics[width=0.5\textwidth]{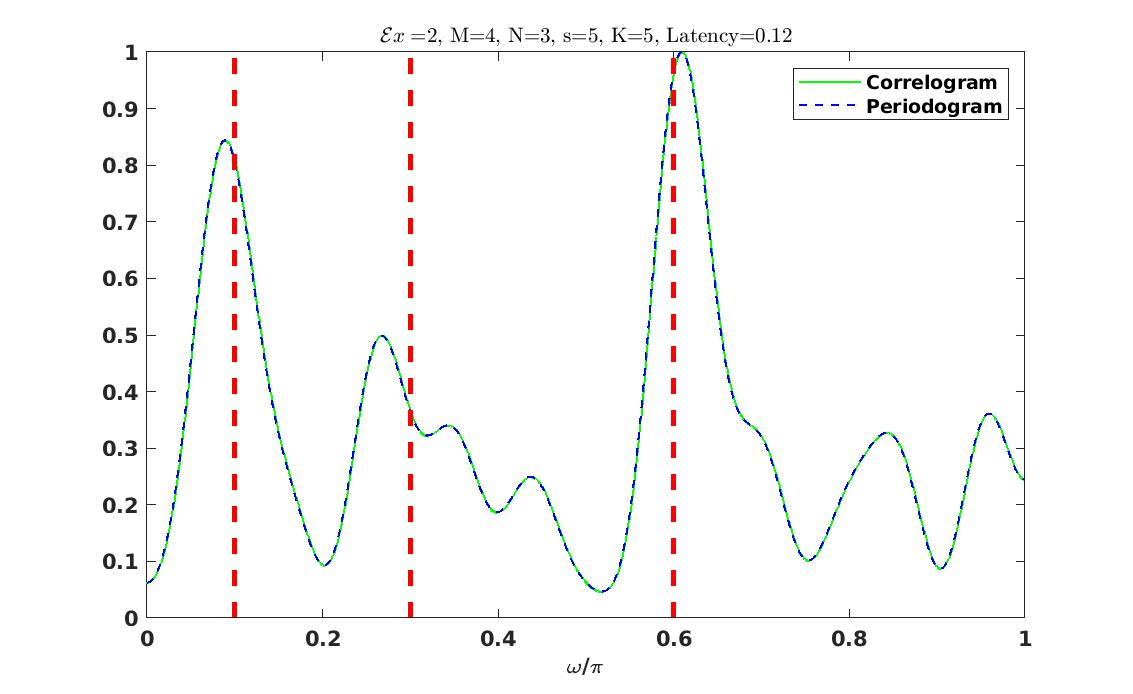}
		\caption{Correlogram spectral estimation for ExSCA with $\mathcal{E}_x=2$, $M=4$, $N=3$, and peaks at locations $[0.1, 0.3, 0.6]$.}
	\label{fig:extreme_Ex2_sim_M4N3}
\end{figure*}
\begin{figure*}[!t]
	\centering
	\includegraphics[width=0.49\textwidth]{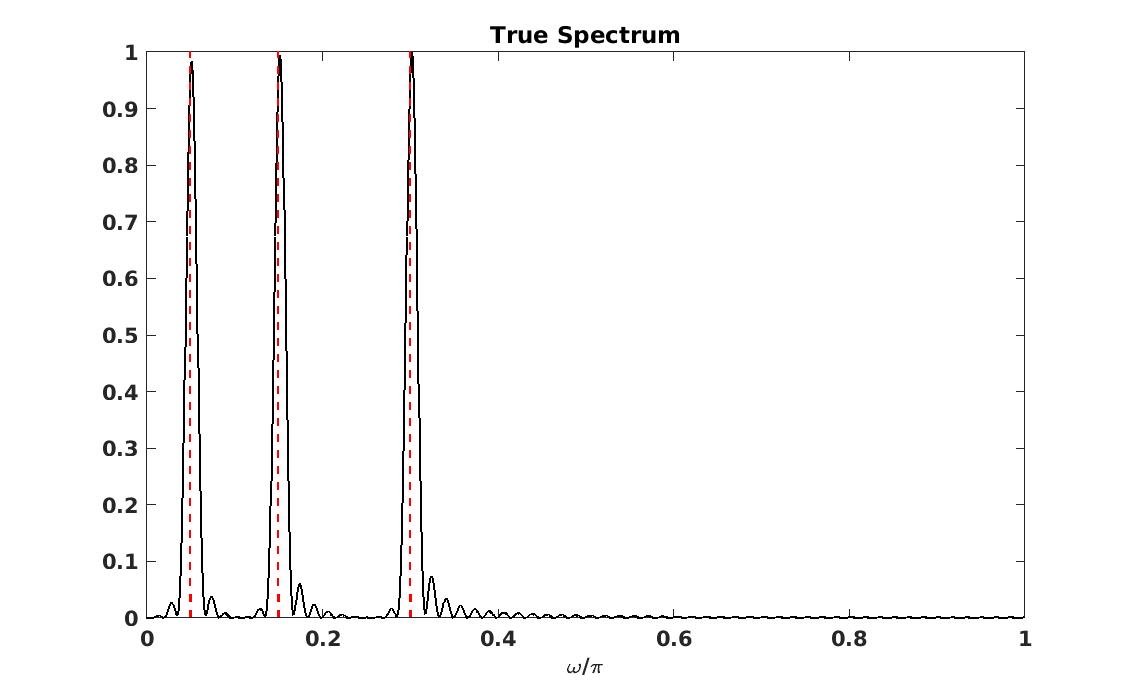}
	\includegraphics[width=0.49\textwidth]{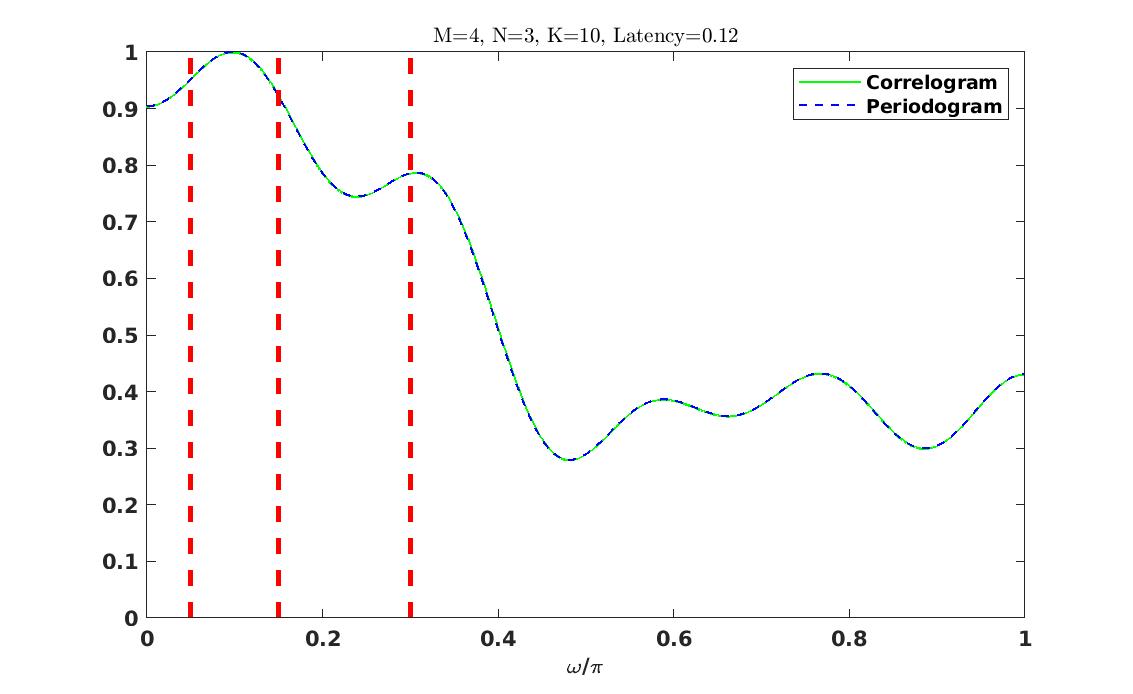}
	\caption{True spectrum (Nyquist) and prototype co-prime correlogram with $M=4$, $N=3$, and peaks at locations $[0.05, 0.15, 0.3]$.}
	\label{fig:extreme_Ex2_sim_true_spect_list2}
\end{figure*}
\begin{figure*}[!t]
	\centering
	\includegraphics[width=0.49\textwidth]{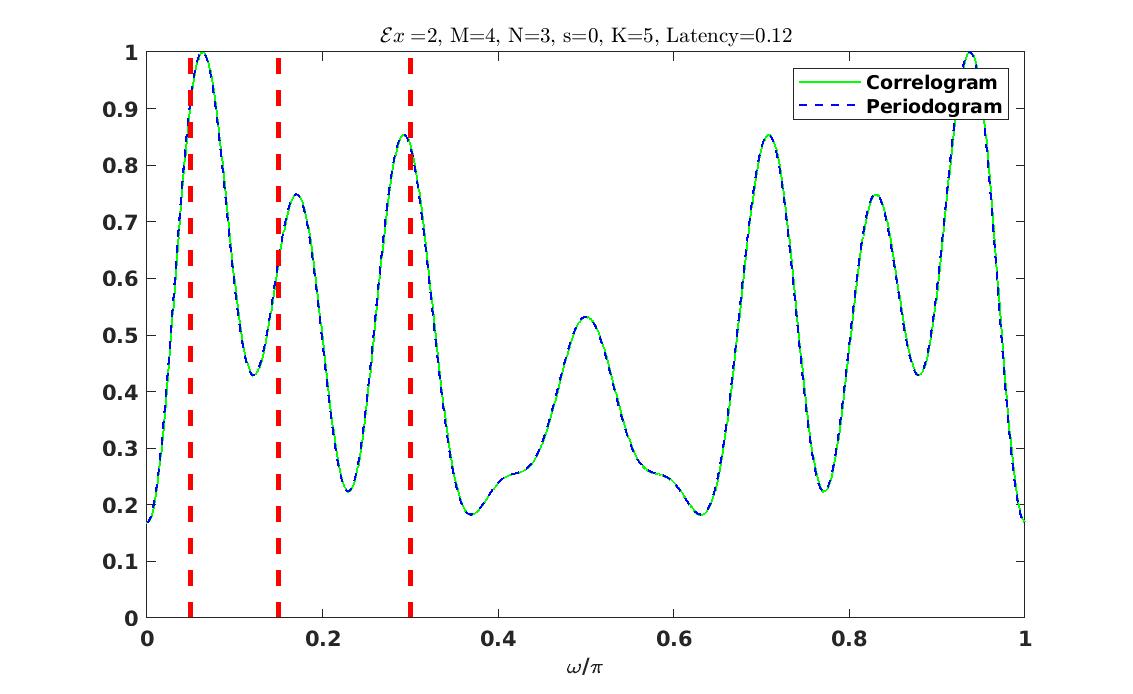}
	\includegraphics[width=0.49\textwidth]{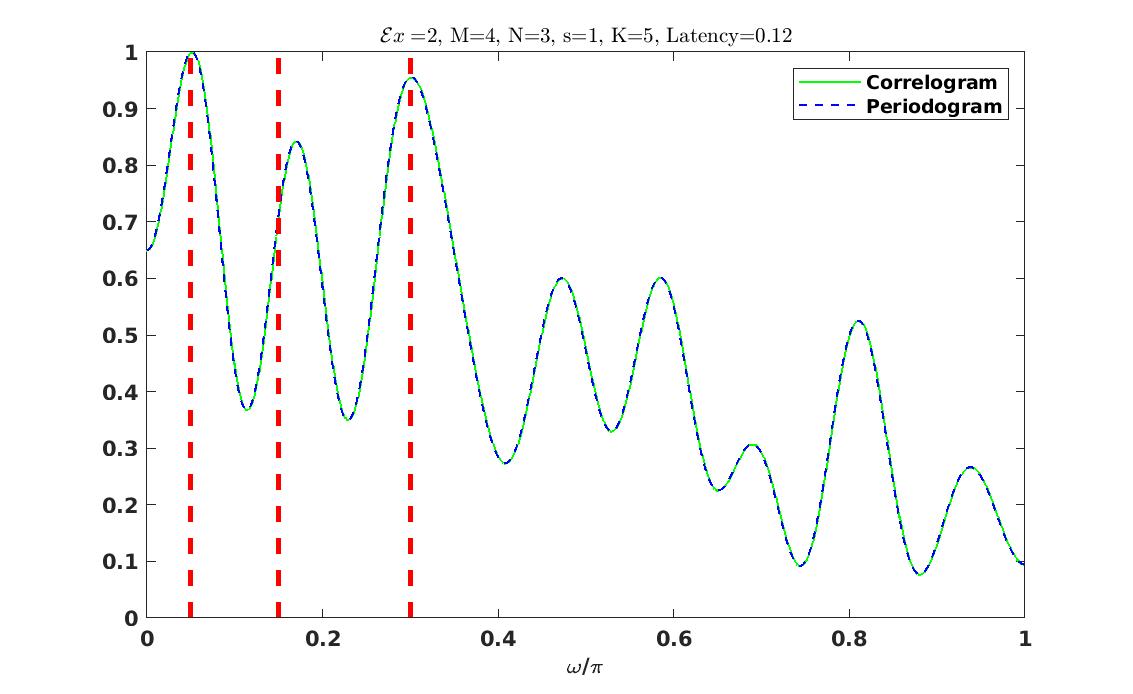}\\
	\includegraphics[width=0.5\textwidth]{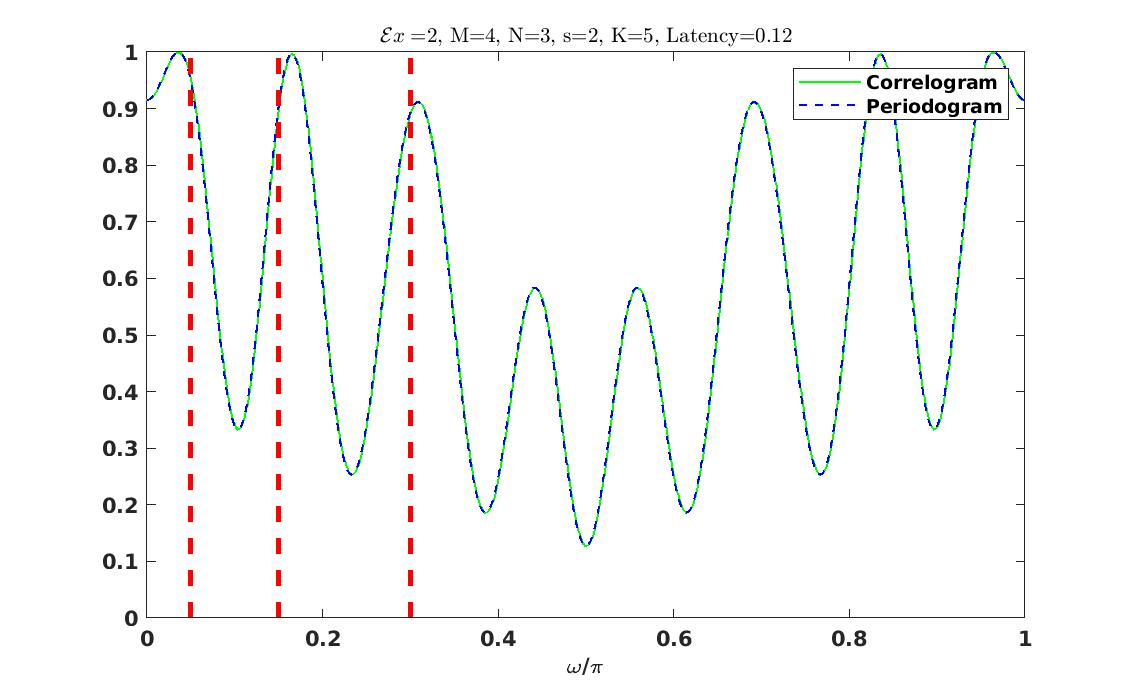}%
	\includegraphics[width=0.5\textwidth]{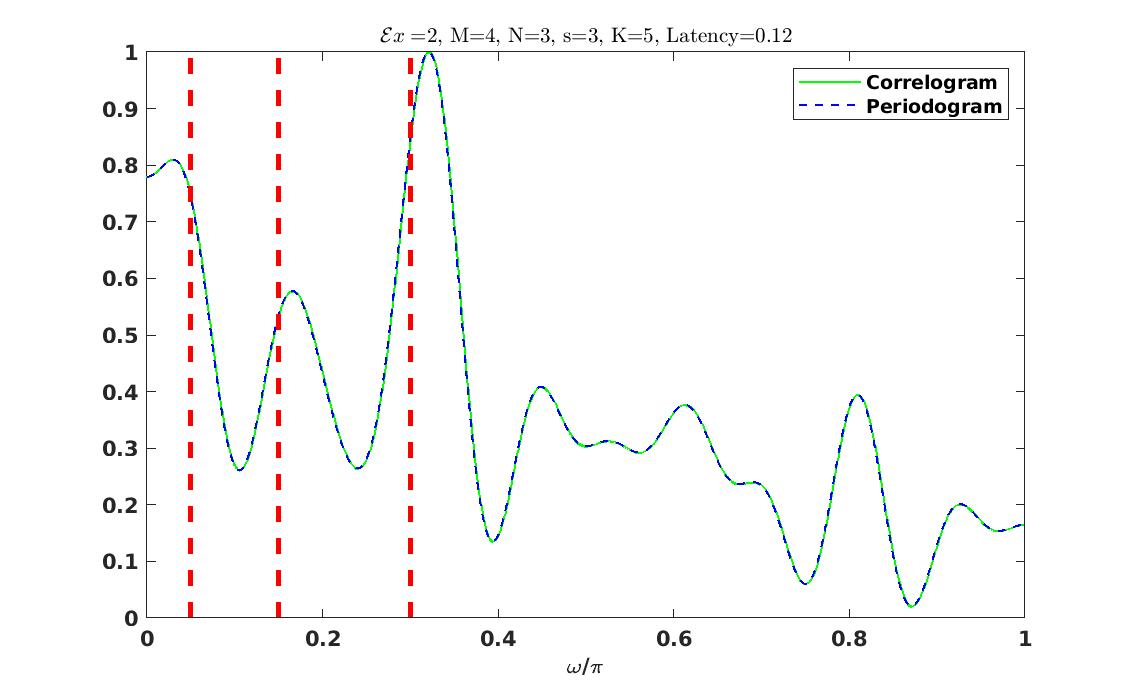}\\
	\includegraphics[width=0.5\textwidth]{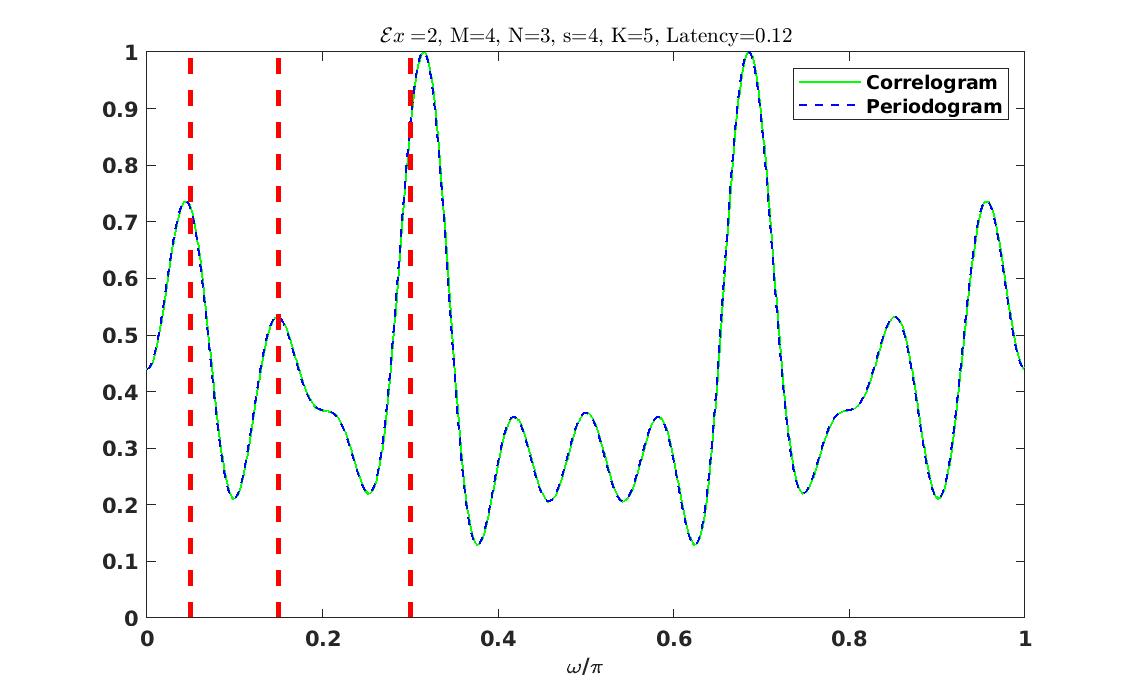}%
	\includegraphics[width=0.5\textwidth]{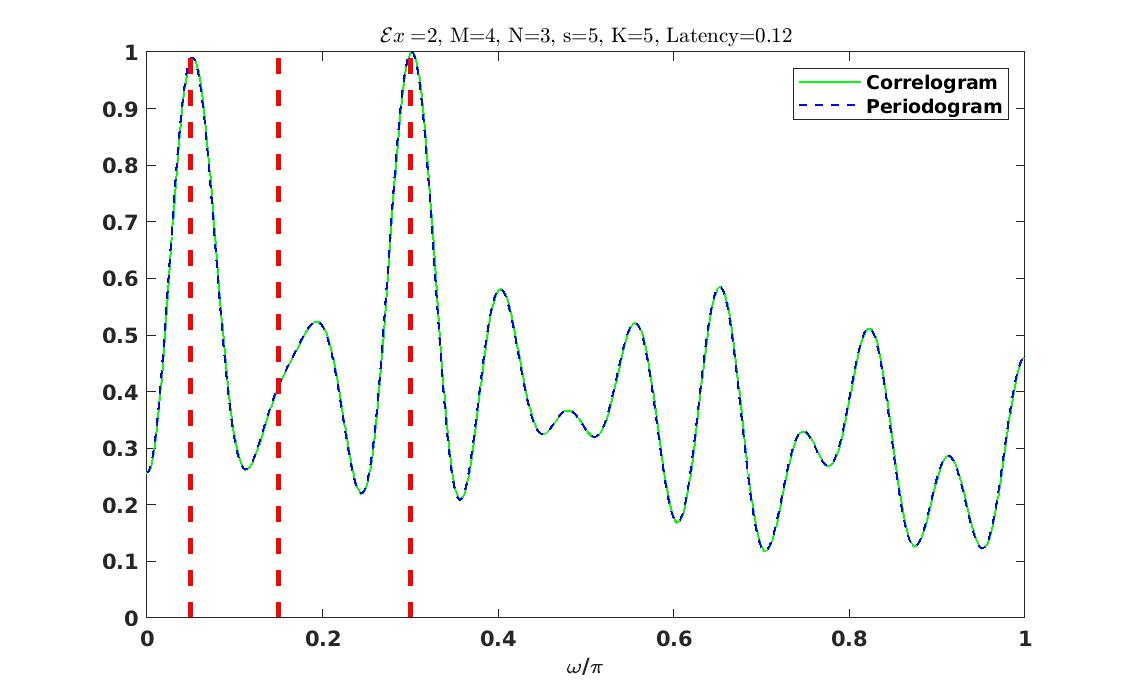}
	\caption{Correlogram spectral estimation for ExSCA with $\mathcal{E}_x=2$, $M=4$, $N=3$, and peaks at locations $[0.05, 0.15, 0.3]$.}
	\label{fig:extreme_Ex2_sim_M4N3_list2}
\end{figure*}
\begin{figure}[!t]
	\centering
	\includegraphics[width=0.49\textwidth]{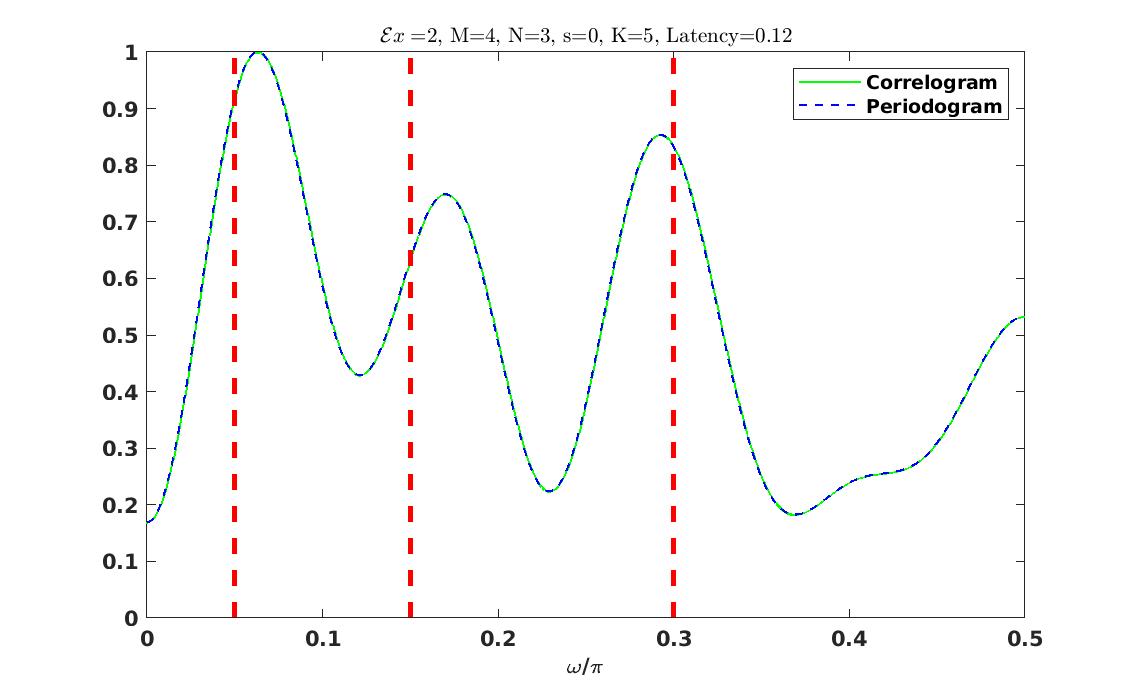}\\
	\includegraphics[width=0.49\textwidth]{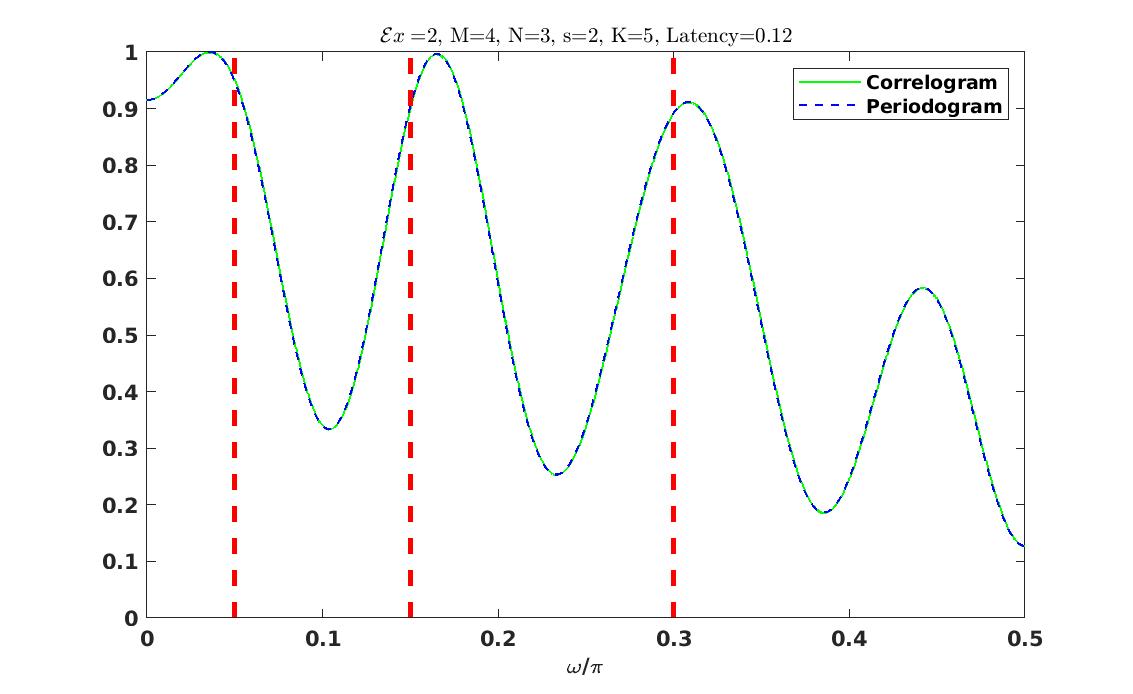}\\
	\includegraphics[width=0.5\textwidth]{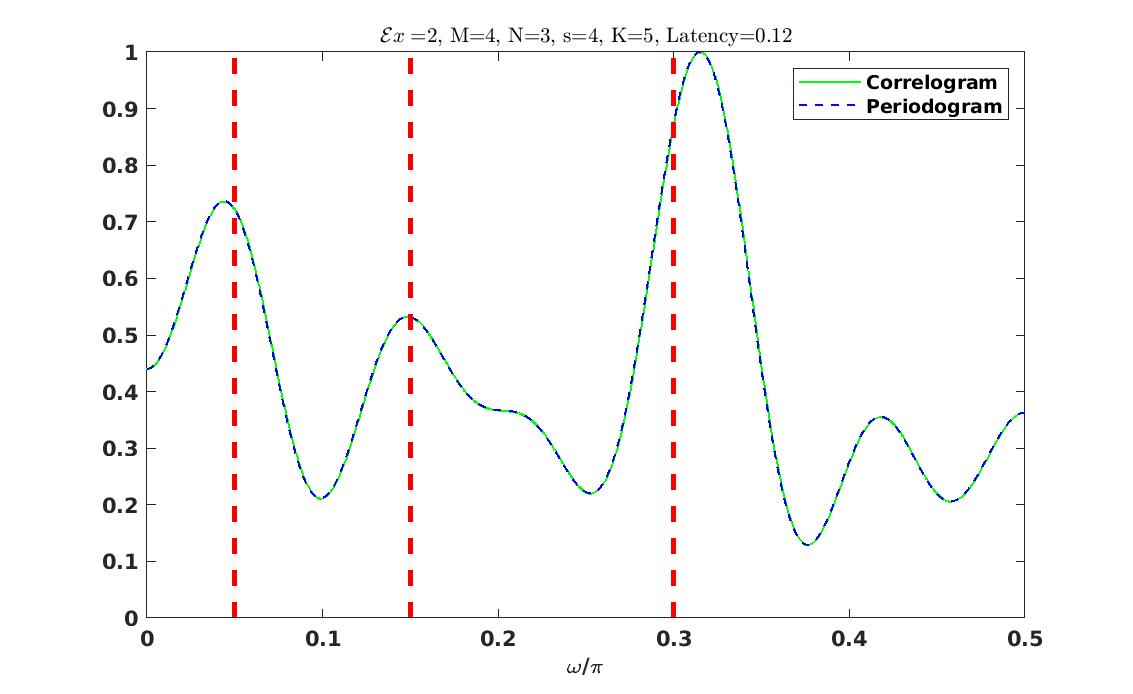}%
	\caption{Correlogram spectral estimation for ExSCA in Fig.~\ref{fig:extreme_Ex2_sim_M4N3_list2} with frequency in range $[0, 0.5]$ for $s=[0, 2, 4]$.}
	\label{fig:extreme_Ex2_sim_M4N3_list2zoom}
\end{figure}
\begin{figure}[!t]
	\centering
	\subfloat[1D ExSCA sampling function: $p^{1D}(k)$]{\includegraphics[width=0.5\textwidth]{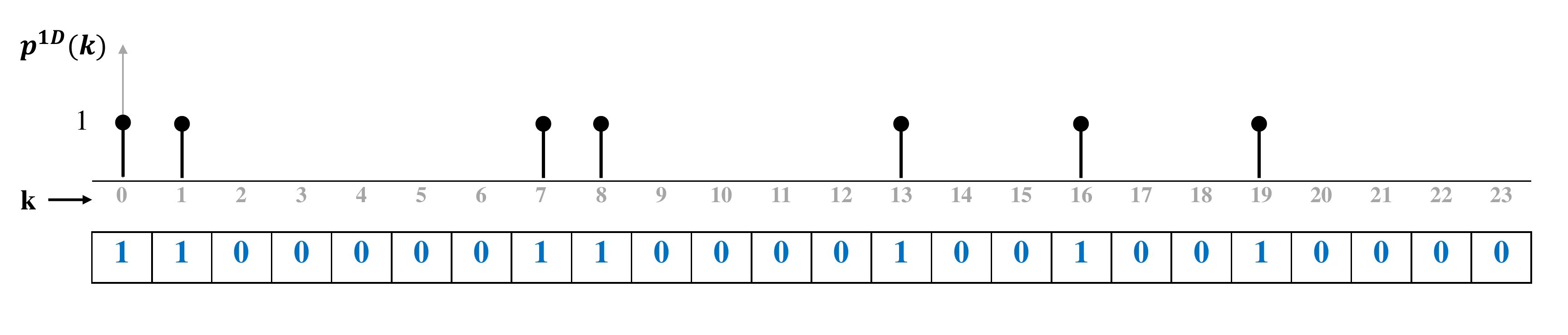}%
		\label{ExSCA_1d_pattern}}
	\hfil
	\subfloat[2D ExSCA sampling function: $p^{2D}(k)=p^{1D}(k_1)\otimes p^{1D}(k_2)$]{\includegraphics[width=0.5\textwidth]{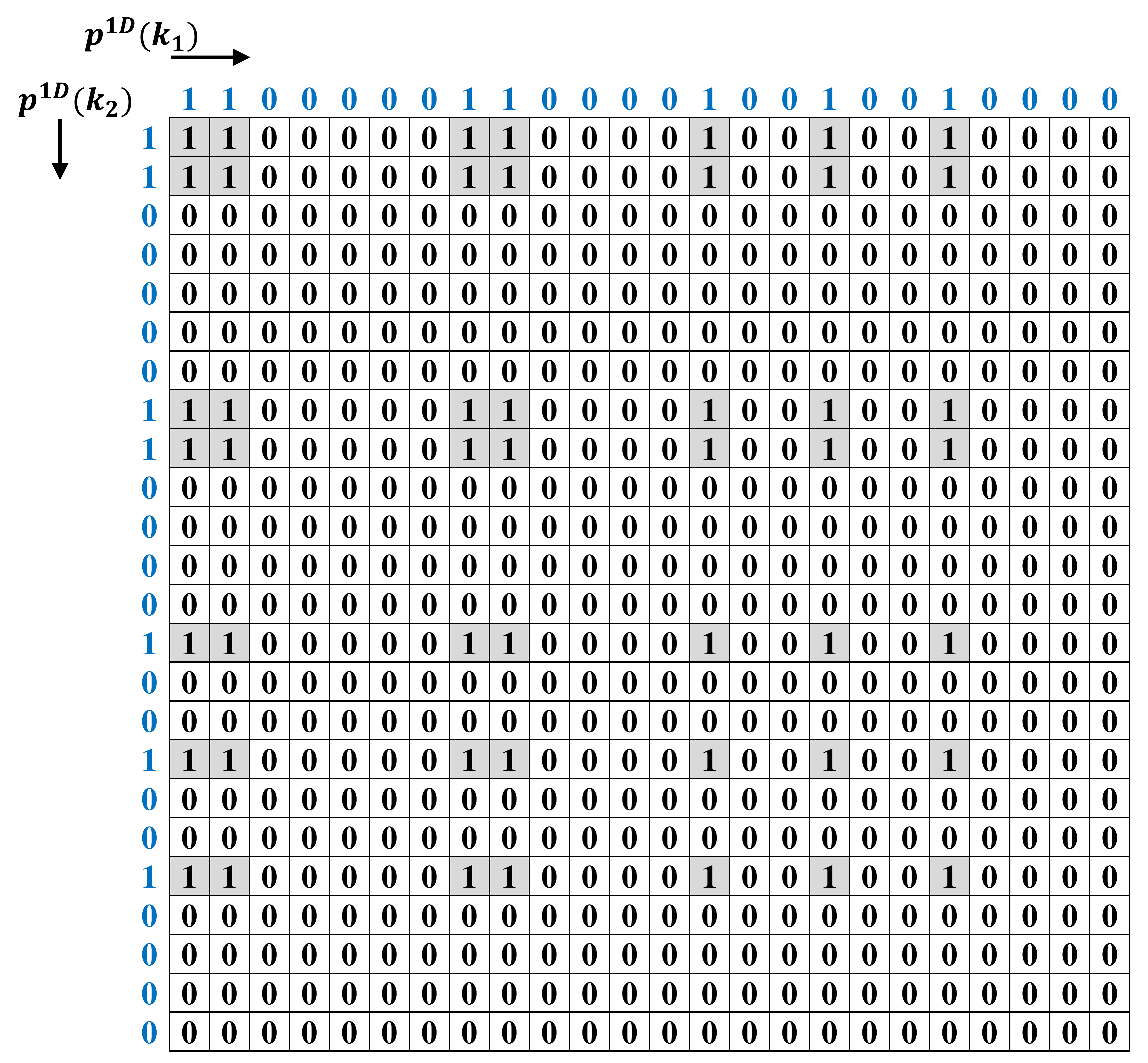}%
		\label{ExSCA_matrix}}
	\hfil
	\subfloat[2D Hybrid ExSCA sampling function: $p^{2D}(k)=p^{1D}(k_1)\otimes p^{1D}(k_2)$]{\includegraphics[width=0.5\textwidth]{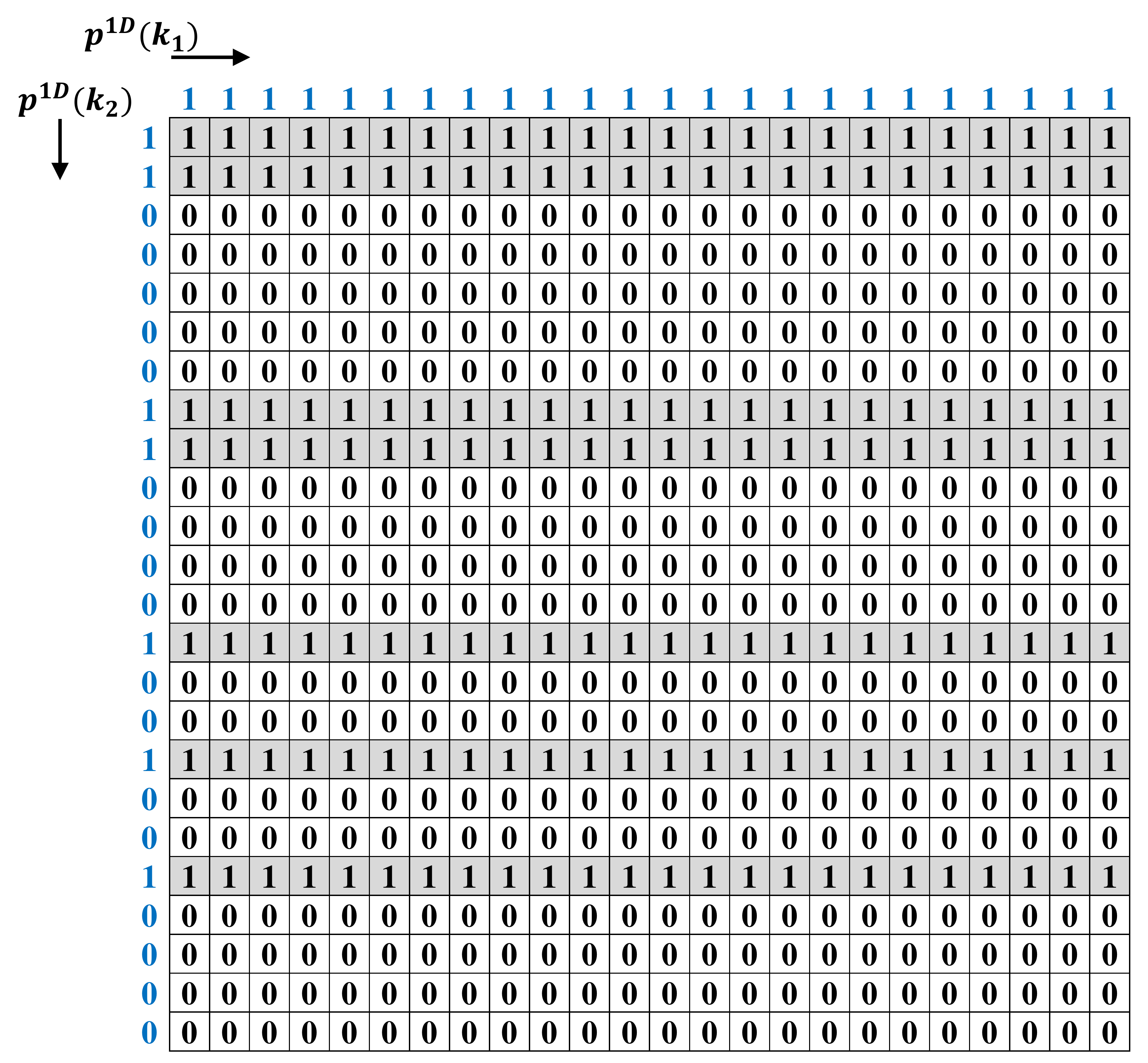}%
	\label{ExSCA_matrixP1P2}}	
	\caption{Example of ExSCA sampling function with $\mathcal{E}_x=2$, $s=1$, $M=4$, and $N=3$.}
	\label{fig:ExSCA_sampling_function}
\end{figure}
\begin{figure*}[!t]
	\centering
		\includegraphics[width=0.5\textwidth]{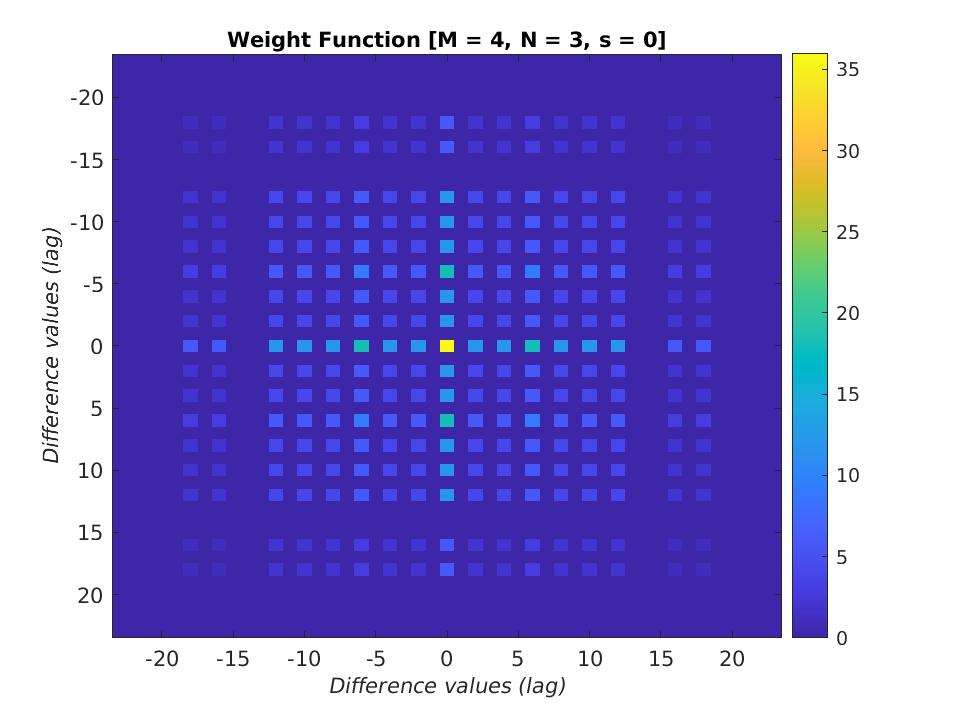}%
		\includegraphics[width=0.5\textwidth]{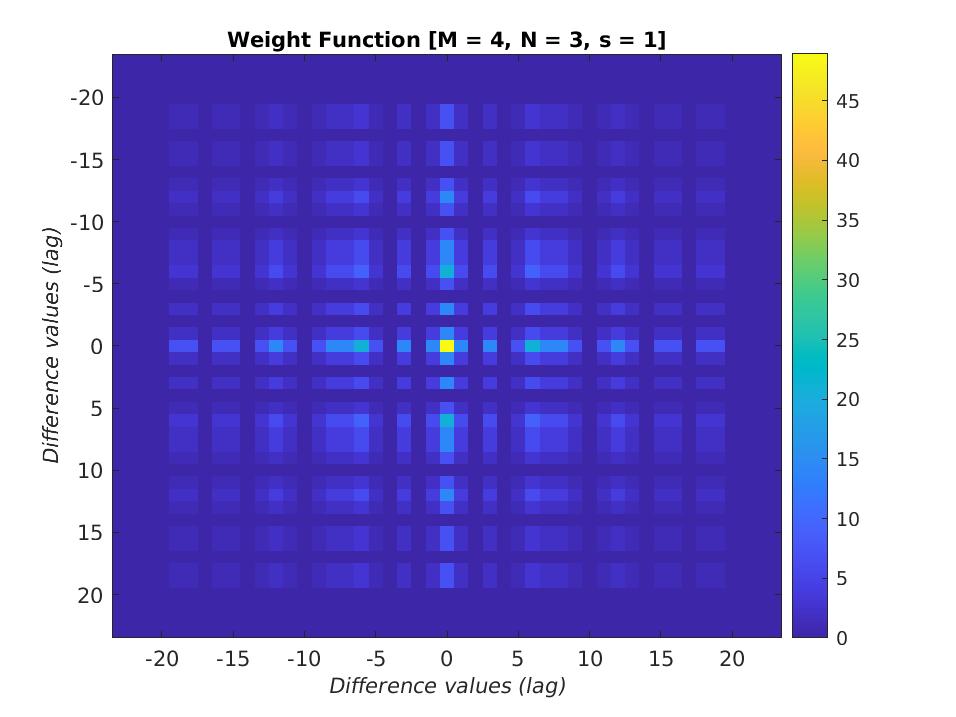}
		
		\includegraphics[width=0.5\textwidth]{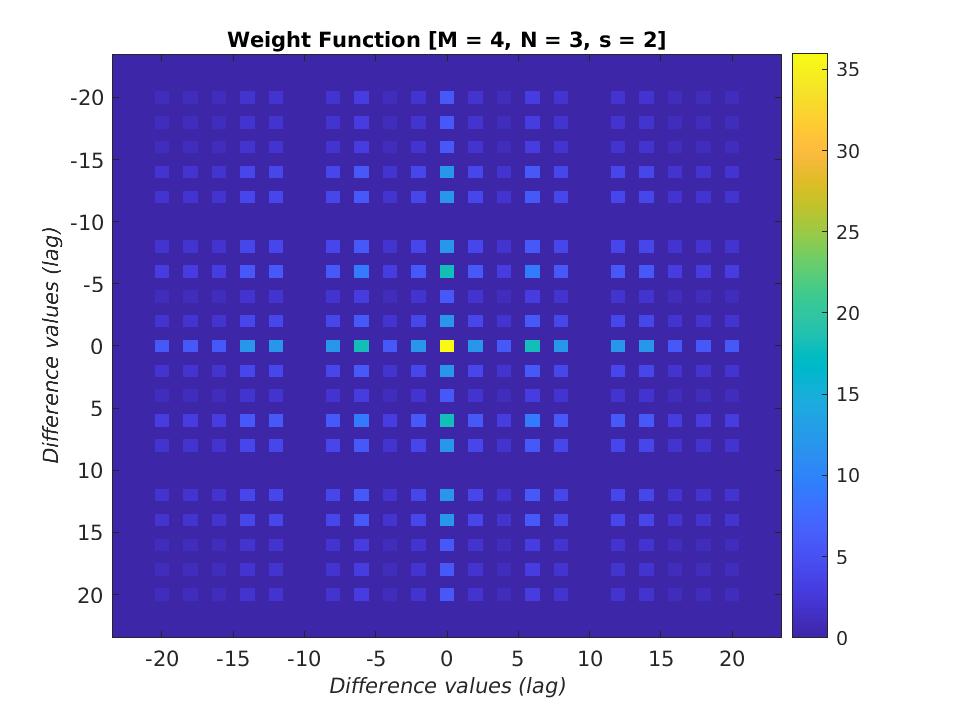}%
		\includegraphics[width=0.5\textwidth]{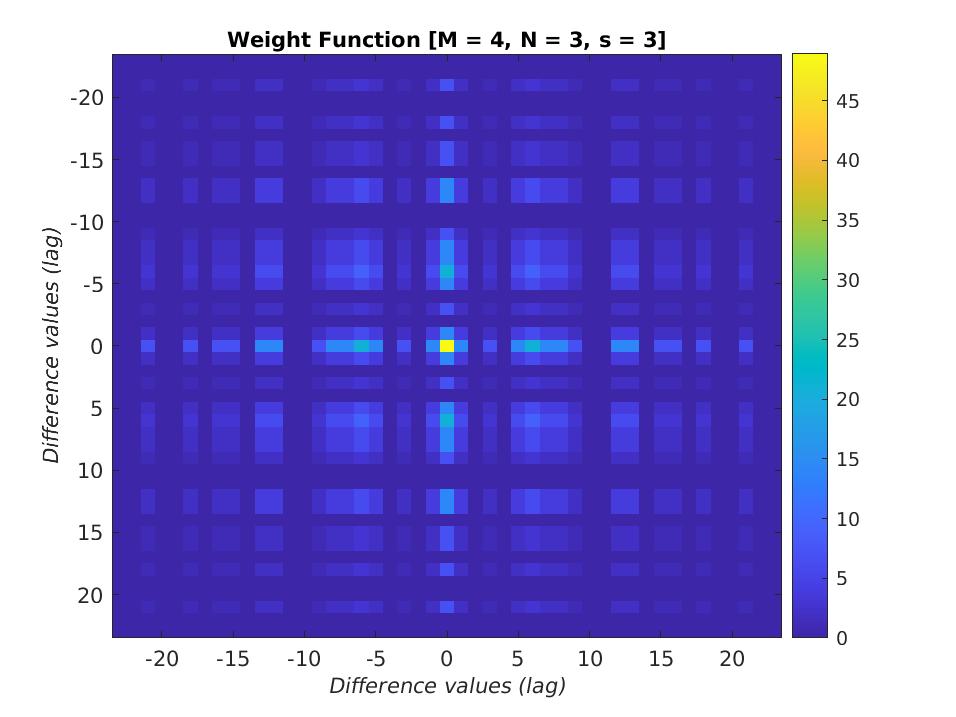}
		\includegraphics[width=0.5\textwidth]{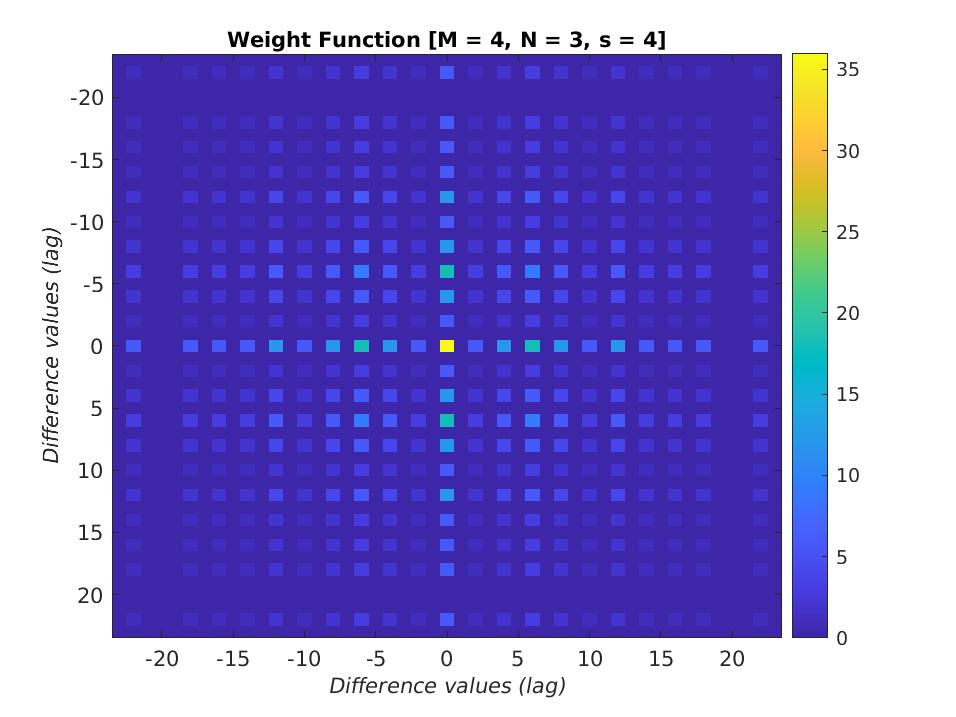}%
		\includegraphics[width=0.5\textwidth]{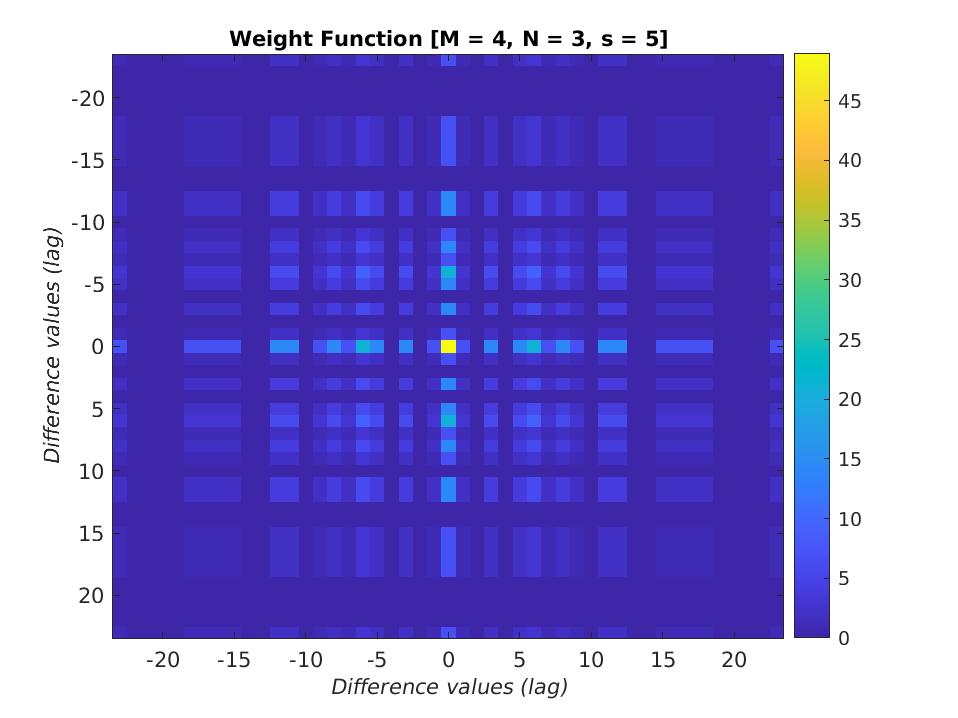}	
	\caption{Example of 2D weight function for ExSCA with $\mathcal{E}_x=2$, $M=4$, $N=3$.}
	\label{fig:2DEx2_weight_M4N3}
\end{figure*}

\begin{figure*}[!t]
	\centering
	\subfloat[$M=4$, $N=3$, $s=0$]{
		\includegraphics[width=0.5\textwidth]{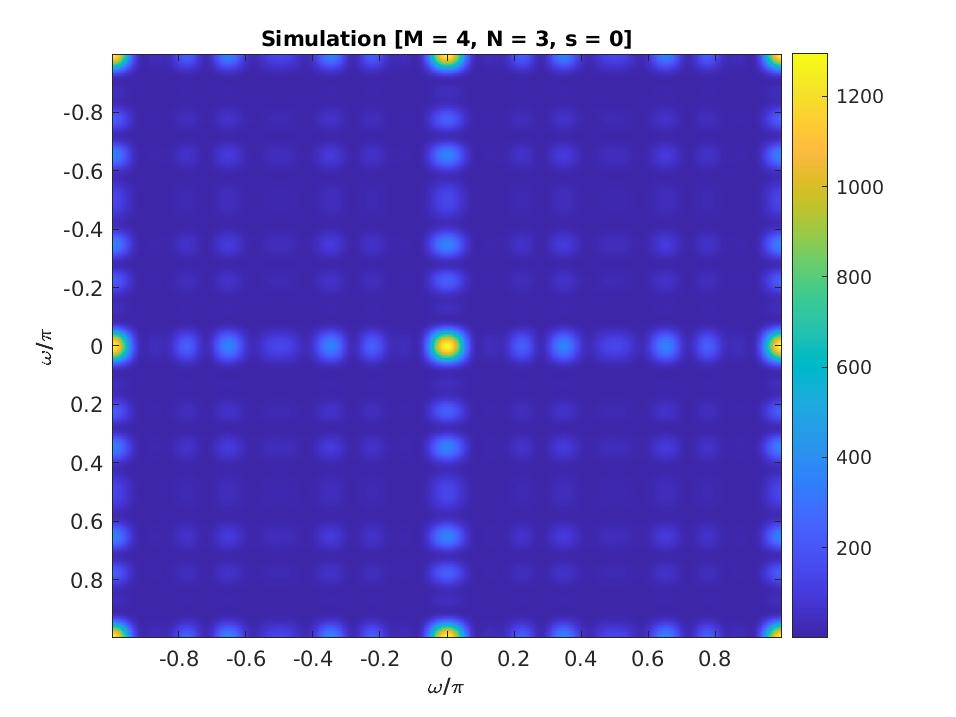}%
		\includegraphics[width=0.5\textwidth]{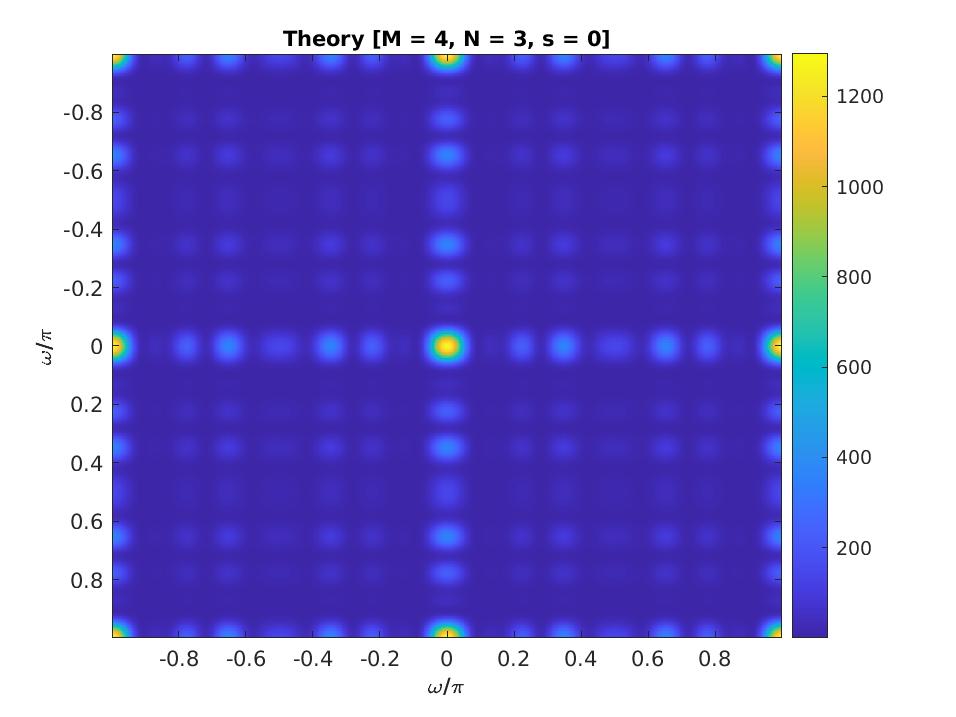}%
		\label{extreme_2DEx2_wt_bias_M4N3s0}}
	\hfil
	\subfloat[$M=4$, $N=3$, $s=1$]{
	\includegraphics[width=0.5\textwidth]{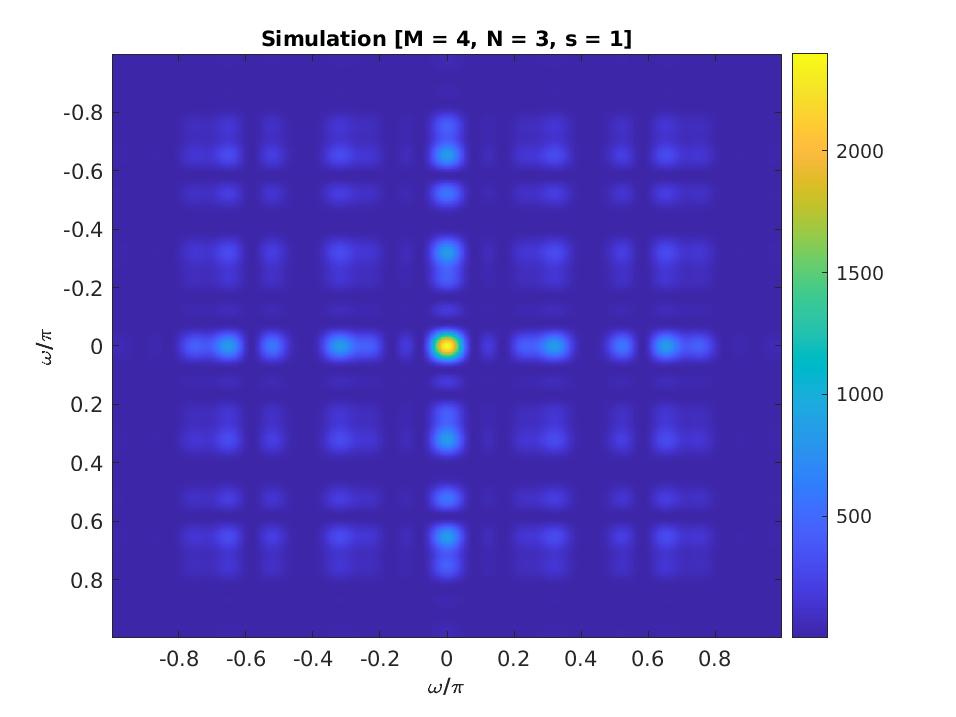}%
	\includegraphics[width=0.5\textwidth]{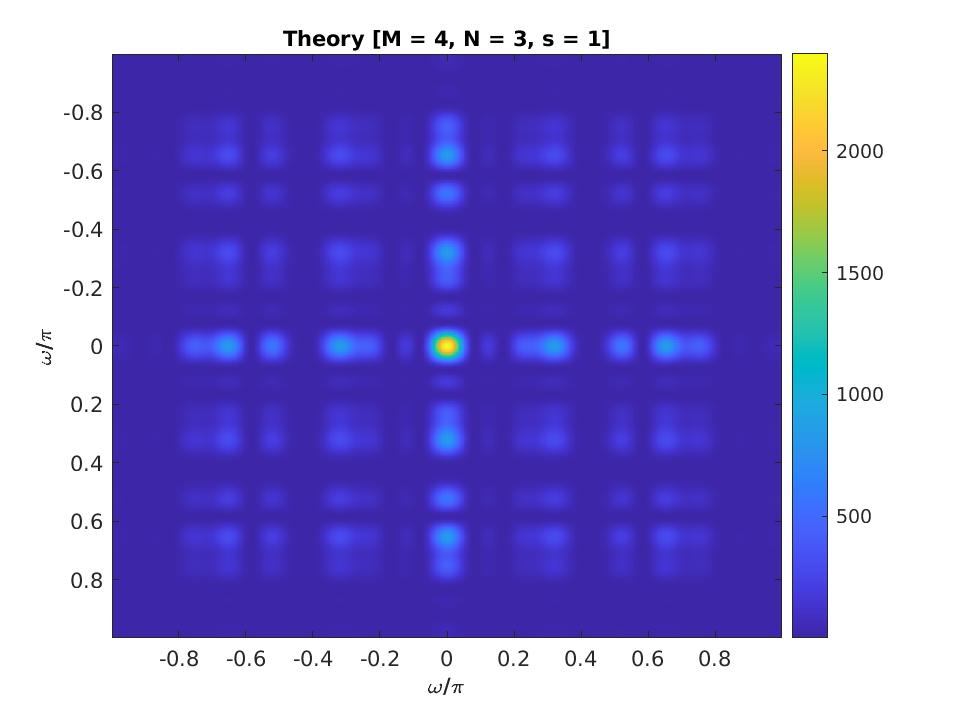}%
	\label{extreme_2DEx2_wt_bias_M4N3s1}}
\hfil
	\subfloat[$M=4$, $N=3$, $s=2$]{
	\includegraphics[width=0.5\textwidth]{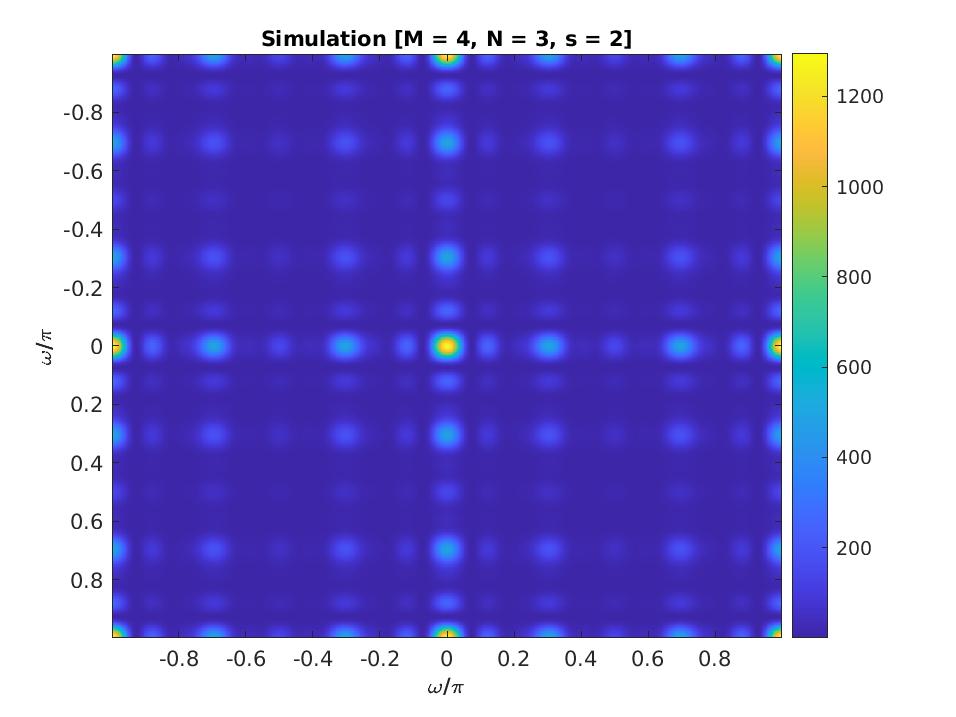}%
	\includegraphics[width=0.5\textwidth]{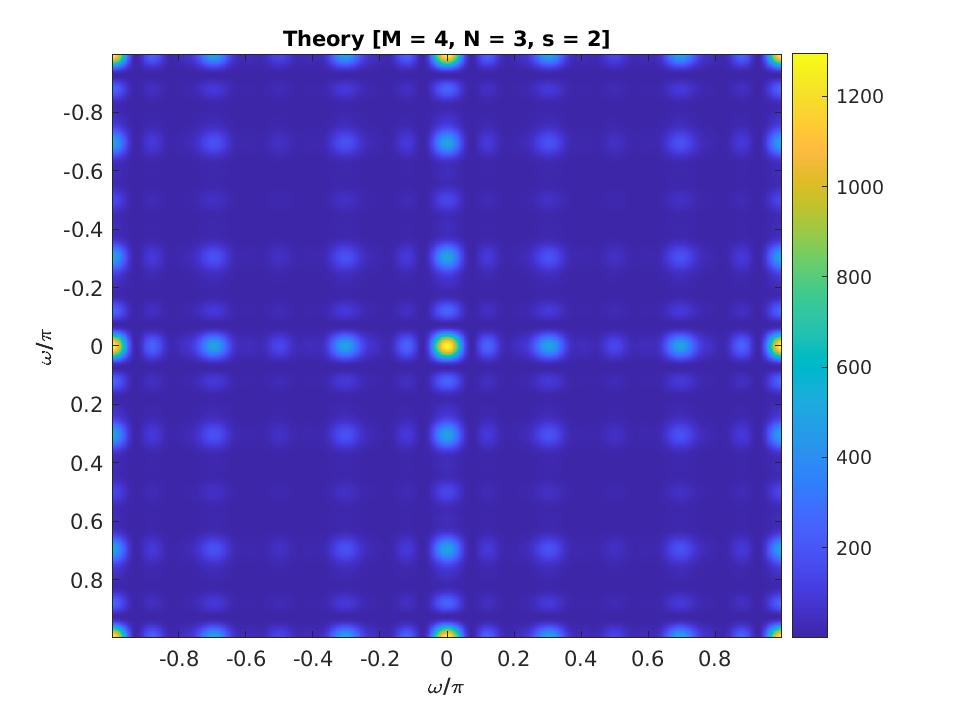}%
	\label{extreme_2DEx2_wt_bias_M4N3s2}}
	\caption{Simulated and theoretical 2D bias window of the correlogram estimator for ExSCA with $\mathcal{E}_x=2$, $M=4$, $N=3$, $s=[0, 2]$.}
\label{fig:extreme_2DEx2_wt_bias_M4N3s02}
\end{figure*}

\begin{figure*}[!t]
	\centering
	\subfloat[$M=4$, $N=3$, $s=3$]{
	\includegraphics[width=0.5\textwidth]{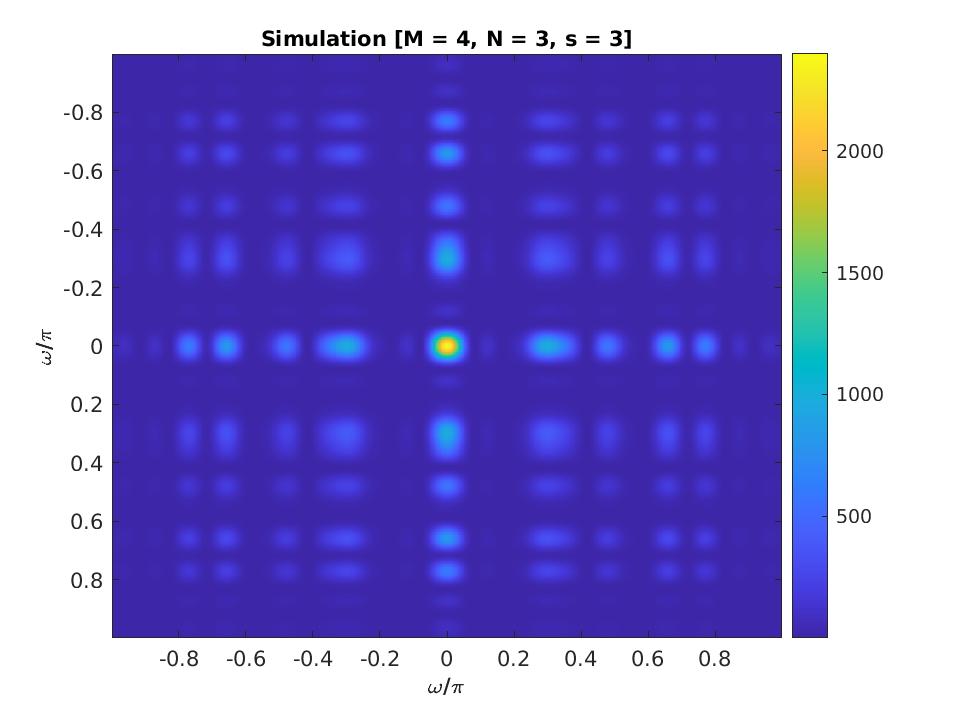}%
	\includegraphics[width=0.5\textwidth]{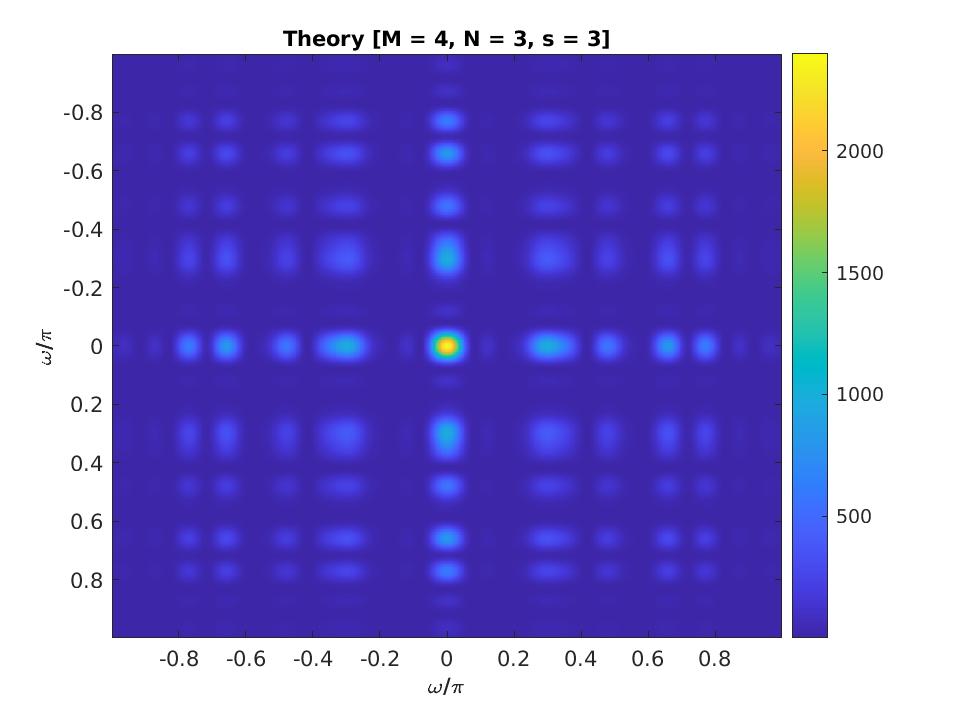}%
	\label{extreme_2DEx2_wt_bias_M4N3s3}}
\hfil
	\subfloat[$M=4$, $N=3$, $s=4$]{
	\includegraphics[width=0.5\textwidth]{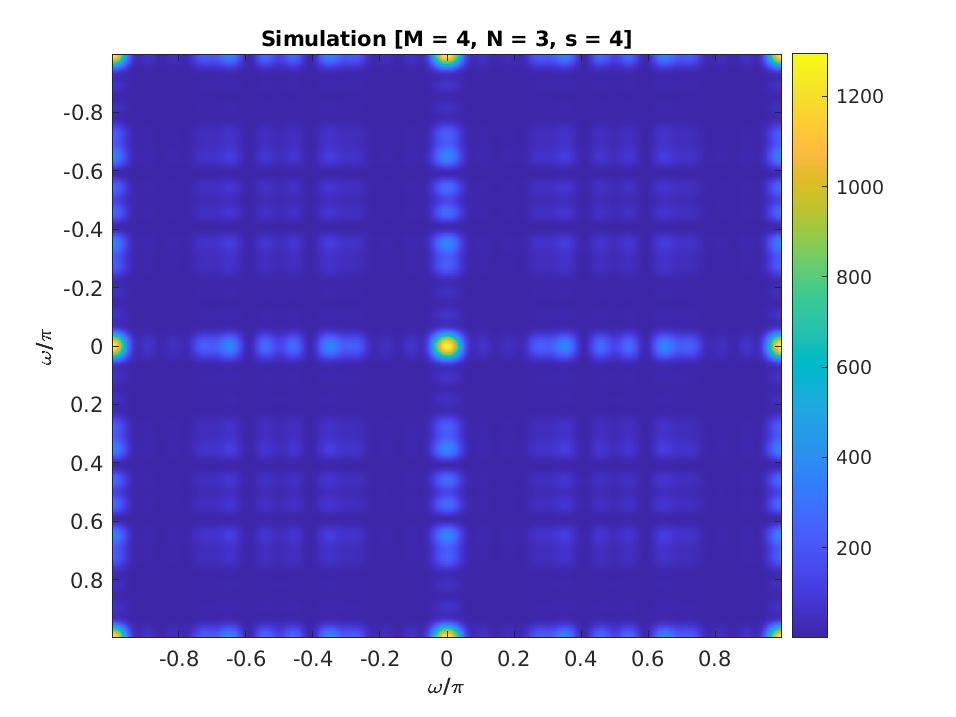}%
	\includegraphics[width=0.5\textwidth]{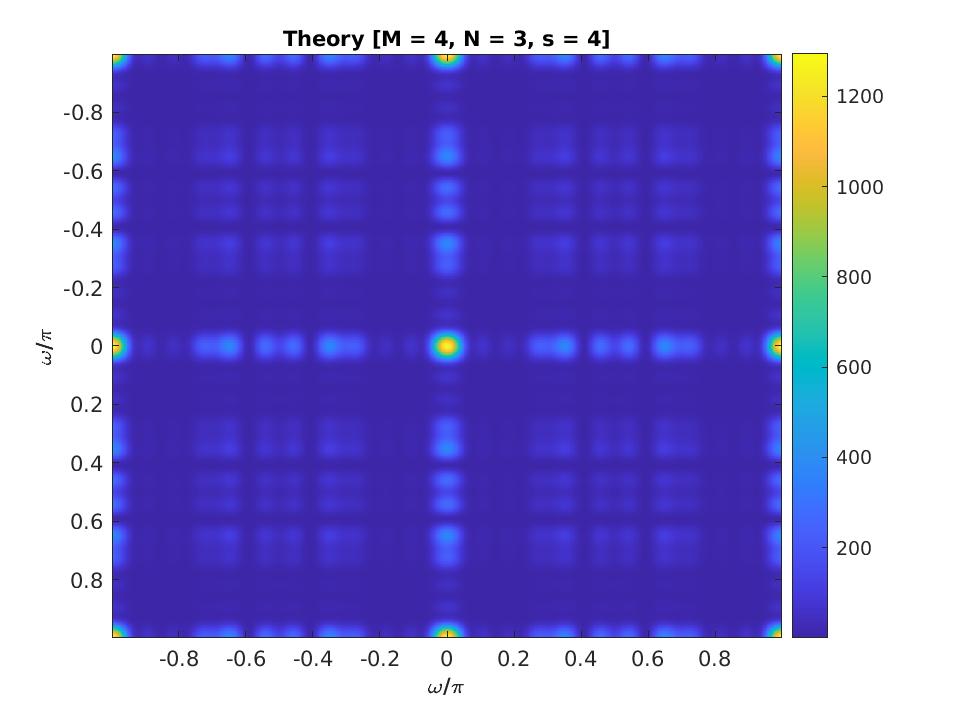}%
	\label{extreme_2DEx2_wt_bias_M4N3s4}}
\hfil
	\subfloat[$M=4$, $N=3$, $s=5$]{
	\includegraphics[width=0.5\textwidth]{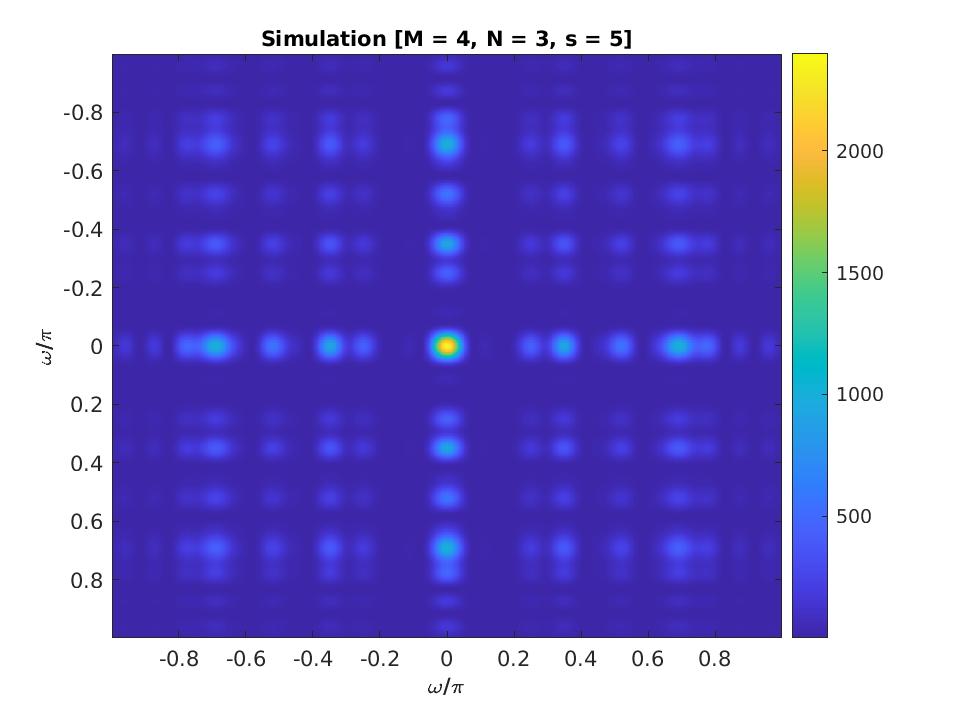}%
	\includegraphics[width=0.5\textwidth]{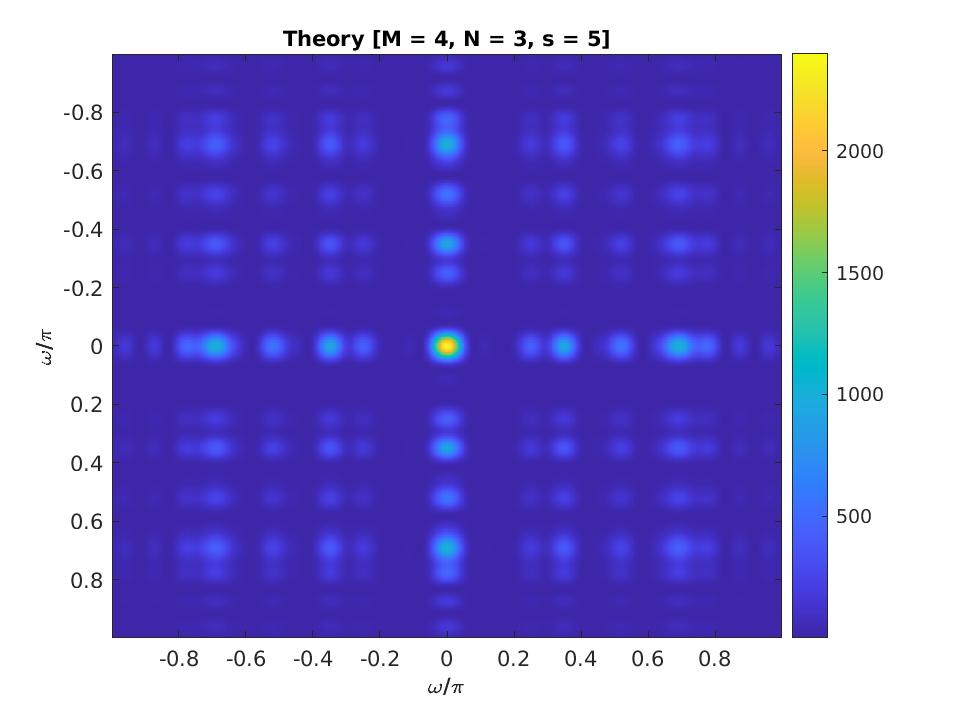}%
	\label{extreme_2DEx2_wt_bias_M4N3s5}}
\hfil
	\caption{Simulated and theoretical 2D bias window of the correlogram estimator for ExSCA with $\mathcal{E}_x=2$, $M=4$, $N=3$, $s=[3, 5]]$.}
\label{fig:extreme_2DEx2_wt_bias_M4N3s35}
\end{figure*}%
\begin{figure*}[!t]
	\centering
	\subfloat[$M=4$, $N=3$, $s=0$]{
		\includegraphics[width=0.5\textwidth]{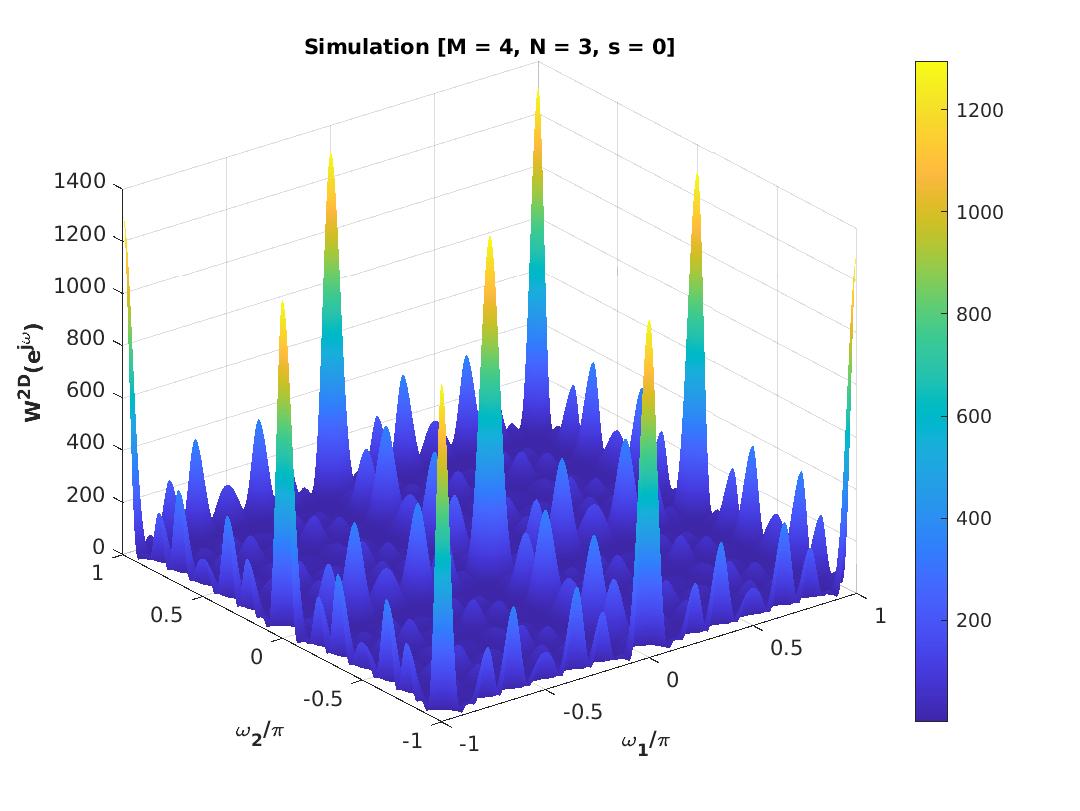}%
		\includegraphics[width=0.5\textwidth]{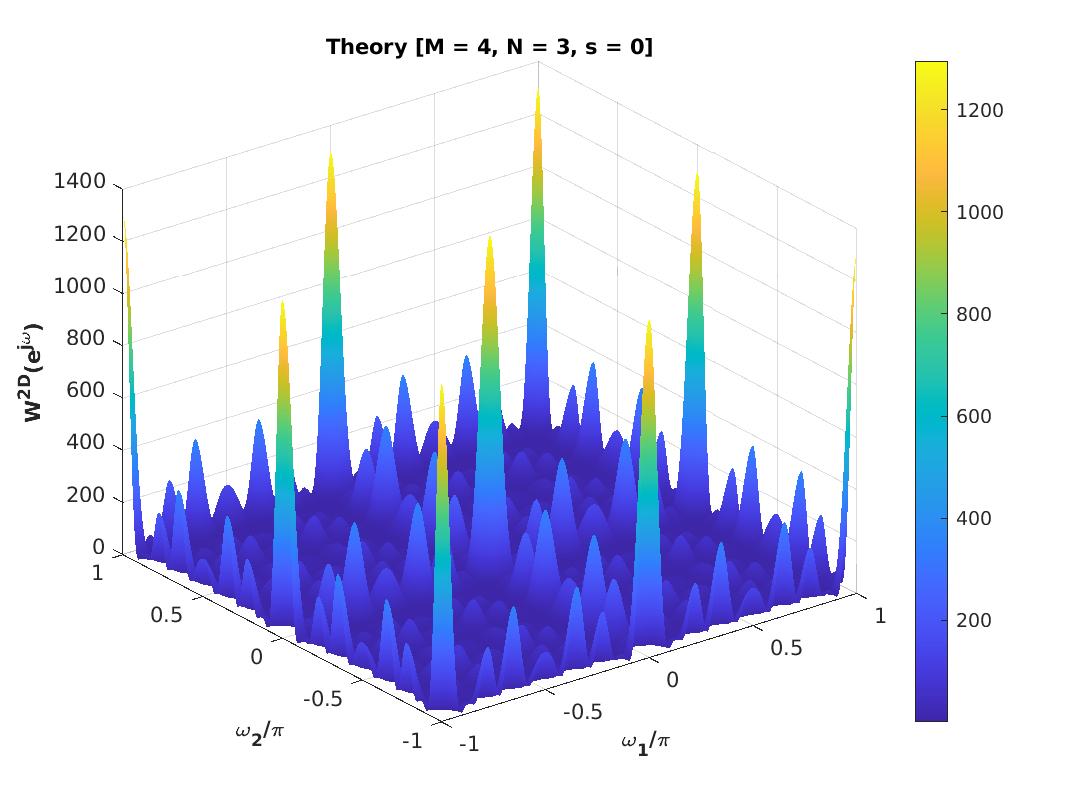}%
		\label{extreme_2DEx2_wt_bias_M4N3s0Surf}}
	\hfil
	\subfloat[$M=4$, $N=3$, $s=1$]{
		\includegraphics[width=0.5\textwidth]{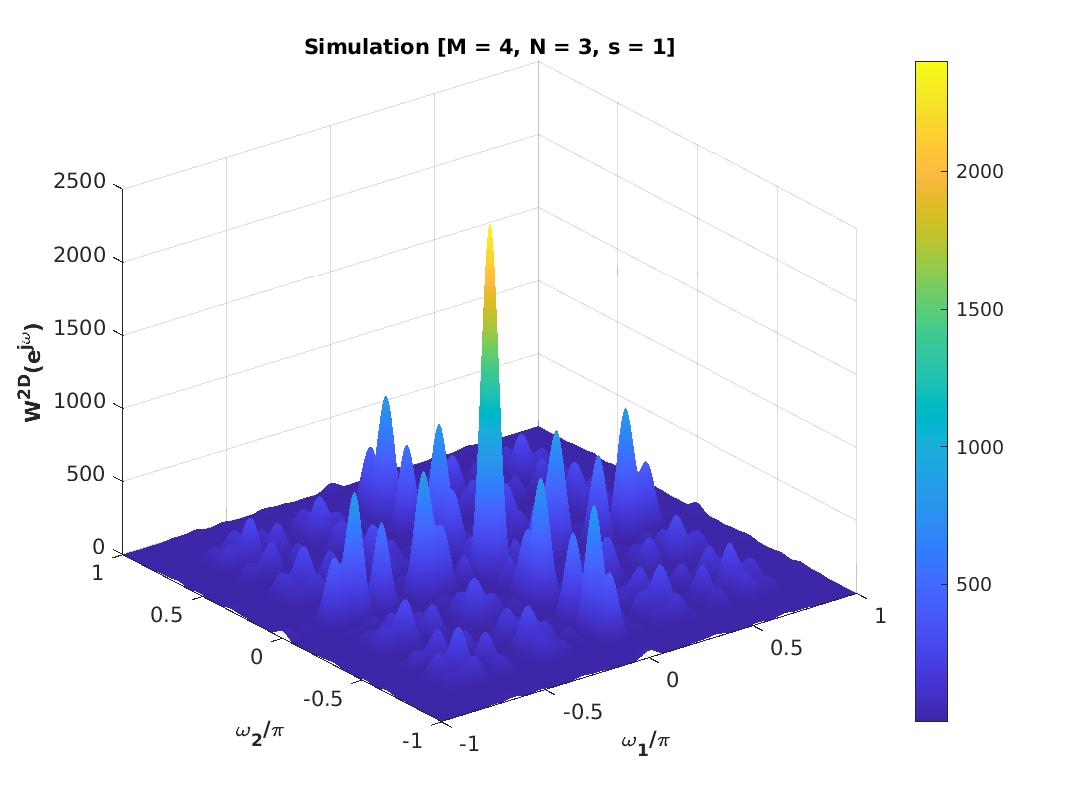}%
		\includegraphics[width=0.5\textwidth]{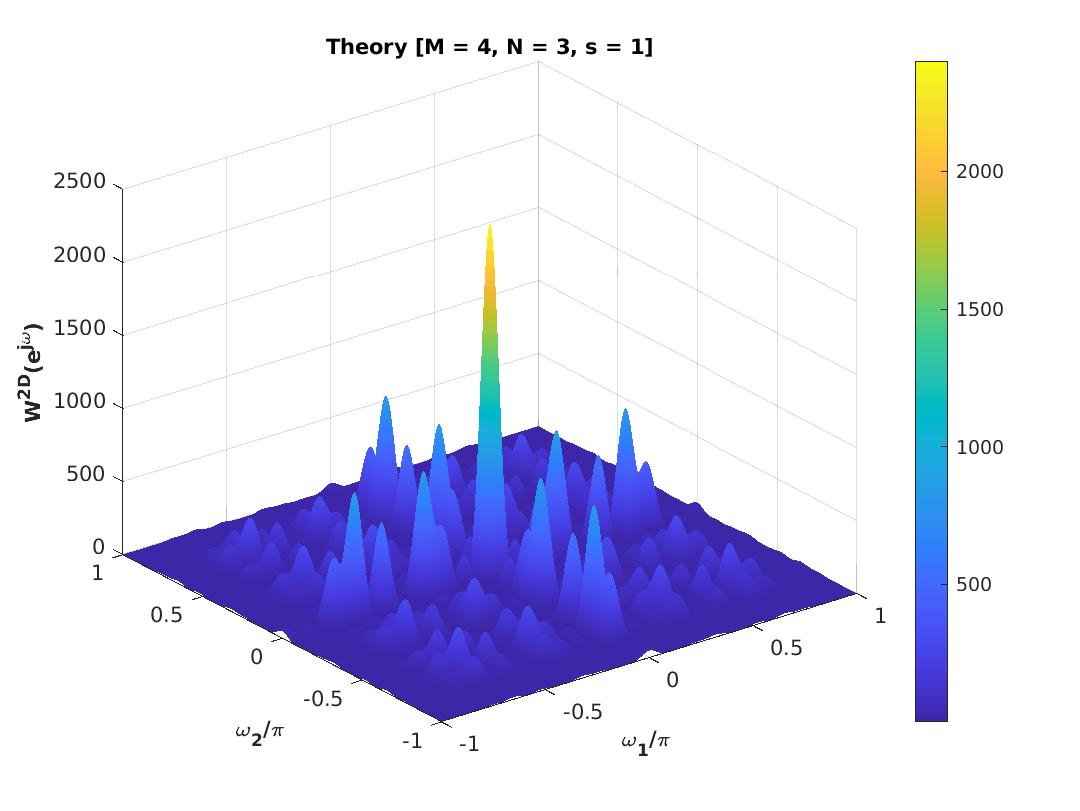}%
		\label{extreme_2DEx2_wt_bias_M4N3s1Surf}}
	\hfil
	\subfloat[$M=4$, $N=3$, $s=2$]{
		\includegraphics[width=0.5\textwidth]{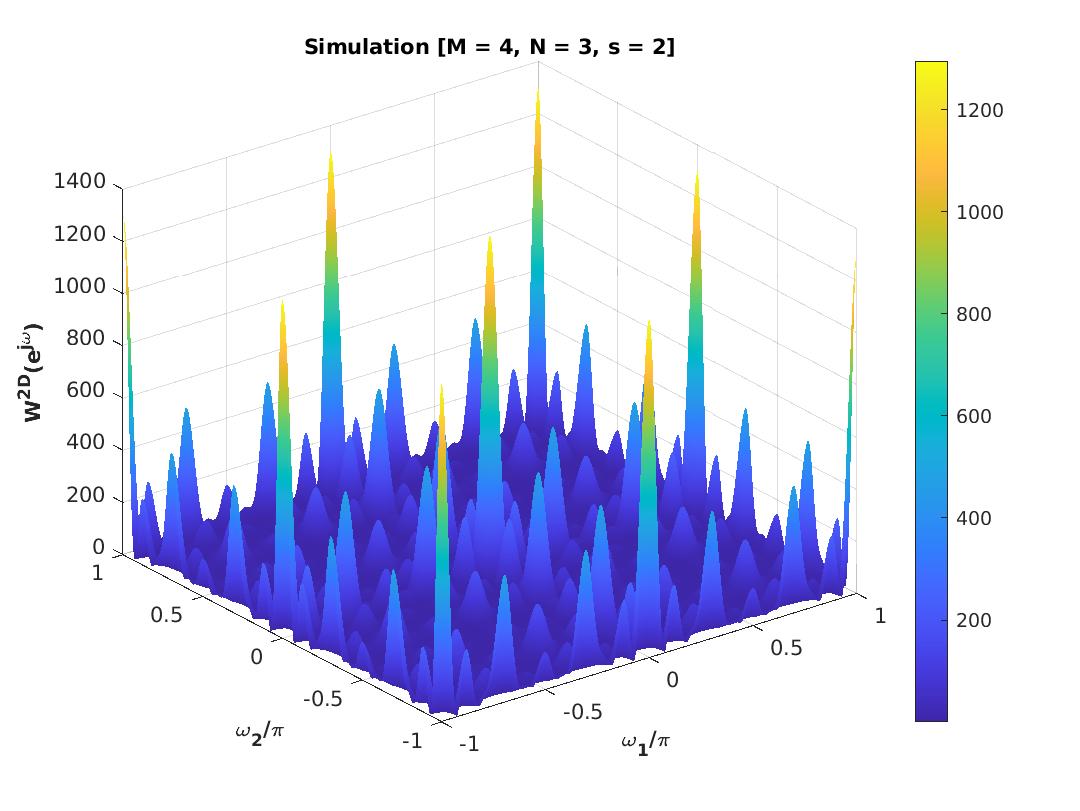}%
		\includegraphics[width=0.5\textwidth]{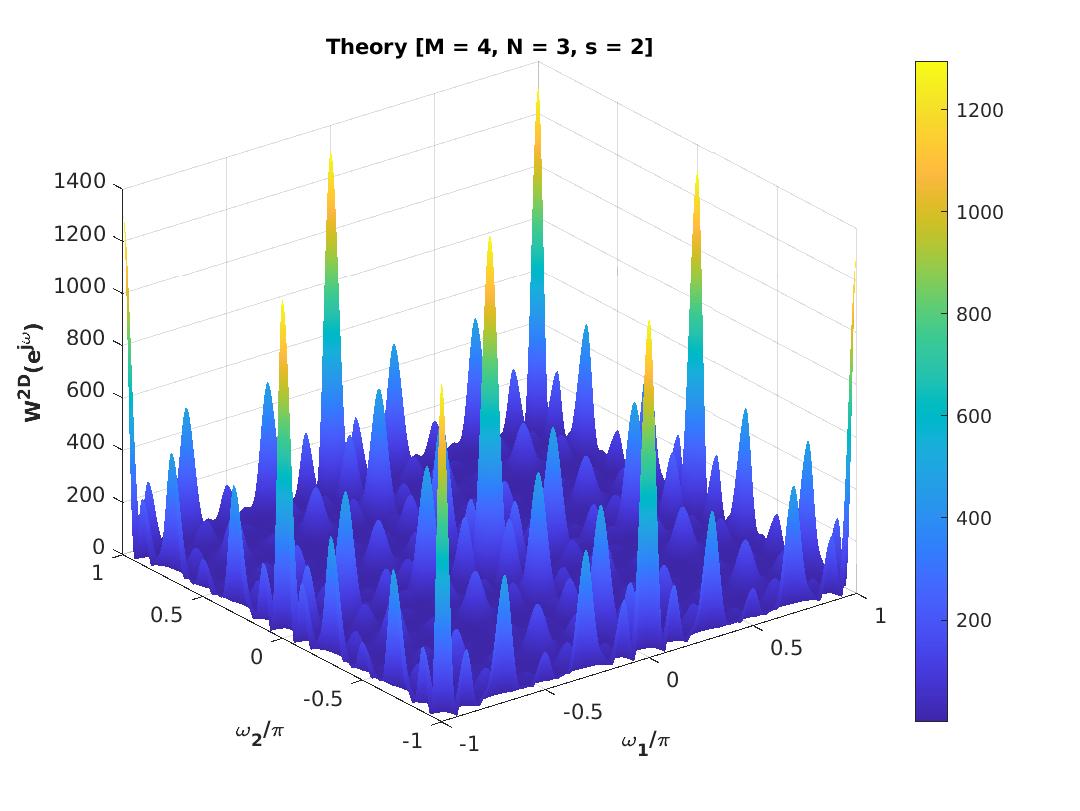}%
		\label{extreme_2DEx2_wt_bias_M4N3s2Surf}}
	\caption{Simulated and theoretical 2D bias window: Surface plot for Fig.~\ref{fig:extreme_2DEx2_wt_bias_M4N3s02}.}
	\label{fig:extreme_2DEx2_wt_bias_M4N3s02Surf}
\end{figure*}

\begin{figure*}[!t]
	\centering
	\subfloat[$M=4$, $N=3$, $s=3$]{
		\includegraphics[width=0.5\textwidth]{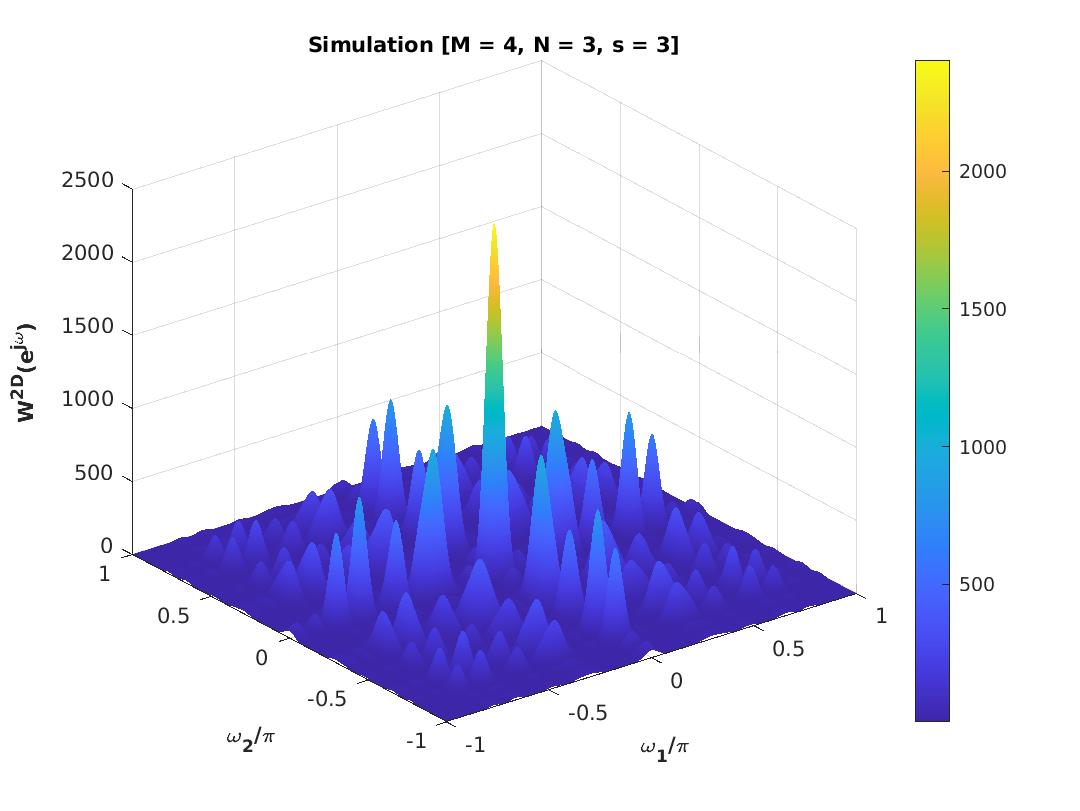}%
		\includegraphics[width=0.5\textwidth]{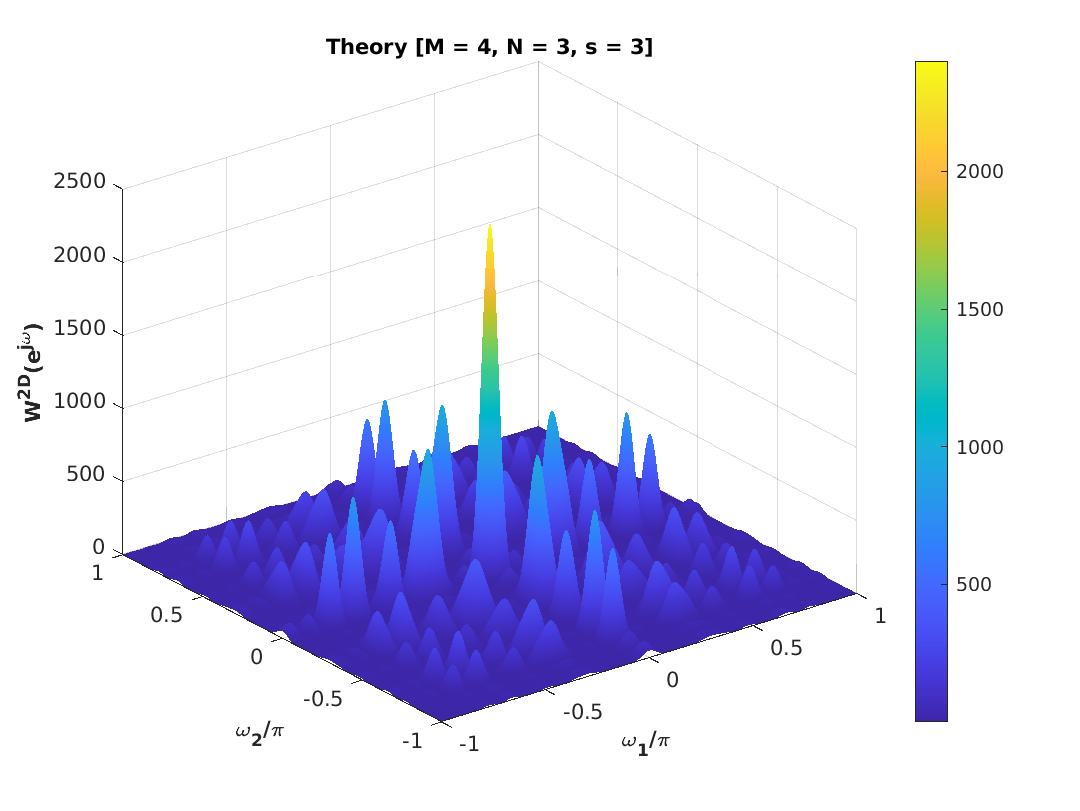}%
		\label{extreme_2DEx2_wt_bias_M4N3s3Surf}}
	\hfil
	\subfloat[$M=4$, $N=3$, $s=4$]{
		\includegraphics[width=0.5\textwidth]{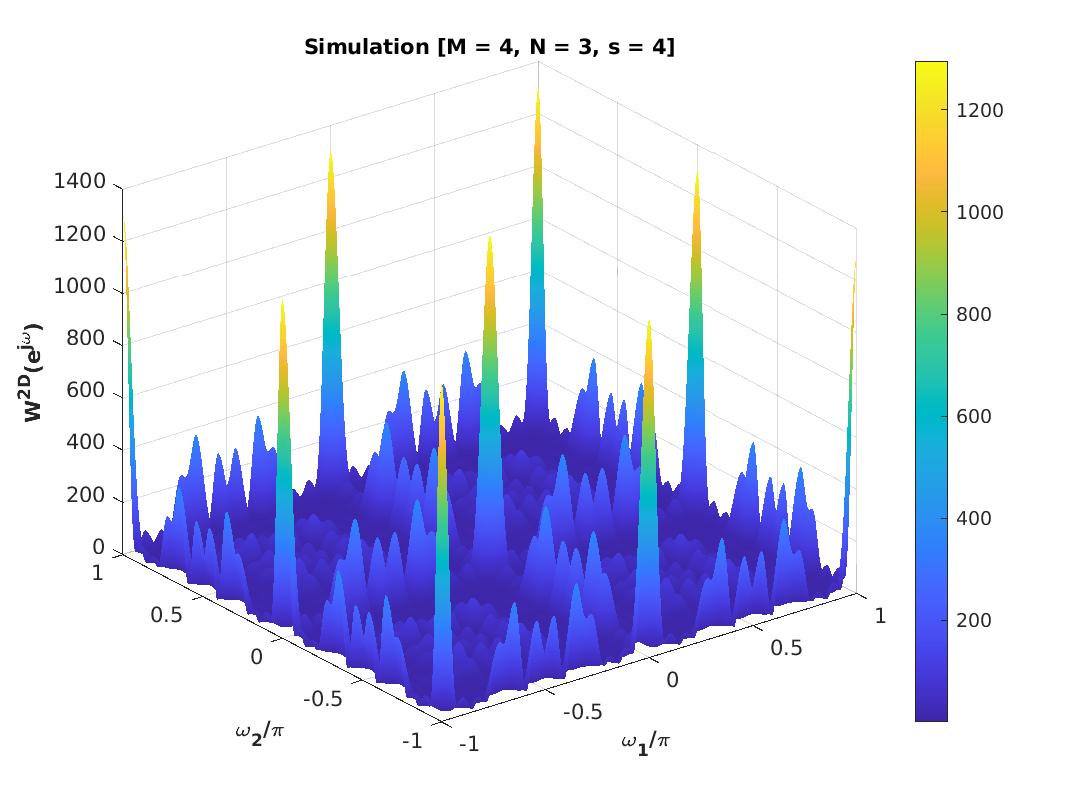}%
		\includegraphics[width=0.5\textwidth]{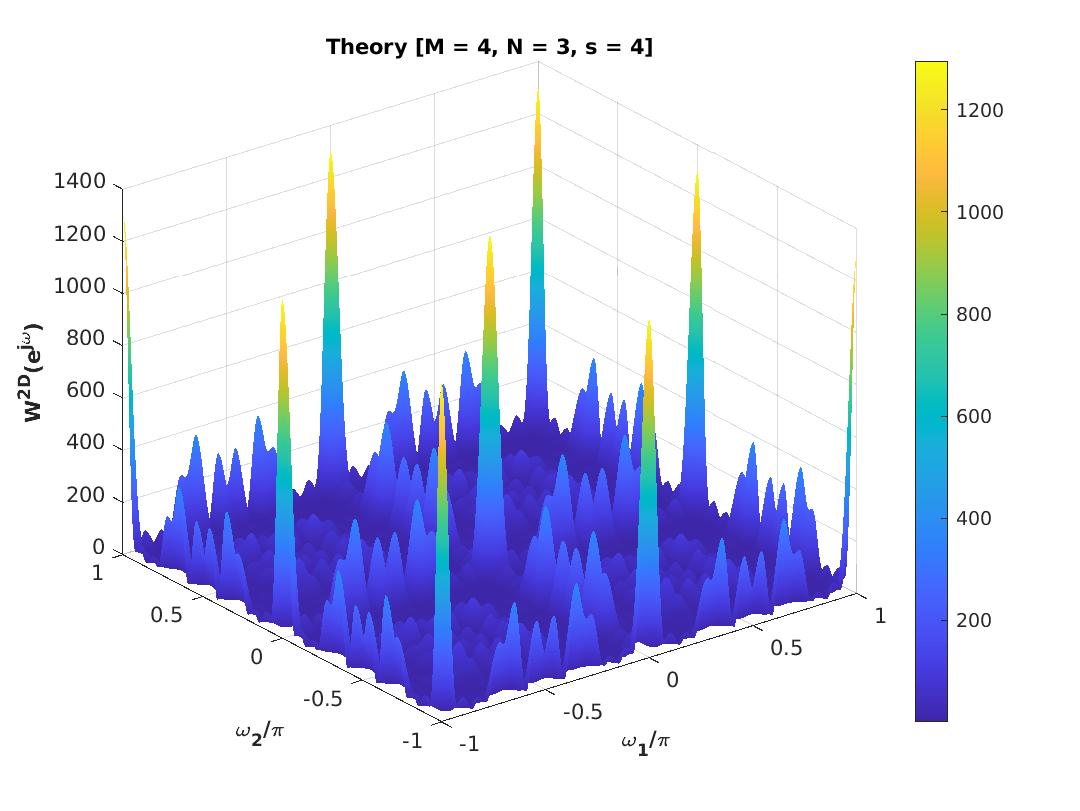}%
		\label{extreme_2DEx2_wt_bias_M4N3s4Surf}}
	\hfil
	\subfloat[$M=4$, $N=3$, $s=5$]{
		\includegraphics[width=0.5\textwidth]{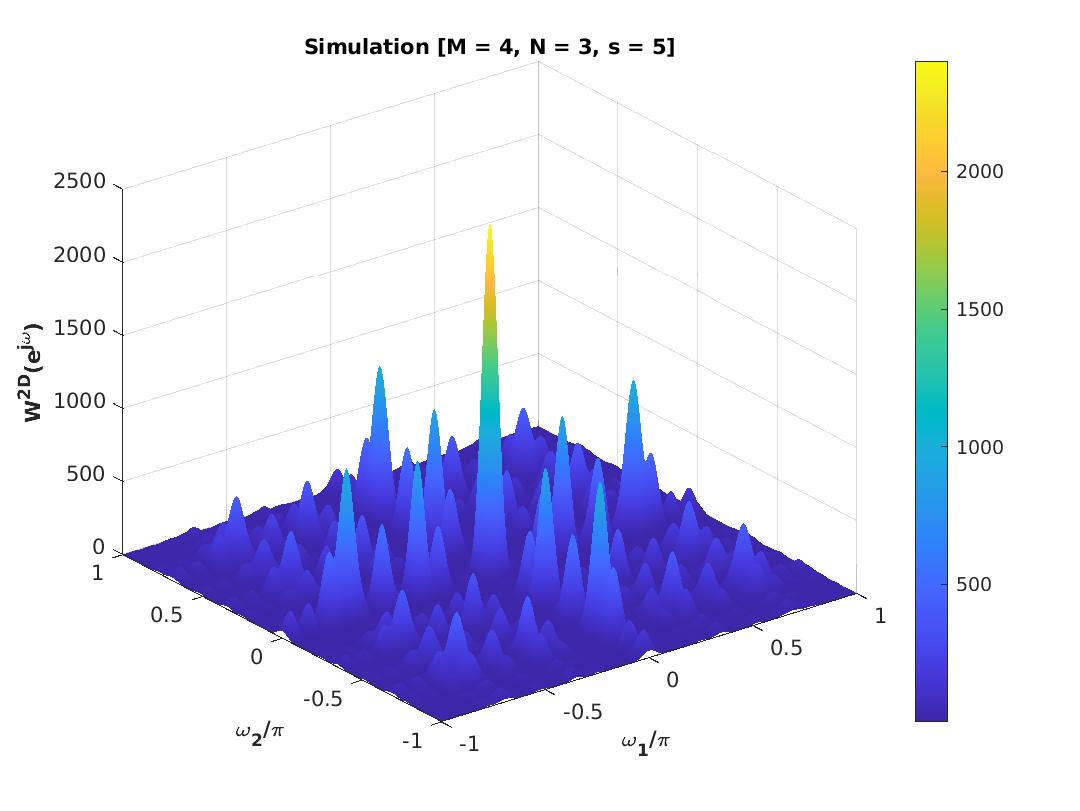}%
		\includegraphics[width=0.5\textwidth]{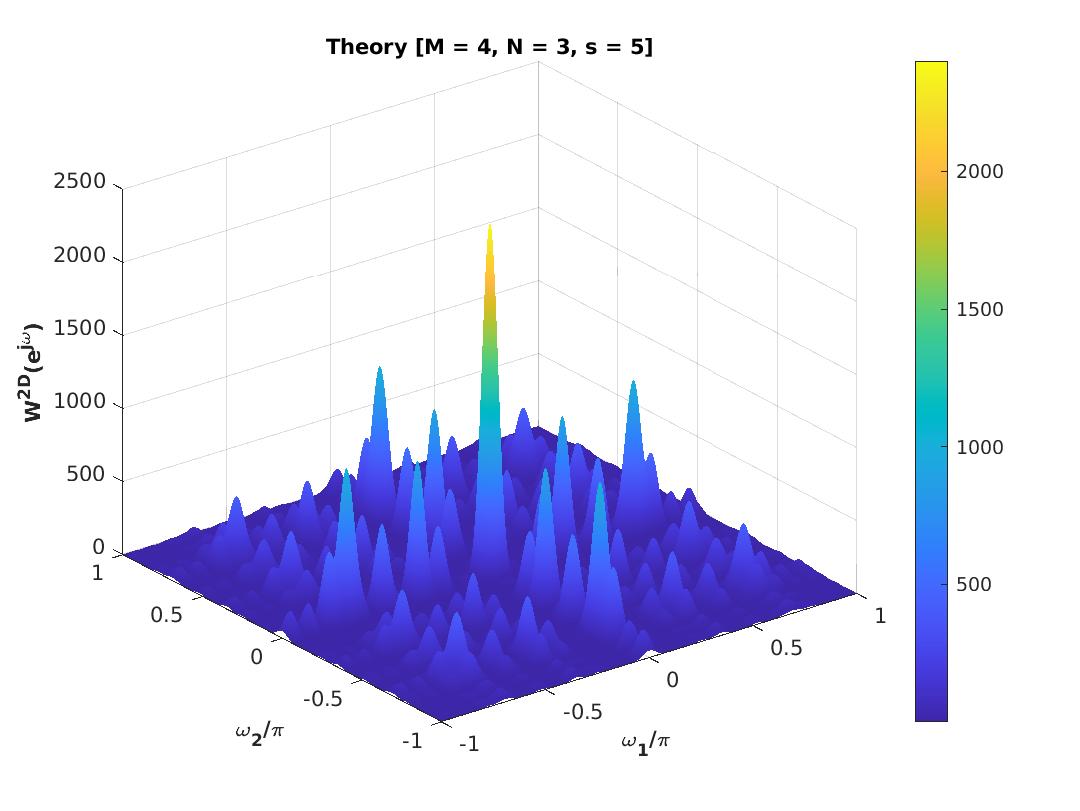}%
		\label{extreme_2DEx2_wt_bias_M4N3s5Surf}}
	\hfil
	\caption{Simulated and theoretical 2D bias window: Surface plot for Fig.~\ref{fig:extreme_2DEx2_wt_bias_M4N3s35}.}
	\label{fig:extreme_2DEx2_wt_bias_M4N3s35Surf}
\end{figure*}%
\begin{figure*}[!t]
	\centering
	\subfloat[Vertical frequencies]{
	\includegraphics[width=0.49\textwidth]{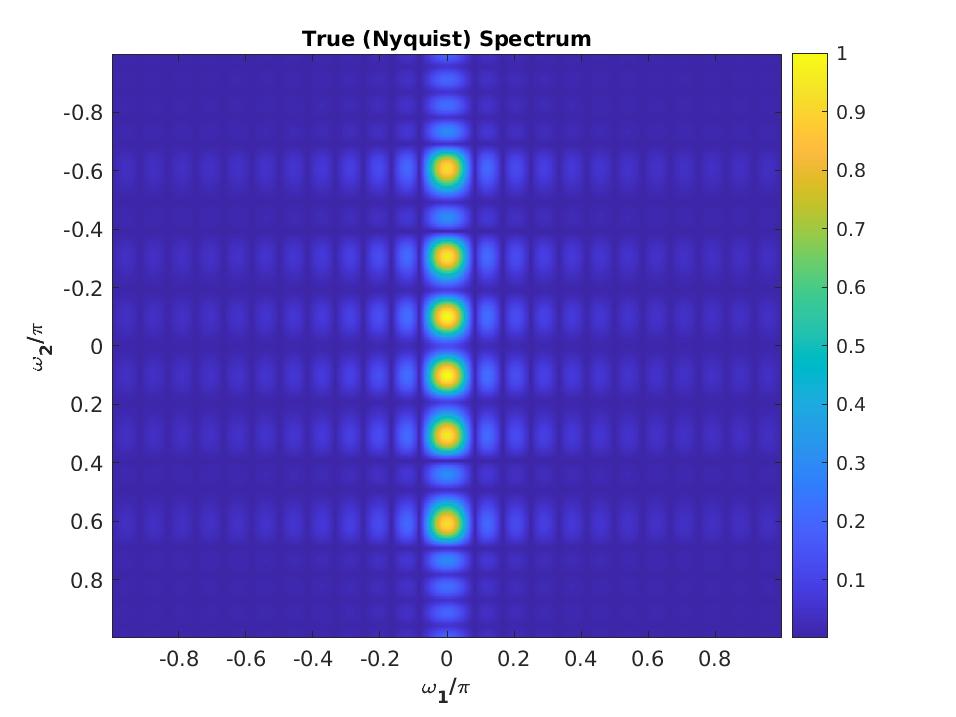}\label{fig:extreme_2D_true_spectV}}
	\subfloat[Horizontal frequencies]{
	\includegraphics[width=0.49\textwidth]{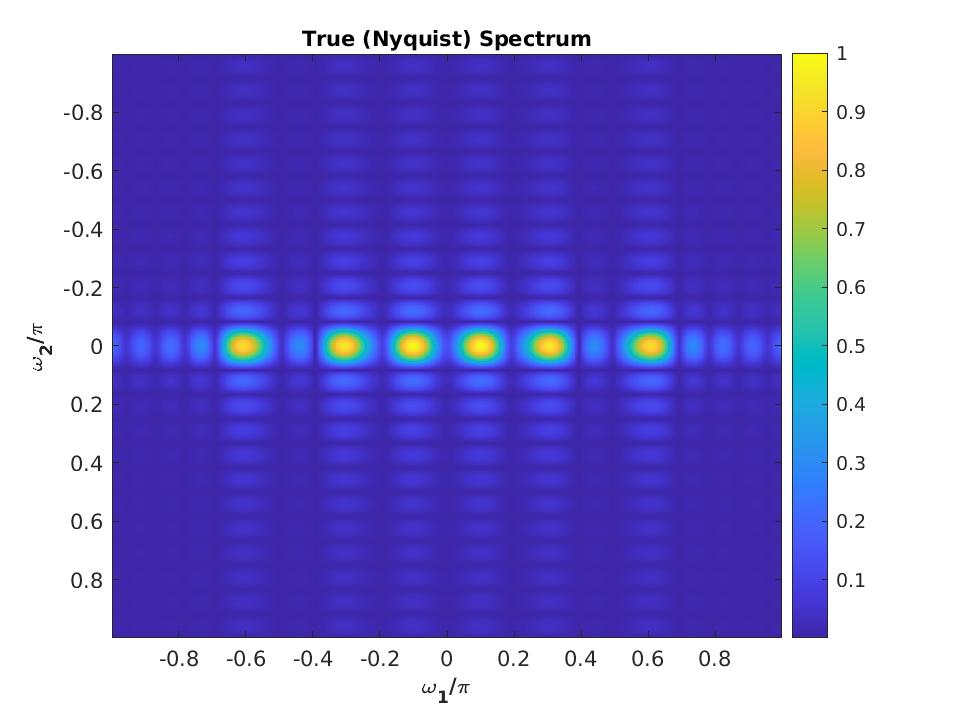}\label{fig:extreme_2D_true_spectH}}\hfil
	\subfloat[Horizontal and vertical frequencies]{
	\includegraphics[width=0.49\textwidth]{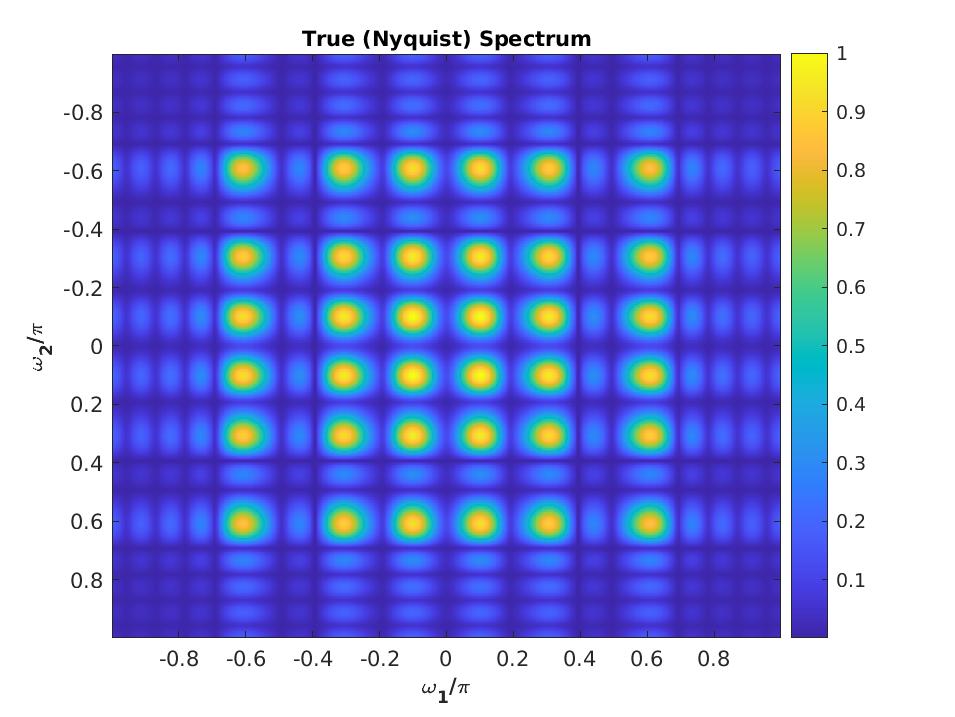}\label{fig:extreme_2D_true_spectVH}}
	\caption{2D true spectrum (Nyquist) with peaks at locations $[0.1, 0.3, 0.6]$ horizontally and vertically.}
	\label{fig:extreme_2D_true_spect}
\end{figure*}
\begin{figure*}[!t]
	\centering
	\includegraphics[width=0.49\textwidth]{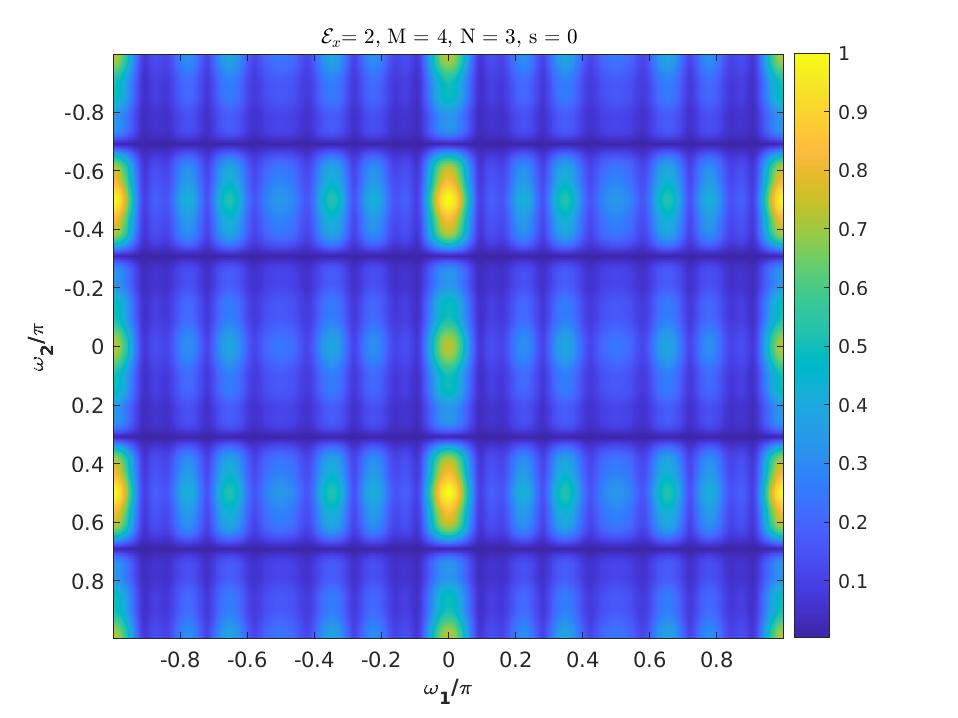}
	\includegraphics[width=0.49\textwidth]{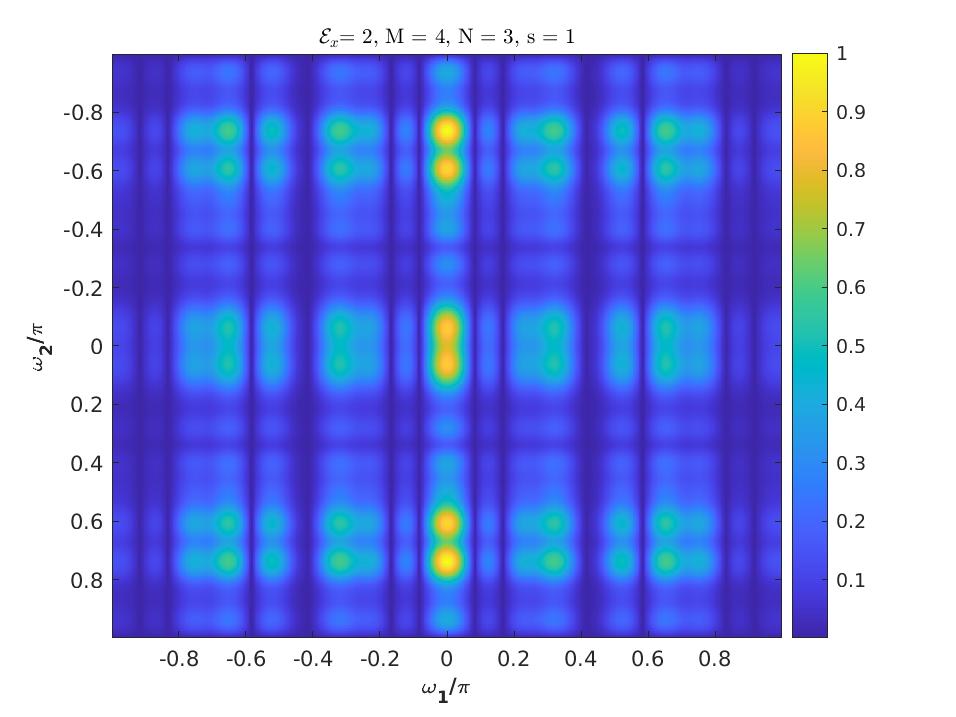}\\
	\includegraphics[width=0.5\textwidth]{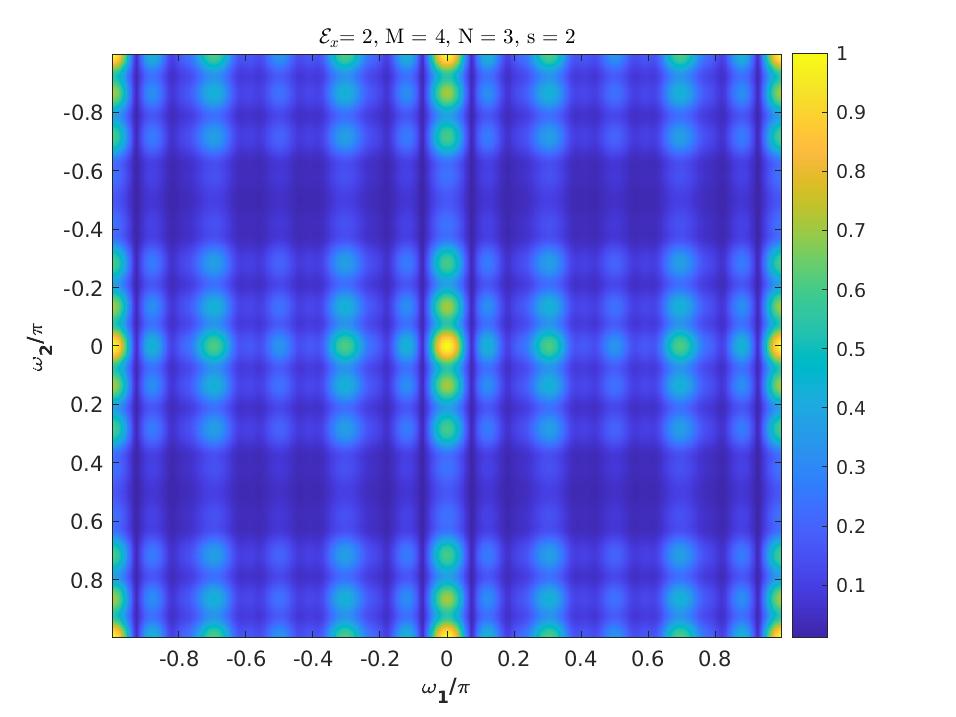}%
	\includegraphics[width=0.5\textwidth]{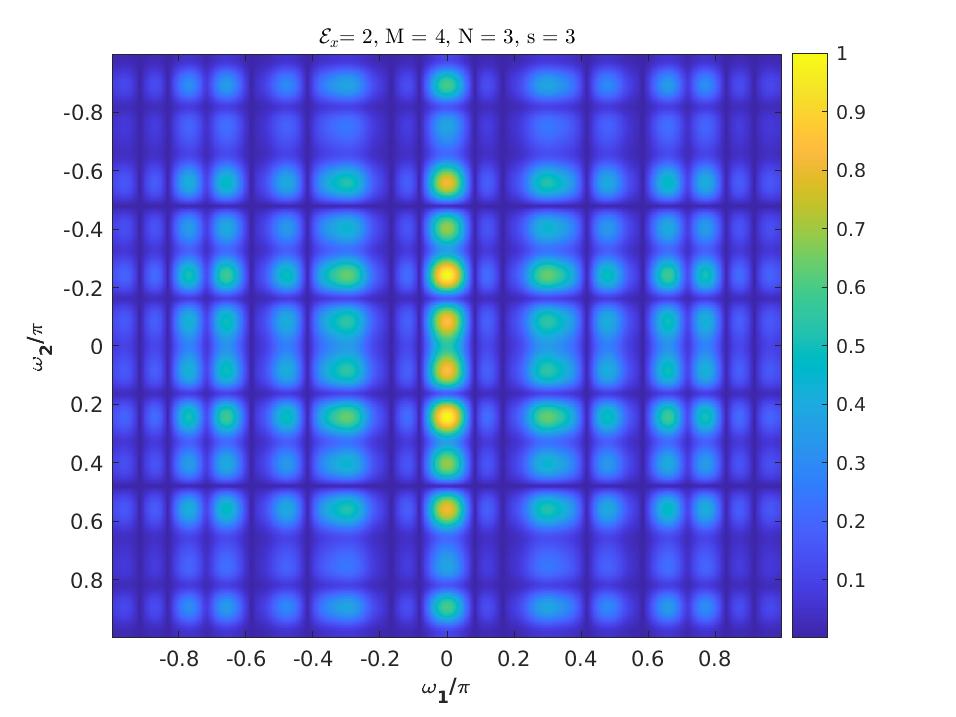}\\
	\includegraphics[width=0.5\textwidth]{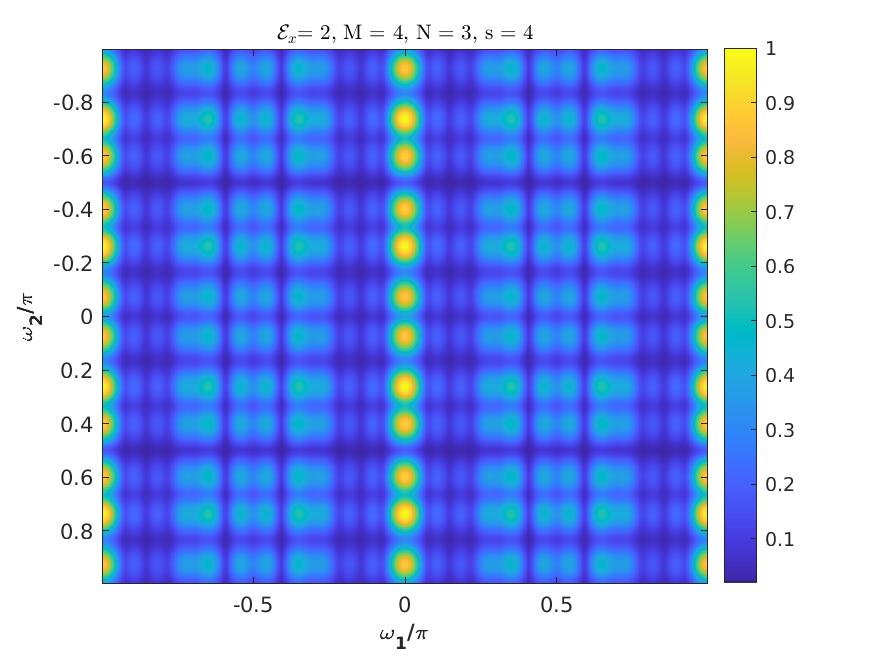}%
	\includegraphics[width=0.5\textwidth]{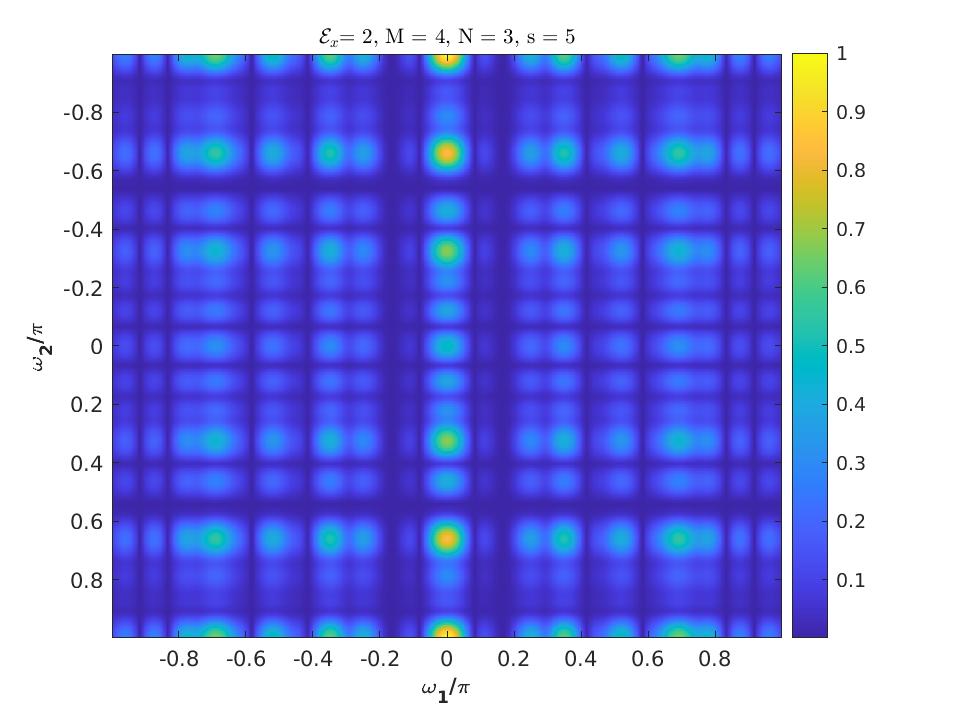}
	\caption{2D ExSCA based spectral estimation for Fig.~\ref{fig:extreme_2D_true_spectV}.}
	\label{fig:extreme_2D_M4N3V}
\end{figure*}

\begin{figure*}[!t]
	\centering
	\includegraphics[width=0.49\textwidth]{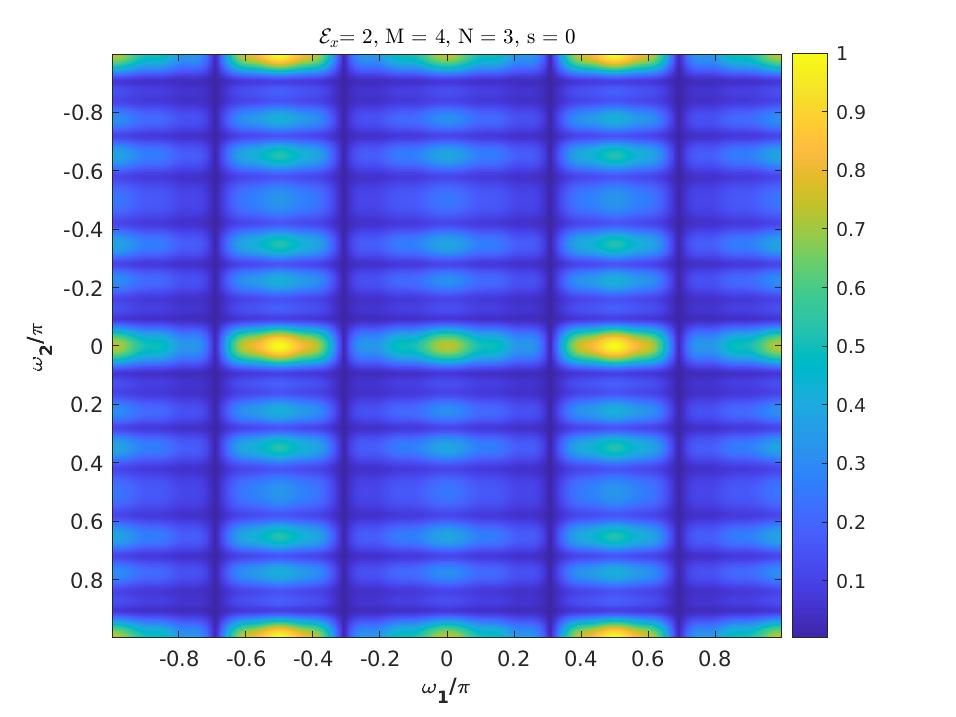}
	\includegraphics[width=0.49\textwidth]{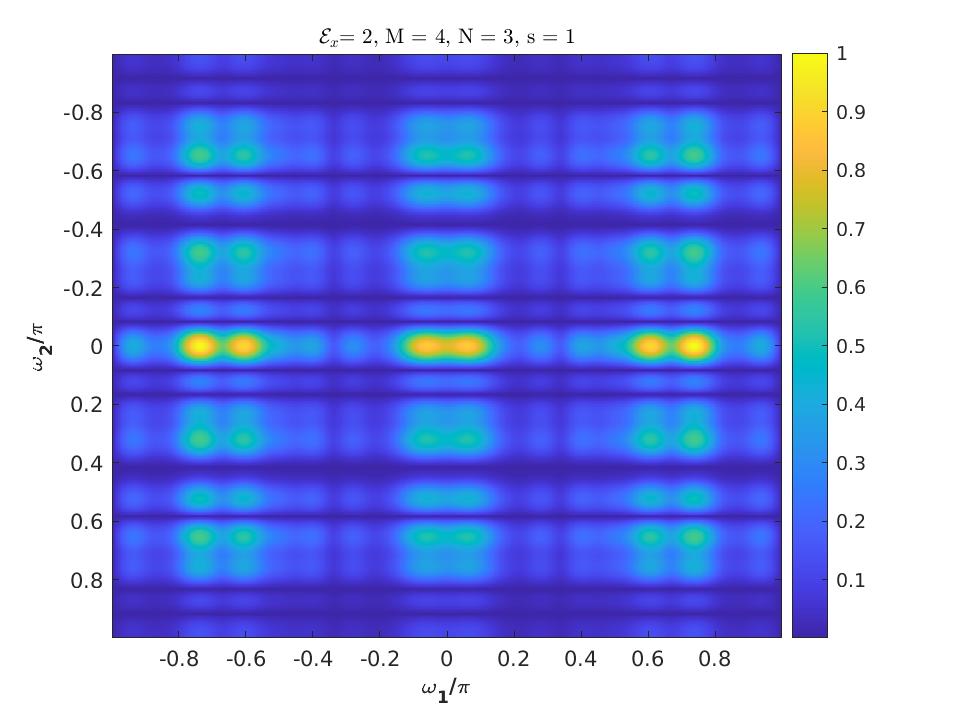}\\
	\includegraphics[width=0.5\textwidth]{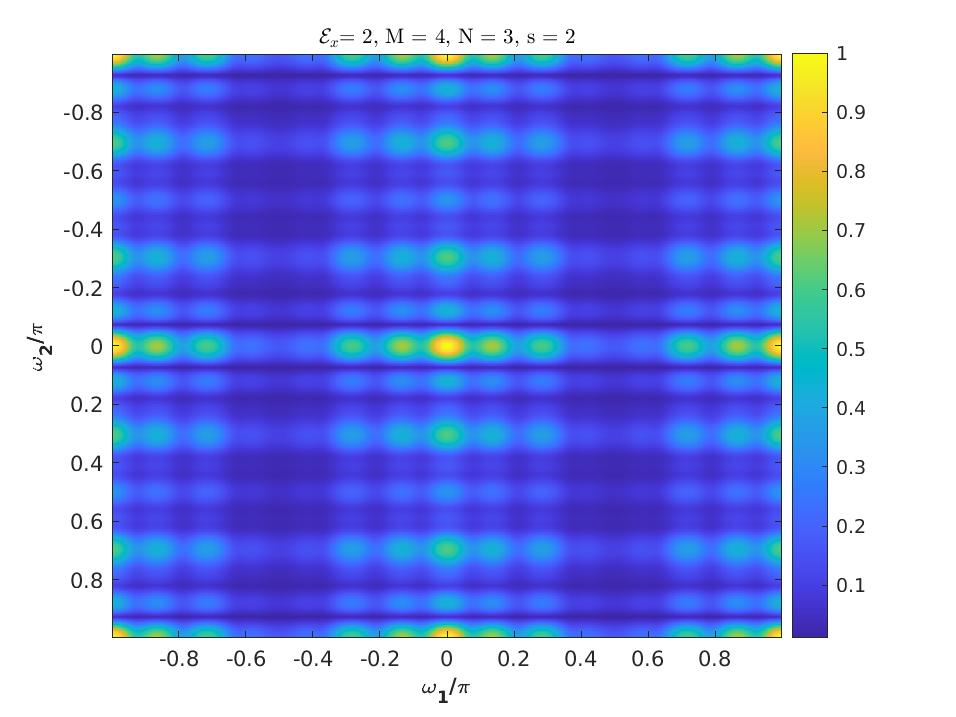}%
	\includegraphics[width=0.5\textwidth]{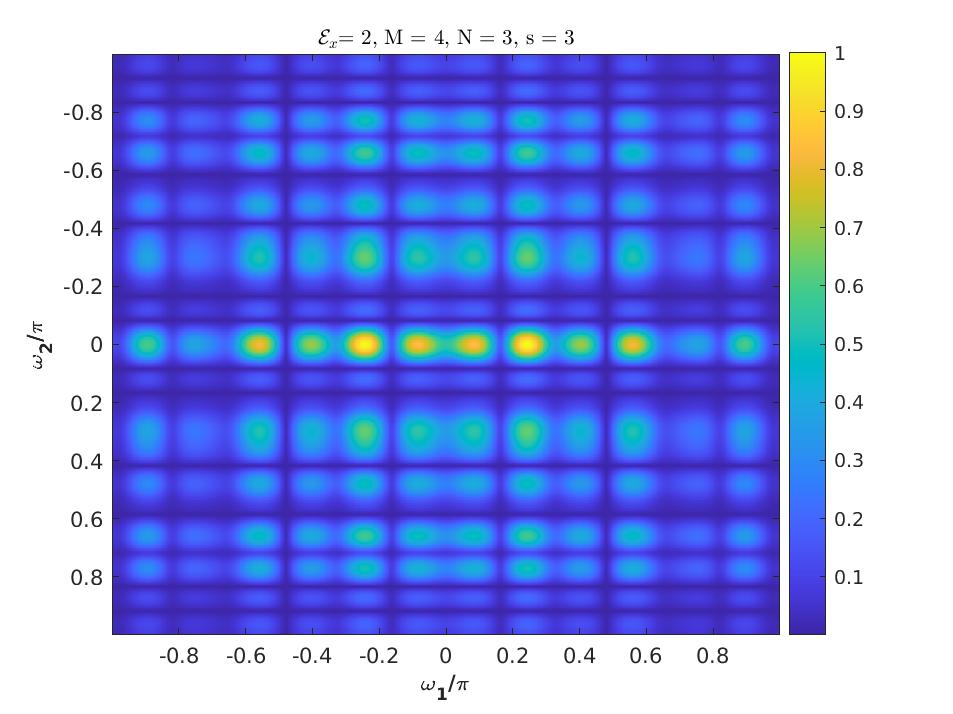}\\
	\includegraphics[width=0.5\textwidth]{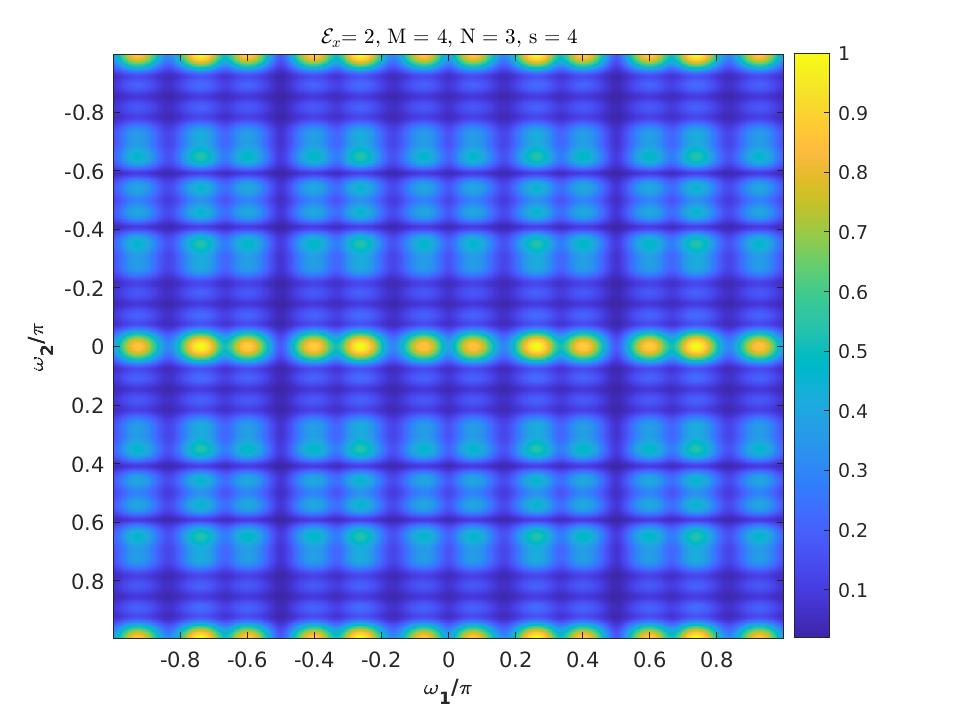}%
	\includegraphics[width=0.5\textwidth]{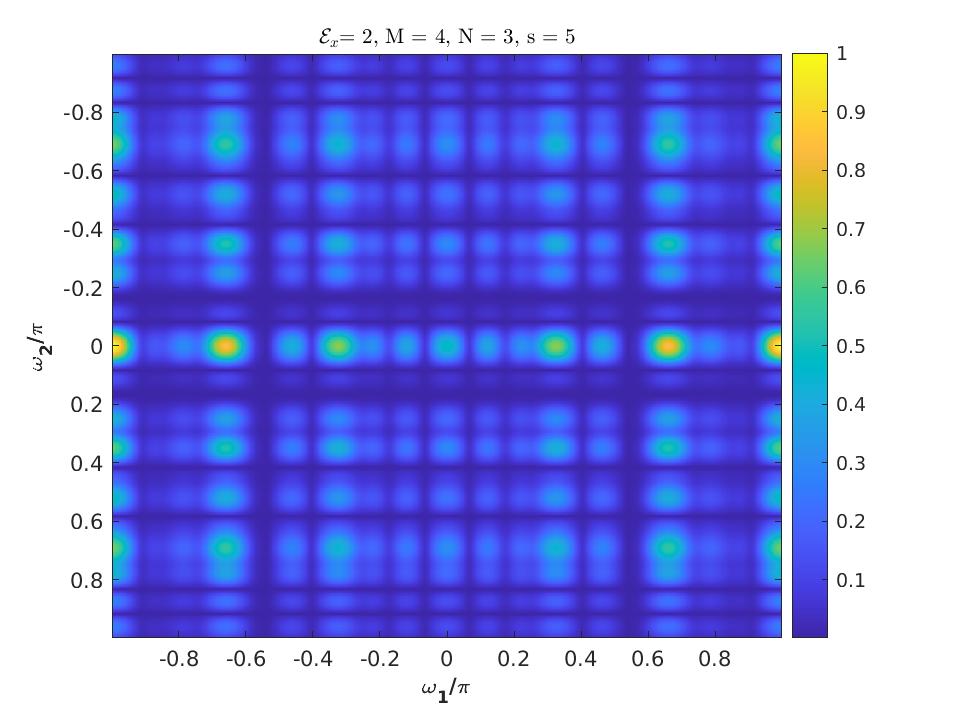}
	\caption{2D ExSCA based spectral estimation for Fig.~\ref{fig:extreme_2D_true_spectH}.}
	\label{fig:extreme_2D_M4N3H}
\end{figure*}

\begin{figure*}[!t]
	\centering
	\includegraphics[width=0.49\textwidth]{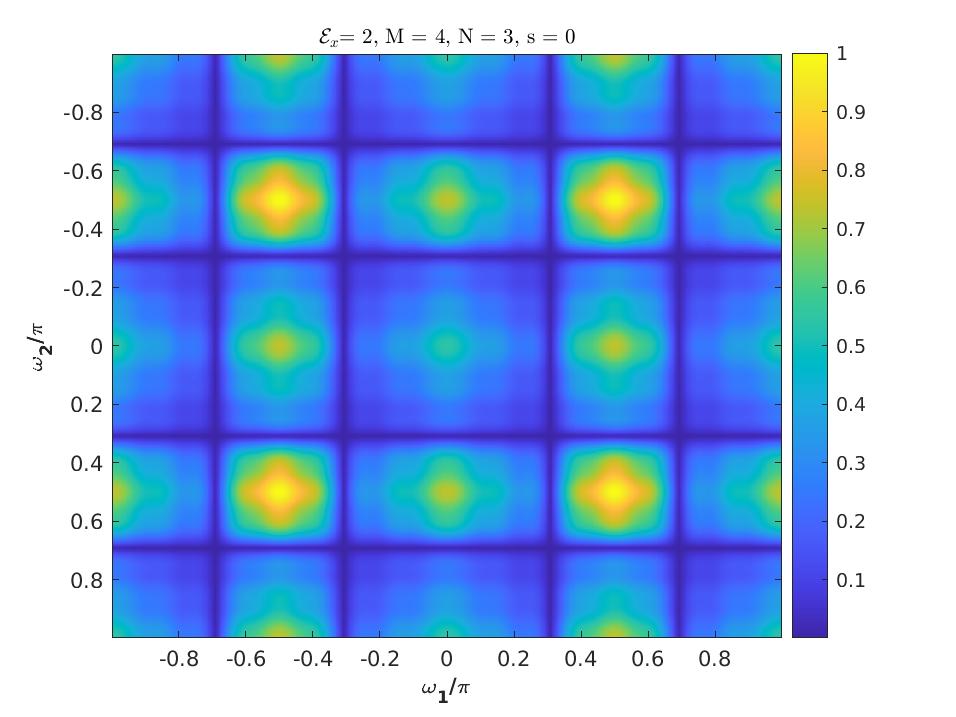}
	\includegraphics[width=0.49\textwidth]{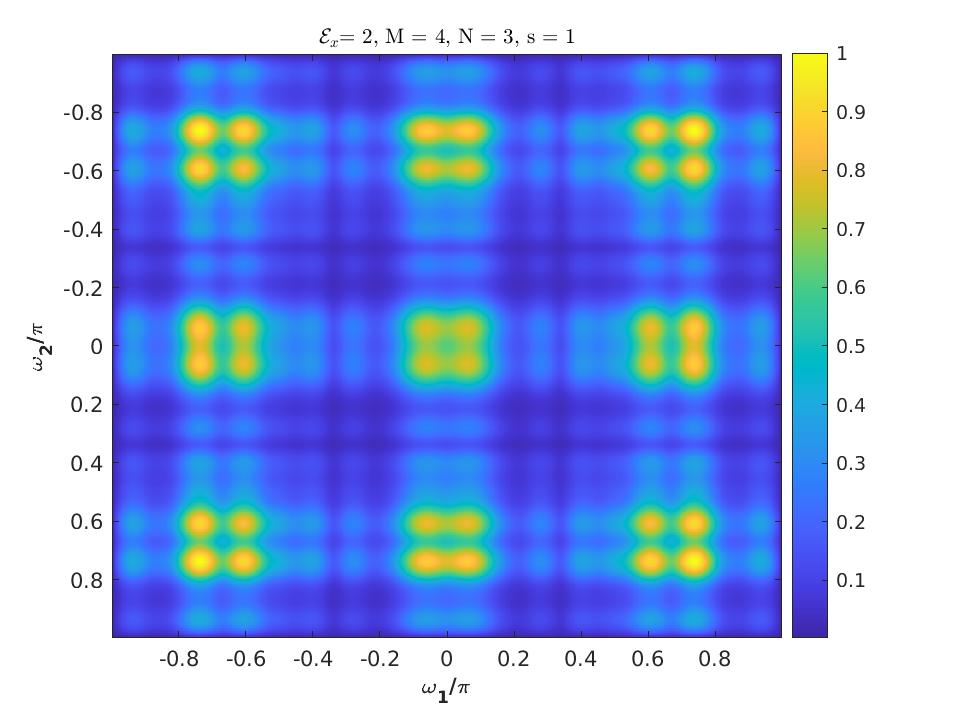}\\
	\includegraphics[width=0.5\textwidth]{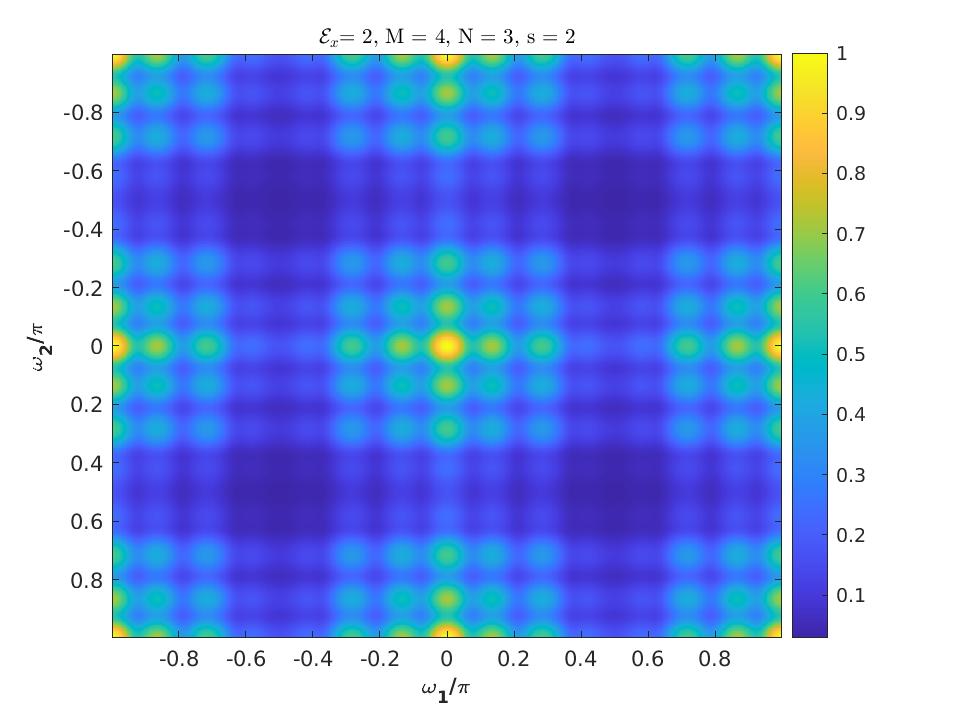}%
	\includegraphics[width=0.5\textwidth]{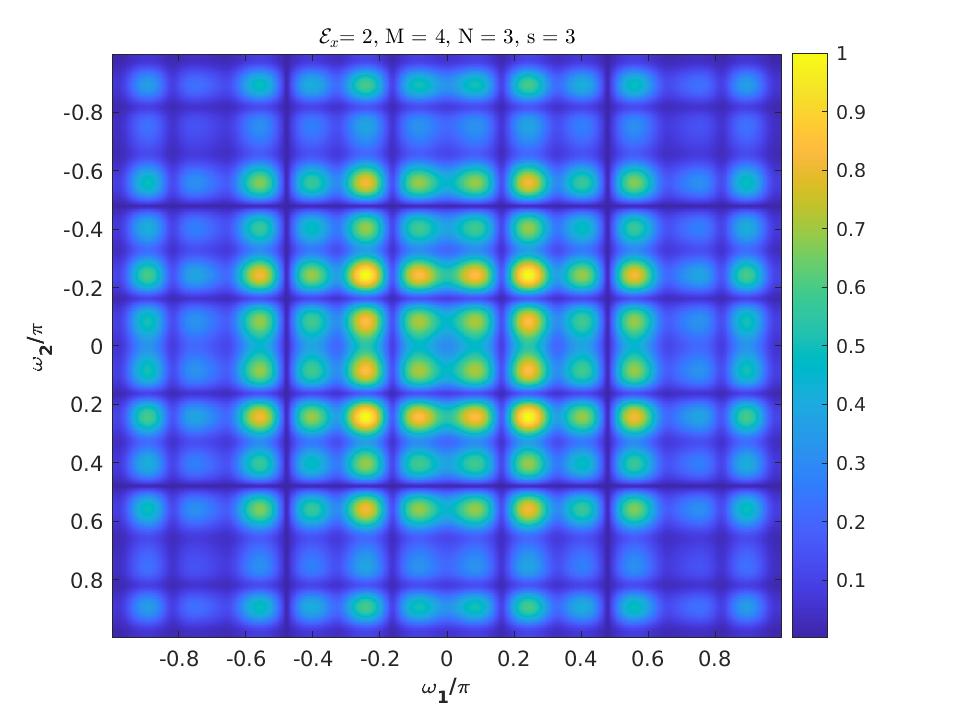}\\
	\includegraphics[width=0.5\textwidth]{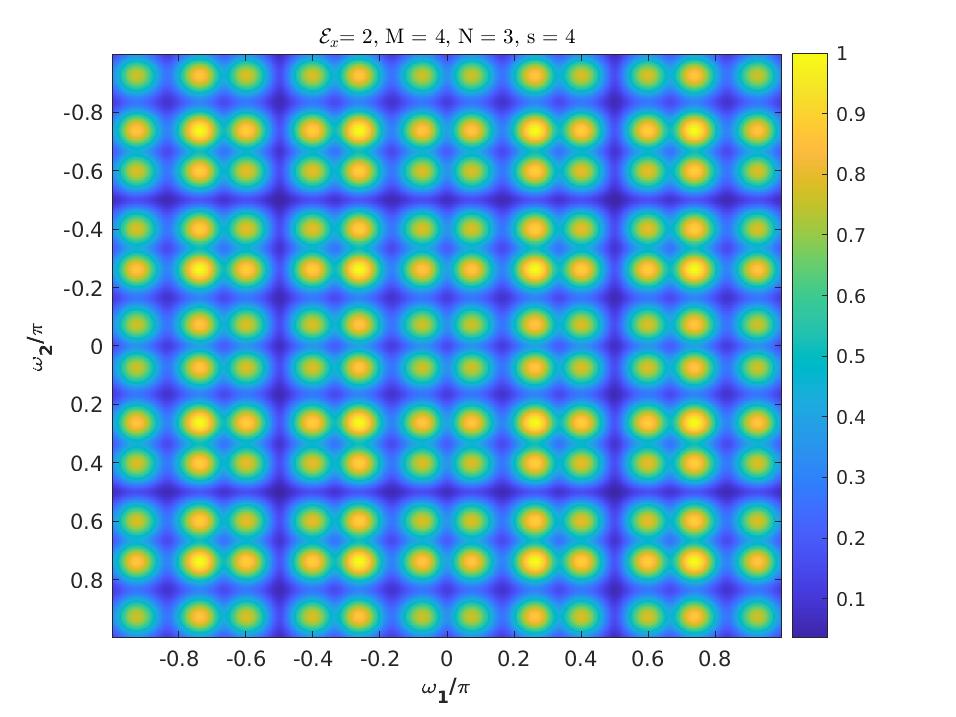}%
	\includegraphics[width=0.5\textwidth]{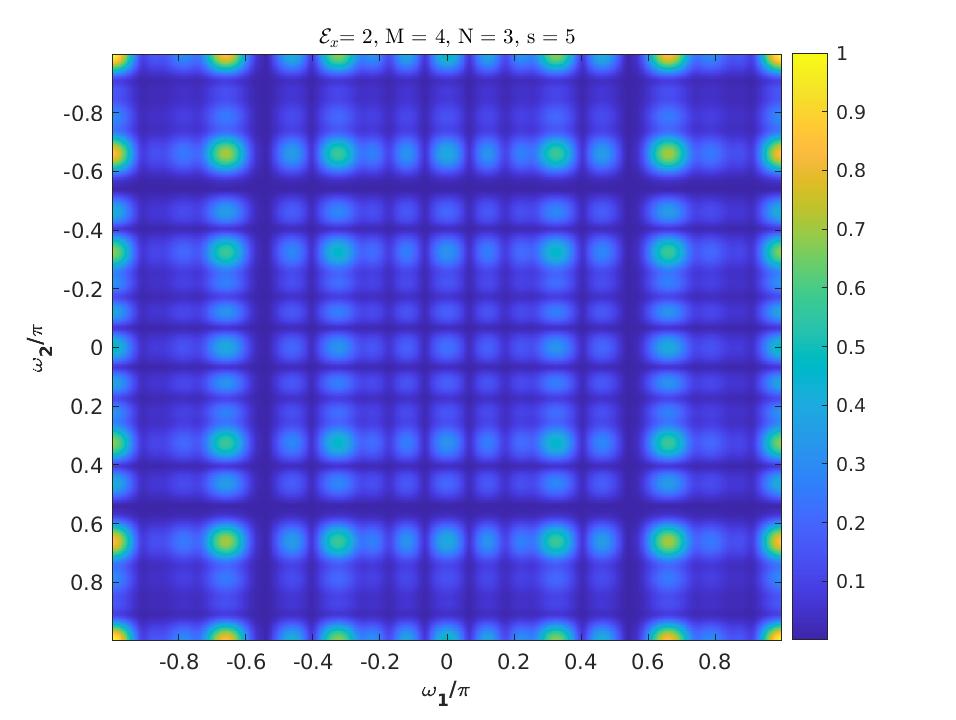}
	\caption{2D ExSCA based spectral estimation for Fig.~\ref{fig:extreme_2D_true_spectVH}.}
	\label{fig:extreme_2D_M4N3HV}
\end{figure*}

Spend some time reflecting on this. You will conclude that the ExSCA with even values of $s$ cannot reconstruct the second order statistics at the Nyquist period $d$. It can only reconstruct at $2d$. 
On the other hand, odd values of $s$ can reconstruct the autocorrelation of the signal at the Nyquist period $d$. Here, the bias window does not have any image of the main lobe. Therefore, odd values of $s$ validates the use of this scheme as an extremely sparse array. While even values of $s$ represents the APCA. 
The above discussion describes the bias of the correlogram estimate for the ExSCA. Now, let us tries to quantify this bias by developing mathematical expressions to describe it as a function of $M$, $N$, and $s$. The closed-form expression for the weight function is given by:
\begin{equation}\label{eq:exsca_entire}
\begin{split}
z(l)=\underbrace{\sum\limits_{n=-(N-1)}^{N-1} (N-\mid n\mid)\delta(l-2Mn)}_\text{A}\\
+\underbrace{\sum\limits_{m=-(M-1)}^{M-1} (M-\mid m\mid)\delta(l-(2Nm))}_\text{B}\\
+\underbrace{\sum\limits_{n=0}^{N-1}\sum\limits_{m=0}^{M-1} \delta(\mid l\mid-\mid 2Mn-(2Nm+s)\mid)}_\text{C}\\
-\left[\sum\limits_{m=0}^{M-1} \delta(\mid l\mid-|2Mn_p-(2Nm+s)|)\right.\\
\underbrace{\left.	+\sum\limits_{n=0}^{N-1} \delta(\mid l\mid-|2Mn-(2Nm_p+s)|)\right]}_\text{D}
\end{split}
\end{equation}
The correlogram bias window is given by:
\begin{equation}\label{eq:FT_exsca_entire}
\begin{split}
 W_{b}(e^{j\omega})=\frac{1}{s_b}\left\{\left |\frac{\sin(\omega MN)}{\sin(\omega M)}\right|^2+\left |\frac{\sin(\omega MN)}{\sin(\omega N)}\right|^2\right .\\
+2\cos \omega (M-N+s) \frac{\sin^2 (\omega MN)}{\sin(\omega M)\sin(\omega N)}\\
-\left[ 2\sin \omega MN \left(\frac{\cos \omega(2Mn_p-MN+N-s)}{\sin \omega N}\right.\right.\\
+\left.\left.\left.\frac{\cos \omega(2Nm_p-MN+M+s)}{\sin \omega M}\right)-1
\right]\right\}
\end{split}
\end{equation}
The above expression can be derived along similar lines as in~\cite{UVD_PHD}. There the significance of the bias window is also described. Note that the convolution of this window with the true power spectrum (or Nyquist power spectrum) represents the ExSCA power spectrum. It had also noted that the bias window represents the covariance if $\omega$ is replaced by $\omega_1-\omega_2$ with a scale factor. It is valid for complex white Gaussian process.
The derived bias window is compared with the simulated bias in Fig.~\ref{fig:extreme_Ex2_wt_bias_M4N3}. It also includes the simulated weight function. Clearly, the derived and simulated bias match well. Note that APCA is a special case of the ExSCA. 
The relationship between APCA and ExSCA is described in Remark 1. In general, d represents the Nyquist distance/ period. Let us define some additional symbols to represent APCA and EXSCA:\\
$d_{A}$ - Nyquist sampling distance/period for APCA.\\
$d_{X}$ - Nyquist sampling distance/period for ExSCA.\\
$f_{s_{A}}=\frac{1}{d_{A}}$ - Nyquist sampling frequency for APCA.\\ $f_{s_{X}}=\frac{1}{d_{X}}$ - Nyquist sampling frequency for ExSCA.\\
$s_{A}$ - an integer shift or pivot selection parameter for APCA.\\
$s_{X}$ - an integer shift or pivot selection parameter for ExSCA.\\
$l_{A}$ - an integer lag or difference value for APCA.\\
$l_{X}$ - an integer lag or difference value for ExSCA. 
%
%
\begin{remark}\label{remarks1}
	ExSCA is equivalent to APCA with $d_{A}=2 d_{X}$, $f_{s_{A}}=\frac{f_{s_{X}}}{2}$, $s_{A}=\frac{s_{X}}{2}$, and $l_{A}=\frac{l_{X}}{2}$. Here, the weight function of APCA is equal to the down-sampled version of the weight function of ExSCA.
\end{remark}
\subsection{Simulation Results}
The simulation model in (Section 4.2.3,~\cite{UVD_PHD}) is used here. Let us consider three spectral peaks at location [0.1, 0.3, 0.6]. Fig.~\ref{fig:extreme_Ex2_sim_true_spect} shows this power spectrum for the Nyquist and the prototype co-prime scheme using the correlogram method. Note that correlogram and periodogram method are equivalent. Fig.~\ref{fig:extreme_Ex2_sim_M4N3} shows the power spectrum for the ExSCA with parameters $\mathcal{E}_x=2$, $M=4$, $N=3$, and $s=[0, 5]$. For even values of s the scheme does not work. This is expected since the bias window has an image of the main-lobe at $\pi$. This gives rise to spurious frequencies and causes aliasing. However, for odd values of s, this scheme works well. Furthermore, it has a better resolution than the prototype co-prime scheme. 

So, are even values of s totally useless? Or is there something more that does not meet the eye? To investigate the even values of s, let us consider another example with spectral peaks at [0.05, 0.15, 0.3]. Fig.~\ref{fig:extreme_Ex2_sim_true_spect_list2} shows the power spectrum for the Nyquist and prototype co-prime scheme. The prototype co-prime scheme fails to estimate the spectral peaks because the main lobe width of the bias window (Fig. 4.3(a) in~\cite{UVD_PHD}) is broad. This reduces the spectral resolution. However, the extremely sparse scheme has a narrow main lobe width (Fig.~\ref{fig:extreme_Ex2_wt_bias_M4N3}). Therefore, the ExSCA scheme works well as shown in Fig.~\ref{fig:extreme_Ex2_sim_M4N3_list2}. It may seem like ExSCA fails for $s=$even. But that's not true. If the desired frequencies are at locations less than 0.5 (normalized frequency), then $s=even$ estimates the spectral peaks in the range [0, 0.5]. The estimated peaks repeat in range [0.5, 1] without aliasing. Fig.~\ref{fig:extreme_Ex2_sim_M4N3_list2zoom} reproduces the results in Fig.~\ref{fig:extreme_Ex2_sim_M4N3_list2} for even values of s with the range limited to [0, 0.5]. The interesting part is that the plots in Fig.~\ref{fig:extreme_Ex2_sim_M4N3_list2zoom} is same as the APCA in Fig.~\ref{Truly_coprime_sim_3peak_M4N3}. This aspect along with Remark~\ref{remarks1} leads to the Remark~\ref{remarks2}:
\begin{remark}\label{remarks2}
	For even values of $s$, ExSCA functions as an APCA. The valid normalized frequency range is [0, 0.5] where $f_{s_{A}}=0.5$.
\end{remark}

Next, let us consider multidimensional ExSCA. 
 
\section{Multidimensional ExSCA}\label{MultiExSCA}
Now, let us discuss multidimensional co-prime arrays. An example of 1D ExSCA combined sampling function is shown in Fig.~\ref{ExSCA_1d_pattern} for $\mathcal{E}_x=2$, $s=1$, $M=4$, and $N=3$. Let us now define a 2D sampling function as $p^{2D}(k)=p^{1D}(k_1)\otimes p^{1D}(k_2)$ where $\otimes$ represents the outer product and $p^{1D}(k)$ is the 1D sampling function. Note that $p^{2D}(k)=p^{2D}(k_1,k_2)$. For ease of representation a single variable $(k)$ is used, with $2D$ as a superscript. In general, the multidimensions could be 3D, 4D, and so on. The sampling pattern as an outer product has the advantage that the weight function as well as the correlogram bias window can be represented as the outer  product of the 1D function. Therefore, the multidimension ExSCA theory relies on the 1D theory. In fact, it is a straightforward extension of the 1D theory.

The idea is explained in Fig.~\ref{fig:ExSCA_sampling_function}. It compares the 1D (Fig.~\ref{ExSCA_1d_pattern}) and 2D (Fig.~\ref{ExSCA_matrix}) sampling function for an example with parameters $\mathcal{E}_x=2$, $M=4$, $N=3$, $s=1$. Note that the 2D weights $z^{2D}(l)$ can be obtained as 2D convolution of the sampling function $p^{2D}(k)$ with its time-reversed version (i.e. autocorrelation). (Refer Fig. 4.6 in~\cite{UVD_PHD} for 1D scenario) Fig.~\ref{fig:2DEx2_weight_M4N3} represents the discrete 2D weight function as a color image for $s=[0, 5]$. The correlogram bias window is the Fourier transform of the weight function. Let us compute the 2D Fourier transform (FFT) of the weight function and refer to it as `Simulated' bias window. Here, the weight function is computed as autocorrelation of the 2D pattern. Next, compute the 2D correlogram bias window as the outer product of the 1D bias window in~\eqref{eq:FT_exsca_entire}. Let us refer to it as `Theoretical' bias window. The simulated and theoretical bias is shown to match in Fig.~\ref{fig:extreme_2DEx2_wt_bias_M4N3s02} and Fig.~\ref{fig:extreme_2DEx2_wt_bias_M4N3s35}. Here, the figures are displayed as images. For better visualization, they have also been displayed as surface plot in Fig.~\ref{fig:extreme_2DEx2_wt_bias_M4N3s02Surf} and Fig.~\ref{fig:extreme_2DEx2_wt_bias_M4N3s35Surf}. Ideally, we wish to have a 2D impulse but practically this may not be possible. Similar to the 1D ExSCA, the 2D ExSCA also has images of the main lobe for $s=even$. This can be observed along the axis as well as the diagonal. However, this represents the 2D APCA with the conditions mentioned in Remarks~\ref{remarks1} and~\ref{remarks2}. 

Let us now consider 2D correlogram/periodogram power spectrum estimation as an application. Fig.~\ref{fig:extreme_2D_true_spect} shows the Nyquist power spectrum. Three scenarios are considered. The first has only vertical frequencies, the second has only horizontal frequencies, and the last scenario has frequencies along different directions. The Nyquist spectrum is considered to be the true spectrum. Therefore, we wish to have a sub-Nyquist system that closely matches the Nyquist estimator. For the 2D ExSCA, power spectrum estimation for vertical frequencies (Fig.~\ref{fig:extreme_2D_true_spectV}) is shown in Fig.~\ref{fig:extreme_2D_M4N3V}. The yellow dots represent the peaks. Observe for $s=3$, these peaks are close to the true vertical frequency peaks i.e. $[-0.6, -0.3, -0.1, 0.1, 0.3, 0.6]$. Therefore, $s=3$ seems to be a good choice for the example considered. Similarly, for the horizontal (Fig.~\ref{fig:extreme_2D_true_spectH}) and mixed frequency directions (Fig.~\ref{fig:extreme_2D_true_spectVH}), 2D ExSCA results are shown in Fig.~\ref{fig:extreme_2D_M4N3H} and Fig.~\ref{fig:extreme_2D_M4N3HV} respectively. Here as well $s=3$ seems good. Similar to the 1D ExSCA, the 2D ExSCA also has spurious peaks for even values of s. Therefore, when s is even, 2D APCA can be realized. The convolution of the ExSCA bias window with the Nyquist (true) spectrum gives the ExSCA spectrum. This theory is explained in~\cite{UVD_PHD}. 

\begin{figure*}[!t]
	\centering
	\includegraphics[width=0.5\textwidth]{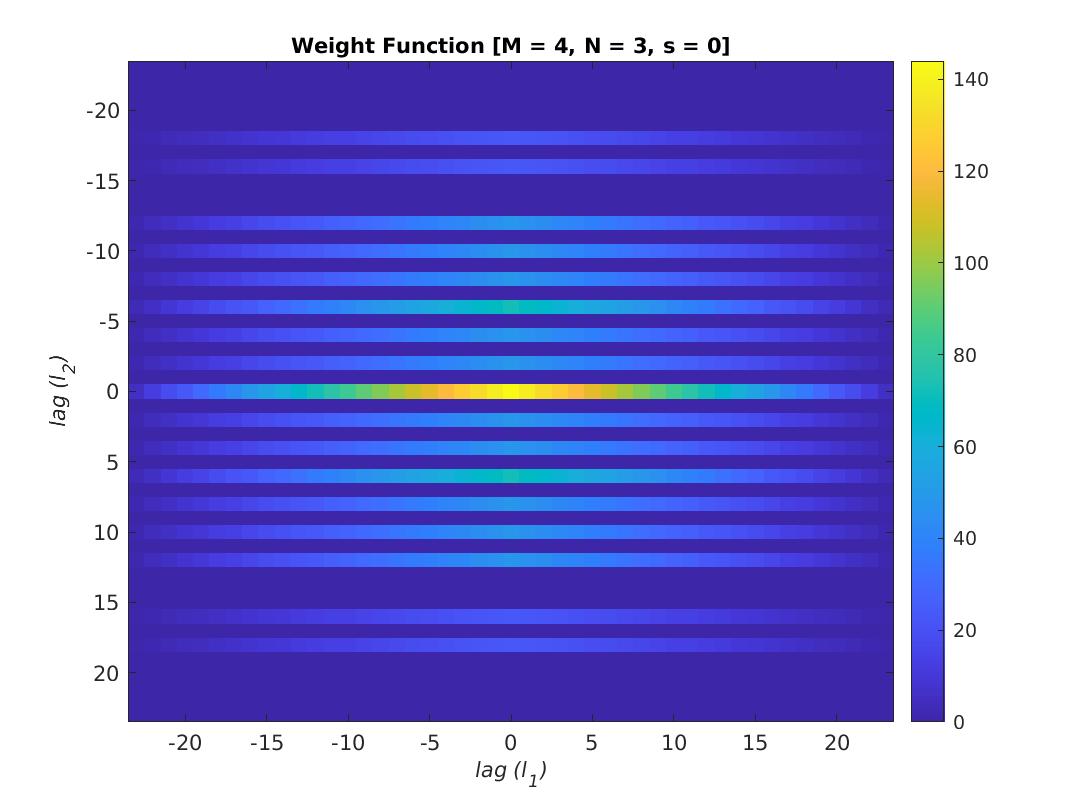}%
	\includegraphics[width=0.5\textwidth]{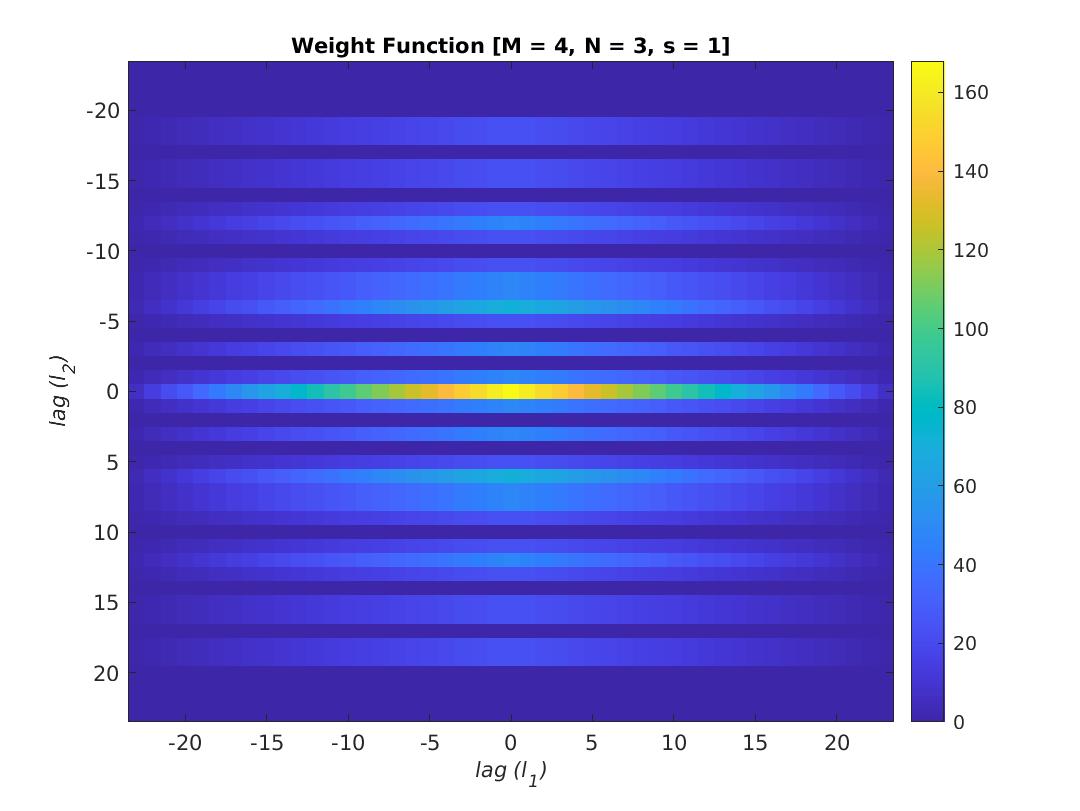}
	
	\includegraphics[width=0.5\textwidth]{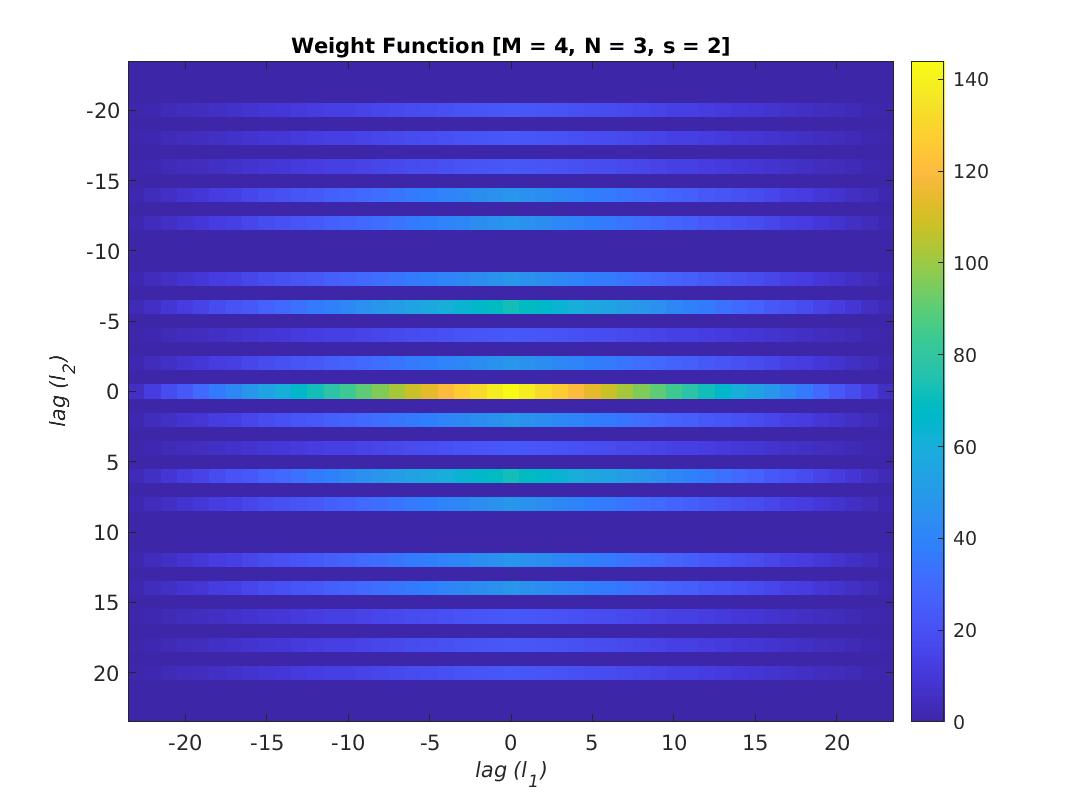}%
	\includegraphics[width=0.5\textwidth]{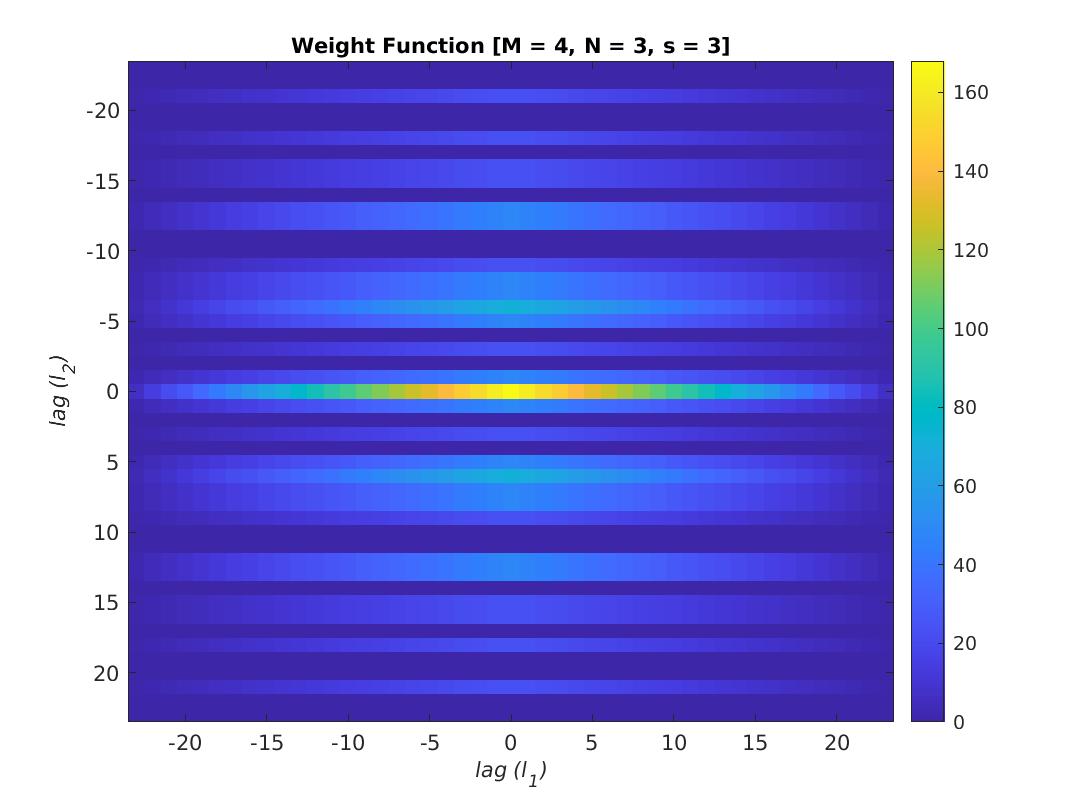}
	\includegraphics[width=0.5\textwidth]{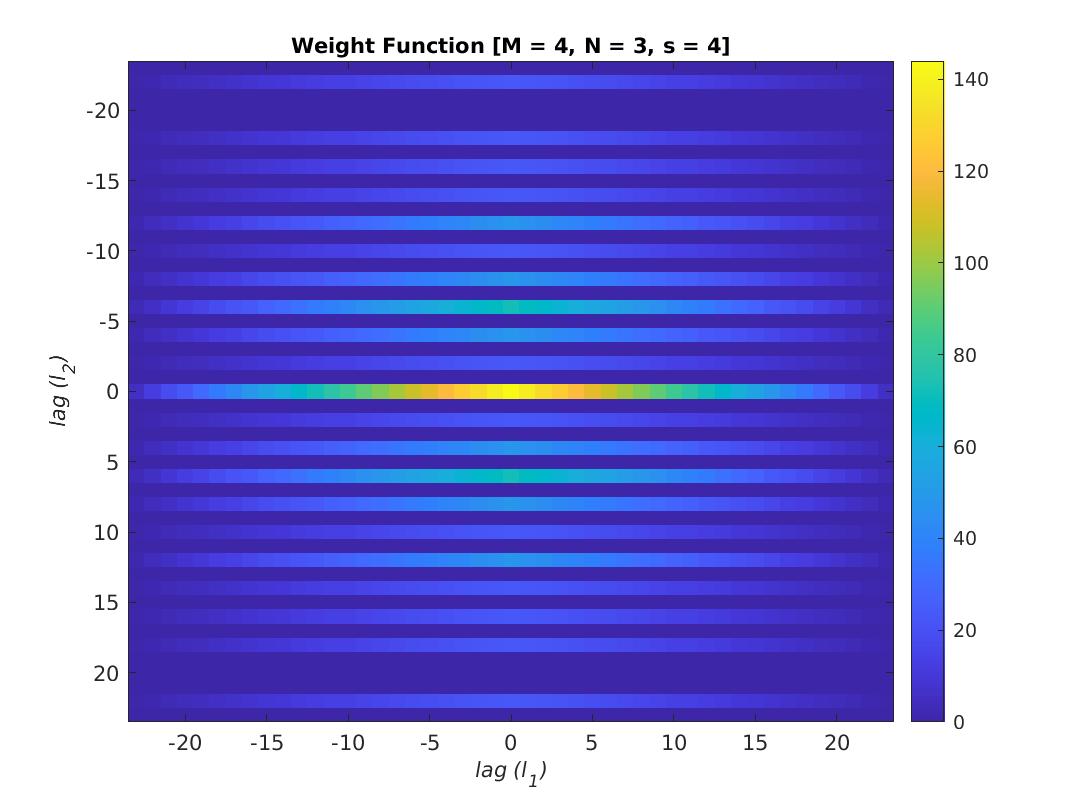}%
	\includegraphics[width=0.5\textwidth]{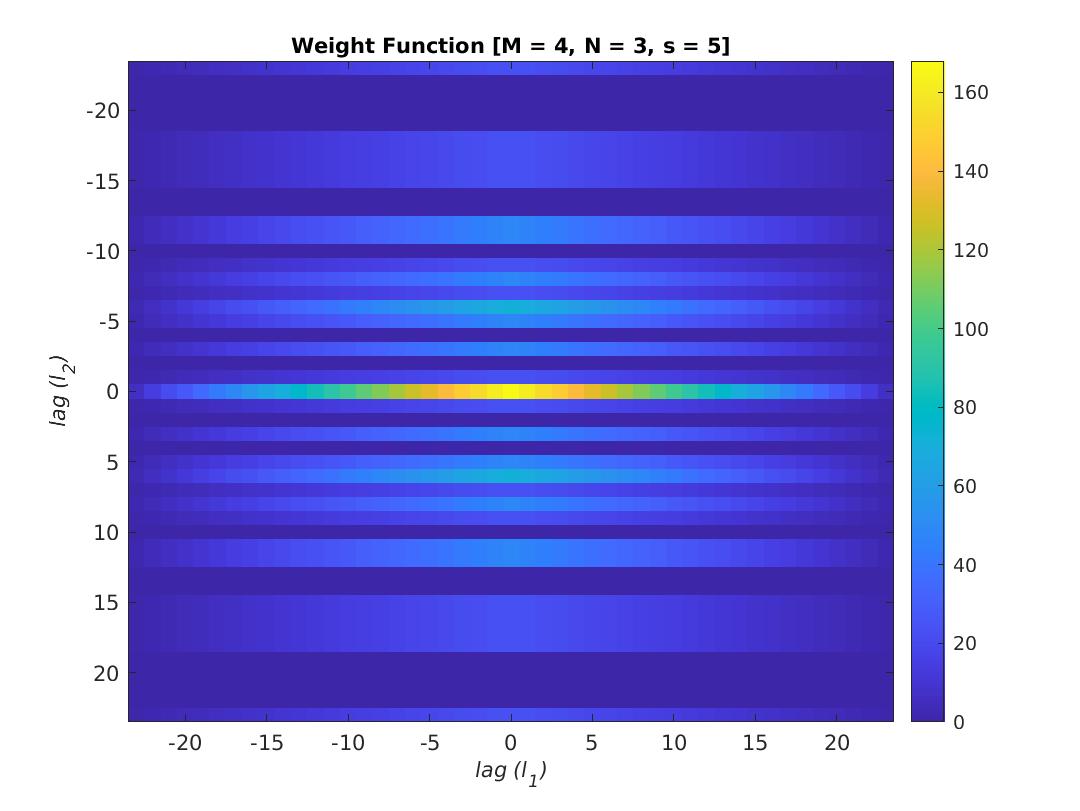}	
	\caption{Example of 2D weight function for Hybrid ExSCA with $\mathcal{E}_x=2$, $M=4$, $N=3$.}
	\label{fig:2DEx2_weight_M4N3P1P2}
\end{figure*}

\begin{figure*}[!t]
	\centering
	\subfloat[$M=4$, $N=3$, $s=0$]{
		\includegraphics[width=0.5\textwidth]{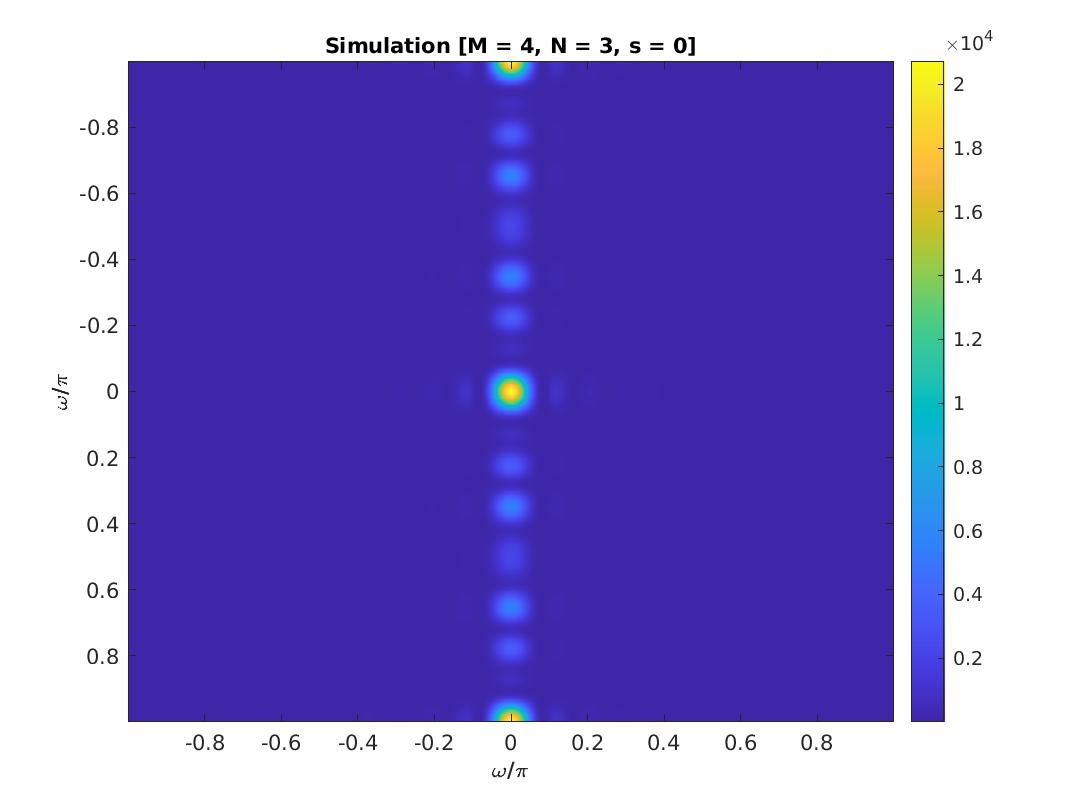}%
		\includegraphics[width=0.5\textwidth]{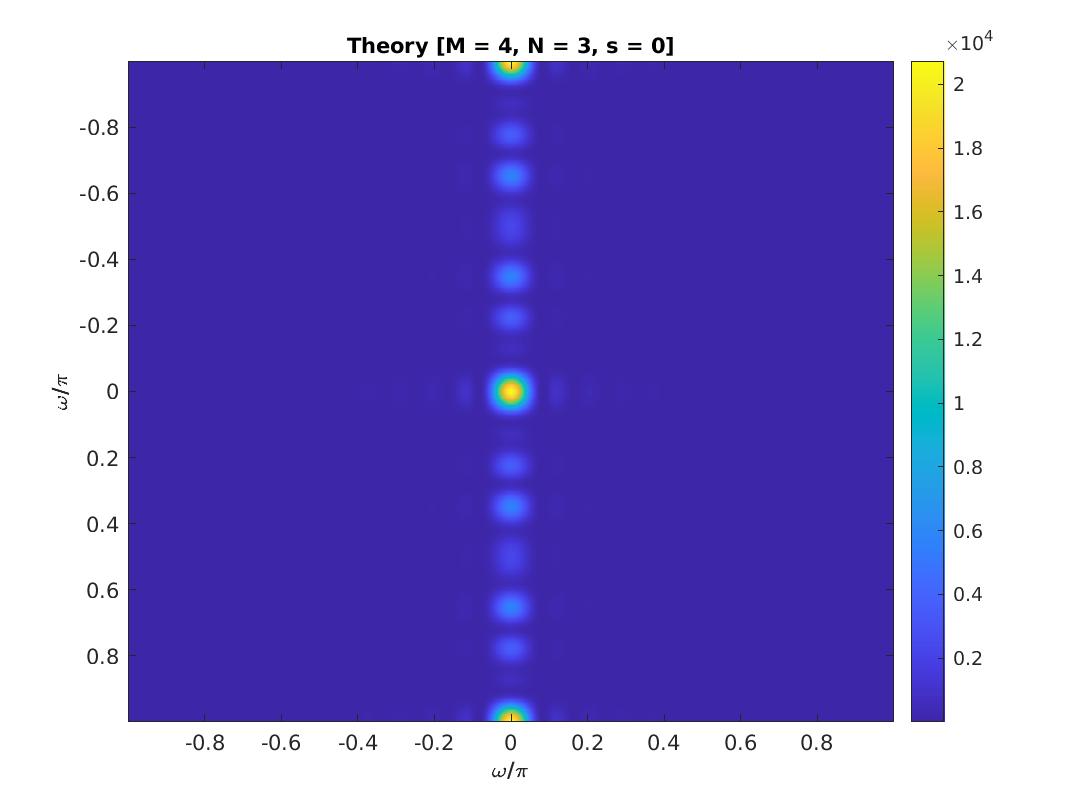}%
		\label{extreme_2DEx2_wt_bias_M4N3s0P1P2}}
	\hfil
	\subfloat[$M=4$, $N=3$, $s=1$]{
		\includegraphics[width=0.5\textwidth]{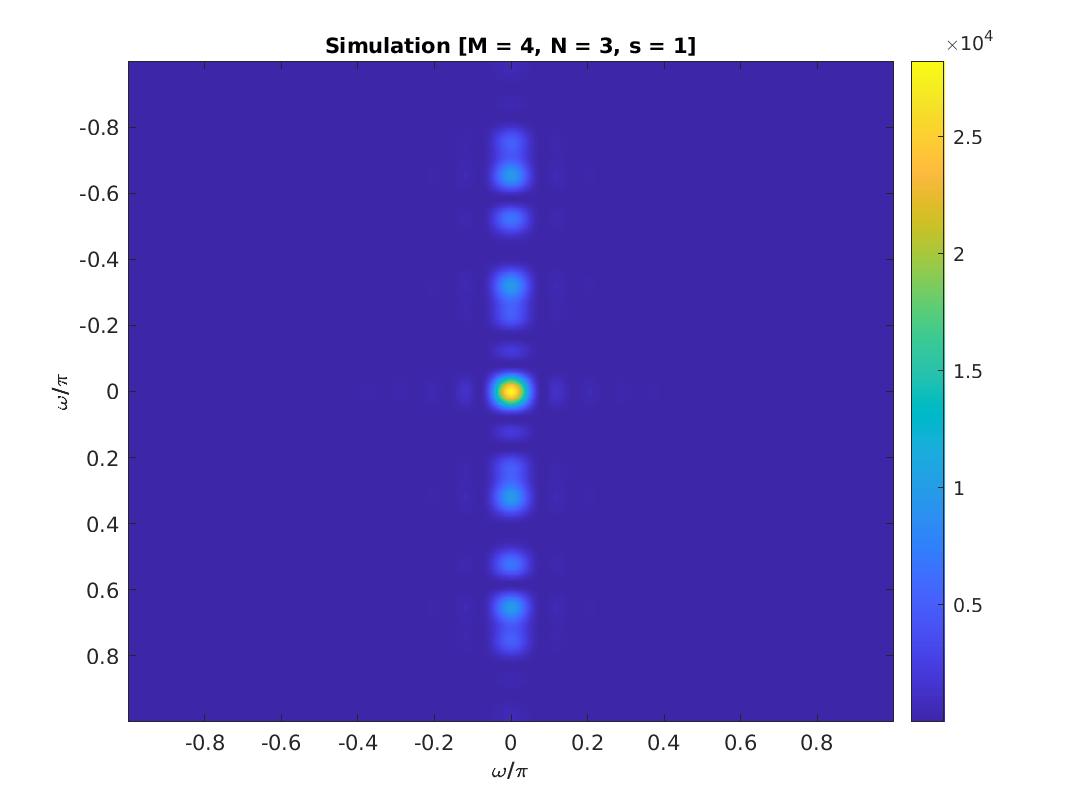}%
		\includegraphics[width=0.5\textwidth]{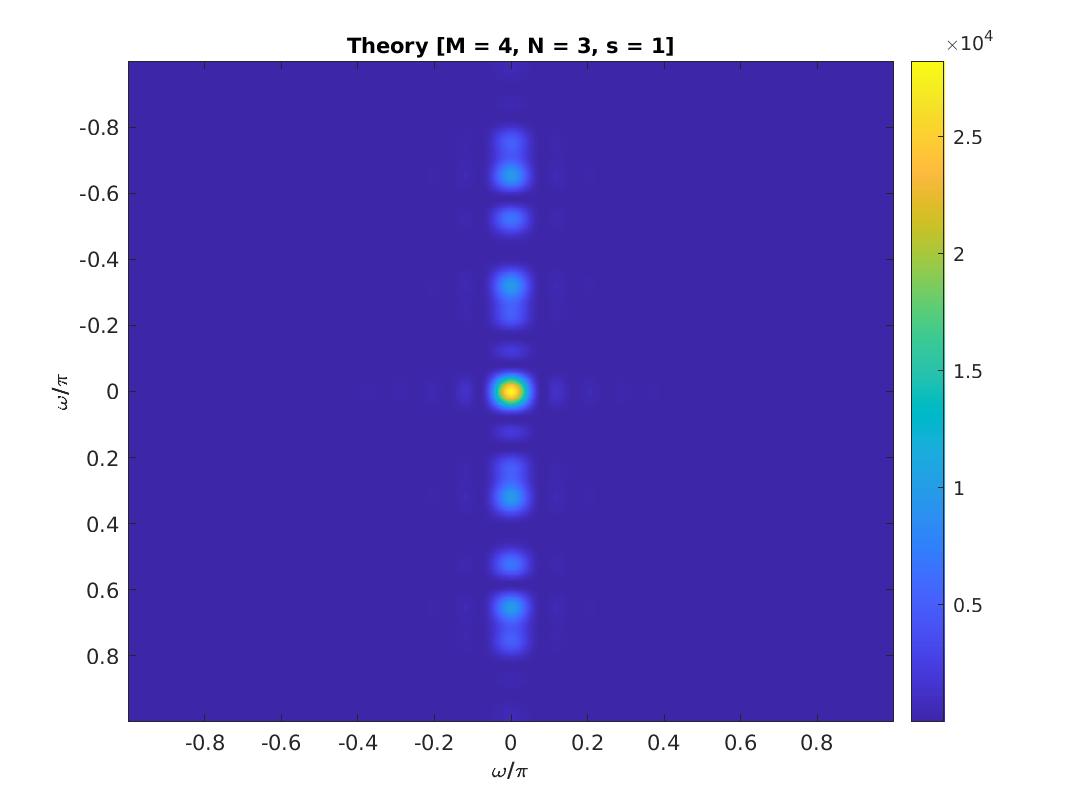}%
		\label{extreme_2DEx2_wt_bias_M4N3s1P1P2}}
	\hfil
	\subfloat[$M=4$, $N=3$, $s=2$]{
		\includegraphics[width=0.5\textwidth]{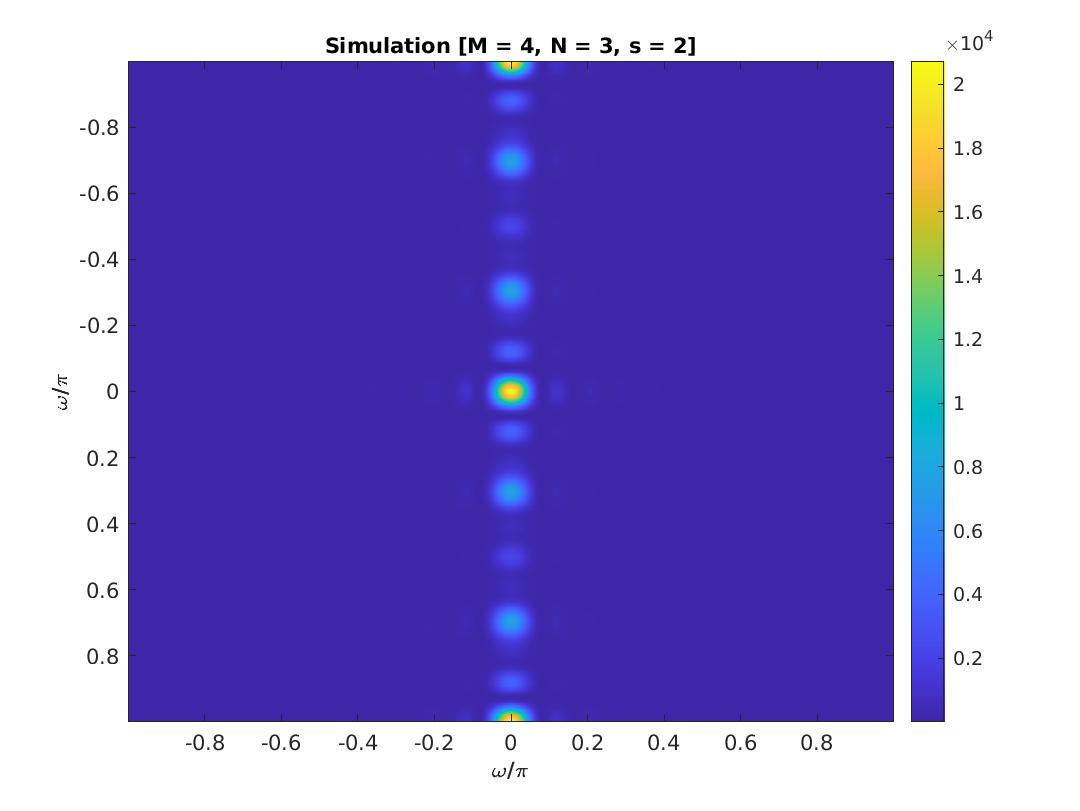}%
		\includegraphics[width=0.5\textwidth]{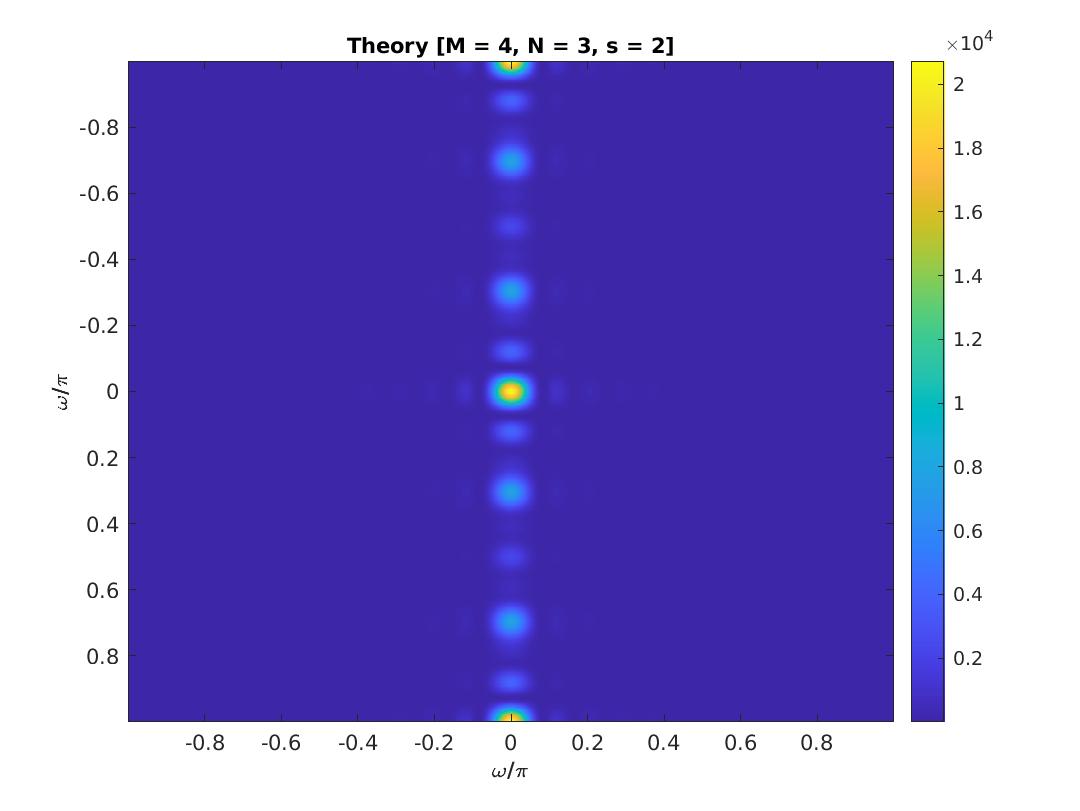}%
		\label{extreme_2DEx2_wt_bias_M4N3s2P1P2}}
	\caption{Simulated and theoretical 2D bias window of the correlogram estimator for Hybrid ExSCA with $\mathcal{E}_x=2$, $M=4$, $N=3$, $s=[0, 2]$.}
	\label{fig:extreme_2DEx2_wt_bias_M4N3s02P1P2}
\end{figure*}

\begin{figure*}[!t]
	\centering
	\subfloat[$M=4$, $N=3$, $s=3$]{
		\includegraphics[width=0.5\textwidth]{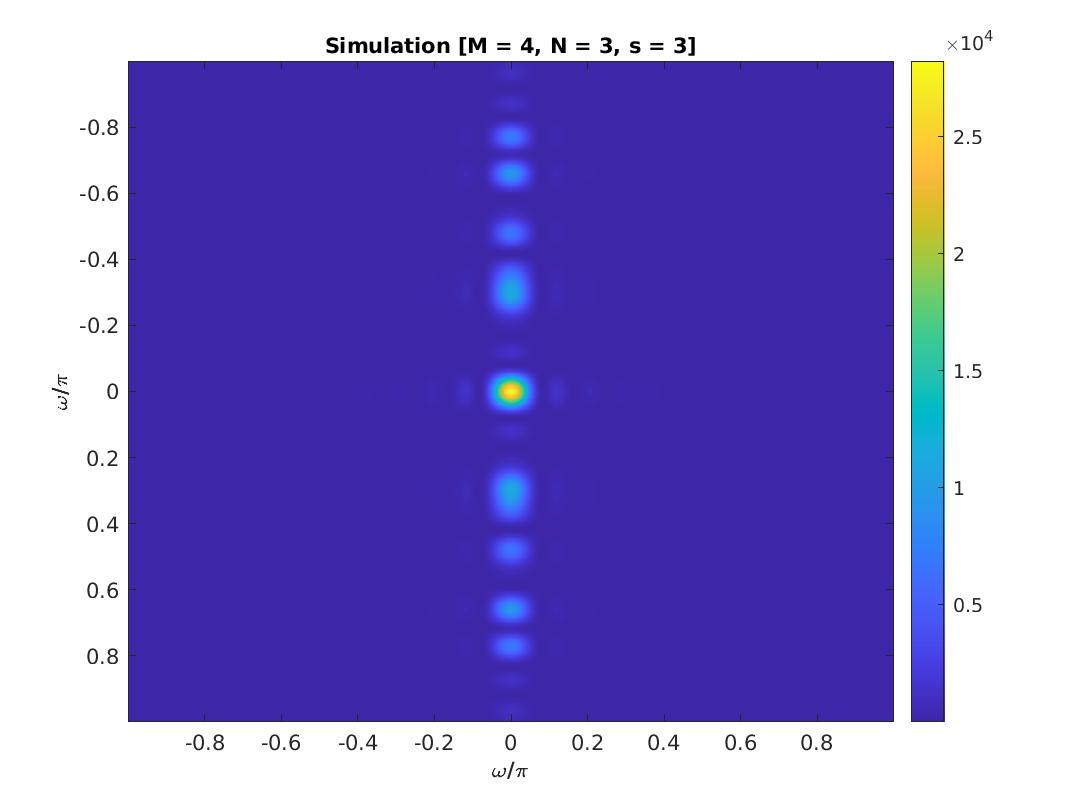}%
		\includegraphics[width=0.5\textwidth]{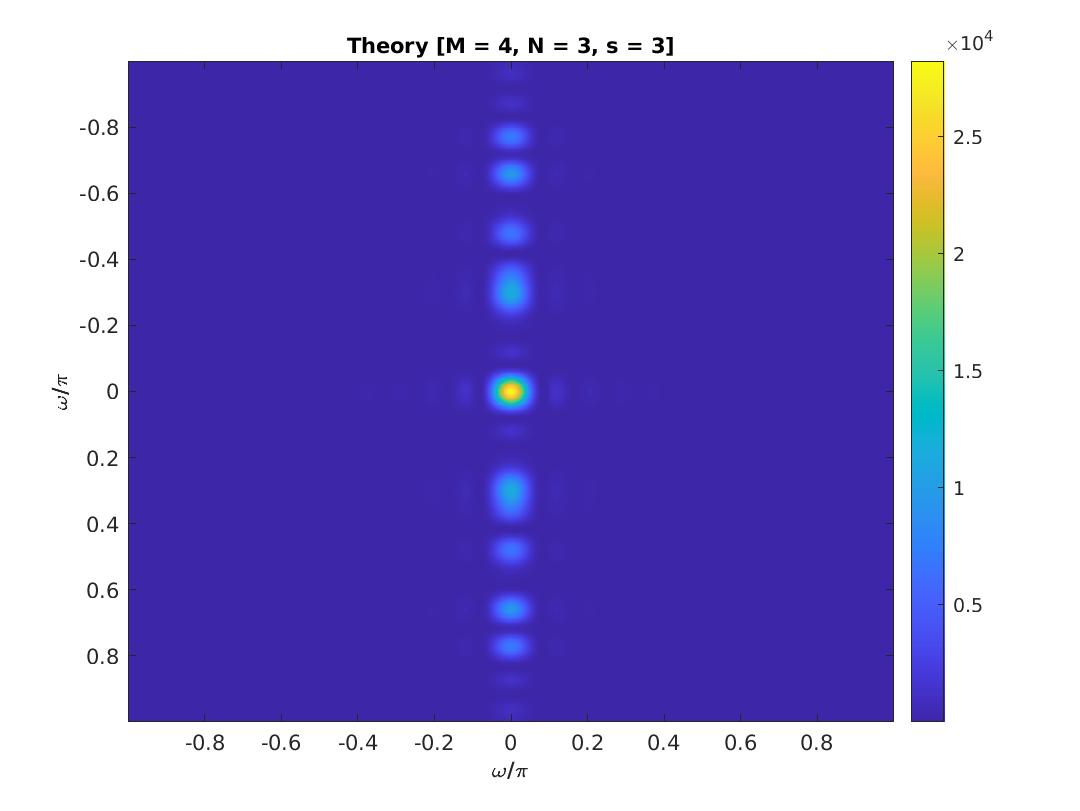}%
		\label{extreme_2DEx2_wt_bias_M4N3s3P1P2}}
	\hfil
	\subfloat[$M=4$, $N=3$, $s=4$]{
		\includegraphics[width=0.5\textwidth]{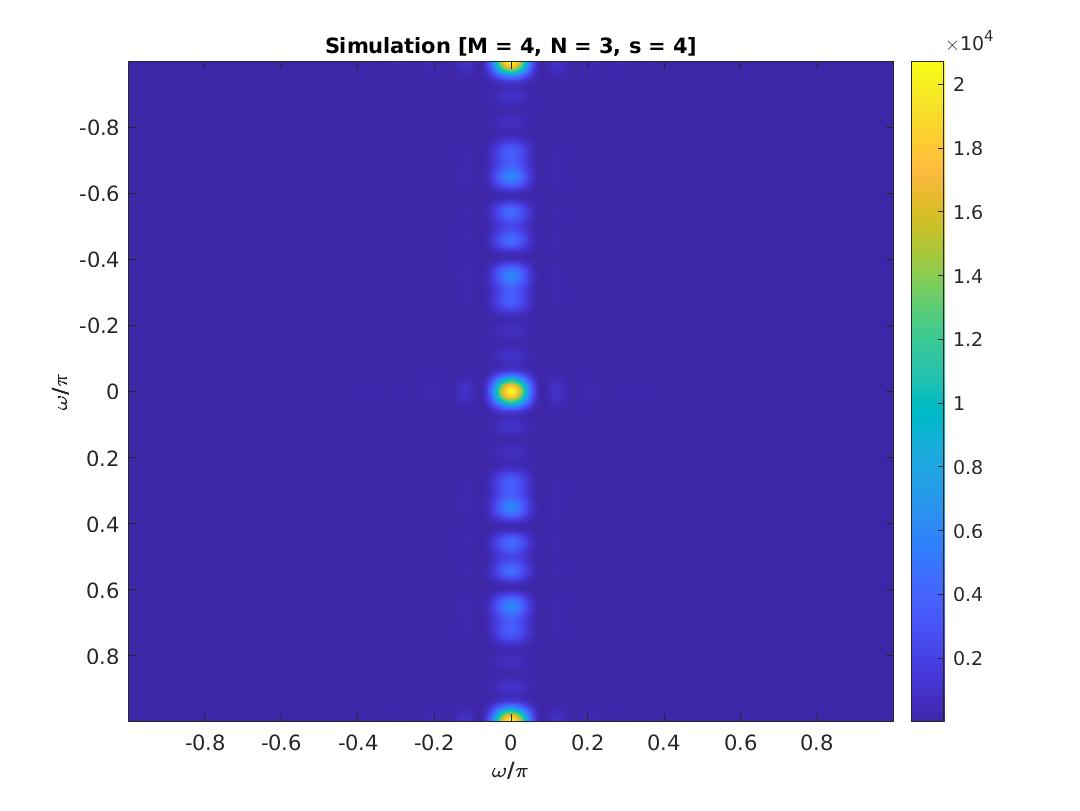}%
		\includegraphics[width=0.5\textwidth]{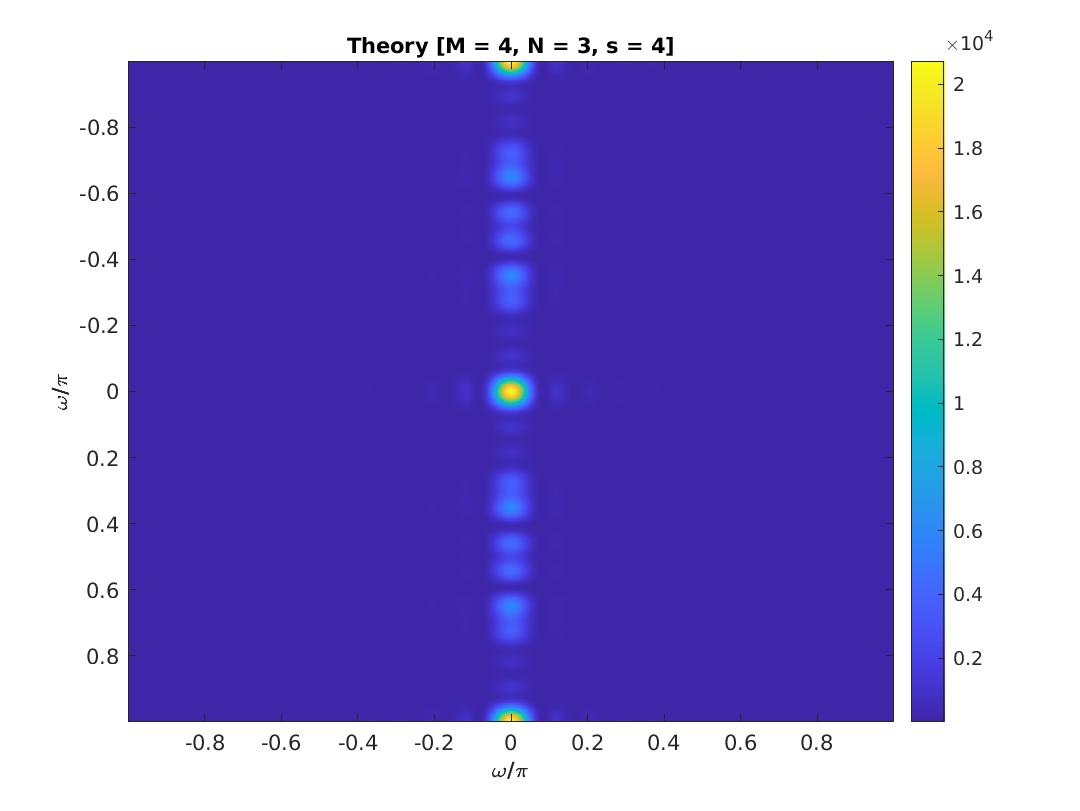}%
		\label{extreme_2DEx2_wt_bias_M4N3s4P1P2}}
	\hfil
	\subfloat[$M=4$, $N=3$, $s=5$]{
		\includegraphics[width=0.5\textwidth]{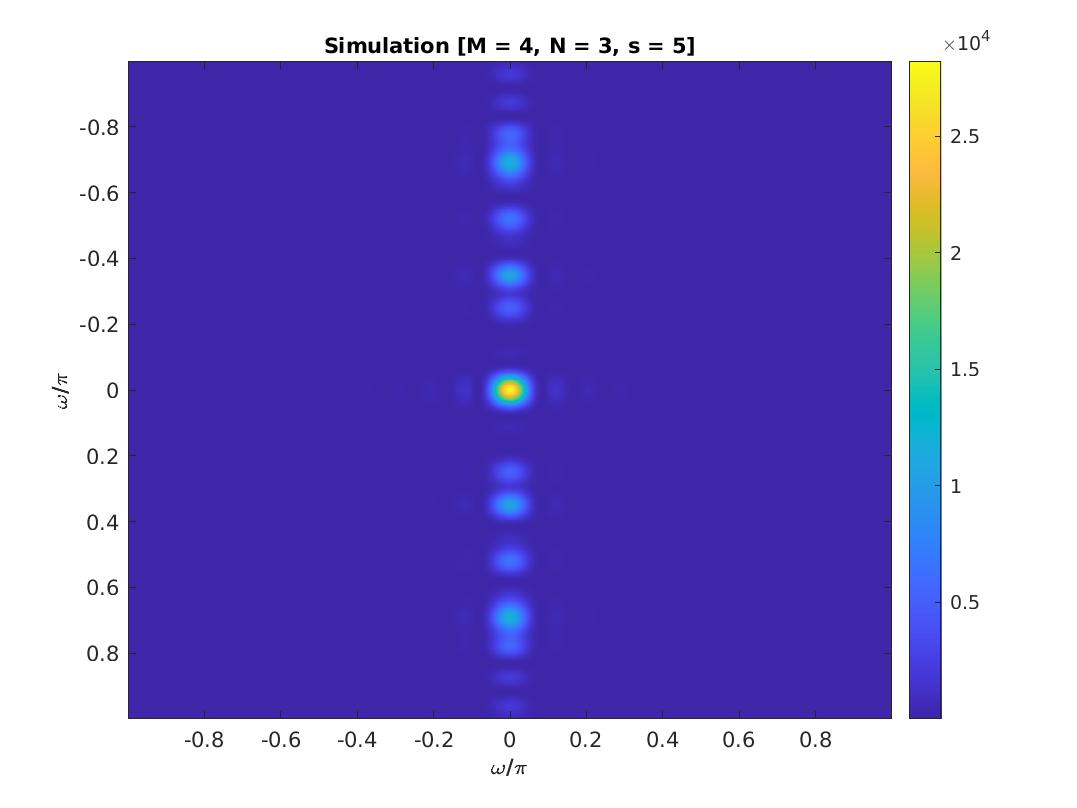}%
		\includegraphics[width=0.5\textwidth]{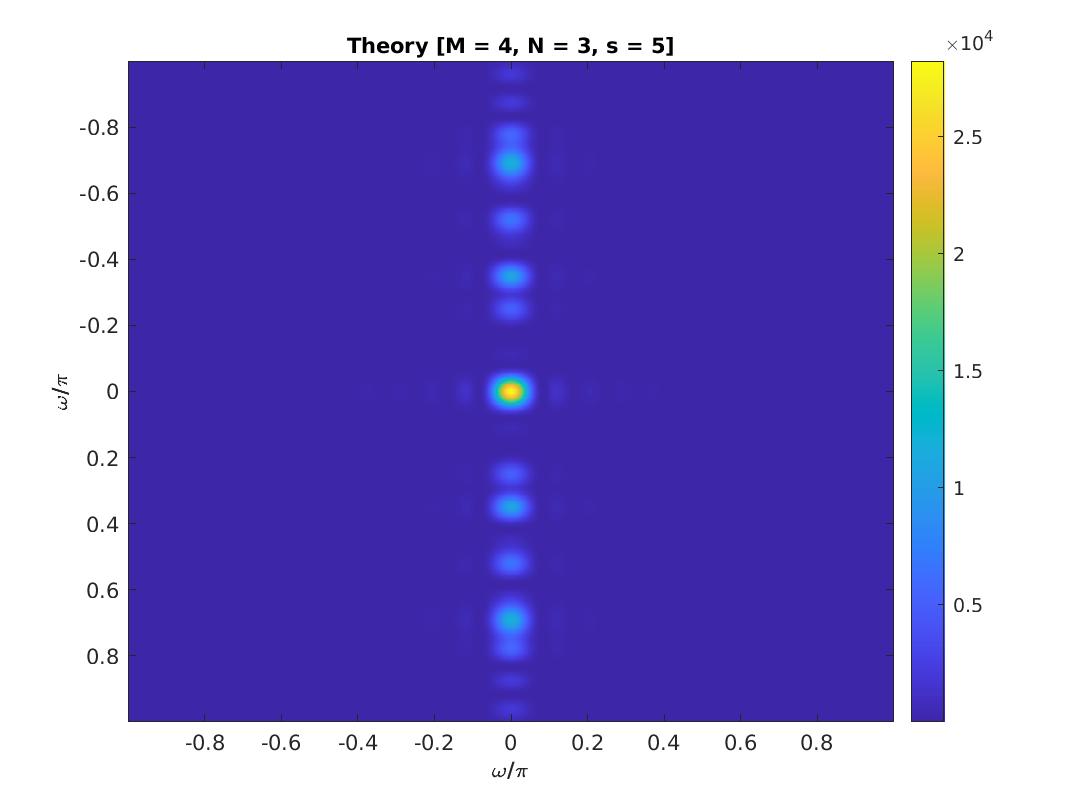}%
		\label{extreme_2DEx2_wt_bias_M4N3s5P1P2}}
	\hfil
	\caption{Simulated and theoretical 2D bias window of the correlogram estimator for Hybrid ExSCA with $\mathcal{E}_x=2$, $M=4$, $N=3$, $s=[3, 5]$.}
	\label{fig:extreme_2DEx2_wt_bias_M4N3s35P1P2}
\end{figure*}

\begin{figure*}[!t]
	\centering
	\subfloat[$M=4$, $N=3$, $s=0$]{
		\includegraphics[width=0.5\textwidth]{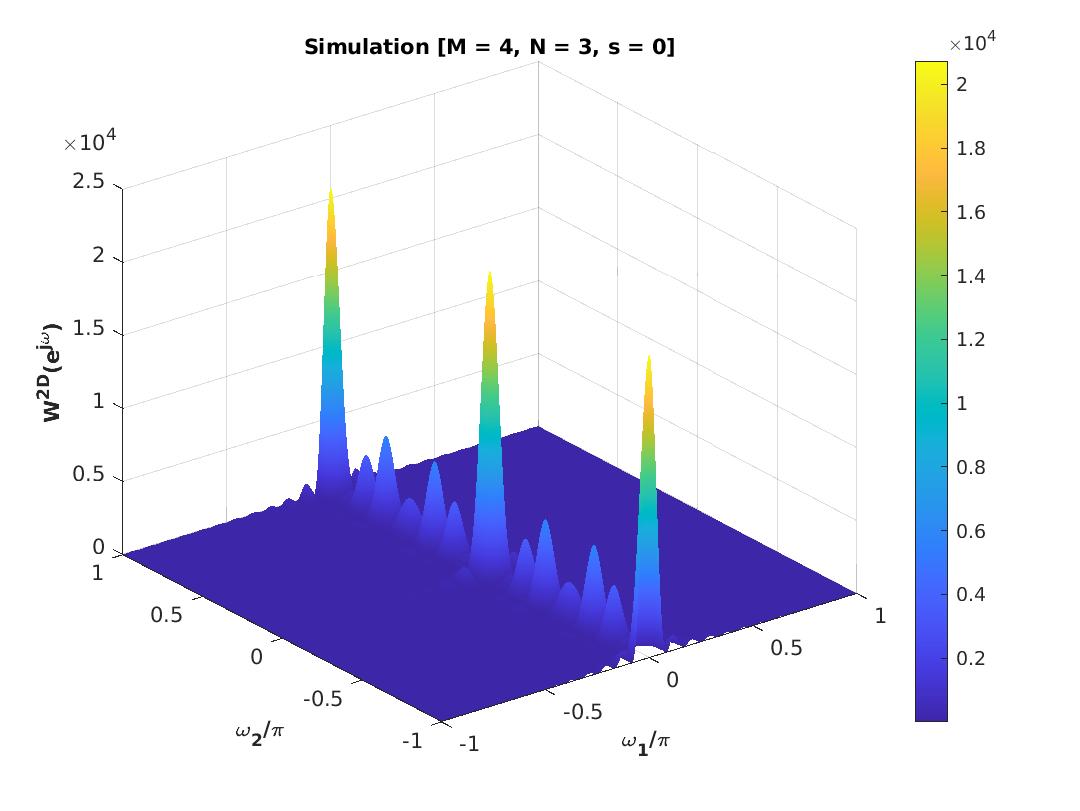}%
		\includegraphics[width=0.5\textwidth]{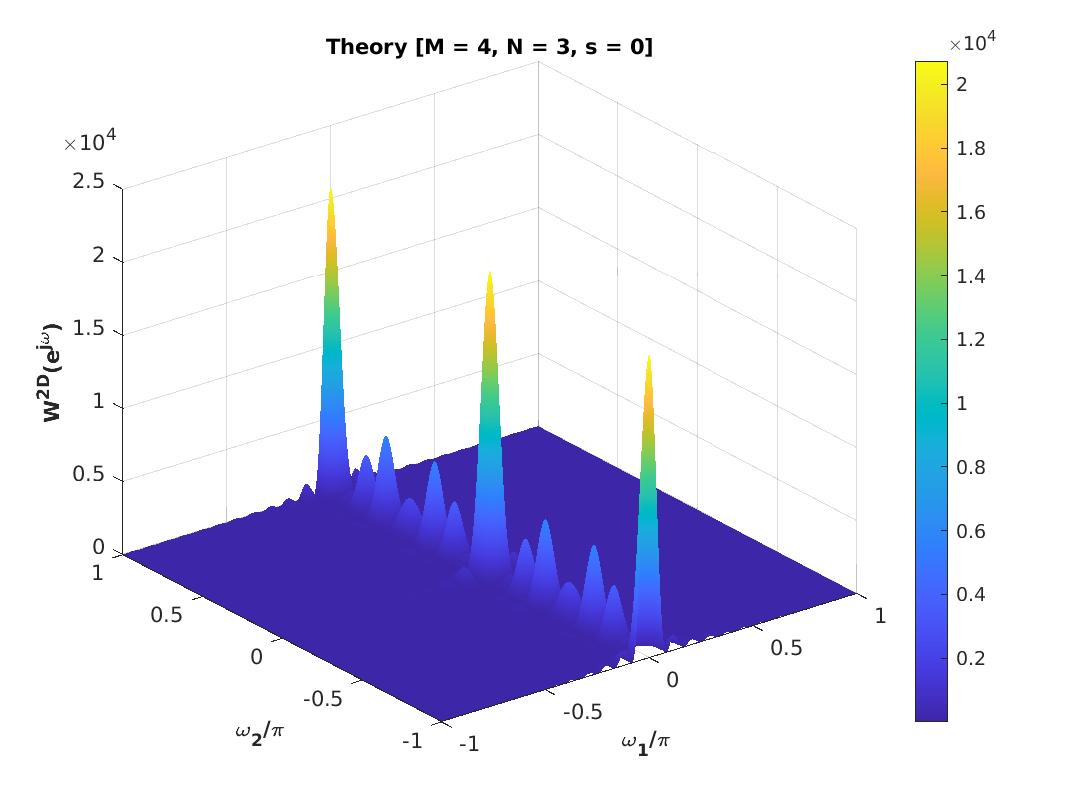}%
		\label{extreme_2DEx2_wt_bias_M4N3s0SurfP1P2}}
	\hfil
	\subfloat[$M=4$, $N=3$, $s=1$]{
		\includegraphics[width=0.5\textwidth]{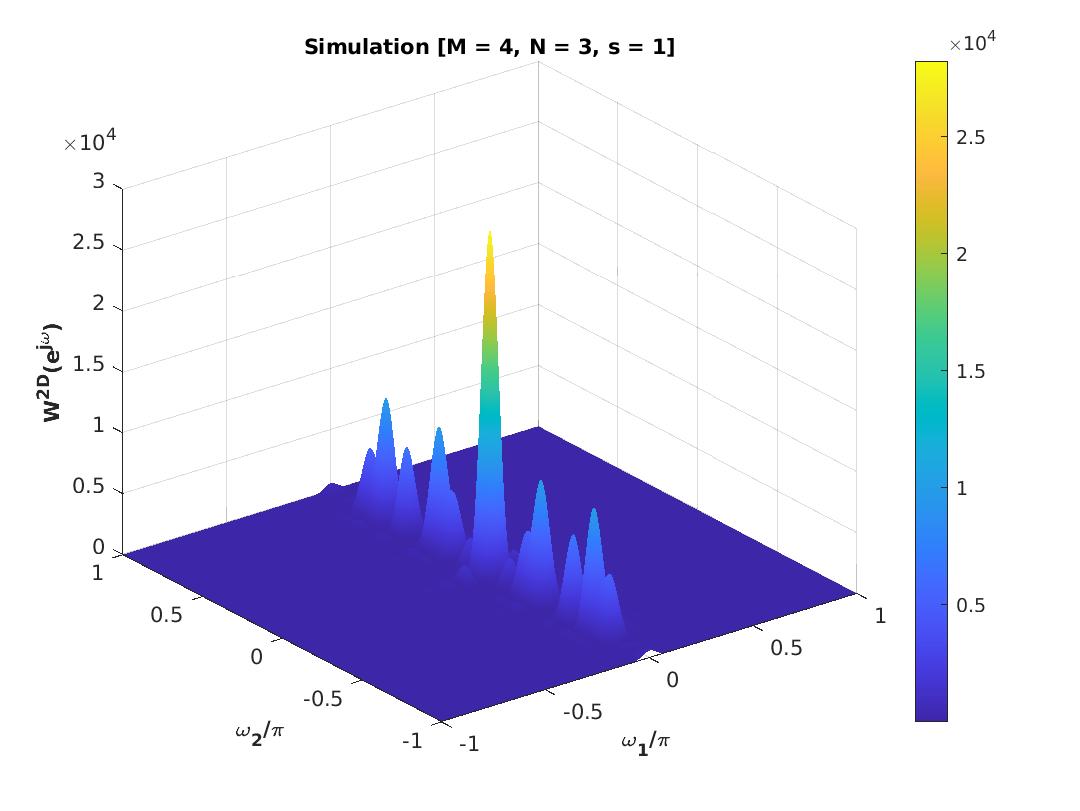}%
		\includegraphics[width=0.5\textwidth]{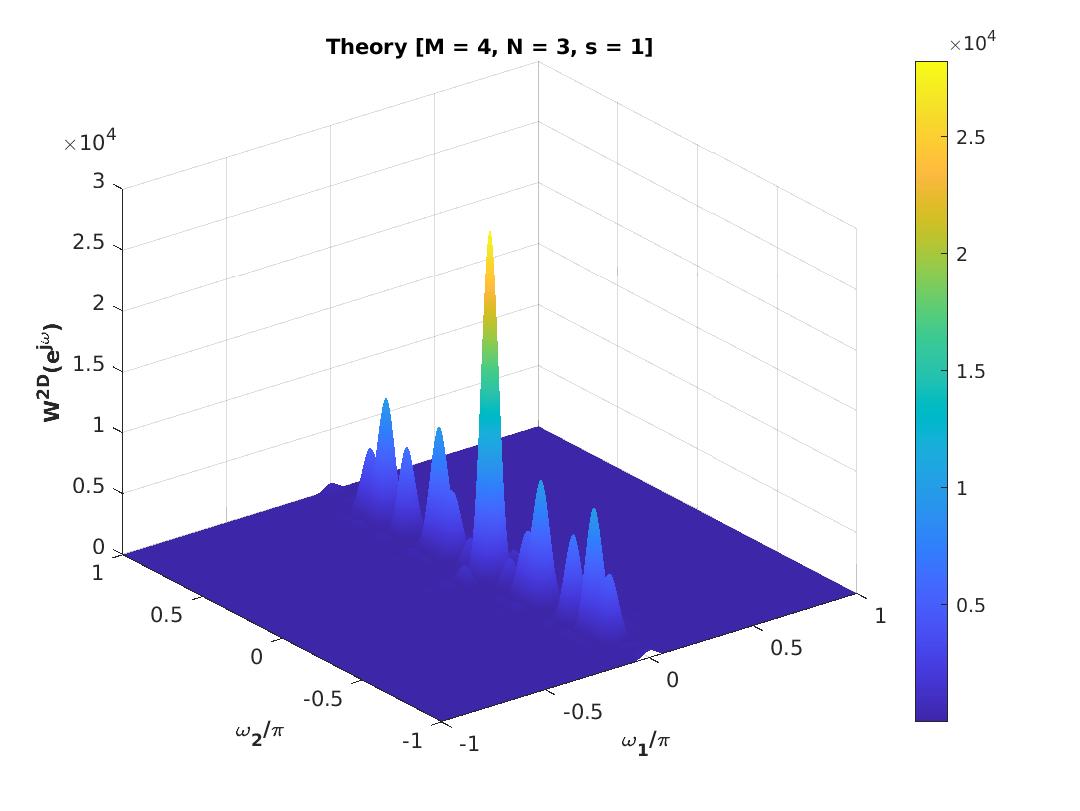}%
		\label{extreme_2DEx2_wt_bias_M4N3s1SurfP1P2}}
	\hfil
	\subfloat[$M=4$, $N=3$, $s=2$]{
		\includegraphics[width=0.5\textwidth]{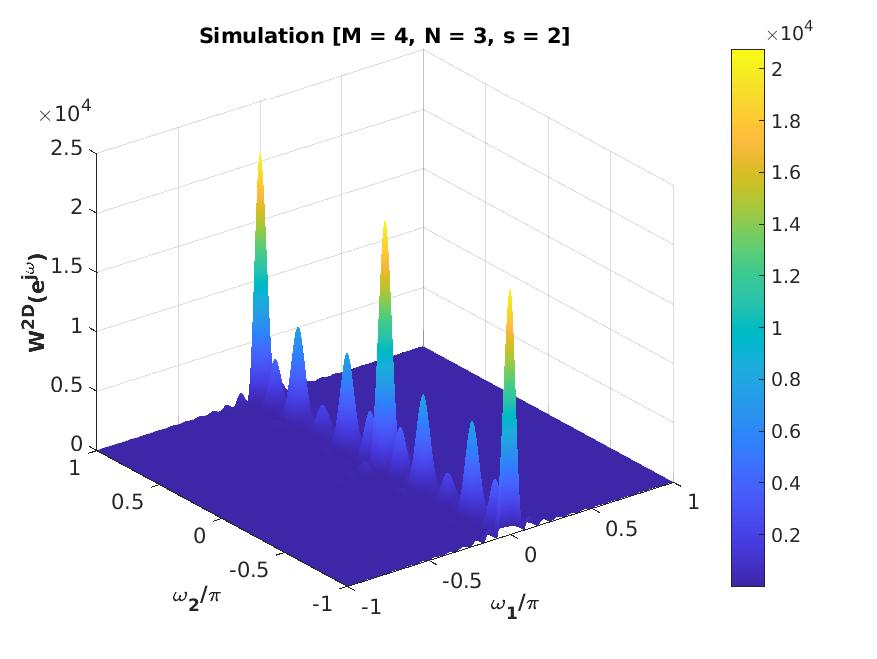}%
		\includegraphics[width=0.5\textwidth]{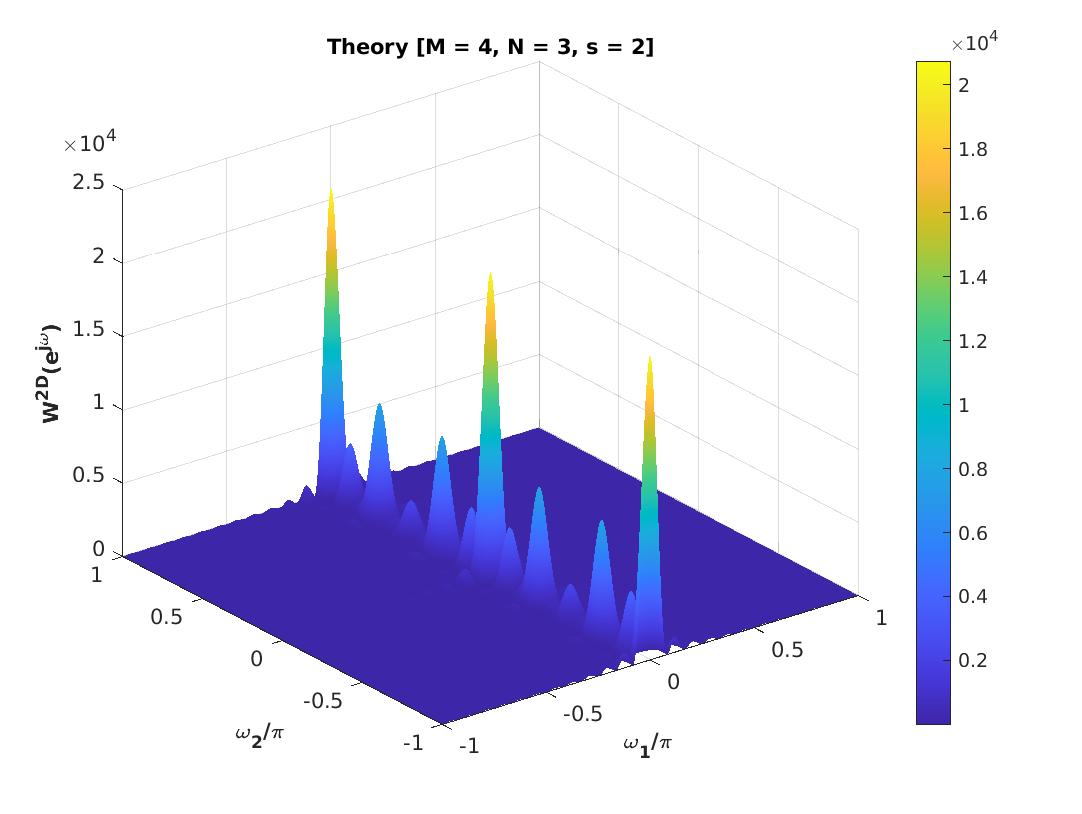}%
		\label{extreme_2DEx2_wt_bias_M4N3s2SurfP1P2}}
	\caption{Simulated and theoretical 2D bias window: Surface plot for Fig.~\ref{fig:extreme_2DEx2_wt_bias_M4N3s02P1P2}.}
	\label{fig:extreme_2DEx2_wt_bias_M4N3s02SurfP1P2}
\end{figure*}

\begin{figure*}[!t]
	\centering
	\subfloat[$M=4$, $N=3$, $s=3$]{
		\includegraphics[width=0.5\textwidth]{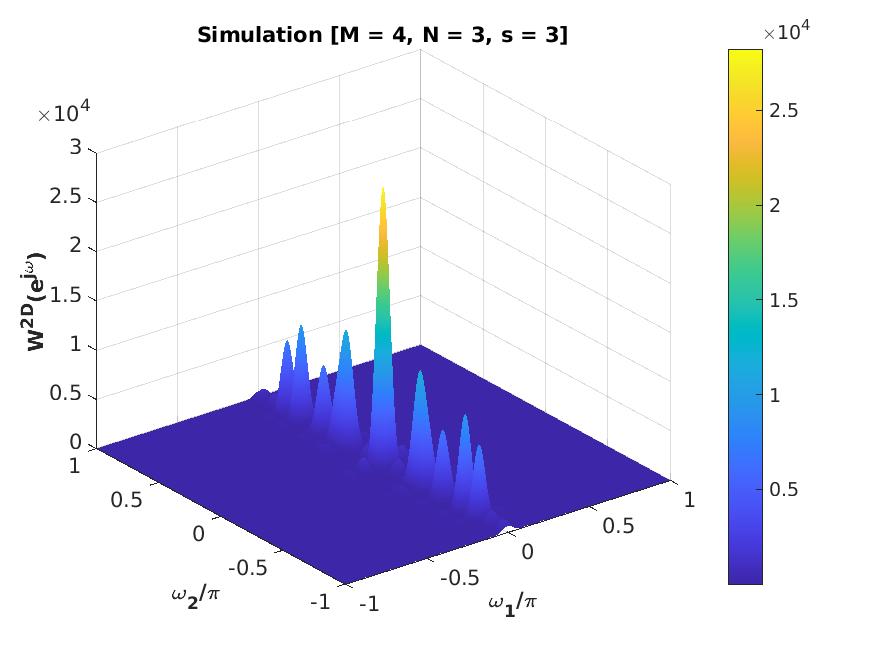}%
		\includegraphics[width=0.5\textwidth]{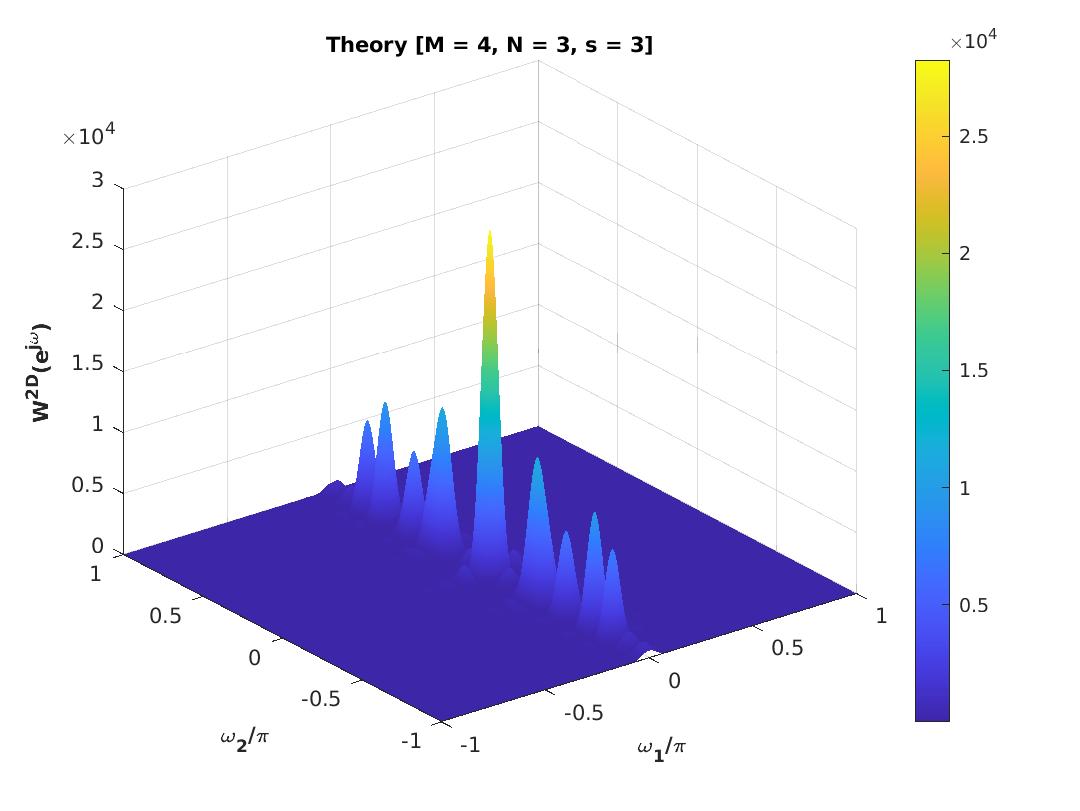}%
		\label{extreme_2DEx2_wt_bias_M4N3s3SurfP1P2}}
	\hfil
	\subfloat[$M=4$, $N=3$, $s=4$]{
		\includegraphics[width=0.5\textwidth]{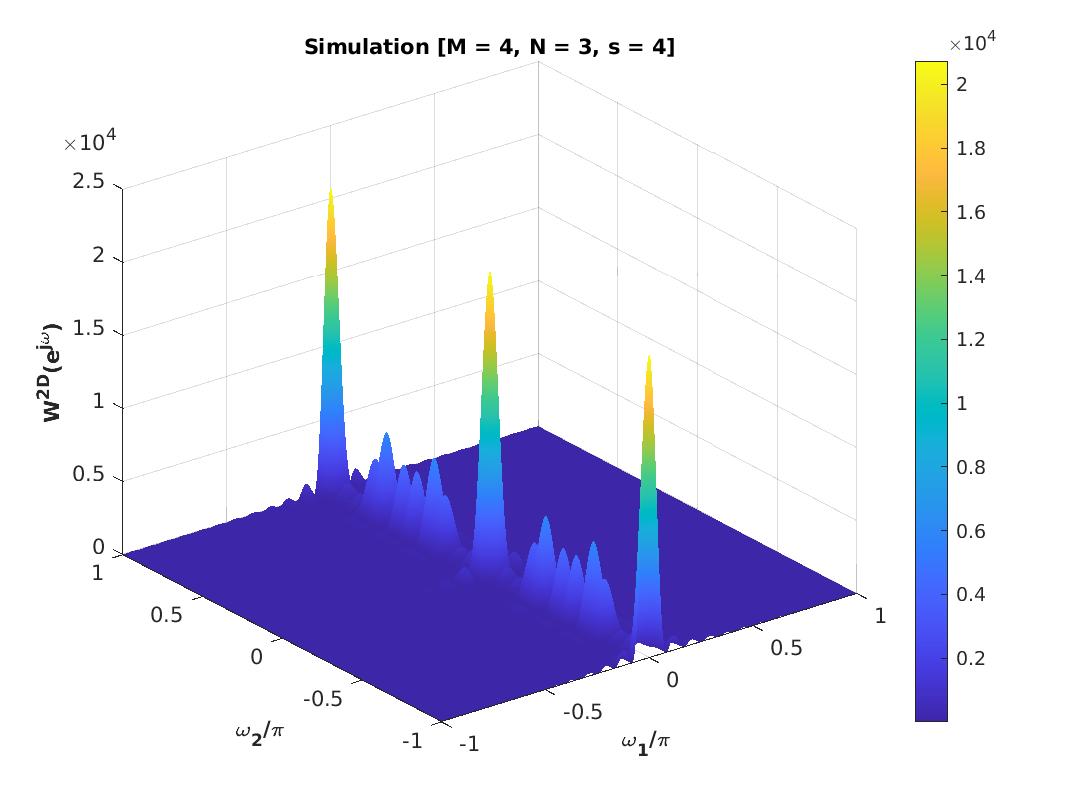}%
		\includegraphics[width=0.5\textwidth]{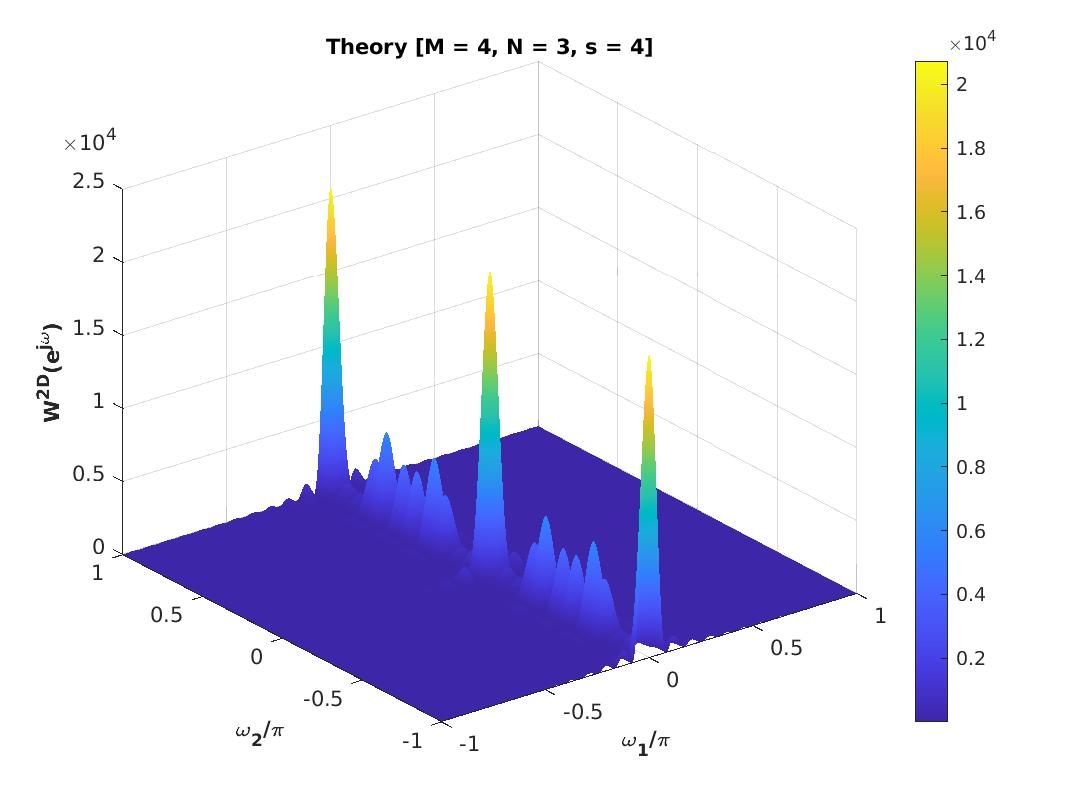}%
		\label{extreme_2DEx2_wt_bias_M4N3s4SurfP1P2}}
	\hfil
	\subfloat[$M=4$, $N=3$, $s=5$]{
		\includegraphics[width=0.5\textwidth]{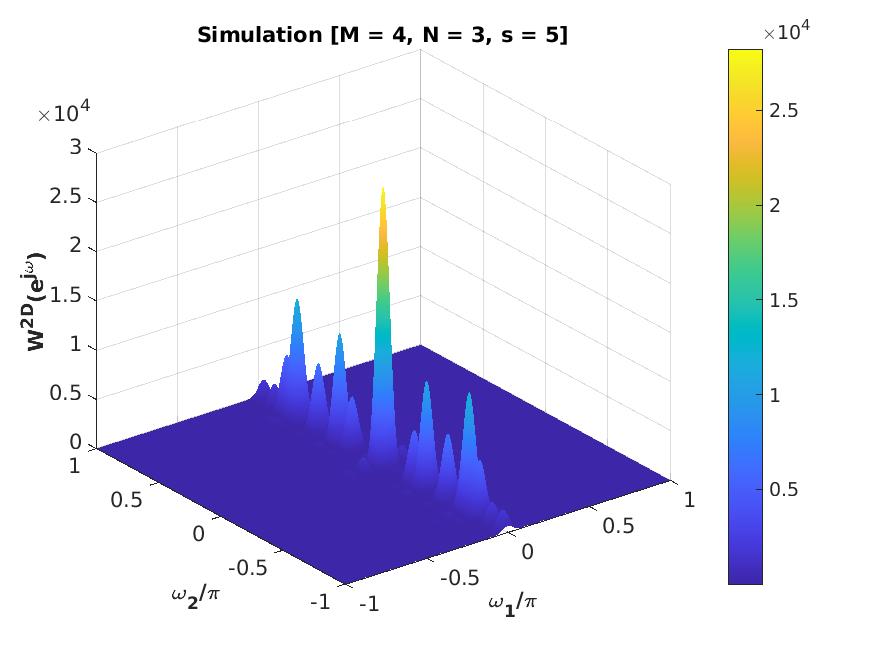}%
		\includegraphics[width=0.5\textwidth]{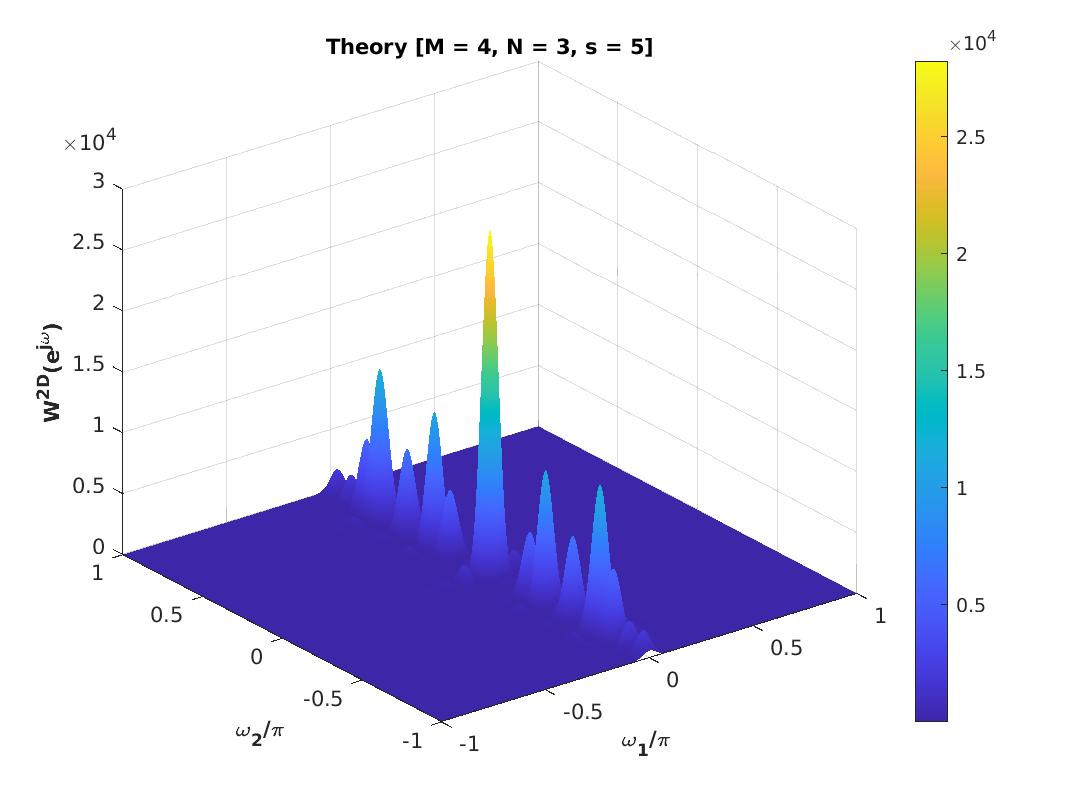}%
		\label{extreme_2DEx2_wt_bias_M4N3s5SurfP1P2}}
	\hfil
	\caption{Simulated and theoretical 2D bias window: Surface plot for Fig.~\ref{fig:extreme_2DEx2_wt_bias_M4N3s35P1P2}.}
	\label{fig:extreme_2DEx2_wt_bias_M4N3s35SurfP1P2}
\end{figure*}

\begin{figure*}[!t]
	\centering
	\includegraphics[width=0.49\textwidth]{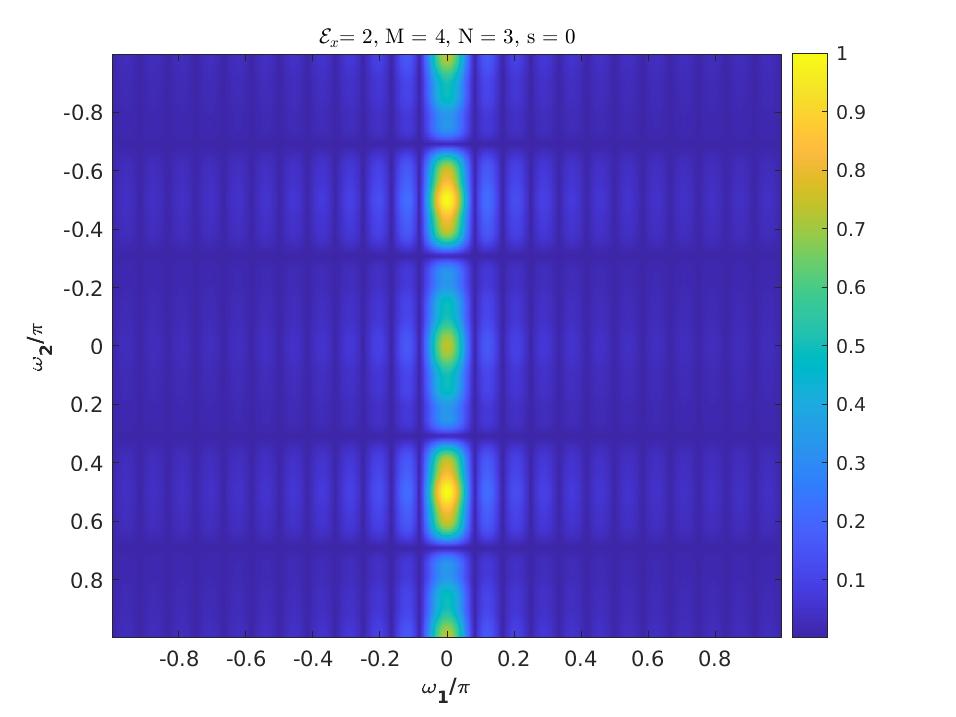}
	\includegraphics[width=0.49\textwidth]{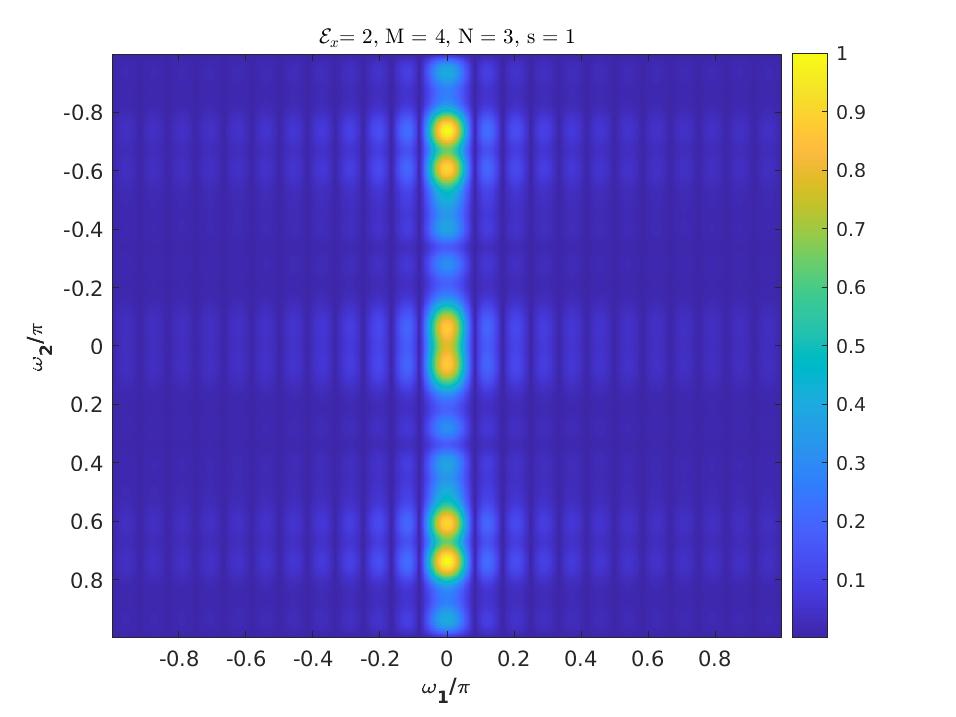}\\
	\includegraphics[width=0.5\textwidth]{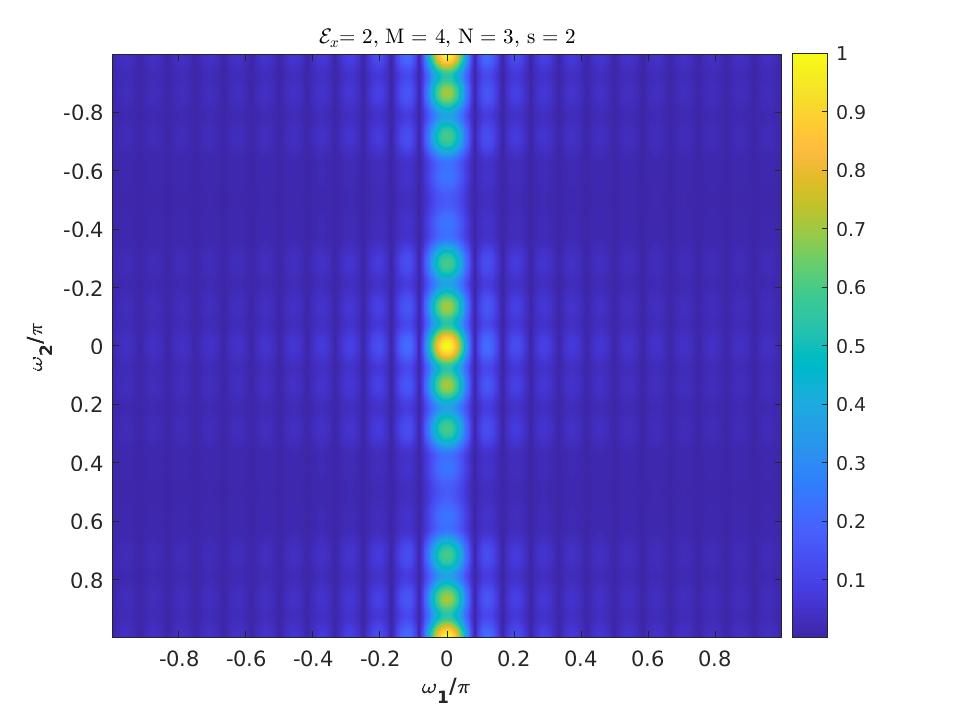}%
	\includegraphics[width=0.5\textwidth]{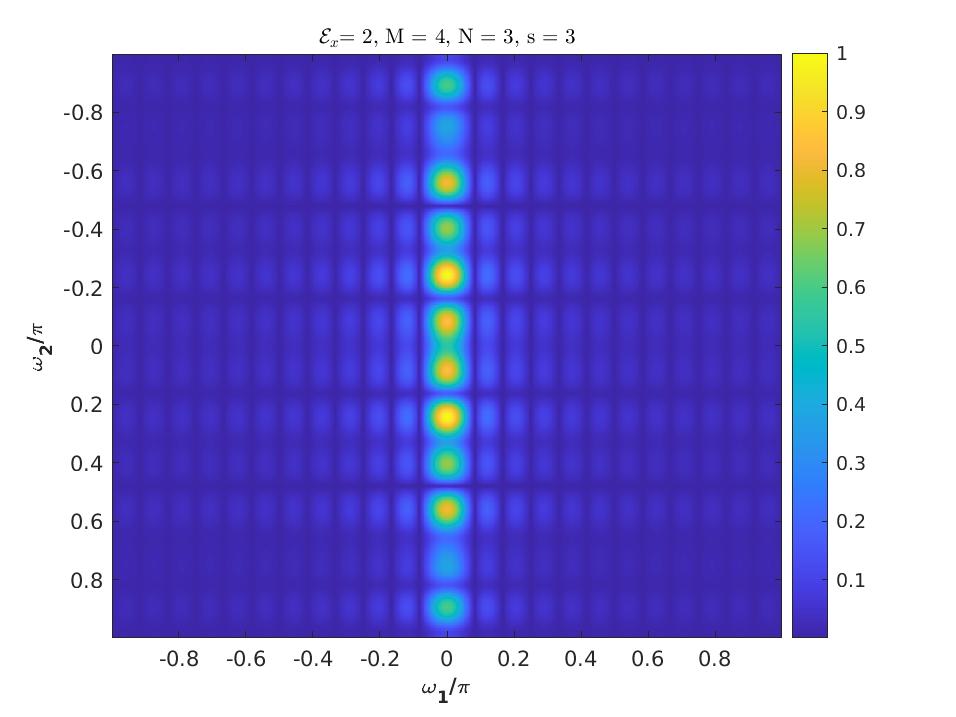}\\
	\includegraphics[width=0.5\textwidth]{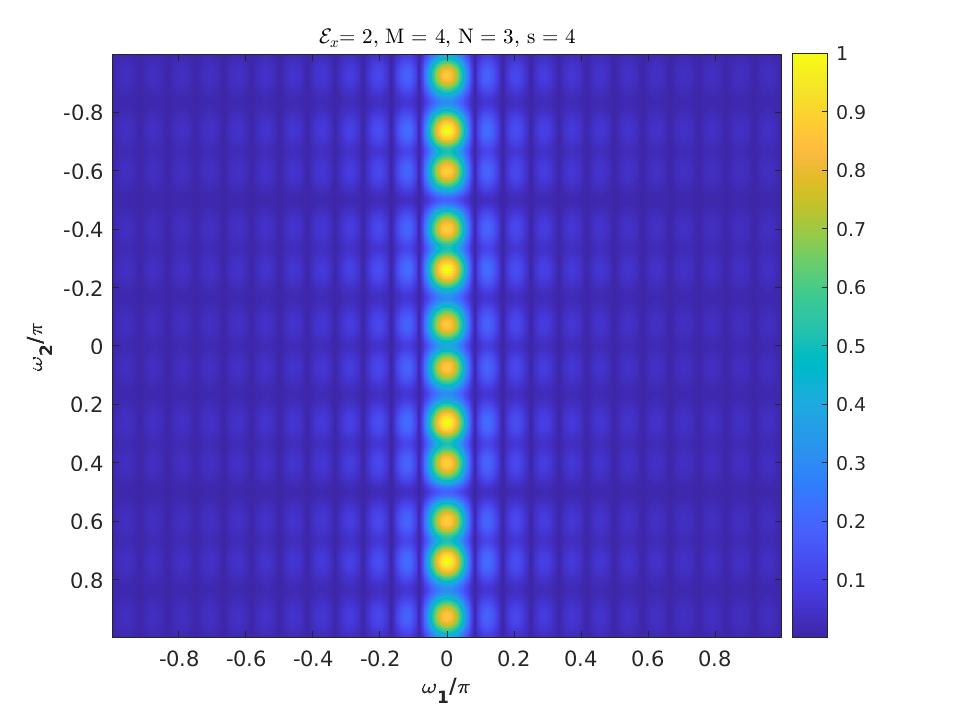}%
	\includegraphics[width=0.5\textwidth]{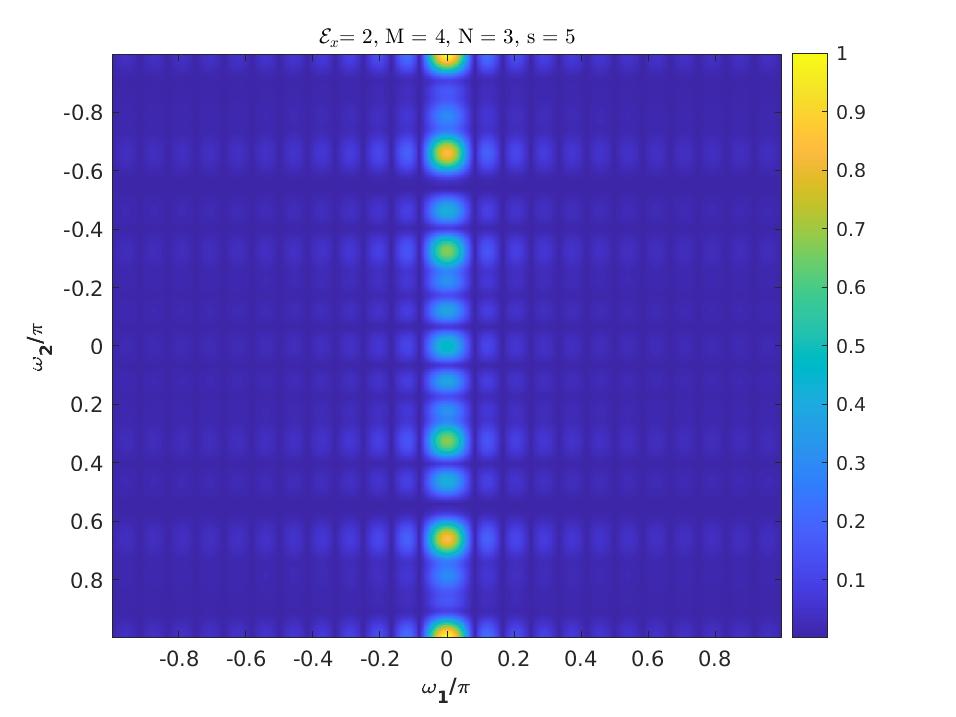}
	\caption{2D Hybrid ExSCA based spectral estimation for Fig.~\ref{fig:extreme_2D_true_spectV}.}
	\label{fig:extreme_2D_M4N3VP1P2}
\end{figure*}

\begin{figure*}[!t]
	\centering
	\includegraphics[width=0.49\textwidth]{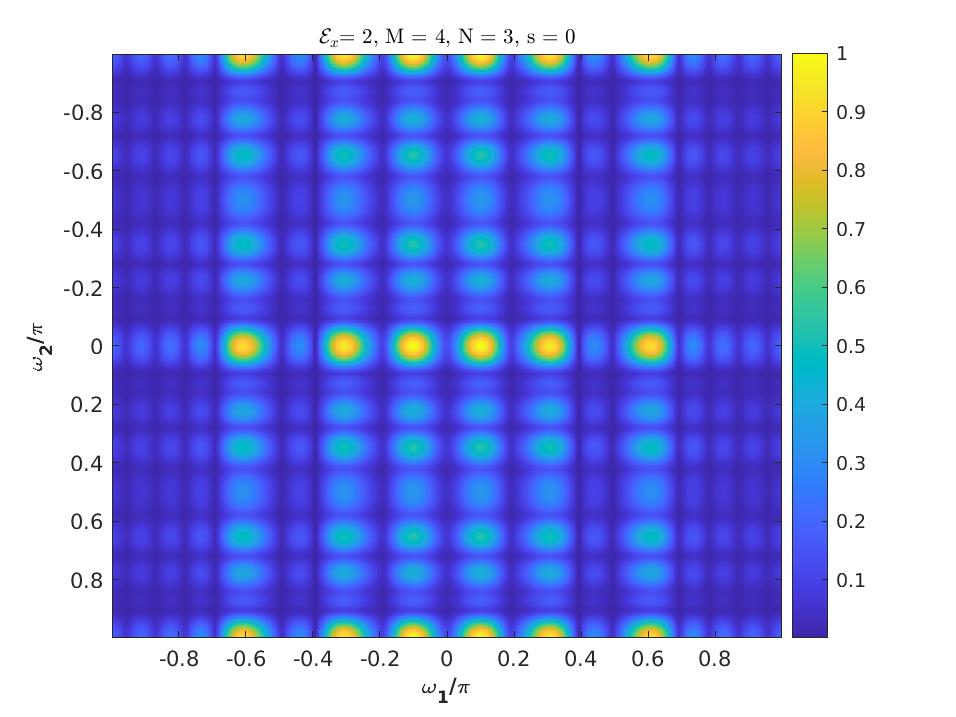}
	\includegraphics[width=0.49\textwidth]{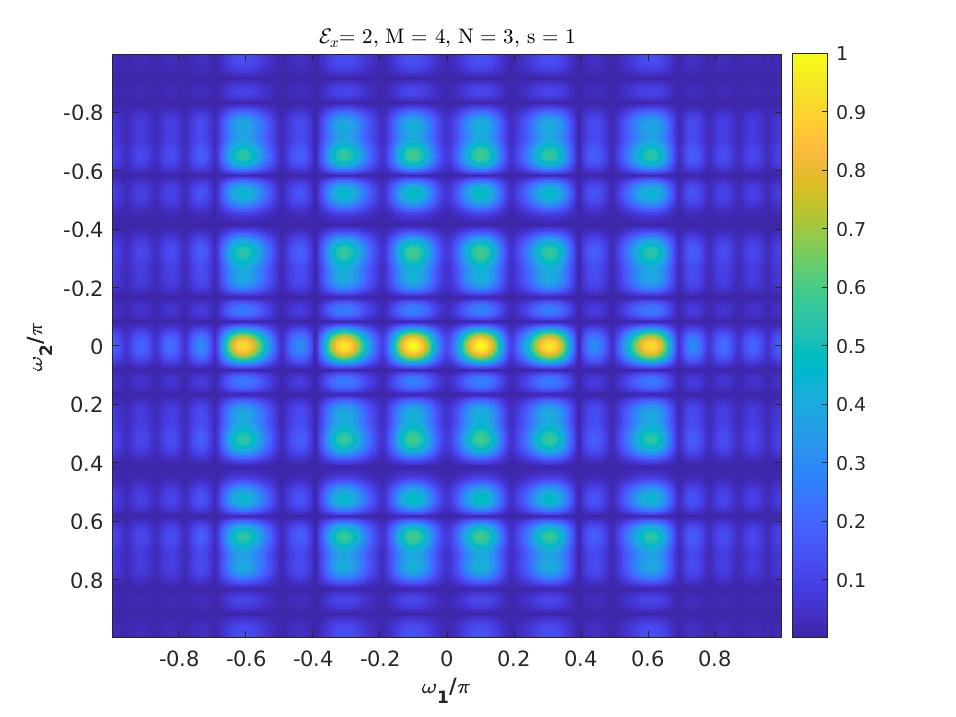}\\
	\includegraphics[width=0.5\textwidth]{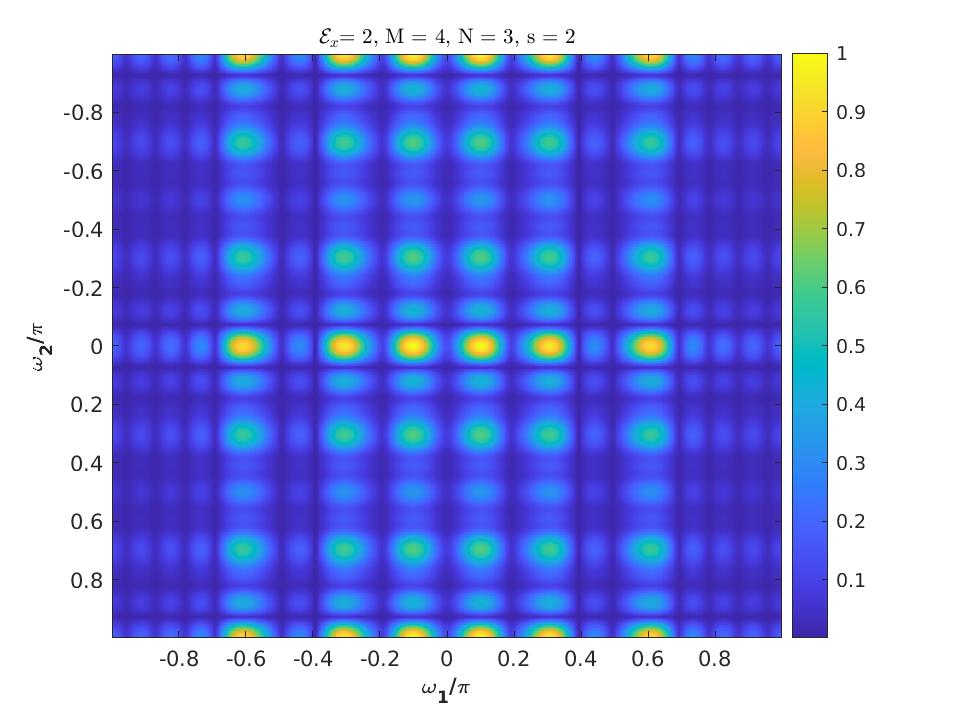}%
	\includegraphics[width=0.5\textwidth]{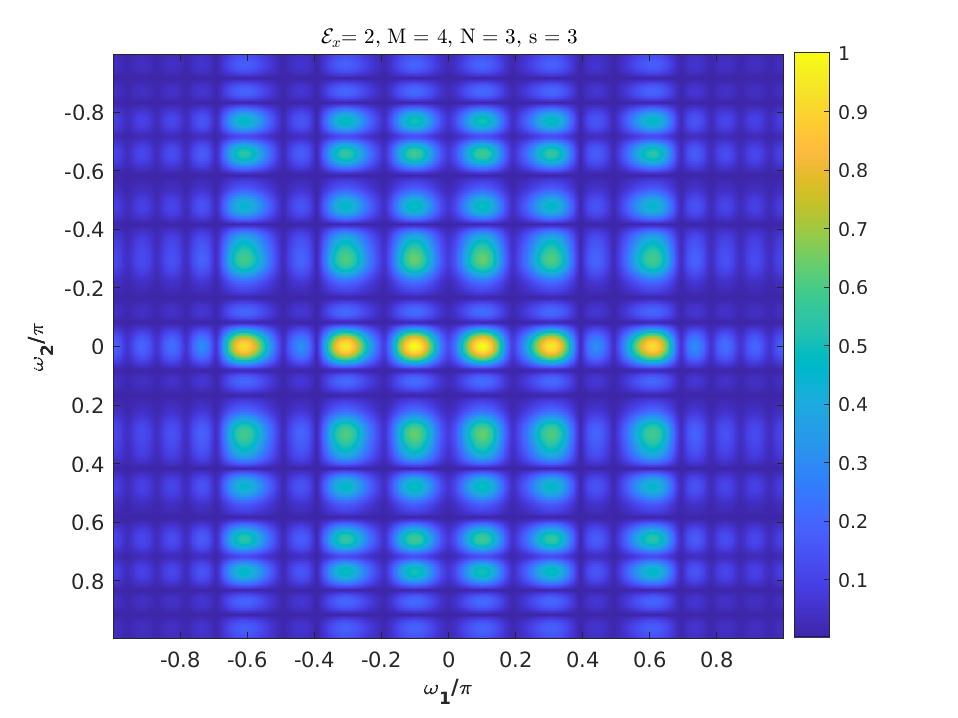}\\
	\includegraphics[width=0.5\textwidth]{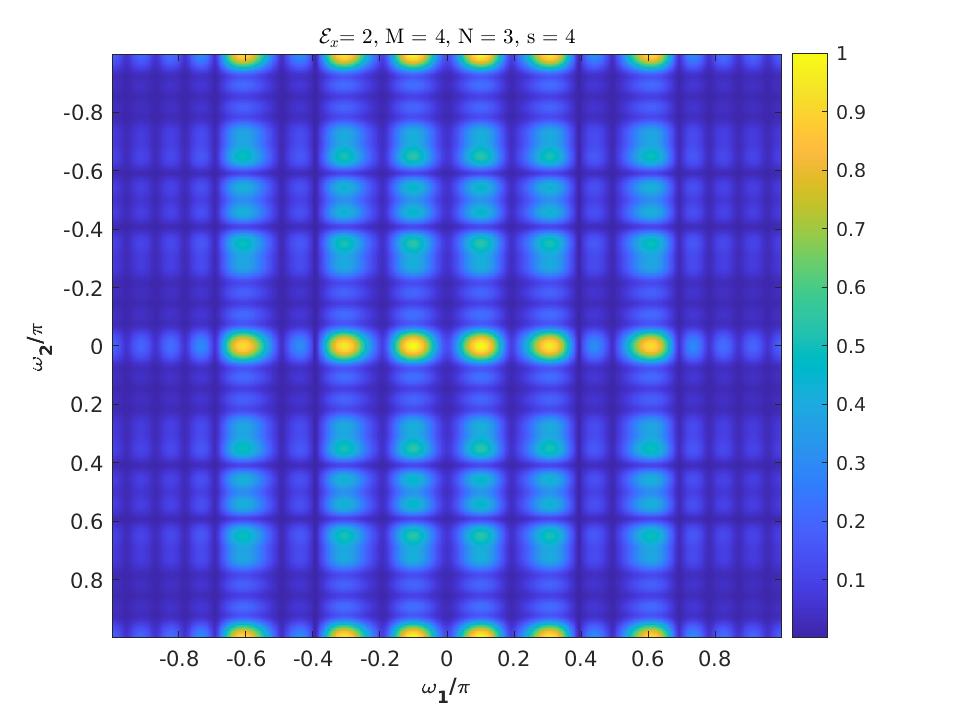}%
	\includegraphics[width=0.5\textwidth]{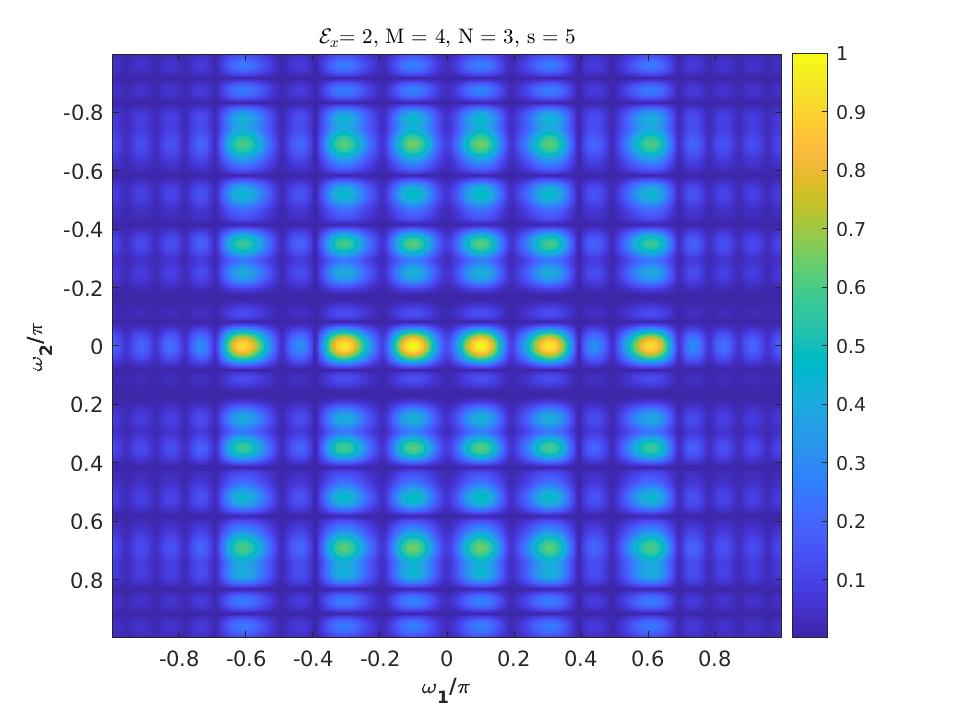}
	\caption{2D Hybrid ExSCA based spectral estimation for Fig.~\ref{fig:extreme_2D_true_spectH}.}
	\label{fig:extreme_2D_M4N3HP1P2}
\end{figure*}

\begin{figure*}[!t]
	\centering
	\includegraphics[width=0.49\textwidth]{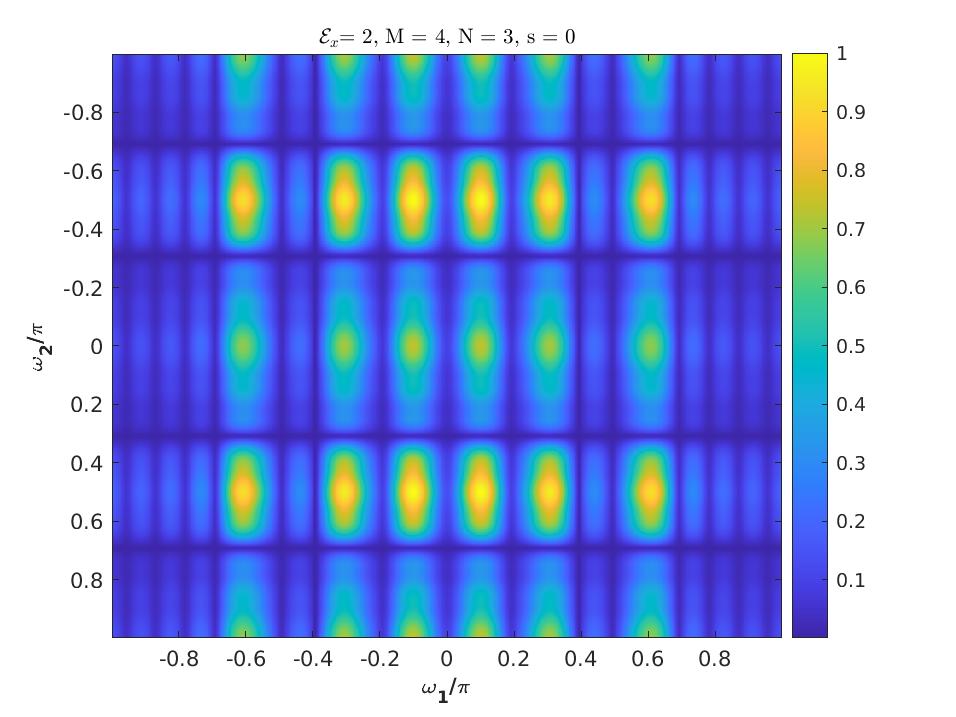}
	\includegraphics[width=0.49\textwidth]{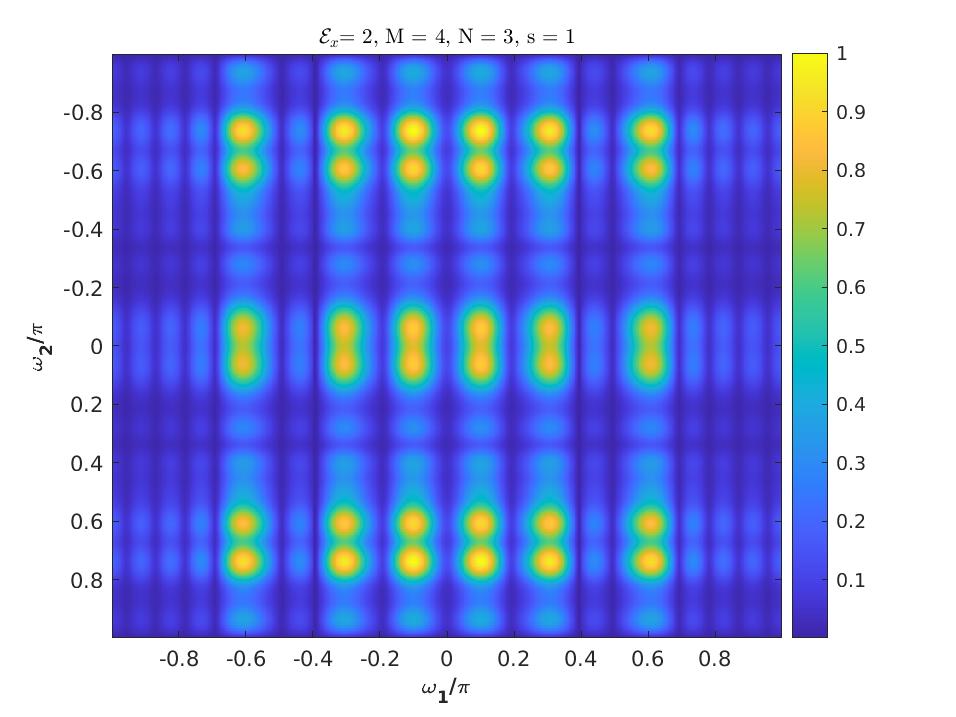}\\
	\includegraphics[width=0.5\textwidth]{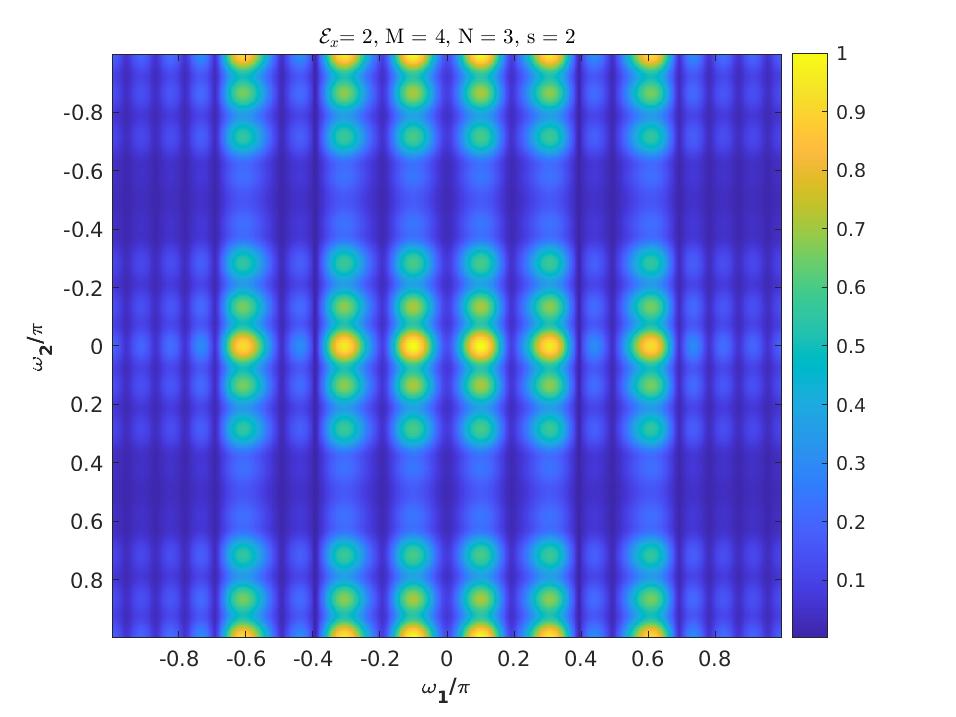}%
	\includegraphics[width=0.5\textwidth]{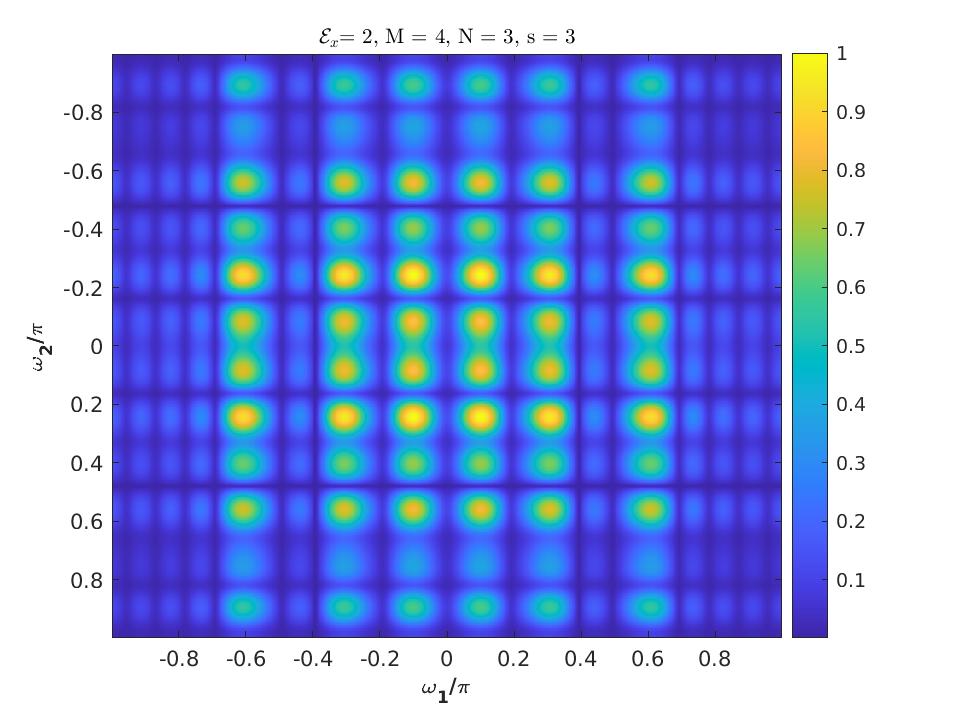}\\
	\includegraphics[width=0.5\textwidth]{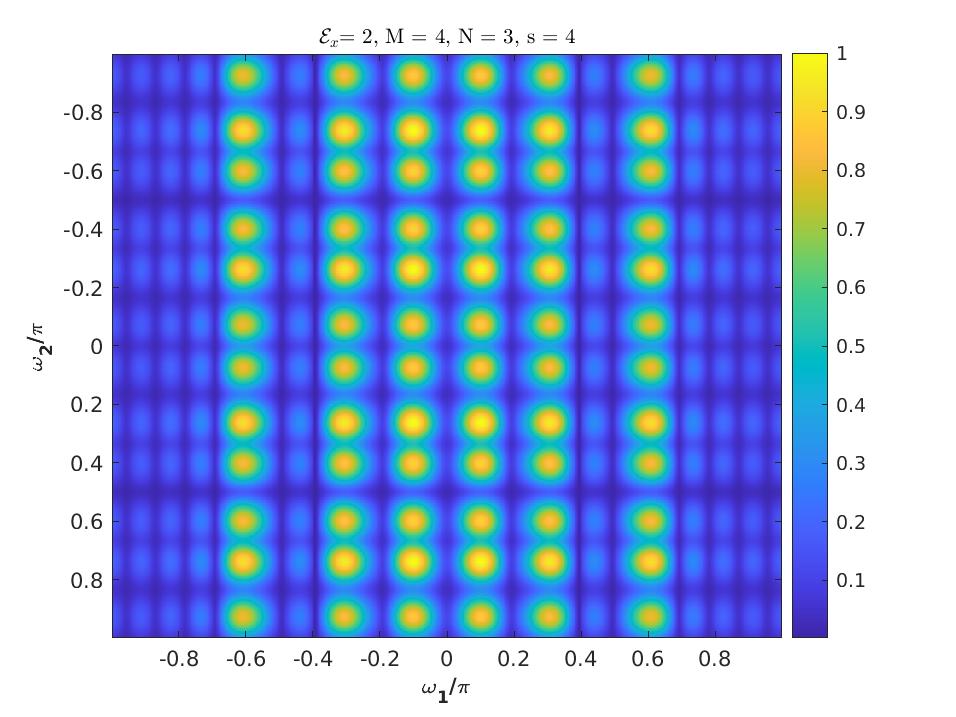}%
	\includegraphics[width=0.5\textwidth]{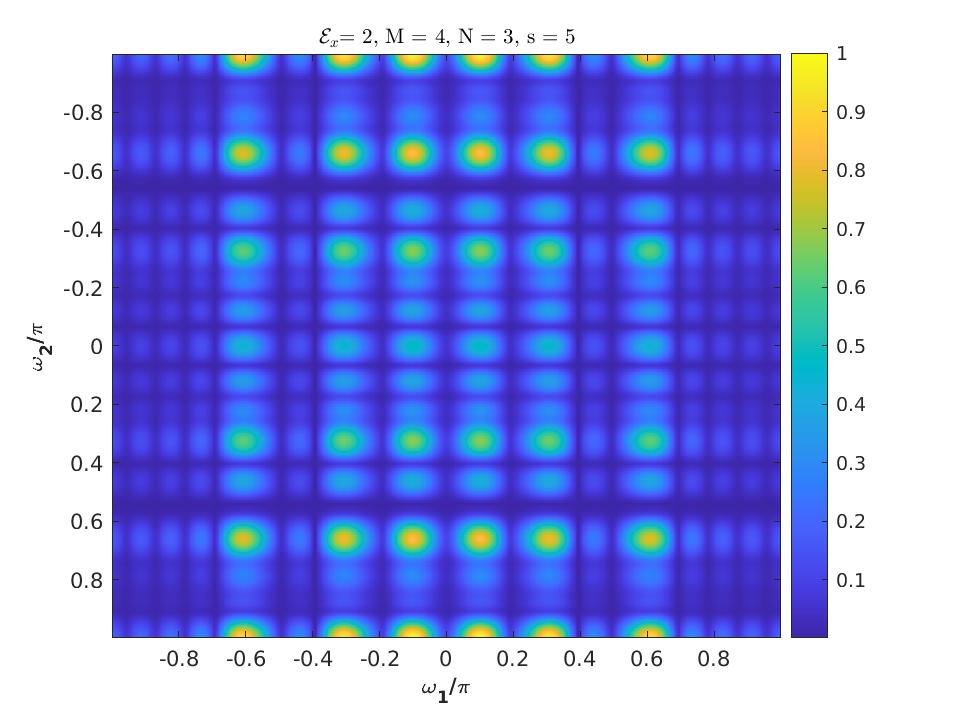}
	\caption{2D Hybrid ExSCA based spectral estimation for Fig.~\ref{fig:extreme_2D_true_spectVH}.}
	\label{fig:extreme_2D_M4N3HVP1P2}
\end{figure*}

There is something more interesting to talk about. Note that in general, two different 1D sampling patterns can be considered to generate the 2D pattern. For example, $p^{1D}(k_1)$ could be a Nyquist sampling function while $p^{1D}(k_2)$ may represent ExSCA. An example of such a sampling pattern is shown in Fig.~\ref{ExSCA_matrixP1P2}. This pattern may be referred to as `Hybrid ExSCA'. Researchers can select some dimensions to represent ExSCA with different parameters, prototype co-prime, any other sub-Nyquist strategies or even Nyquist sampling. For the hybrid ExSCA example in Fig.~\ref{ExSCA_matrixP1P2}, the 2D hybrid weight function is shown in Fig.~\ref{fig:2DEx2_weight_M4N3P1P2}. The corresponding bias window is shown in Fig.~\ref{fig:extreme_2DEx2_wt_bias_M4N3s02P1P2} and Fig.~\ref{fig:extreme_2DEx2_wt_bias_M4N3s35P1P2}. For better visualization, the surface plot is shown in Fig.~\ref{fig:extreme_2DEx2_wt_bias_M4N3s02SurfP1P2} and Fig.~\ref{fig:extreme_2DEx2_wt_bias_M4N3s35SurfP1P2}. Here, the `Theory' is the outer product of~\eqref{eq:FT_exsca_entire} i.e. ExSCA and the Fourier transform of triangular function (2.4.15)~\cite{7.1} i.e. Nyquist bias window. Note the bias distortion is large along the vertical axis. The convolution of this bias window with the Nyquist (true) spectrum may produce lesser distortion. Let us check if the hybrid system can estimate the spectrum. Simulation results for the hybrid system is shown in Fig.~\ref{fig:extreme_2D_M4N3VP1P2}, Fig.~\ref{fig:extreme_2D_M4N3HP1P2}, and Fig.~\ref{fig:extreme_2D_M4N3HVP1P2} for examples Fig.~\ref{fig:extreme_2D_true_spectV}, Fig.~\ref{fig:extreme_2D_true_spectH}, and Fig.~\ref{fig:extreme_2D_true_spectVH} respectively. The hybrid ExSCA works for certain values of $s$. Appropriate parameters $(M, N, s)$ will have to be investigated. To explain the concept only two dimensions are considered here. However, several dimensions can be considered. In general:\\
Sampling pattern:
	\begin{eqnarray}
\nonumber	p^{2D}(k)&=&p^{1D}(k_1)\otimes p^{1D}(k_2)\\
\nonumber	p^{\eta D}(k)&=&p^{(\eta -1)D}(k)\otimes p^{1D}(k_{\eta})\\
	&=&p^{1D}(k_1)\otimes p^{1D}(k_2)\otimes \dots \otimes p^{1D}(k_\eta)
	\end{eqnarray}
Weight function:	
	\begin{eqnarray}
\nonumber	z^{2D}(l)&=&z^{1D}(l_1)\otimes z^{1D}(l_2)\\
\nonumber	z^{\eta D}(l)&=&z^{(\eta -1)D}(l)\otimes z^{1D}(l_{\eta})\\
&=&z^{1D}(l_1)\otimes z^{1D}(l_2)\otimes \dots \otimes z^{1D}(l_\eta)
\end{eqnarray}
Bias window:
	\begin{eqnarray}
\nonumber	 W^{2D}_{b}(e^{j\omega})&=&W^{1D}_{b}(e^{j\omega_1})\otimes W^{1D}_{b}(e^{j\omega_2})\\
	W^{\eta D}_{b}(e^{j\omega})&=&W^{(\eta -1)D}_{b}(e^{j\omega})\otimes W^{1D}_{b}(e^{j\omega_\eta})\\
\nonumber&=&W^{1D}_{b}(e^{j\omega_1})\otimes W^{1D}_{b}(e^{j\omega_2}) \dots \otimes W^{1D}_{b}(e^{j\omega_\eta})
\end{eqnarray}

Note once again that the 2D or $\eta$D bias window should be as close as possible to an impulse function. Next, let us discuss a generalized ExSCA sampling strategy.


\section{Generalized Extremely Sparse Co-Prime Arrays}
\label{GenExSCA}
\begin{figure*}[!t]
	\centering
	\includegraphics[width=0.99\textwidth]{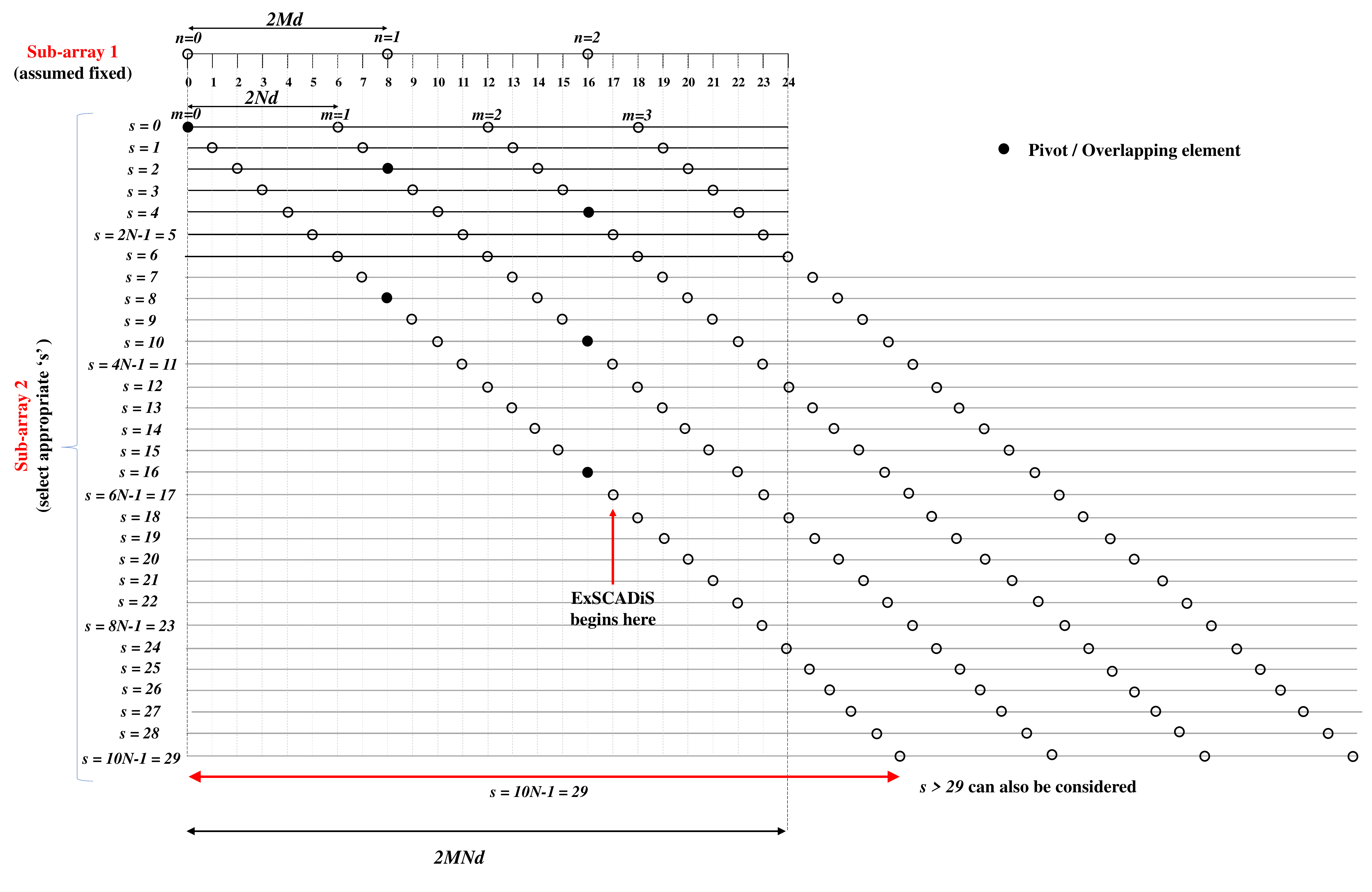}%
	\caption{Extremely sparse co-prime array with arbitrary value of $s$. ExSCADiS is a special case.}
	\label{fig:ExSCADiS}
\end{figure*}

\begin{figure*}[!t]
	\centering
	\includegraphics[width=0.99\textwidth]{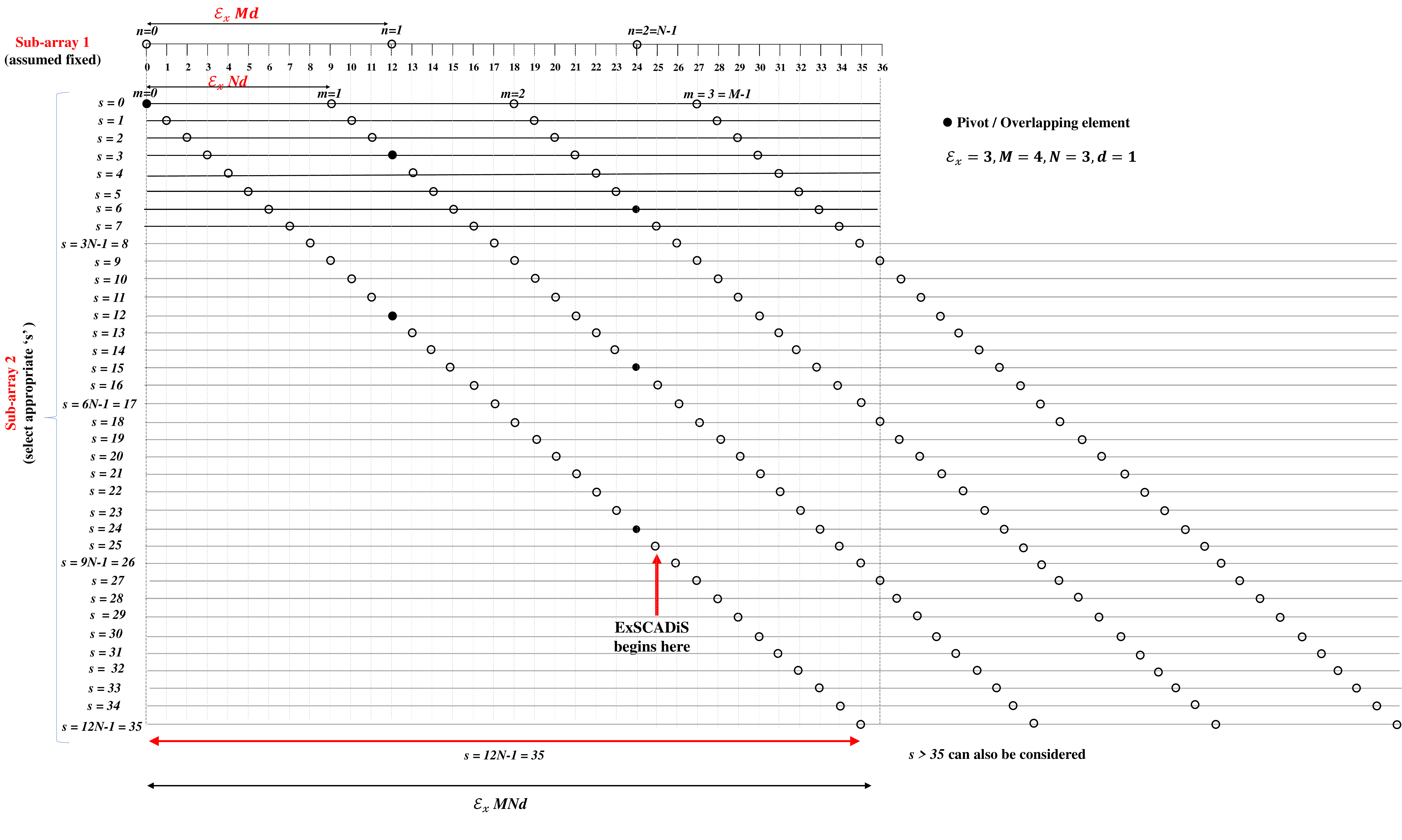}%
	\caption{Extremely sparse co-prime array with arbitrary value of $s$ and $\mathcal{E}_{x}=3$. ExSCADiS is a special case.}
	\label{fig:ExSCADiS_Ex3}
\end{figure*}

This section will consider arbitrary values of shift, $s$ and sparity factor $\mathcal{E}_x$.
In the discussion carried out so far, the value of shift `$s$' was restricted to the range $0\leq s \leq \mathcal{E}_{x}N-1$, where $\mathcal{E}_{x}=2$. This section presents the structure for arbitrary values of shift $s$. Fig.~\ref{fig:ExSCADiS} describes this structure. Note that for $s>16$ i.e. $s>2M(N-1)$, the second sub-array is beyond the last element of the first sub-array. This concept is referred to as Extremely Sparse Co-prime Array with Displaced Sub-arrays (ExSCADiS). Next let us consider an arbitrary sparsity factor $\mathcal{E}_{x}$. Fig.~\ref{fig:ExSCADiS_Ex3} describes the structure with $\mathcal{E}_{x}=3$. ExSCADiS is possible for $s>\mathcal{E}_{x}M(N-1)$. Let us pause for a moment, and reflect on what we know and where we are going. 
Based on prior discussion and specifically the work in \cite{UVD_PHD}, following comments are reiterated:
\begin{enumerate}
	\item The worth of the correlogram spectral estimator largely depends on the bias window.
	\item Under the assumption of complex Gaussian white process, the covariance expression is similar to the bias window expression.
	\item A sampling pattern may be selected such that the correlogram bias window is as close as possible to an impulse function.
	\item It would ensure that convolution operation between bias window and true (Nyquist) spectrum is approximately equal to the true spectrum itself.
\end{enumerate}

Therefore, let us only focus on the bias window expression. Below is the generalized ExSCA weight function:
\begin{equation}\label{eq:exsca_gen}
\begin{split}
&z(l)=\underbrace{\sum\limits_{i=1}^{q}\sum\limits_{n=-(r_iN_i-1)}^{r_iN_i-1} (r_iN_i-\mid n_i\mid)\delta(l-\mathcal{E}_i\breve{M}_in_i)}_\text{A}\\
&+\sum\limits_{i=1}^{q}\sum\limits_{k=i+1}^{q}\sum\limits_{n_i=0}^{r_iN_i-1}\sum\limits_{n_k=0}^{r_kN_k-1} \delta(\mid l\mid-\\
&\underbrace{	\mid \mathcal{E}_i\breve{M}_in_i+s_i-(\mathcal{E}_k\breve{M}_kn_k+s_k)\mid)}_\text{B}\\
\end{split}
\end{equation}
The correlogram bias window is given by:
\begin{equation}\label{eq:FT_exsca_gen}
\begin{split}
&W_{b}(e^{j\omega})=\frac{1}{s_b}\left\{\sum\limits_{i=1}^{q}\left |\frac{\sin(\frac{\omega \mathcal{E}_i\breve{M}_ir_iN_i}{2})}{\sin(\frac{\omega \mathcal{E}_i\breve{M}_i}{2})}\right|^2\right .\\
&+\sum\limits_{i=1}^{q}\sum\limits_{k=i+1}^{q}\left[\frac{\sin(\frac{\omega \mathcal{E}_i\breve{M}_ir_iN_i}{2})\sin(\frac{\omega \mathcal{E}_k\breve{M}_kr_kN_k}{2})}{\sin(\frac{\omega \mathcal{E}_i\breve{M}_i}{2})\sin(\frac{\omega \mathcal{E}_k\breve{M}_k}{2})}\right.\\
& 2 \cos \left(\omega\left[\frac{\mathcal{E}_i\breve{M}_ir_iN_i-\mathcal{E}_k\breve{M}_kr_kN_k-(\mathcal{E}_i\breve{M}_i-\mathcal{E}_k\breve{M}_k)}{2}\right.\right.\\
&\left.\left.\left.\left.+(s_i-s_k)\right]\right)\right]\right\}
\end{split}
\end{equation}

\begin{figure*}[!t]
	\centering
	\includegraphics[width=0.49\textwidth]{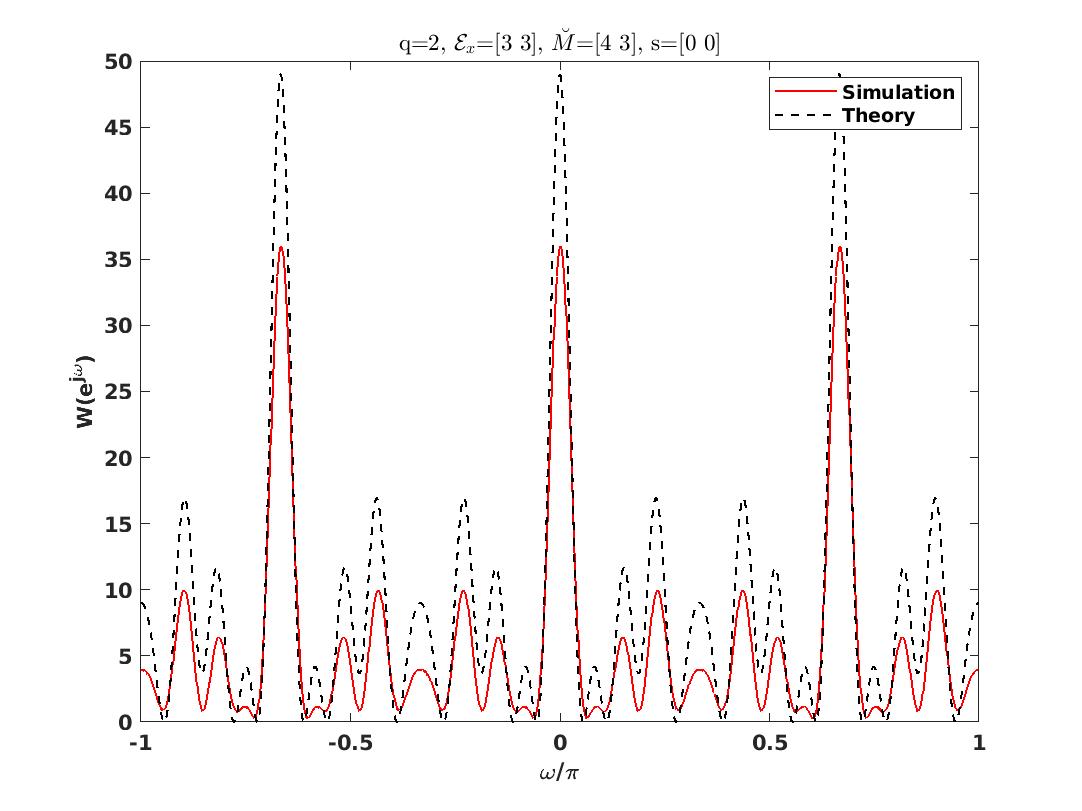}
	\includegraphics[width=0.49\textwidth]{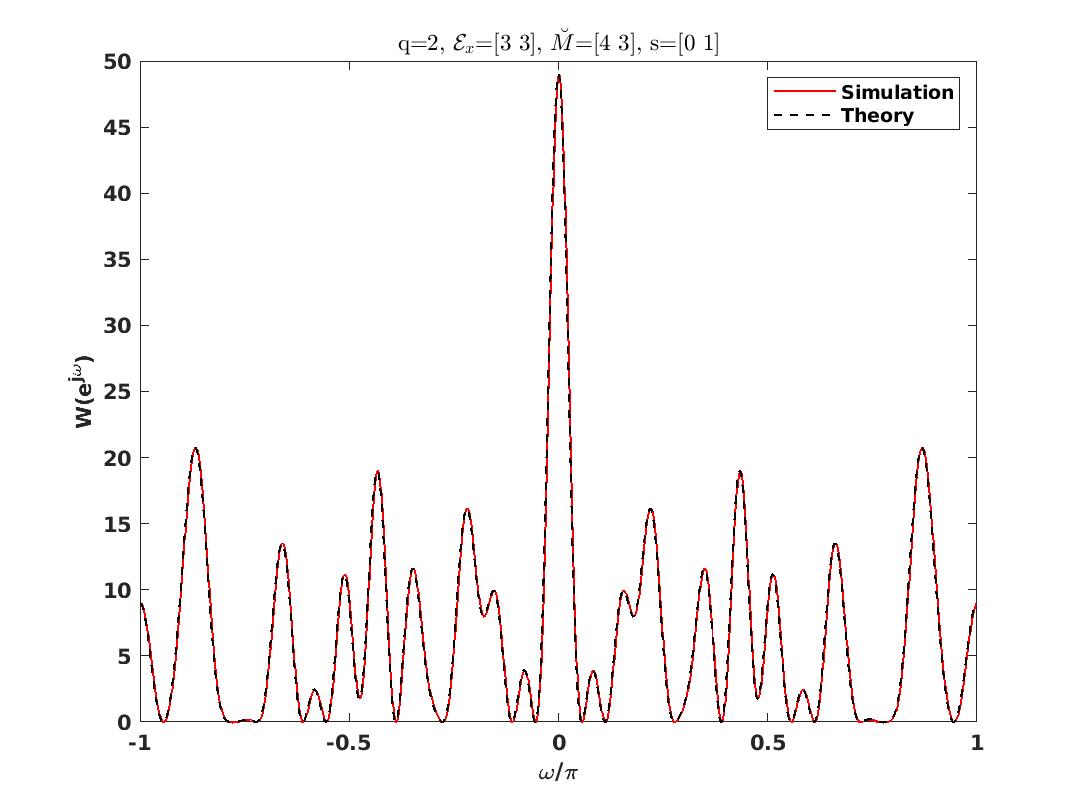}\\
	\includegraphics[width=0.5\textwidth]{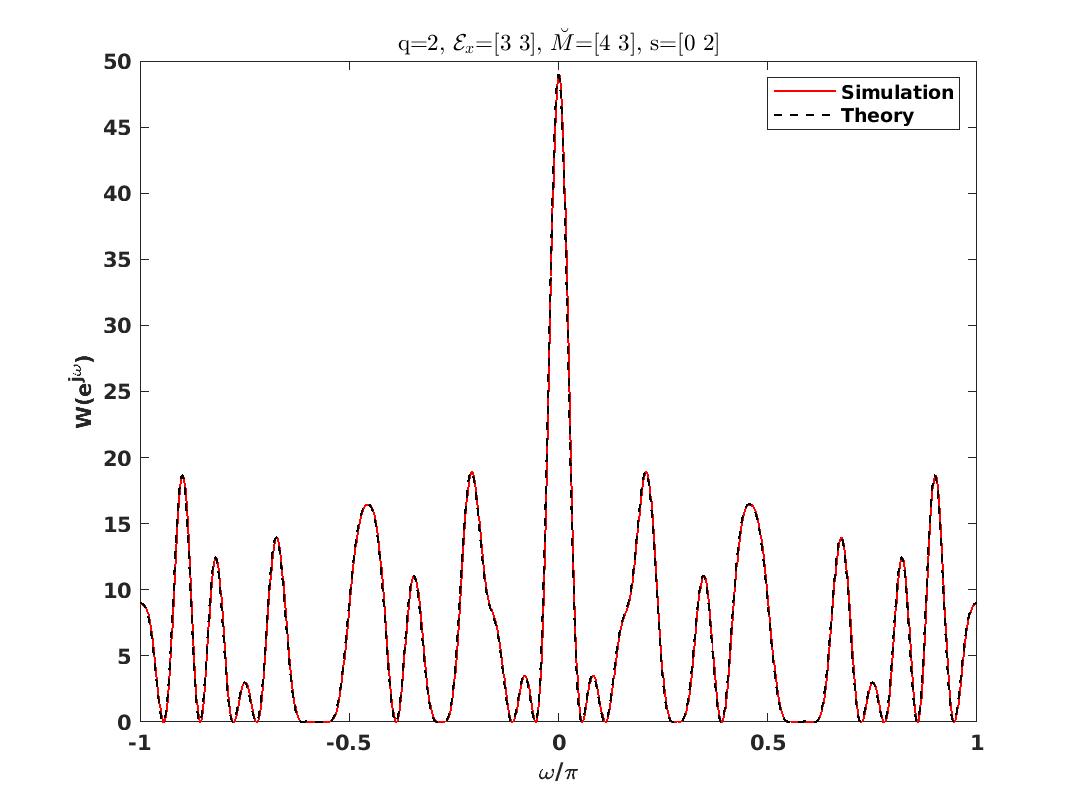}%
	\includegraphics[width=0.5\textwidth]{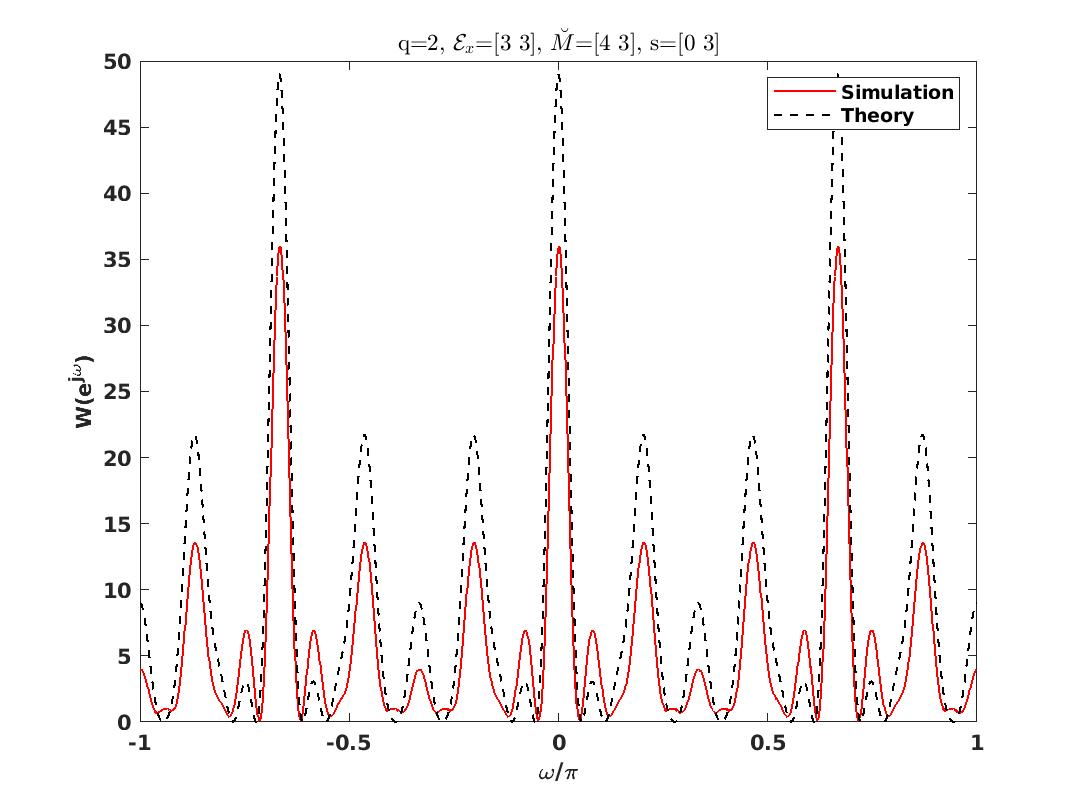}\\
	\includegraphics[width=0.5\textwidth]{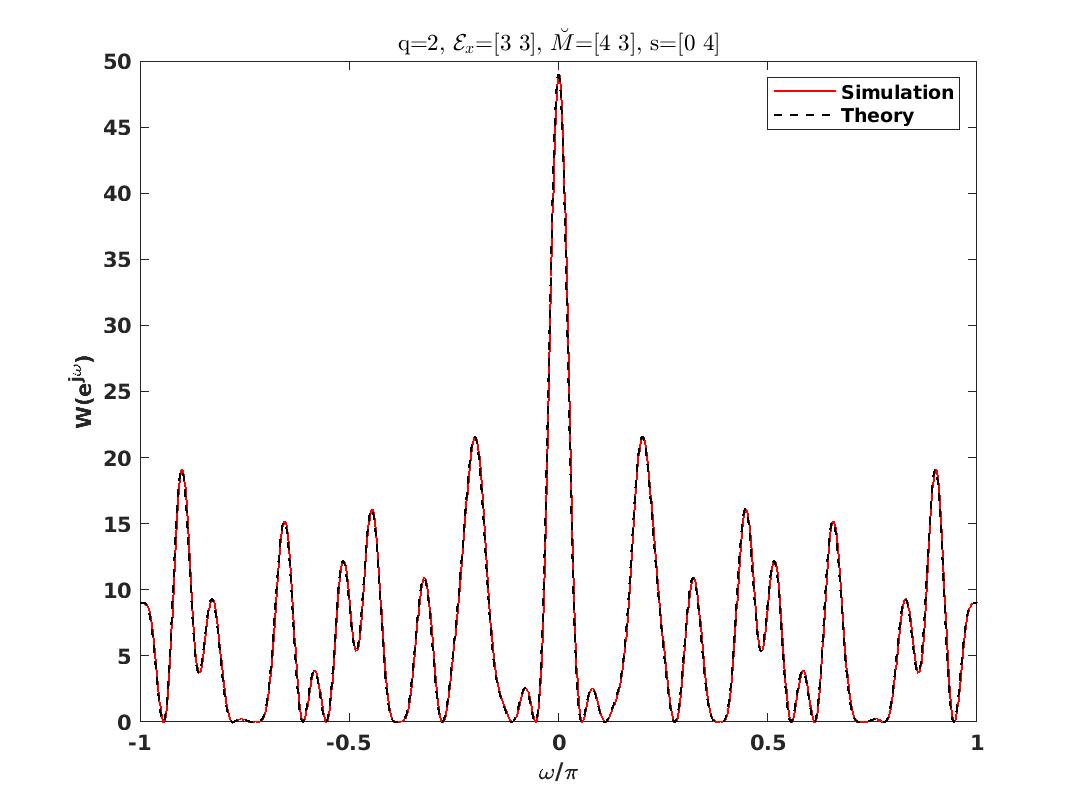}%
	\includegraphics[width=0.5\textwidth]{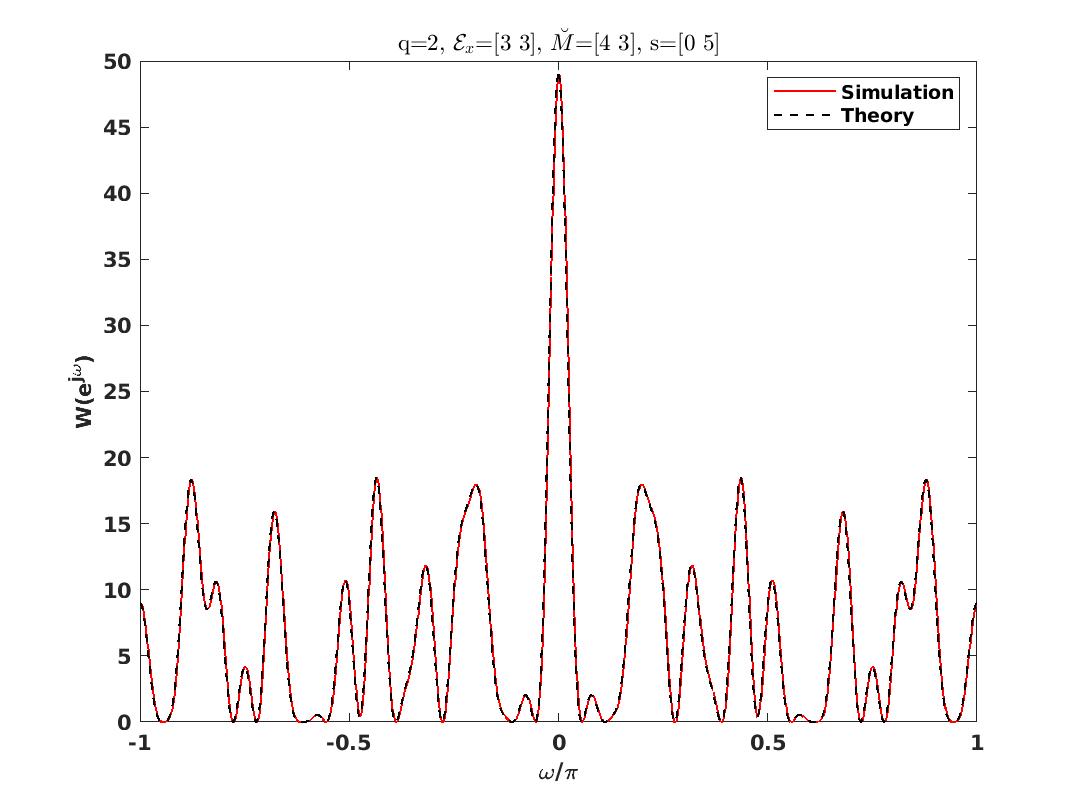}
	\caption{Bias window for generalized ExSCA as in  Fig.~\ref{fig:ExSCADiS_Ex3} for $0\leq s\leq5$.}
	\label{fig:gen_ExSCA_Ex3s0_5}
\end{figure*}

\begin{figure*}[!t]
	\centering
	\includegraphics[width=0.49\textwidth]{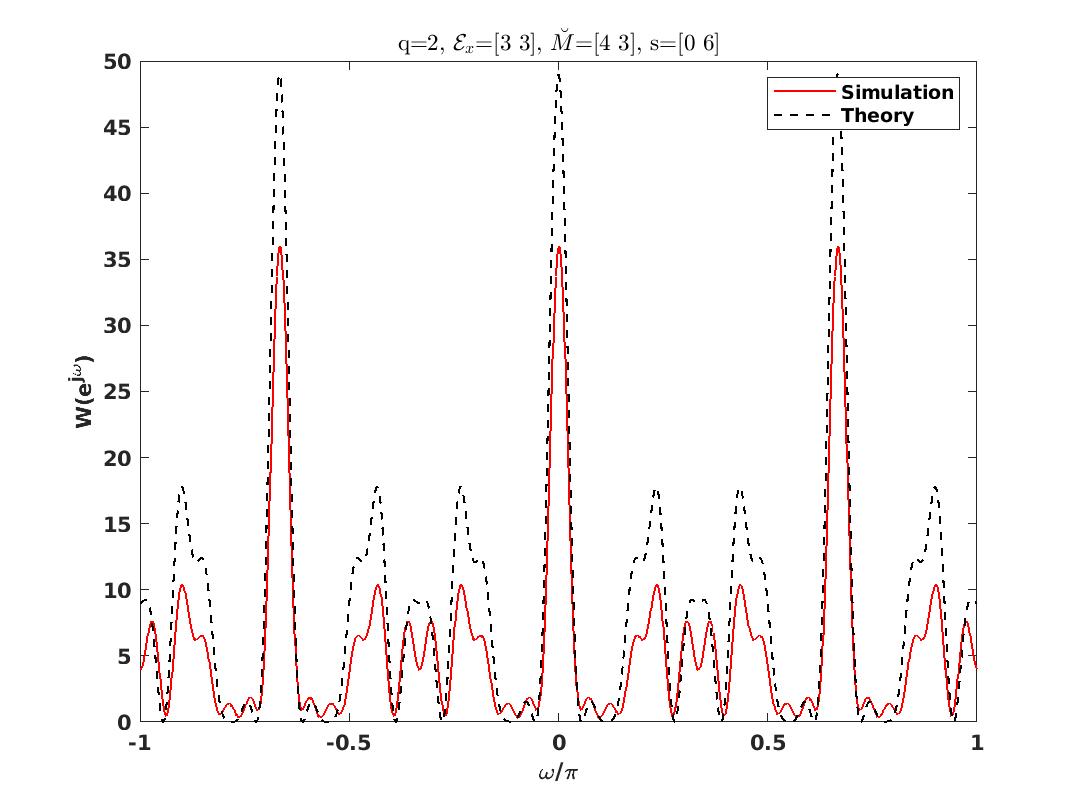}
	\includegraphics[width=0.49\textwidth]{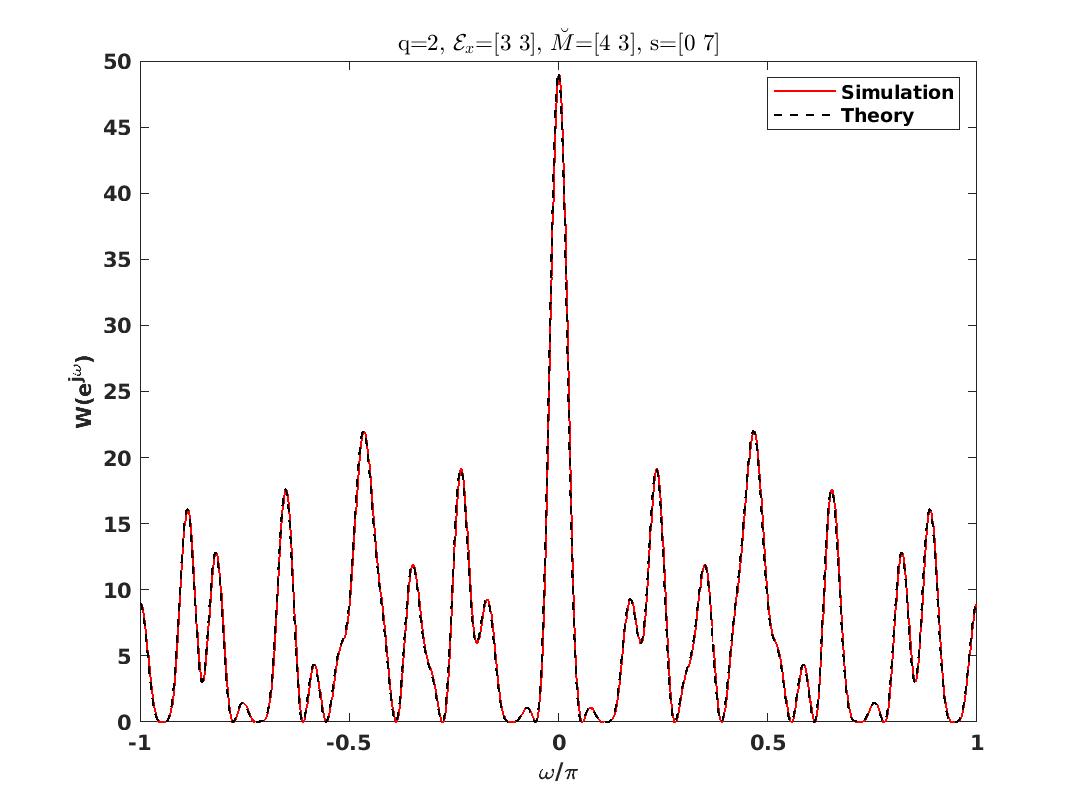}\\
	\includegraphics[width=0.5\textwidth]{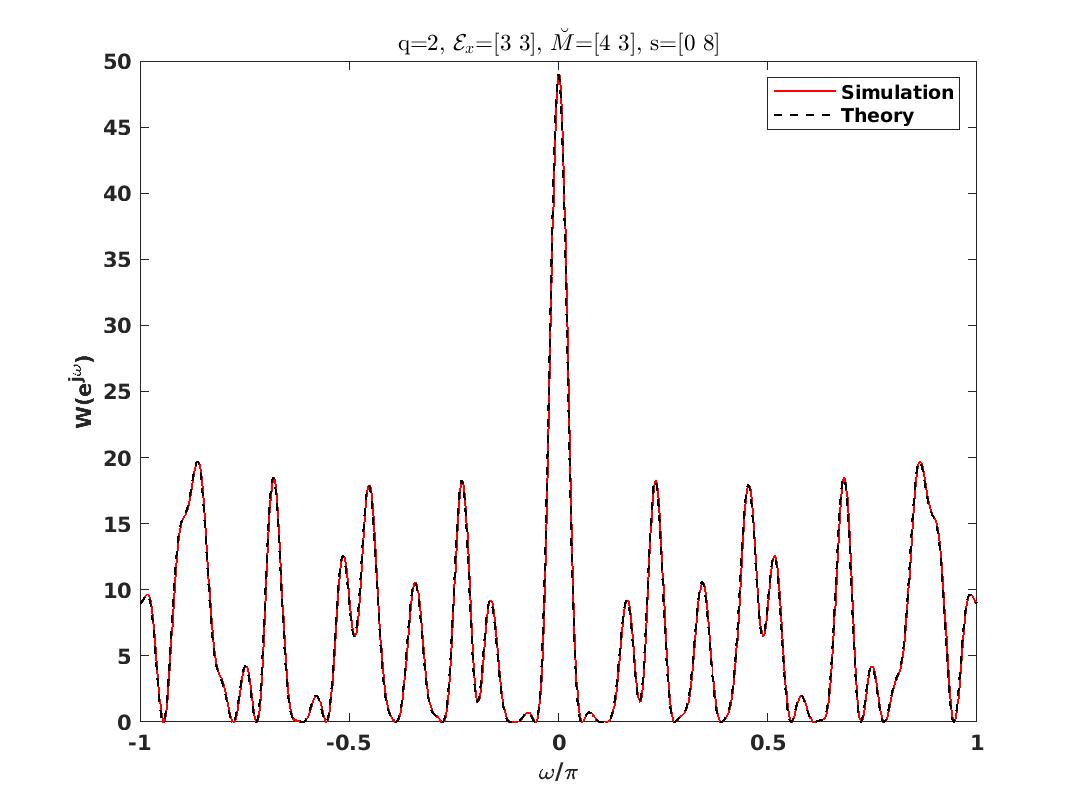}%
	\includegraphics[width=0.5\textwidth]{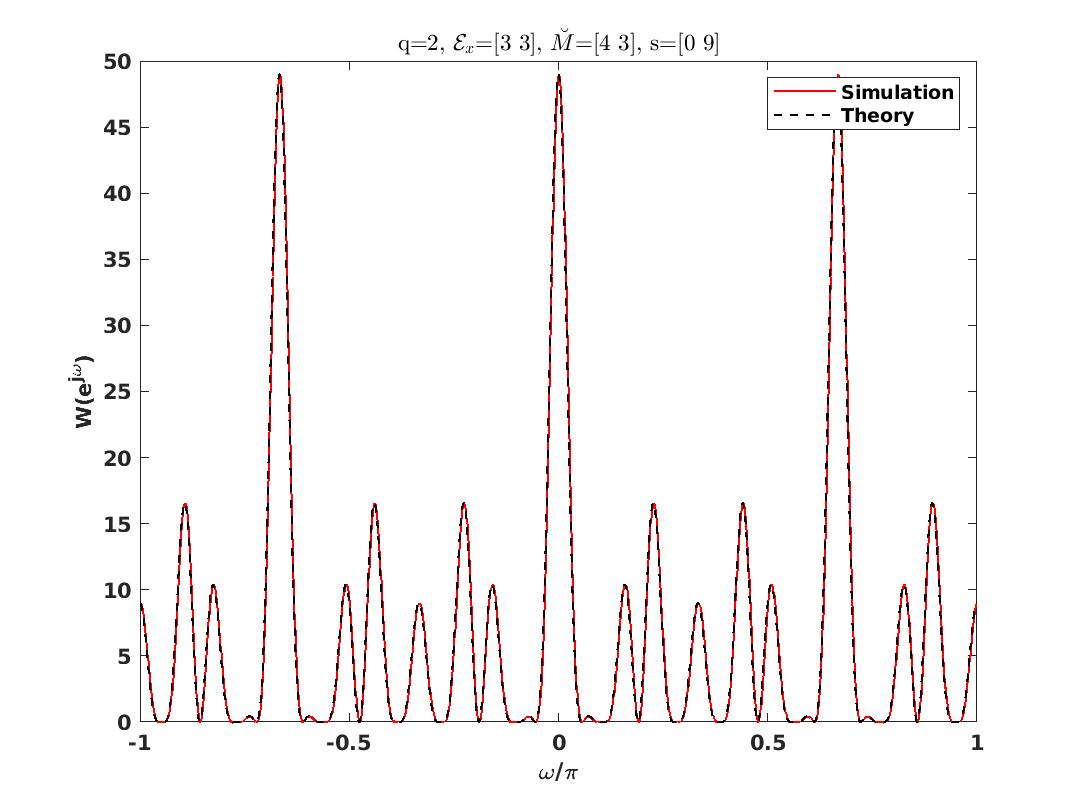}\\
	\includegraphics[width=0.5\textwidth]{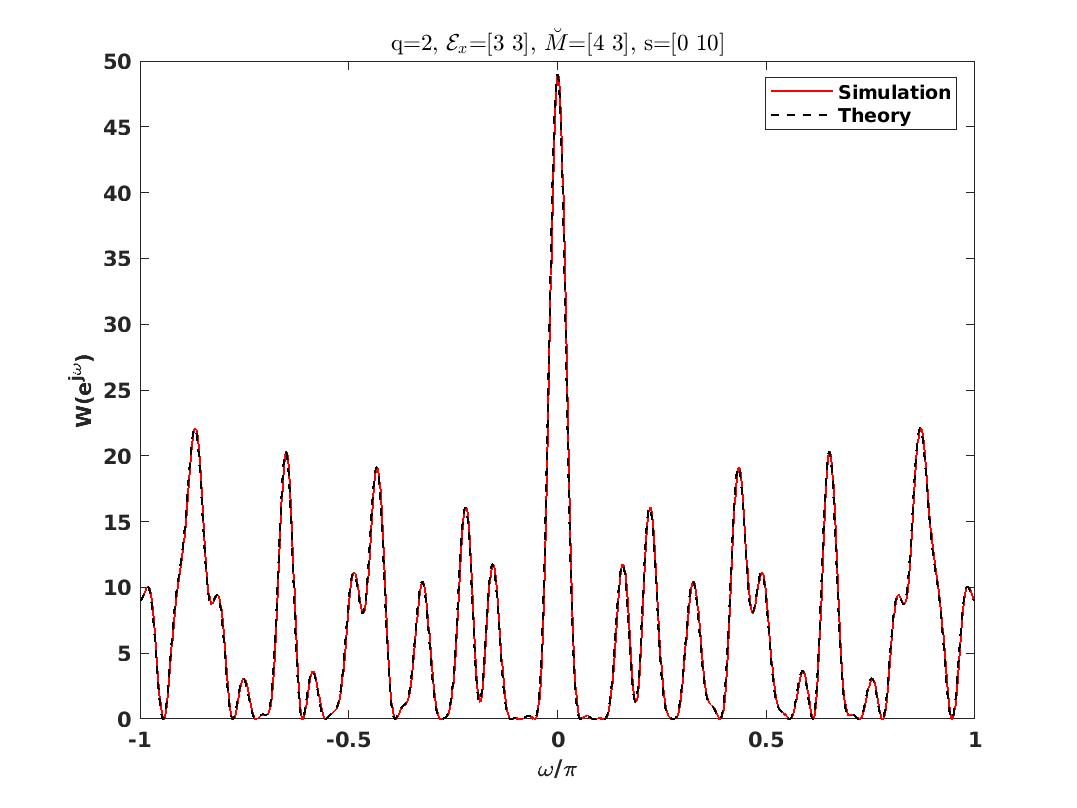}%
	\includegraphics[width=0.5\textwidth]{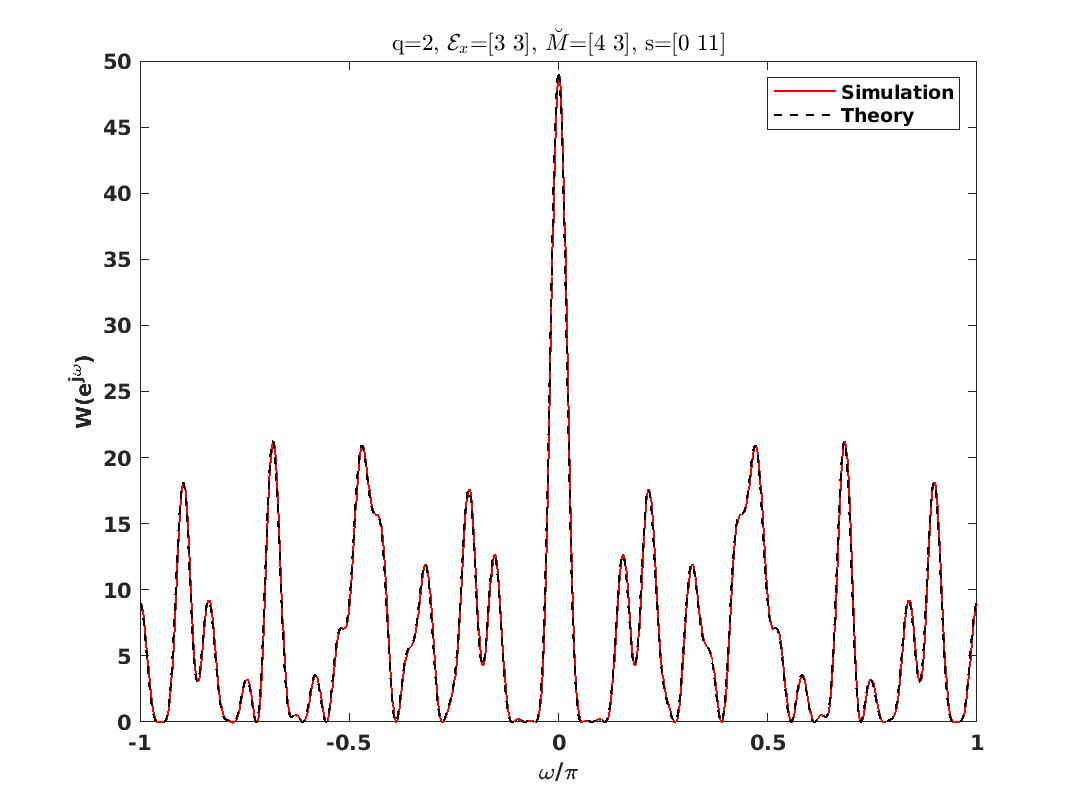}
	\caption{Bias window for generalized ExSCA as in  Fig.~\ref{fig:ExSCADiS_Ex3} for $6\leq s\leq 11$.}
	\label{fig:gen_ExSCA_Ex3s6_11}
\end{figure*}

It is valid for all the generalized sampling strategies provided the sub-arrays do not have overlapping elements/samples. The overlapping case can be easily incorporated for speciﬁc scenarios as described for the prototype ExSCA. Fig.~\ref{fig:gen_ExSCA_Ex3s0_5} and Fig.~\ref{fig:gen_ExSCA_Ex3s6_11} displays the bias window for the struture shown in Fig.~\ref{fig:ExSCADiS_Ex3} with different values of $s$. Note that the proposed expression closely matches the simulation except for $s=$ 0, 3, and 6 since they have overlapping elements. $q=2$ implies the structure has two sub-arrays.
 
\begin{figure*}[!t]
	\centering
	\includegraphics[width=0.99\textwidth]{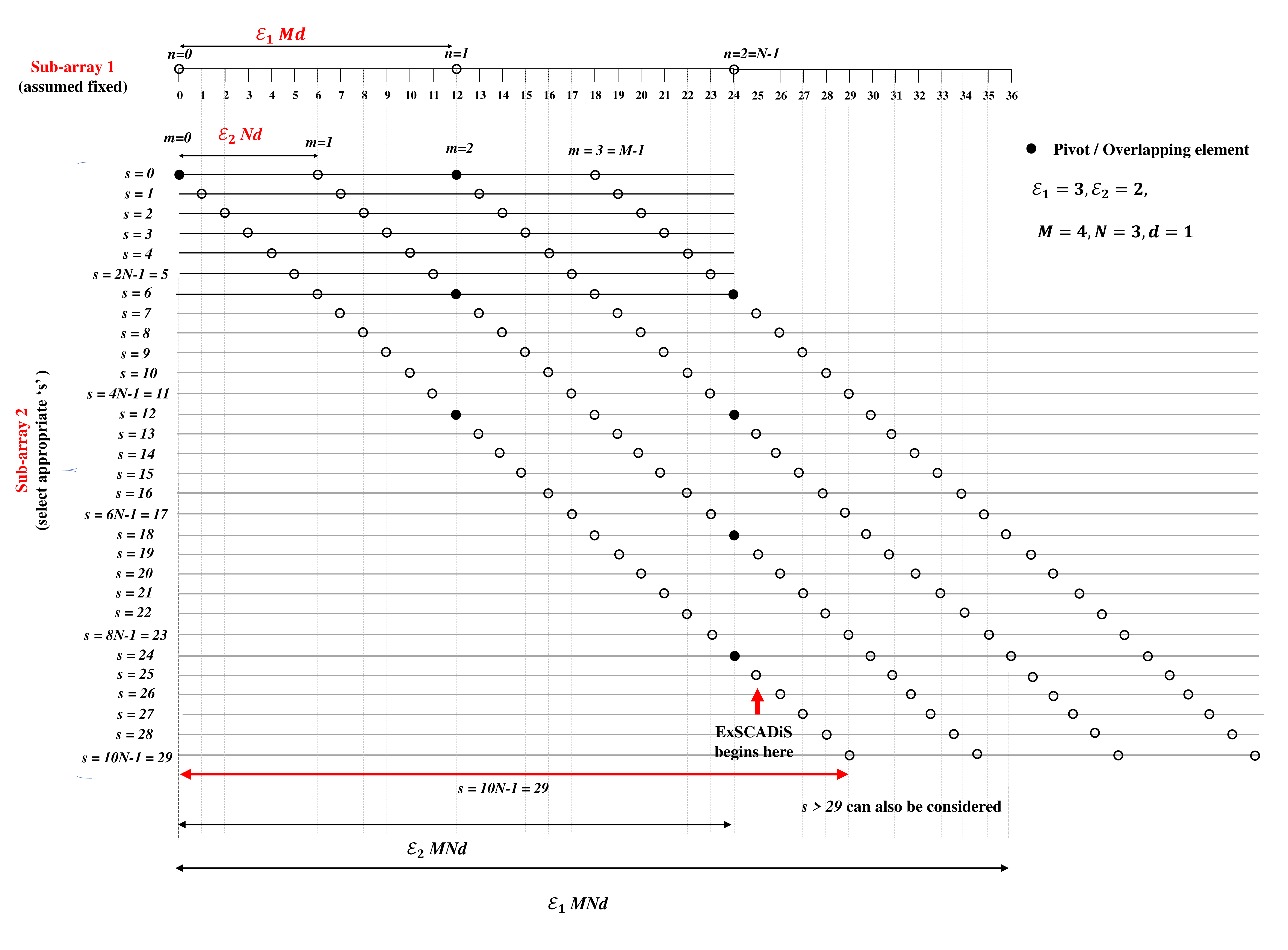}%
	\caption{Extremely sparse co-prime array with arbitrary value of $s$, $\mathcal{E}_{1}=3$, and $\mathcal{E}_{2}=2$. ExSCADiS is a special case.}
	\label{fig:ExSCADiS_E1E2}
\end{figure*}

\begin{figure*}[!t]
	\centering
	\includegraphics[width=0.49\textwidth]{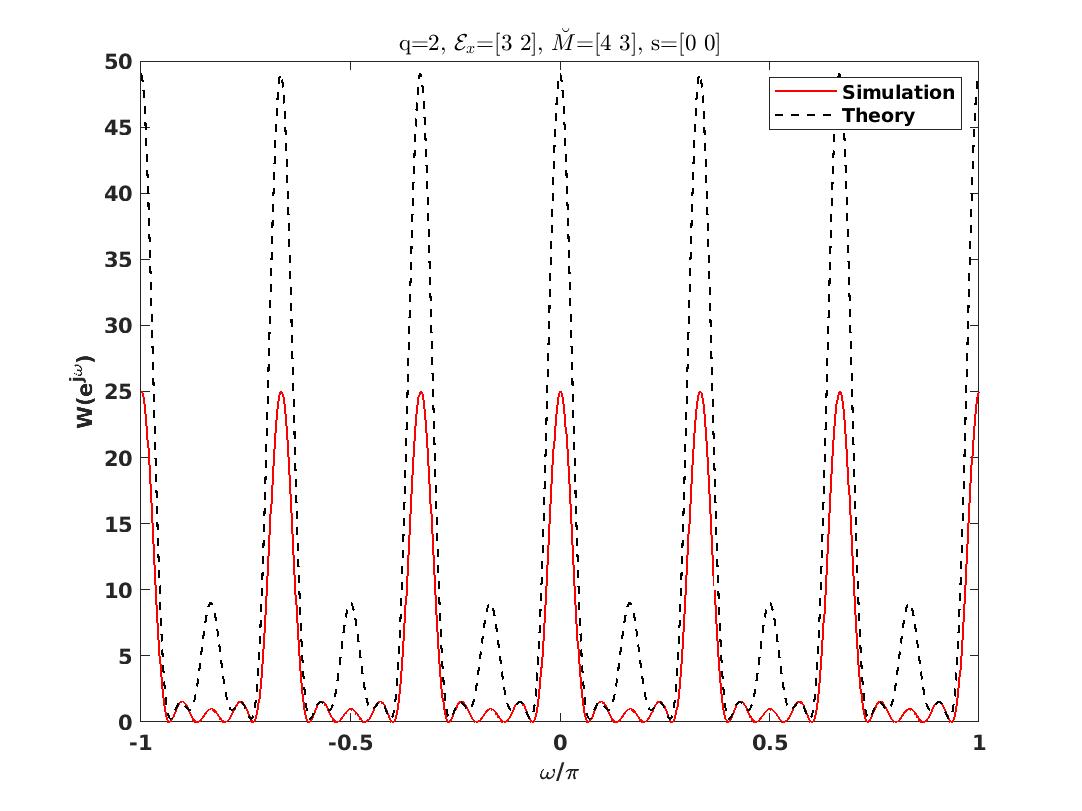}
	\includegraphics[width=0.49\textwidth]{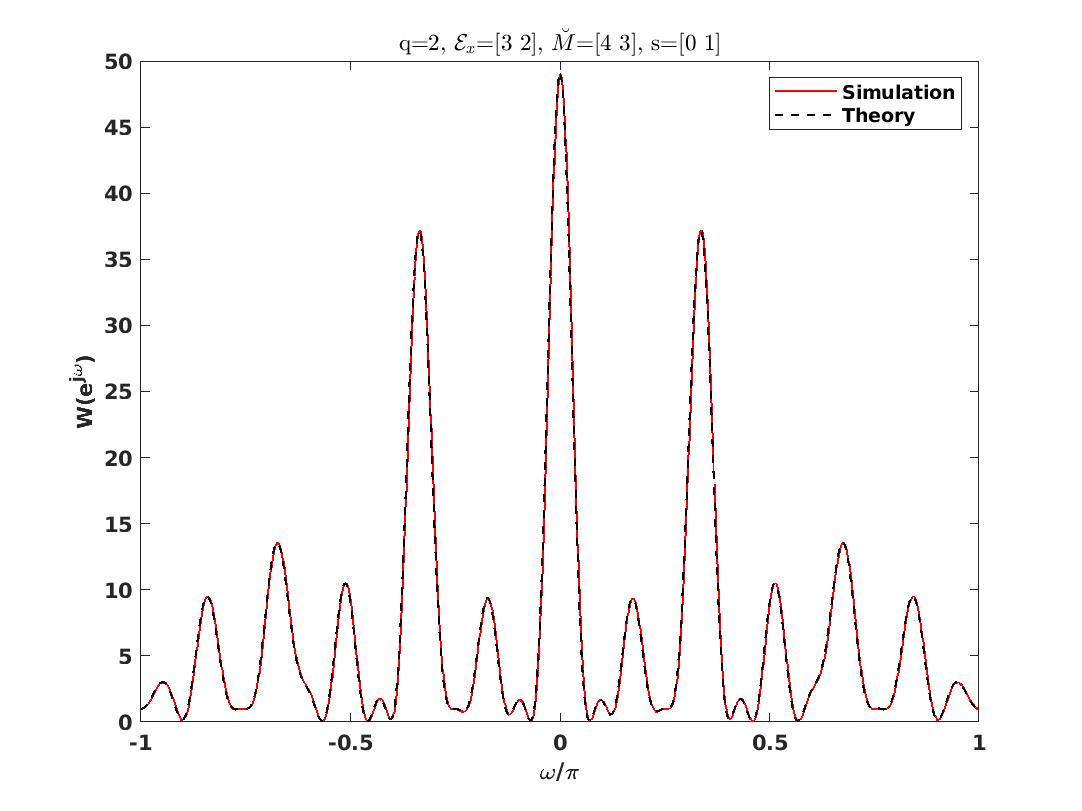}\\
	\includegraphics[width=0.5\textwidth]{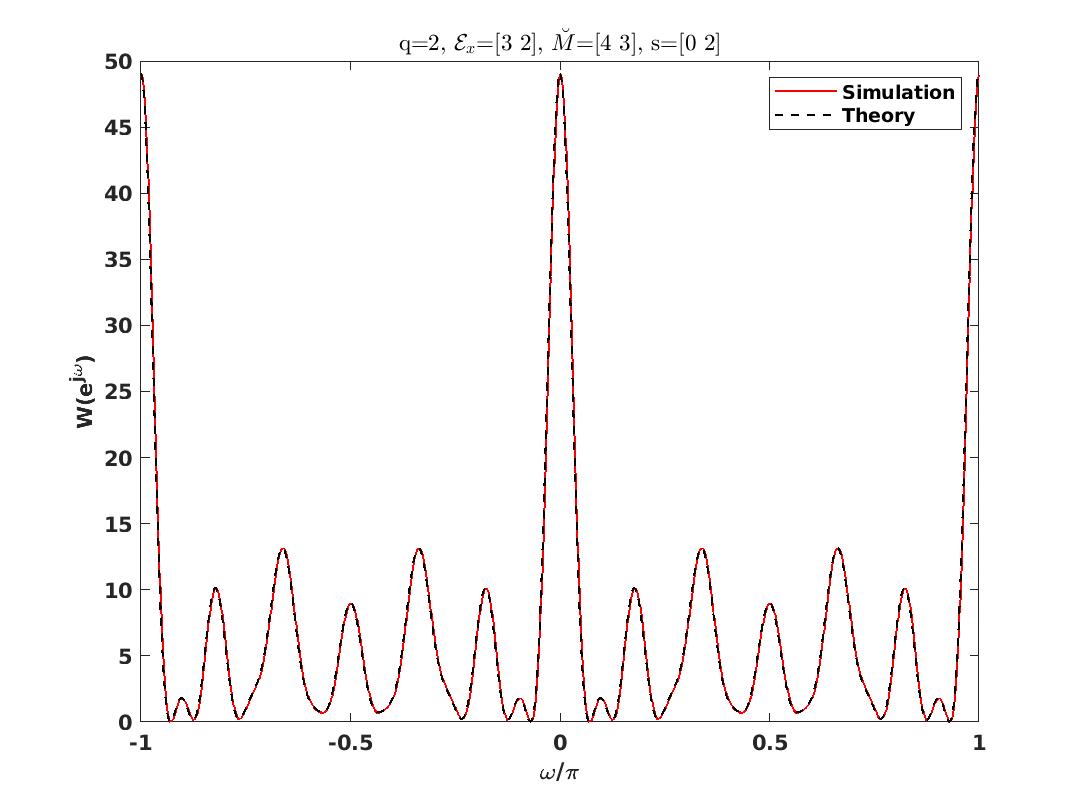}%
	\includegraphics[width=0.5\textwidth]{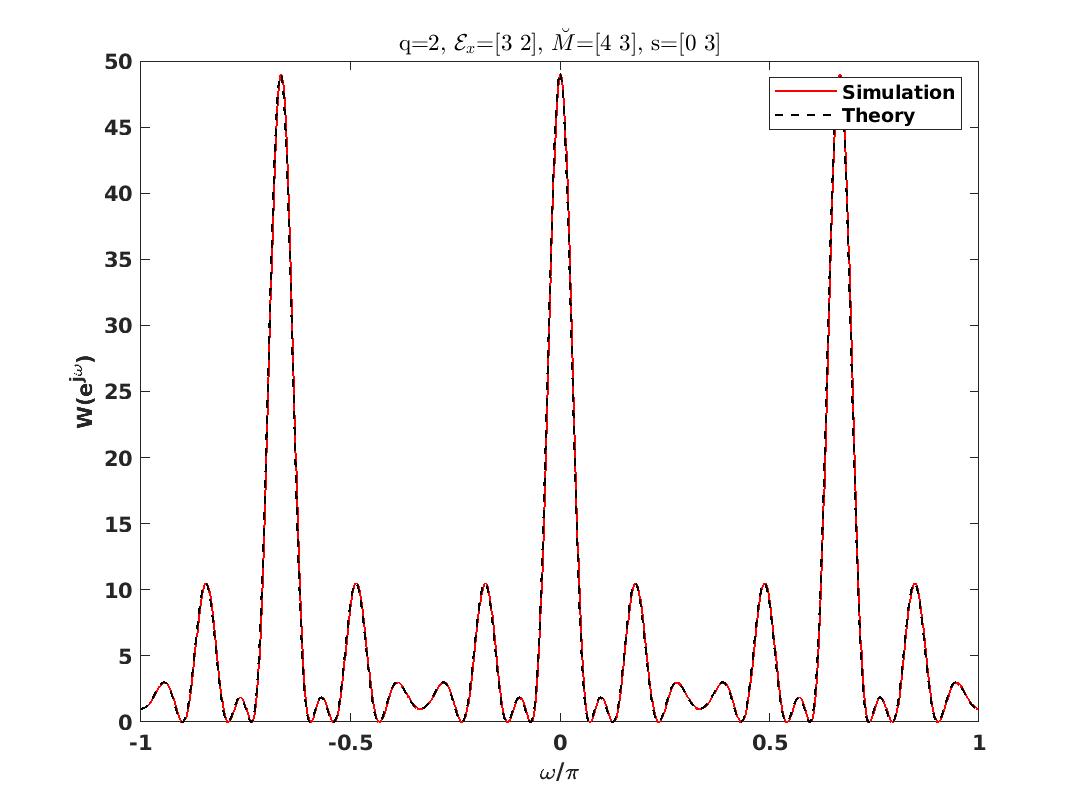}\\
	\includegraphics[width=0.5\textwidth]{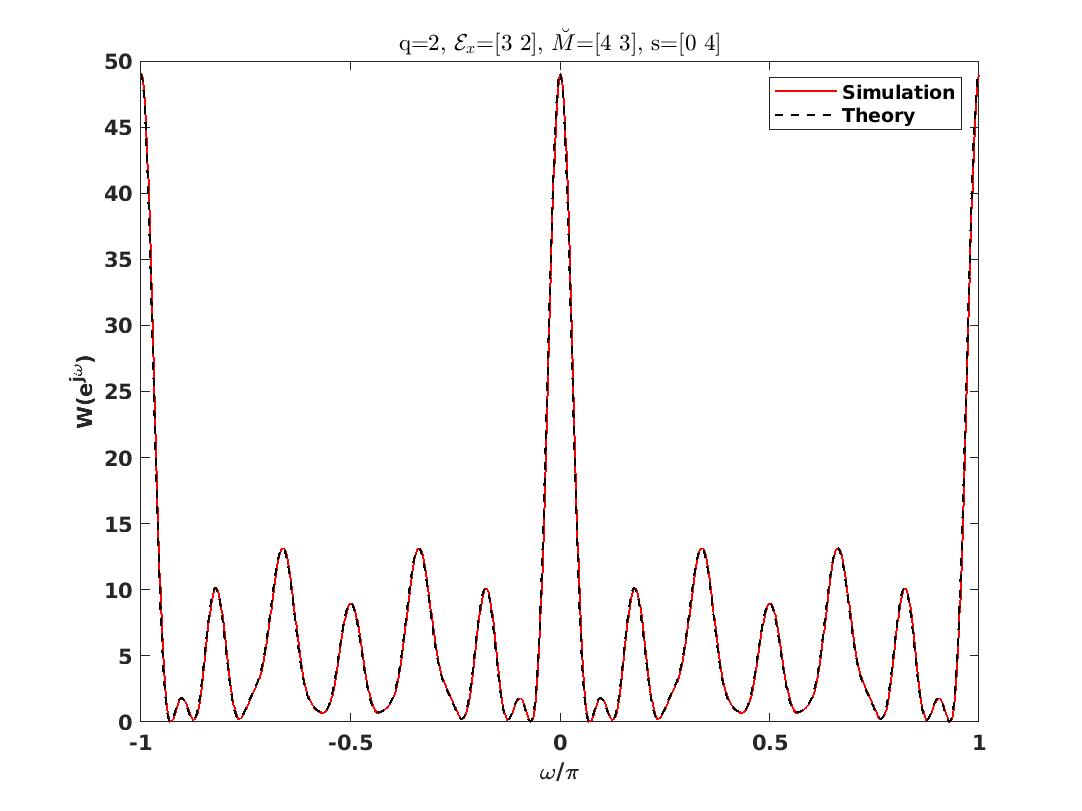}%
	\includegraphics[width=0.5\textwidth]{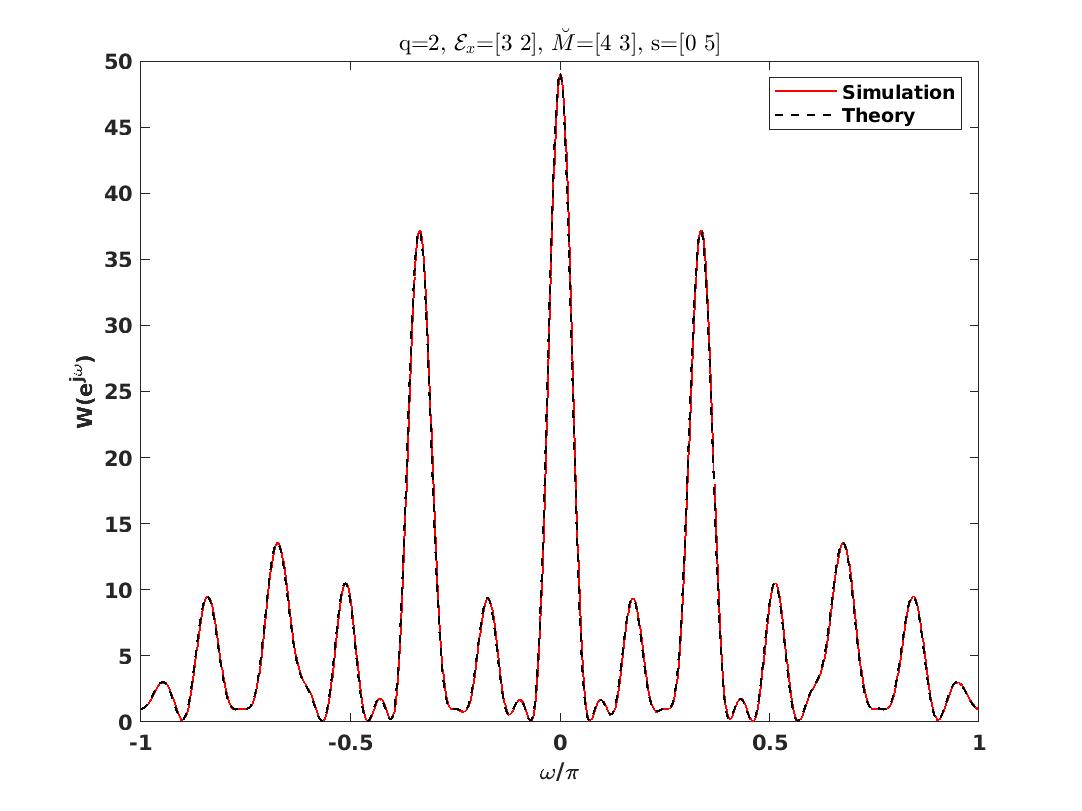}
	\caption{Bias window for generalized ExSCA as in  Fig.~\ref{fig:ExSCADiS_E1E2} for $0\leq s\leq5$.}
	\label{fig:gen_ExSCA_E3E2s0_5}
\end{figure*}

\begin{figure*}[!t]
	\centering
	\includegraphics[width=0.49\textwidth]{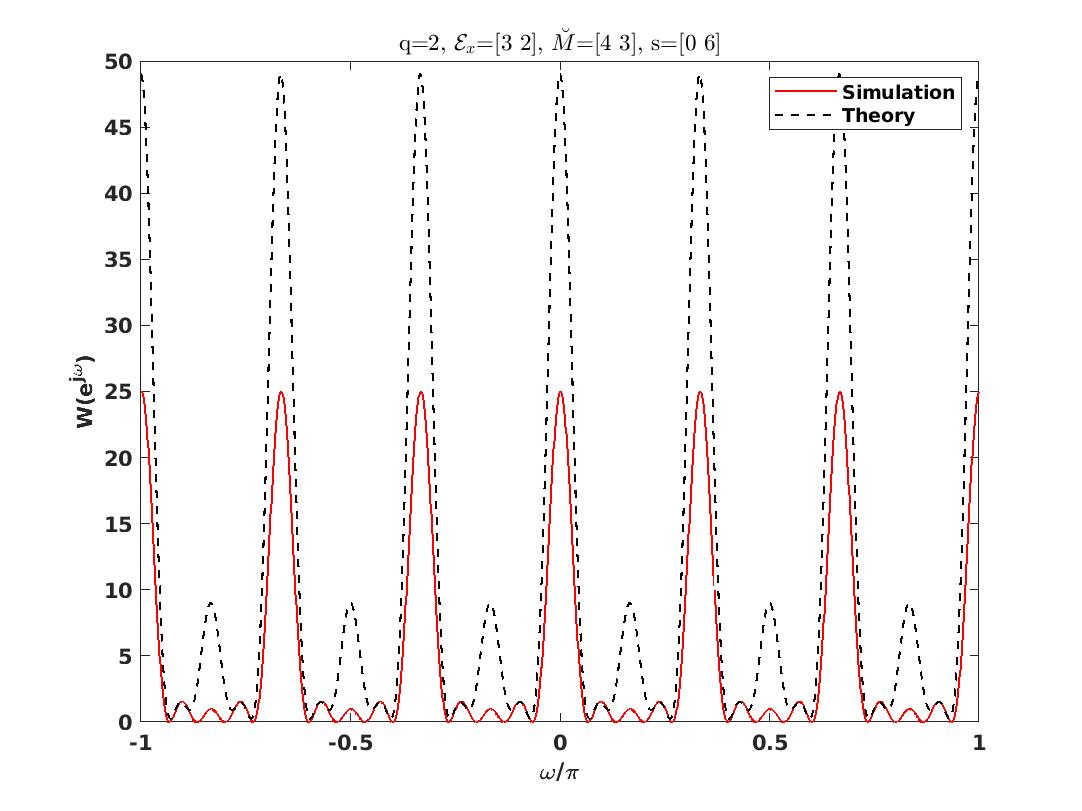}
	\includegraphics[width=0.49\textwidth]{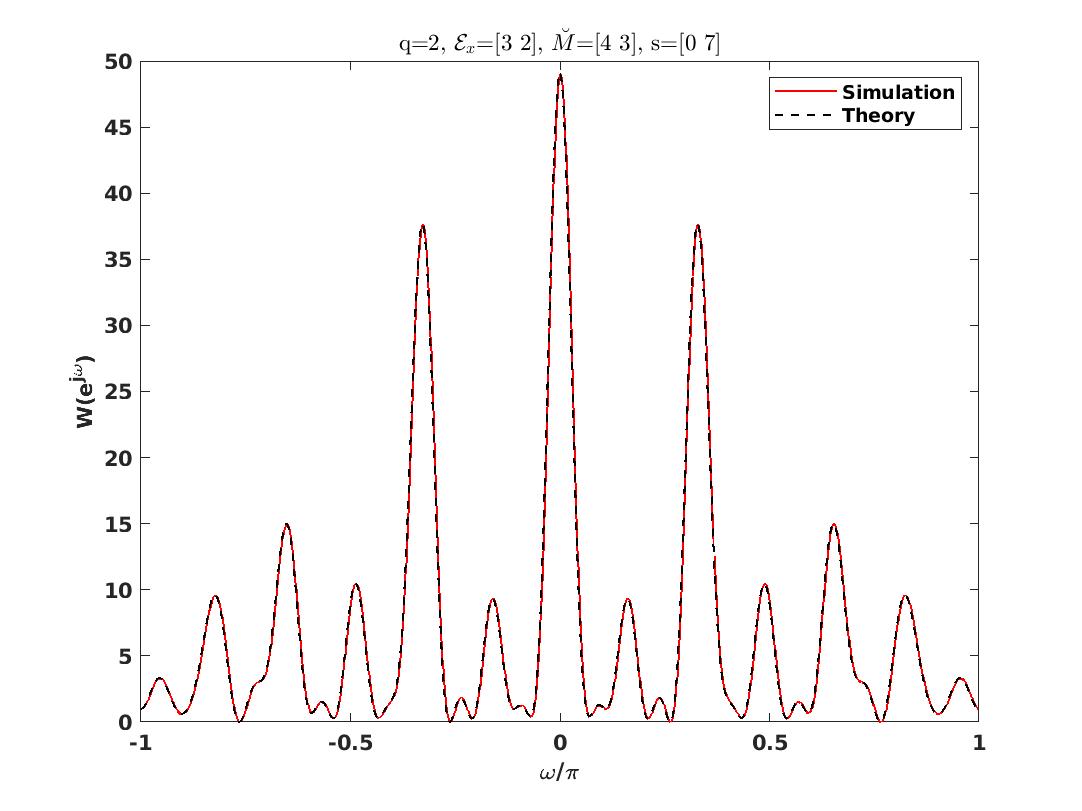}\\
	\includegraphics[width=0.5\textwidth]{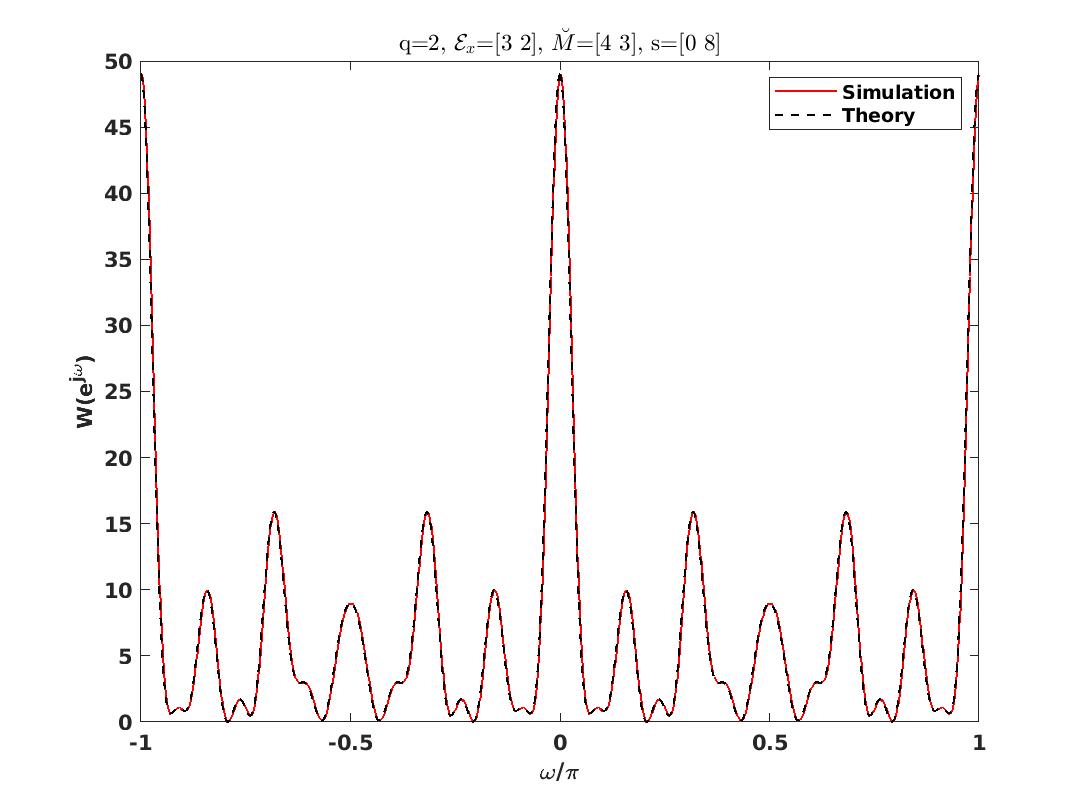}%
	\includegraphics[width=0.5\textwidth]{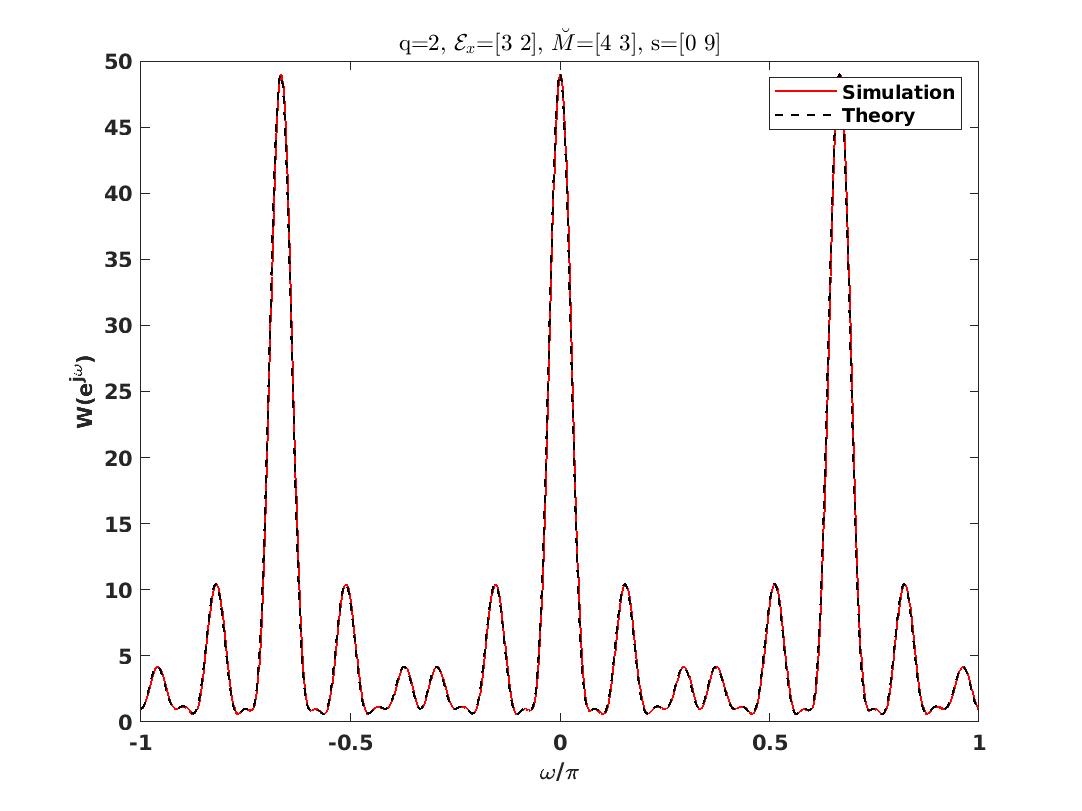}\\
	\includegraphics[width=0.5\textwidth]{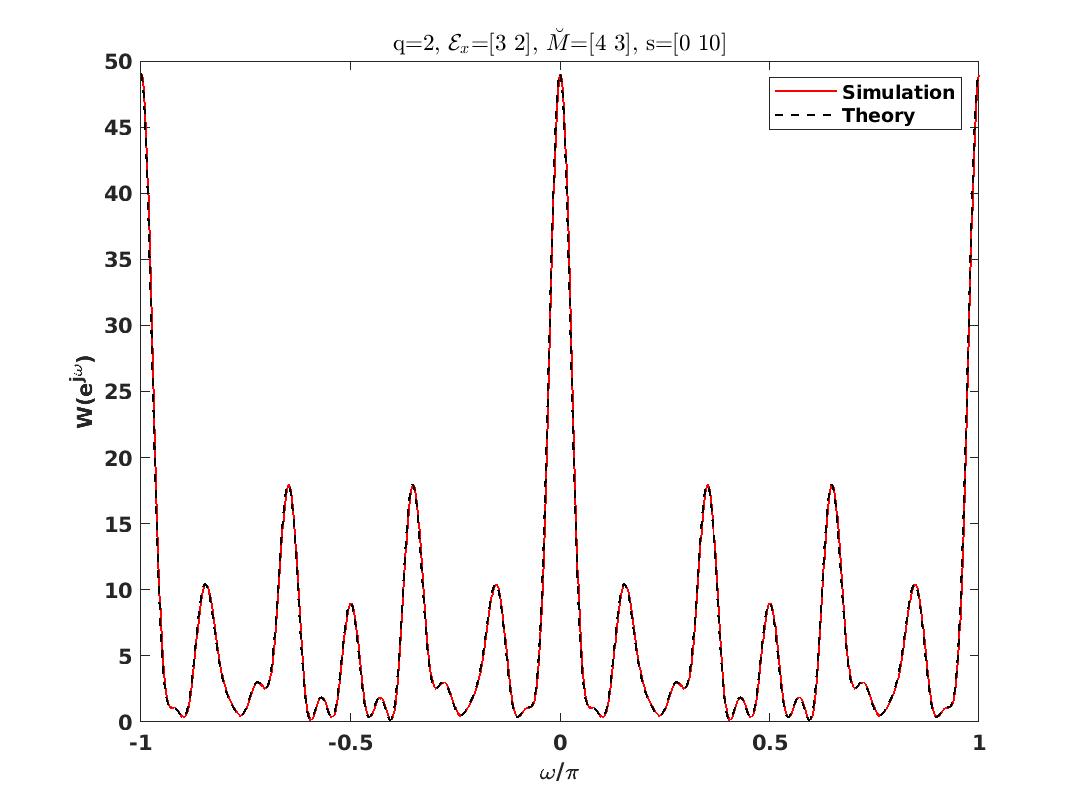}%
	\includegraphics[width=0.5\textwidth]{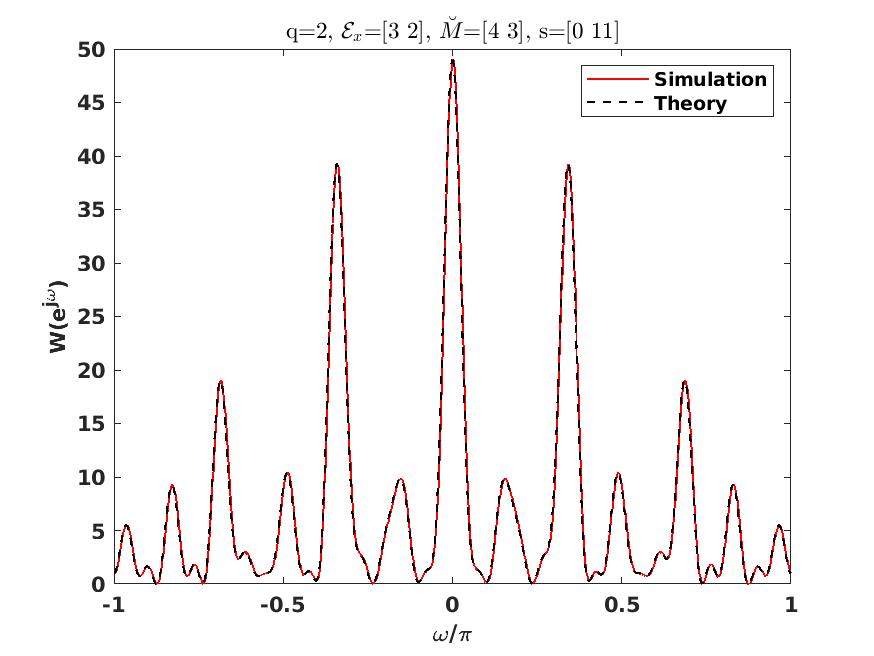}
	\caption{Bias window for generalized ExSCA as in  Fig.~\ref{fig:ExSCADiS_E1E2} for $6\leq s\leq 11$.}
	\label{fig:gen_ExSCA_E3E2s6_11}
\end{figure*}

Next note that we can also design an array to have different sparsity factors. For example, in Fig.~\ref{fig:ExSCADiS_E1E2}, the first sub-array has $\mathcal{E}_{1}=3$ and the second sub-array has $\mathcal{E}_{2}=2$. Fig.~\ref{fig:gen_ExSCA_E3E2s0_5} and Fig.~\ref{fig:gen_ExSCA_E3E2s6_11} shows the bias window for this scenario. It closely matches the expression except for $s=$ 0 and 6 since they have overlapping elements.

\begin{figure*}[!t]
	\centering
	\includegraphics[width=0.99\textwidth]{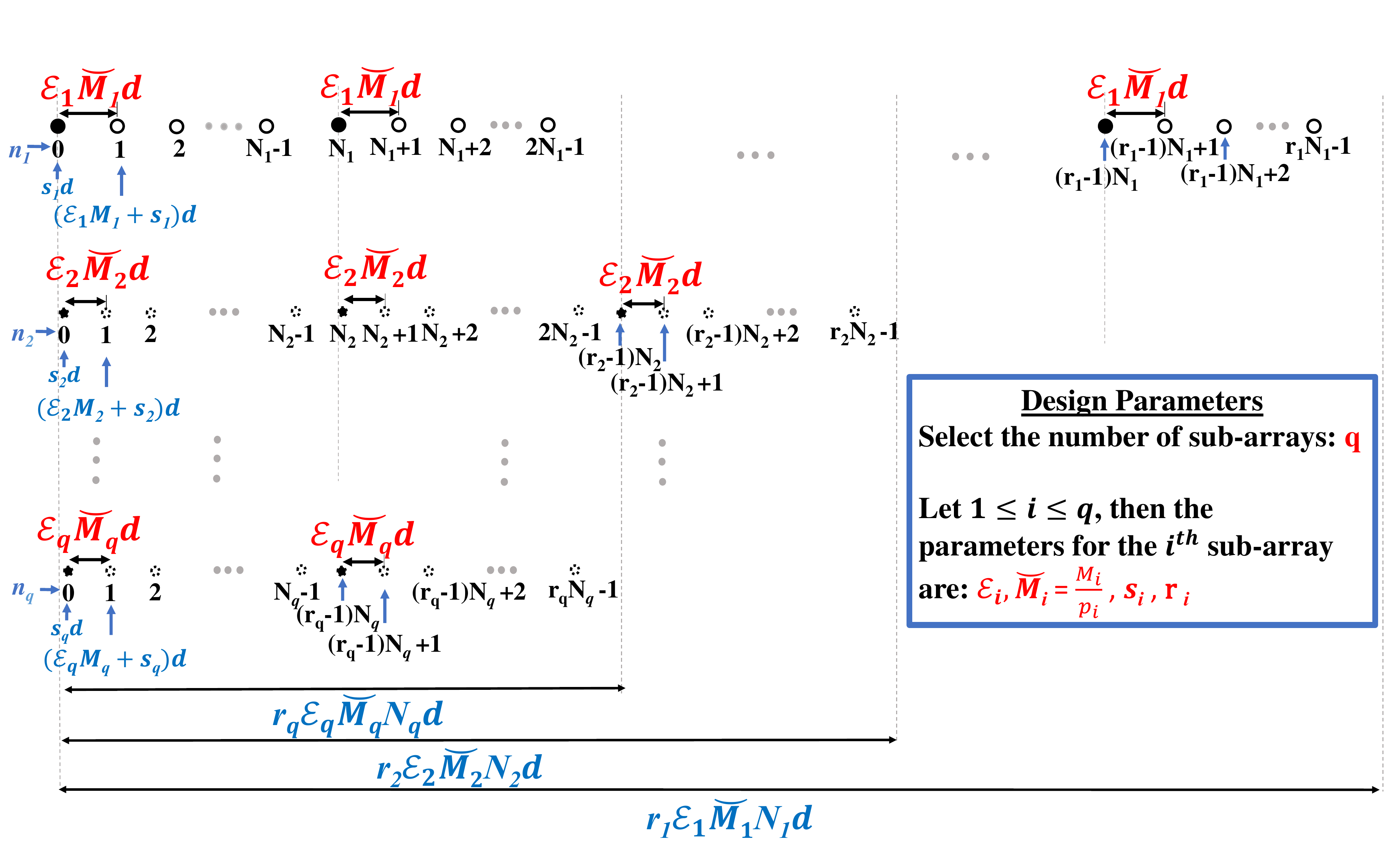}%
	\caption{Generalized extremely sparse co-prime array.}
	\label{fig:ExSCA_structure_ntuple_multi}
\end{figure*}

Finally, a generalized ExSCA is presented. It has several design parameters as shown in Fig.~\ref{fig:ExSCA_structure_ntuple_multi}. Multiple periods, multiple levels/sub-arrays (also known as n-tuple), CACIS, CADiS, APCA, nested arrays, extended co-prime, prototype co-prime, etc. are special cases of the generalized Extremely Sparse Co-Prime Arrays. It may be noted that if we replace $d$ with $\frac{d}{q}$ we obtain the Super-Nyquist Co-Prime Scheme. It was the motivating factor in the ExSCA design. Before moving forward, let us redefine the parameters for convenience. The number of sub-arrays is $q$. Till now, we had considered $q=2$. Each of the sub-array can be shifted by $s=[s_1, s_2, \dots s_q]$. Previously, we had considered $q=2$ with one array fixed and second array shifted by constant $s$, i.e. $[0, s]$. Furthermore, since $q=2$ we had only two integers $(M, N)$ representing the sub-array inter-element spacing. For a general value $q$, it may be convenient to define the sub-array inter-element spacing as $(M_1, M_2, \dots M_q)$ (\textit{otherwise we would run-out-of alphabets $(M, N, O, \dots)$}). Let us also incorporate the compression factor as in CACIS for each sub-array i.e. $p=[p_1, p_2, \dots p_q]$. Therefore, the inter-element spacing can now be compressed to give $(\breve{M}_1, \breve{M}_2, \dots \breve{M}_q)$ where  $\breve{M}_i=\frac{M_i}{p_i}$ and $1\leq i\leq q$. Note that when $p_i=M_i$ nested array configurations are possible. $p_i=1$ gives the prototype style structures without compression. We will also redefine $N$ to represent the number of sensors/antennas/samples/elements in each sub-array i.e. $N=[N_1, N_2, \dots N_q]$. In addition, different sparsity factors can be used for each sub-array i.e. $\mathcal{E}_x=[\mathcal{E}_1, \mathcal{E}_2, \dots \mathcal{E}_q]$. Furthermore, different multiple periods can be considered for each sub-array i.e. $r=[r_1, r_2, \dots r_q]$.

\begin{figure*}[!t]
	\centering
	\includegraphics[width=0.49\textwidth]{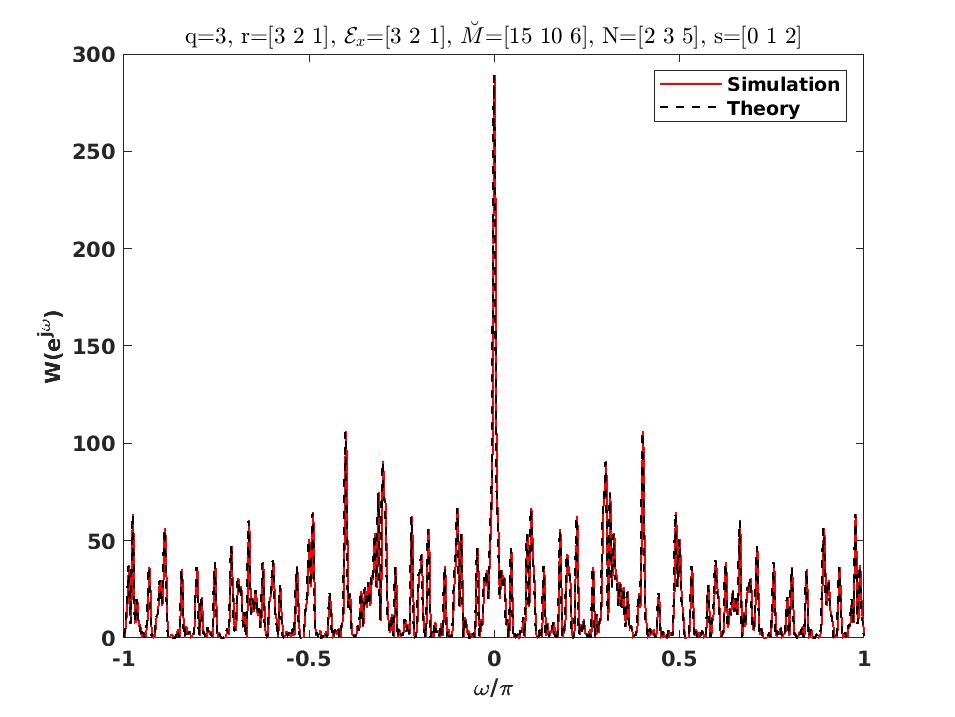}
	\includegraphics[width=0.49\textwidth]{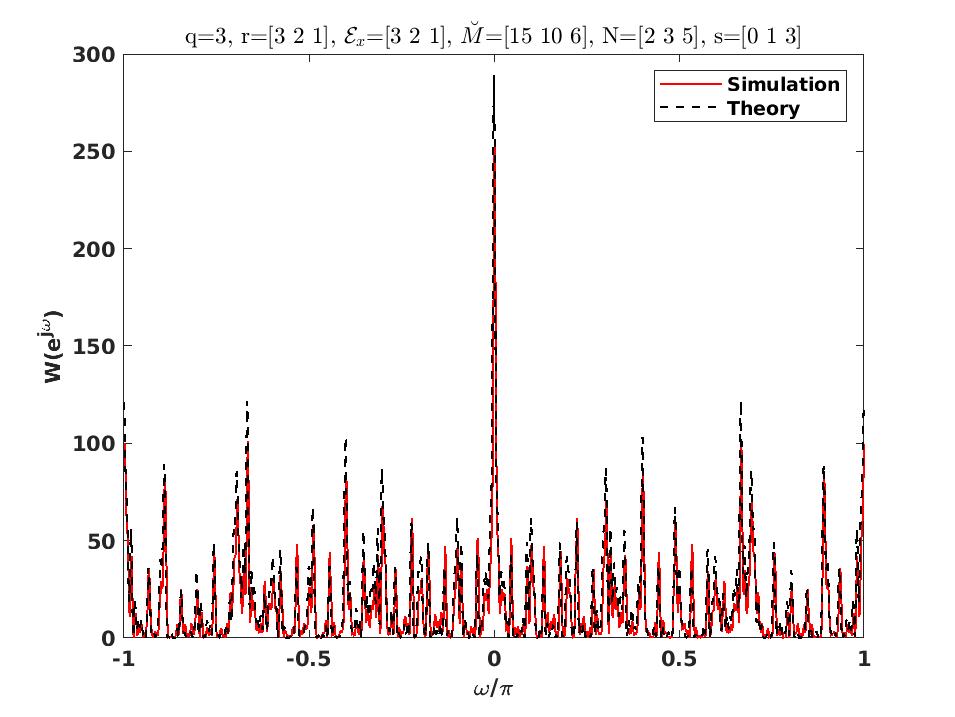}\\
	\includegraphics[width=0.5\textwidth]{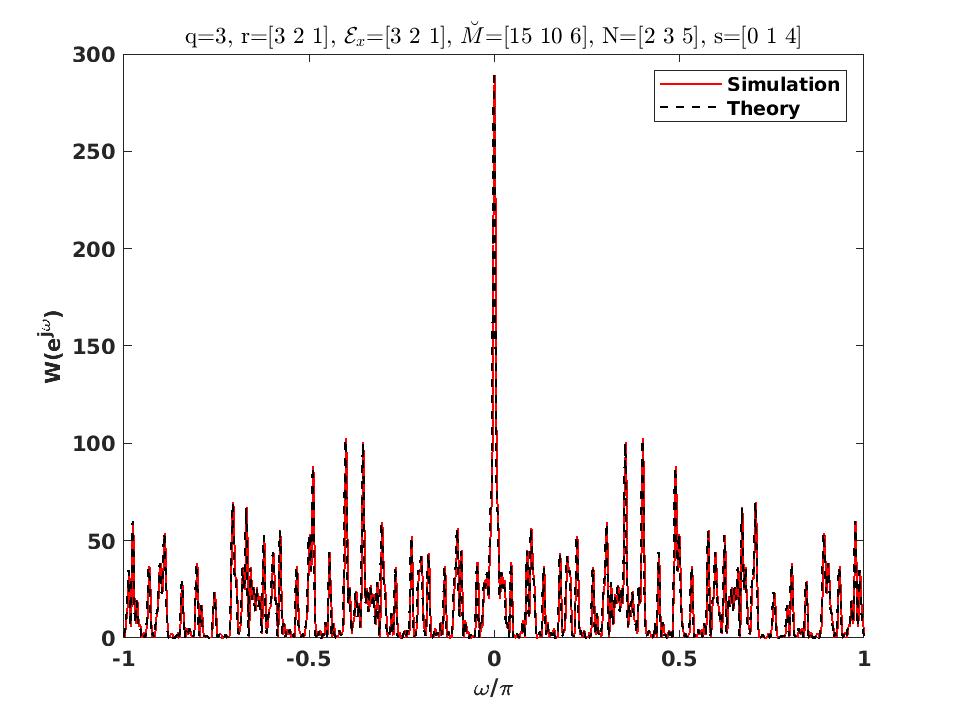}%
	\includegraphics[width=0.5\textwidth]{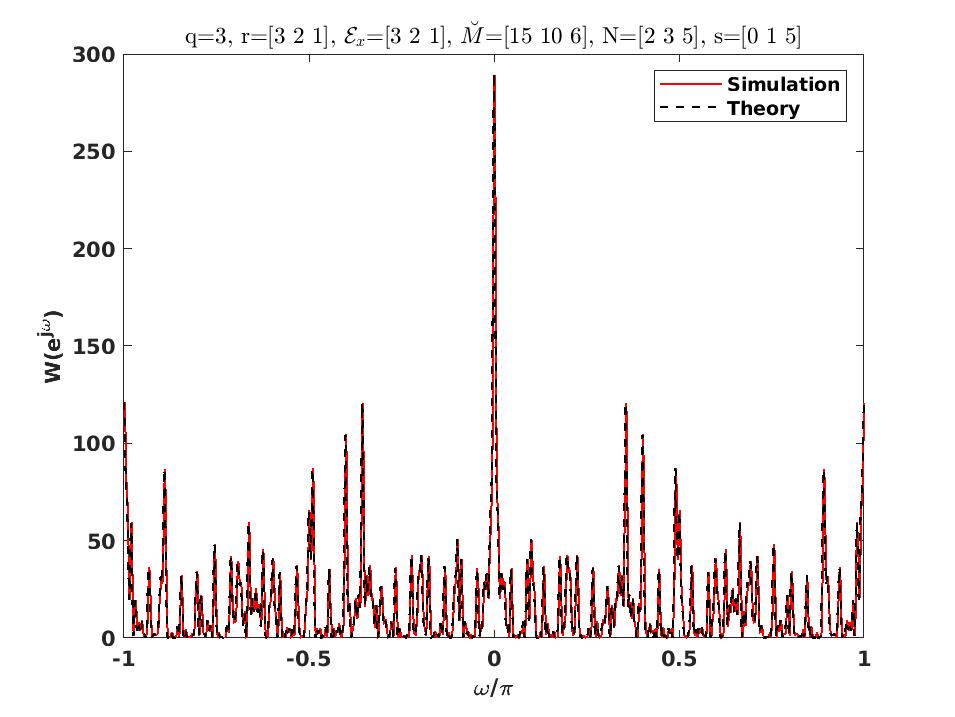}\\
	\includegraphics[width=0.5\textwidth]{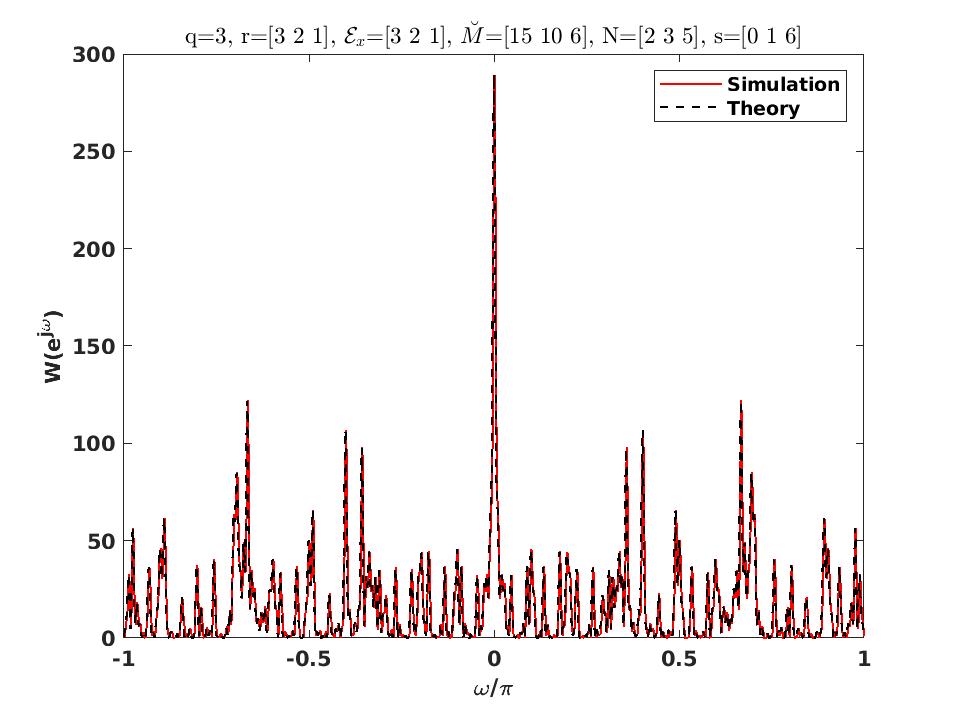}%
	\includegraphics[width=0.5\textwidth]{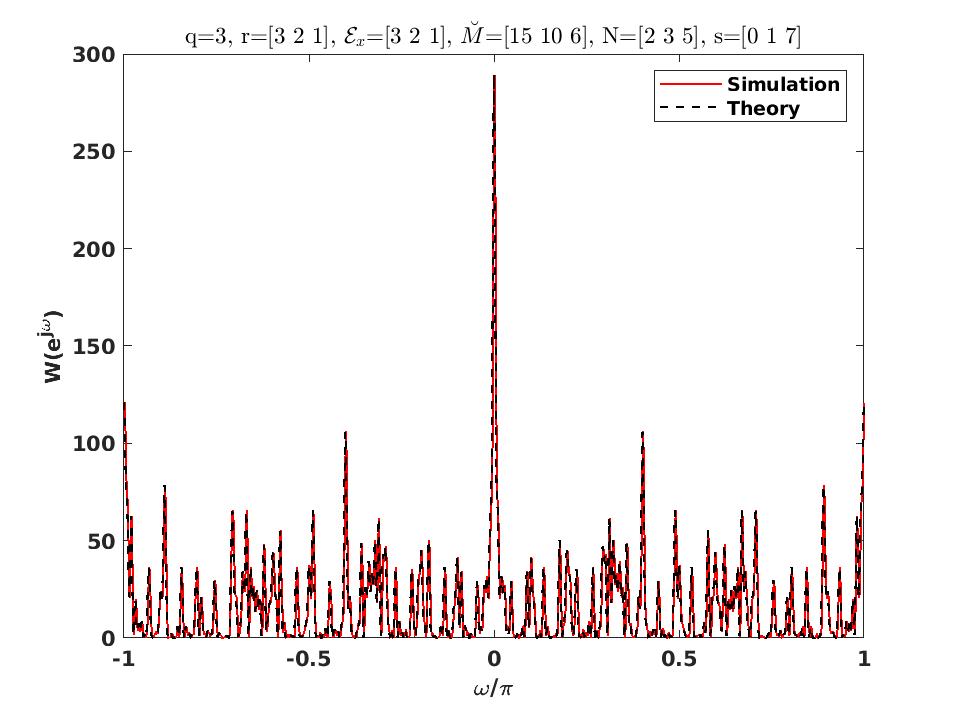}
	\caption{Bias window for generalized ExSCA as in  Fig.~\ref{fig:ExSCADiS_E1E2} for $2\leq s_3\leq7$.}
	\label{fig:gen_ExSCA_set4_s2_7}
\end{figure*}

\begin{figure*}[!t]
	\centering
	\includegraphics[width=0.49\textwidth]{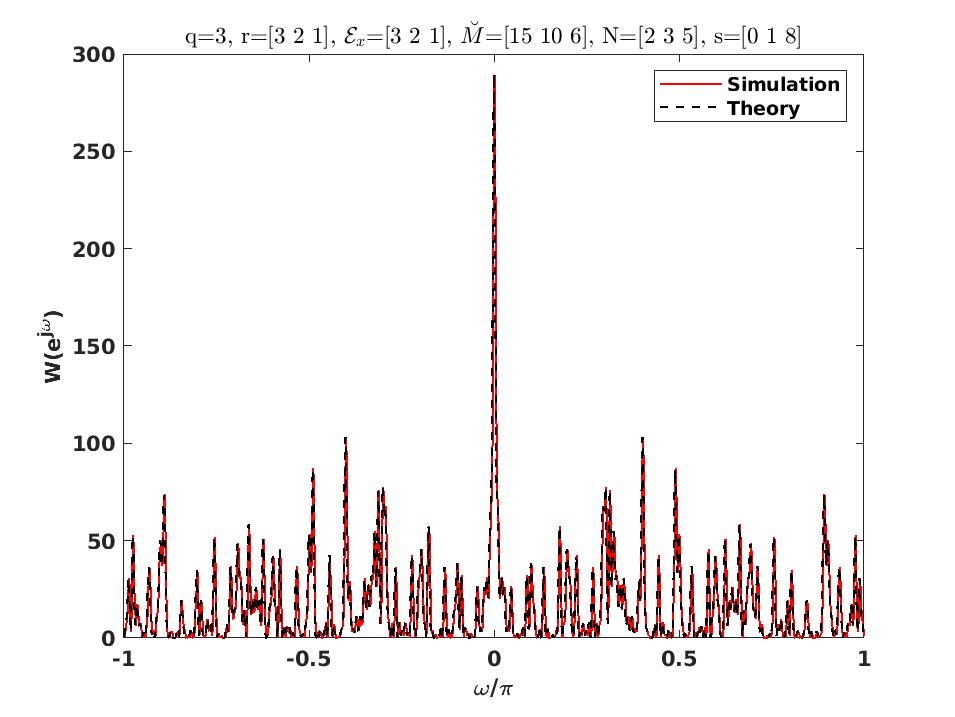}
	\includegraphics[width=0.49\textwidth]{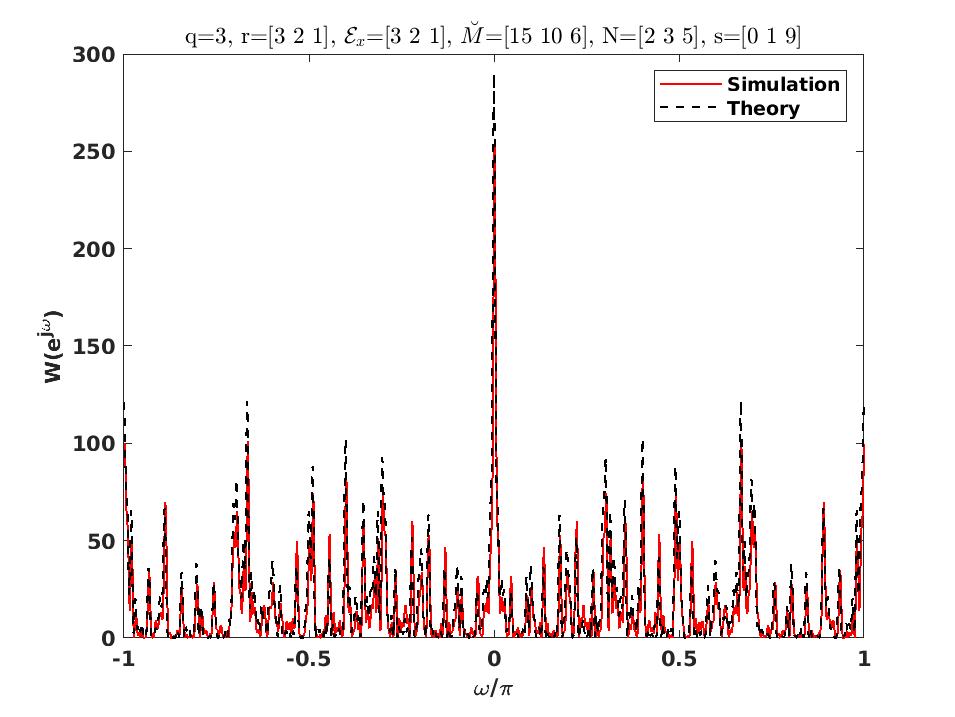}\\
	\includegraphics[width=0.5\textwidth]{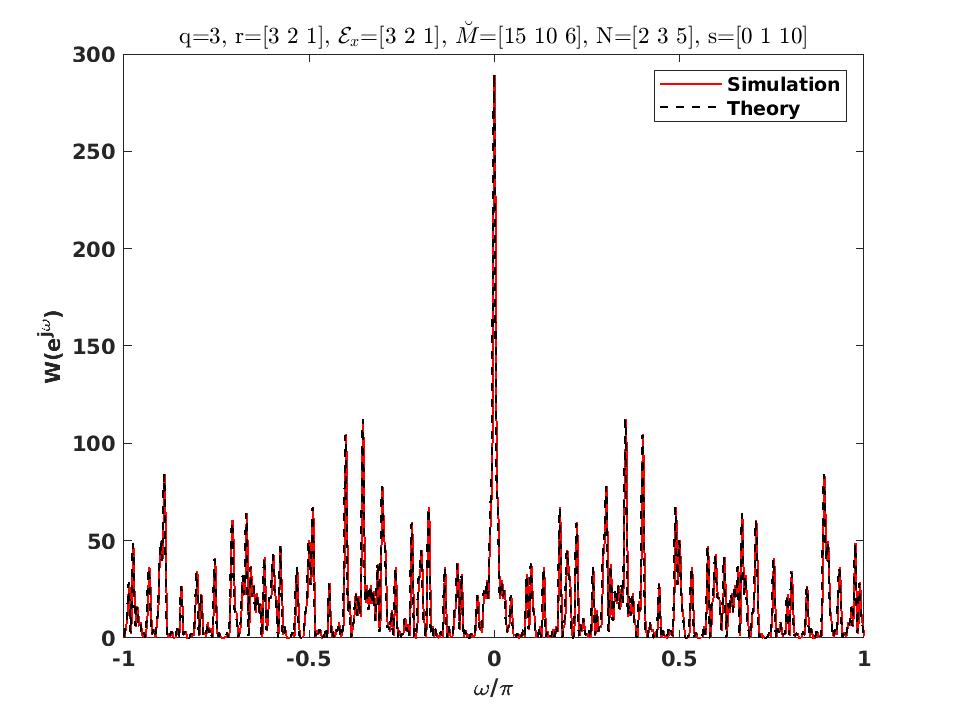}%
	\includegraphics[width=0.5\textwidth]{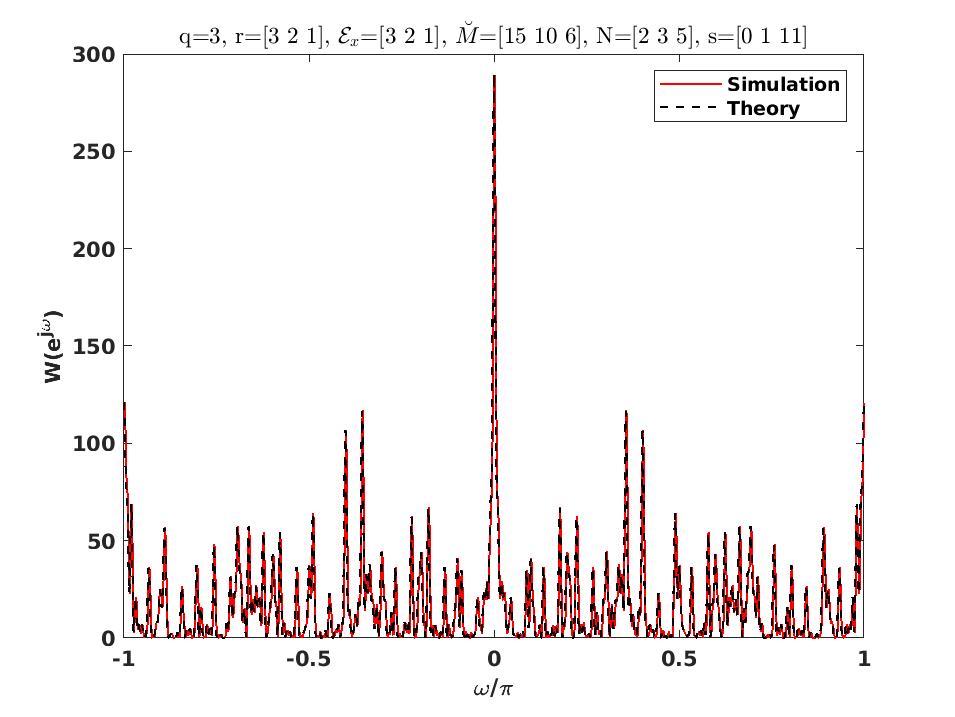}\\
	\includegraphics[width=0.5\textwidth]{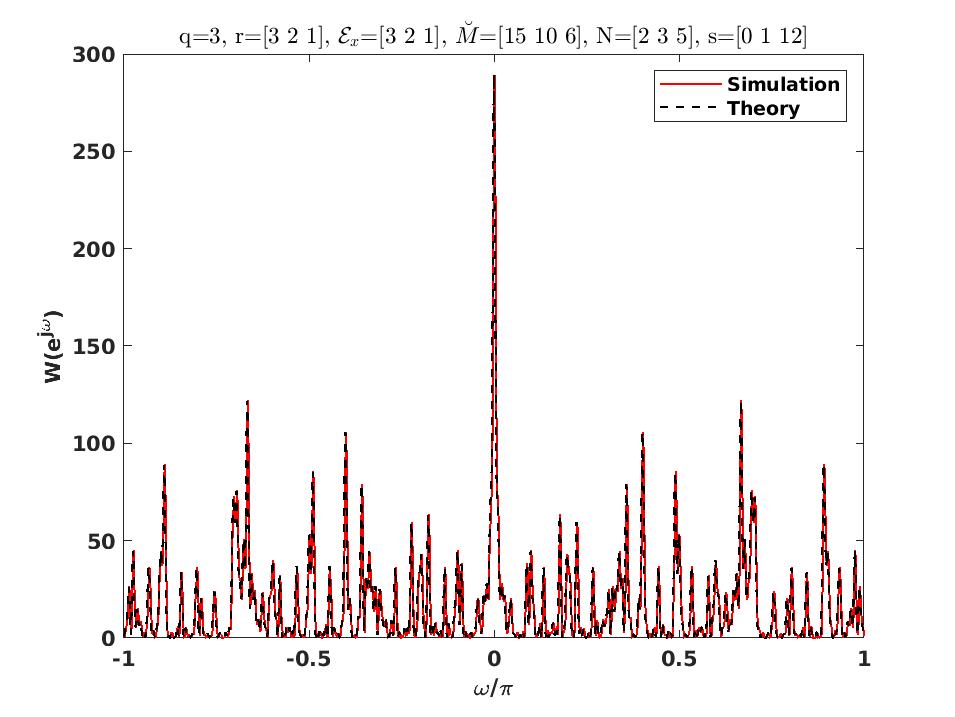}%
	\includegraphics[width=0.5\textwidth]{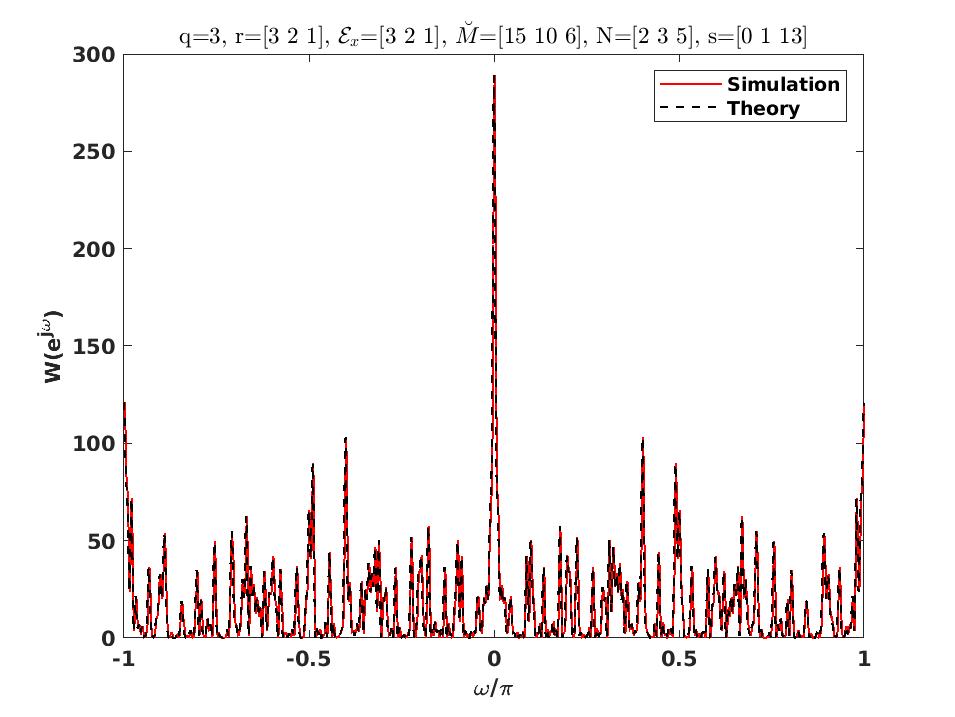}
	\caption{Bias window for generalized ExSCA as in  Fig.~\ref{fig:ExSCADiS_E1E2} for $8\leq s_3\leq 13$.}
	\label{fig:gen_ExSCA_set4_s8_13}
\end{figure*}

We also wish to show that the bias window expression in~\eqref{eq:FT_exsca_gen} is valid (except when overlapping elements are present). Consider an example of the generalized ExSCA with parameters $q=3$, $r=[3, 2, 1]$, $\mathcal{E}_x=[3, 2, 1]$, $\breve{M}=[15, 10, 6]$ with $p=[1, 1, 1]$ (no compression), $N=[2, 3, 5]$ (sensors per sub-array) and $s=[0, 1, 2]$. Let us vary the third shift parameter from $s_3=2$ to $s_3=13$. Fig.~\ref{fig:gen_ExSCA_set4_s2_7} and Fig.~\ref{fig:gen_ExSCA_set4_s8_13} shows that the expression closely matches the simulation. It definitely fails for the case when the sensors overlap. Consider another example with the same parameters but now introduce compression i.e. $p=[3, 1, 1]$. Therefore, $\breve{M}=[5, 10, 6]$. The bias window expression closely matches the simulation (except for overlapping sensors). Generalized ExSCA when combined with other sampling strategies provides a generalized hybrid ExSCA as in Section~\ref{MultiExSCA}. It was shown that the 1D theory is important since the 2D or $\eta$D theory is derived from the 1D-theory.
\begin{figure*}[!t]
	\centering
	\includegraphics[width=0.49\textwidth]{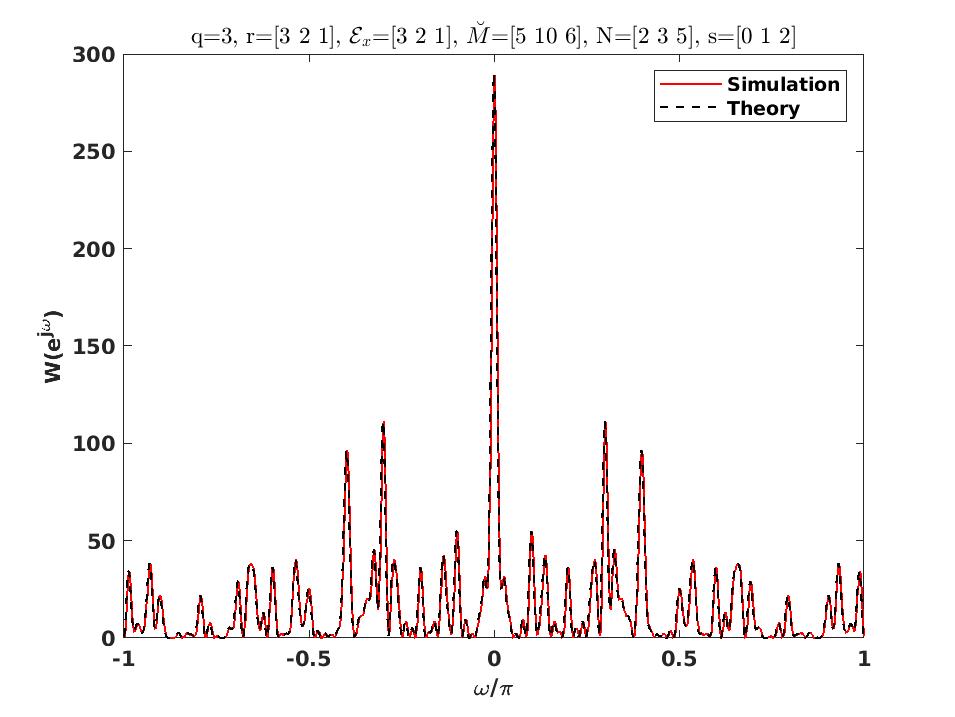}
	\includegraphics[width=0.49\textwidth]{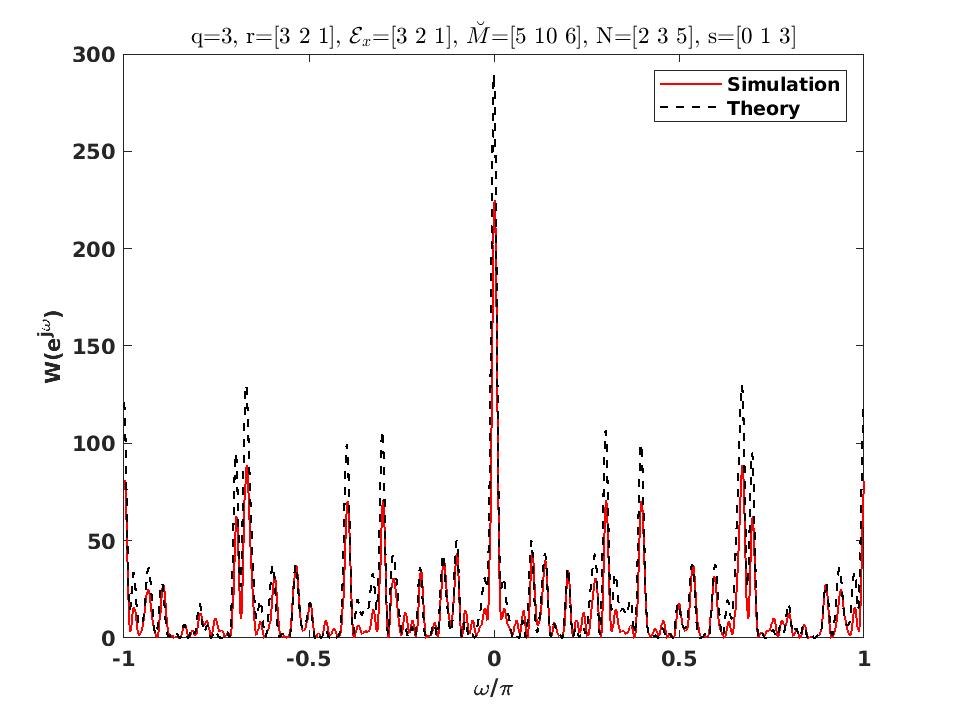}\\
	\includegraphics[width=0.5\textwidth]{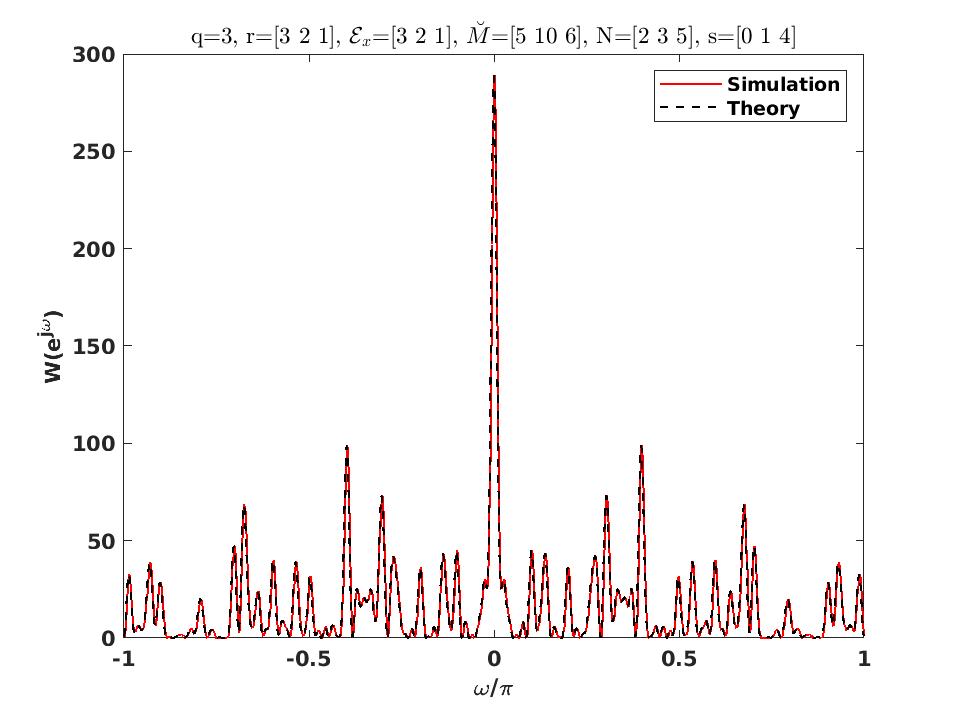}%
	\includegraphics[width=0.5\textwidth]{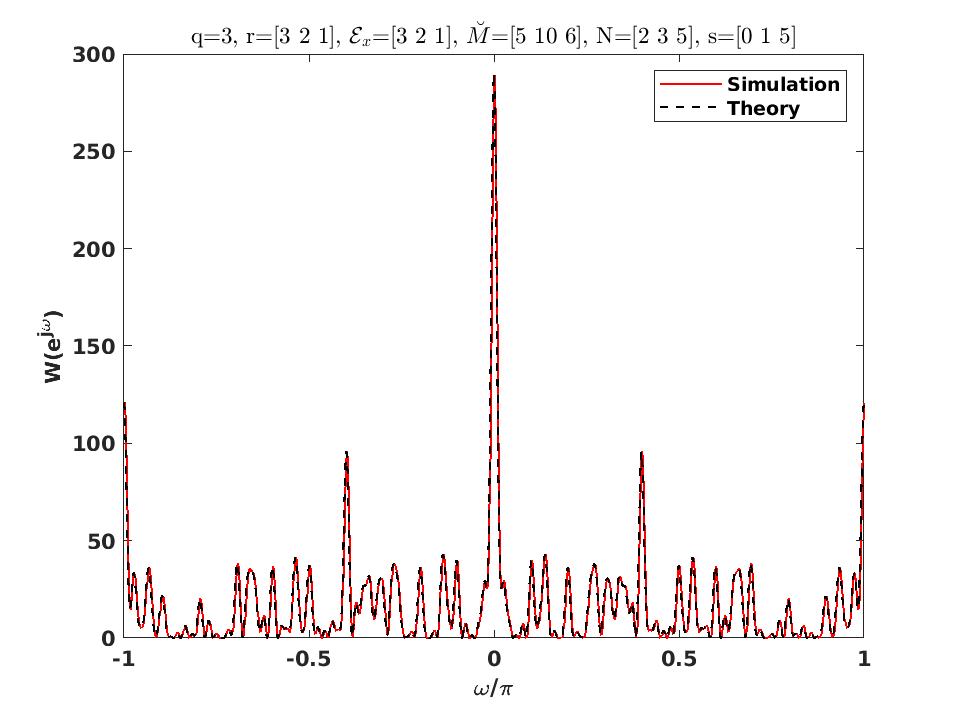}\\
	\includegraphics[width=0.5\textwidth]{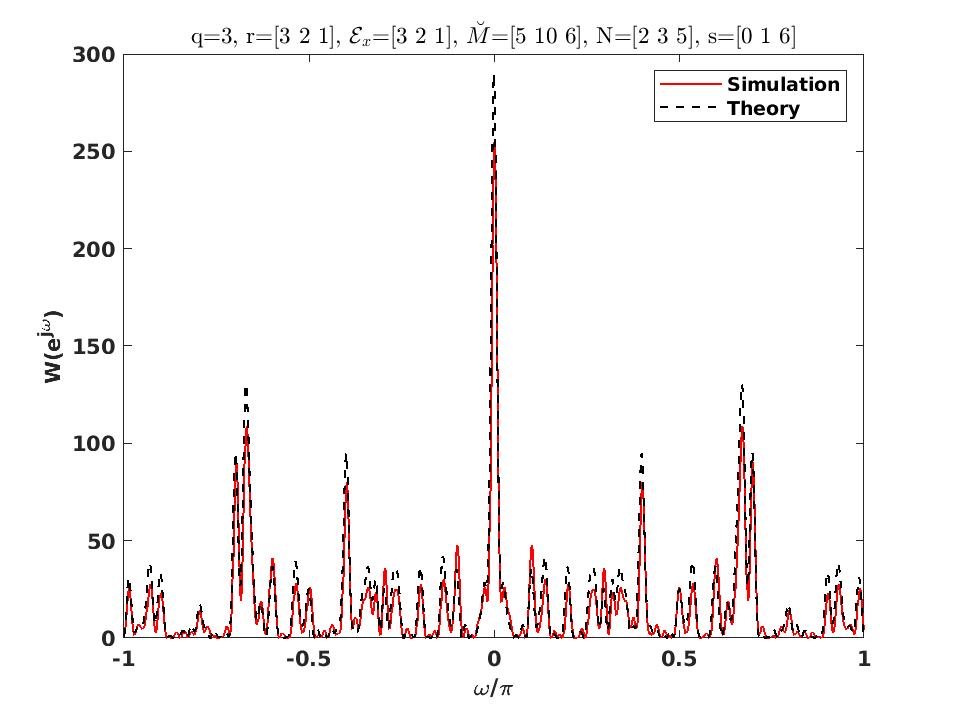}%
	\includegraphics[width=0.5\textwidth]{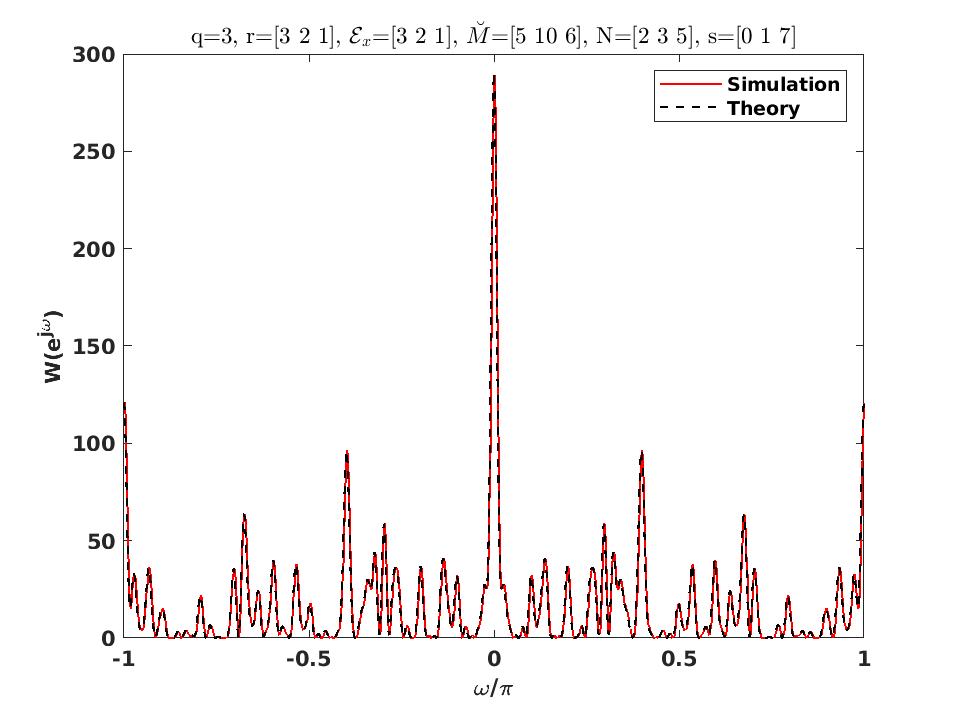}
	\caption{Bias window for generalized ExSCA as in  Fig.~\ref{fig:ExSCADiS_E1E2} for $2\leq s_3\leq7$.}
	\label{fig:gen_ExSCA_set5_s2_7}
\end{figure*}
\begin{figure*}[!t]
	\centering
	\includegraphics[width=0.49\textwidth]{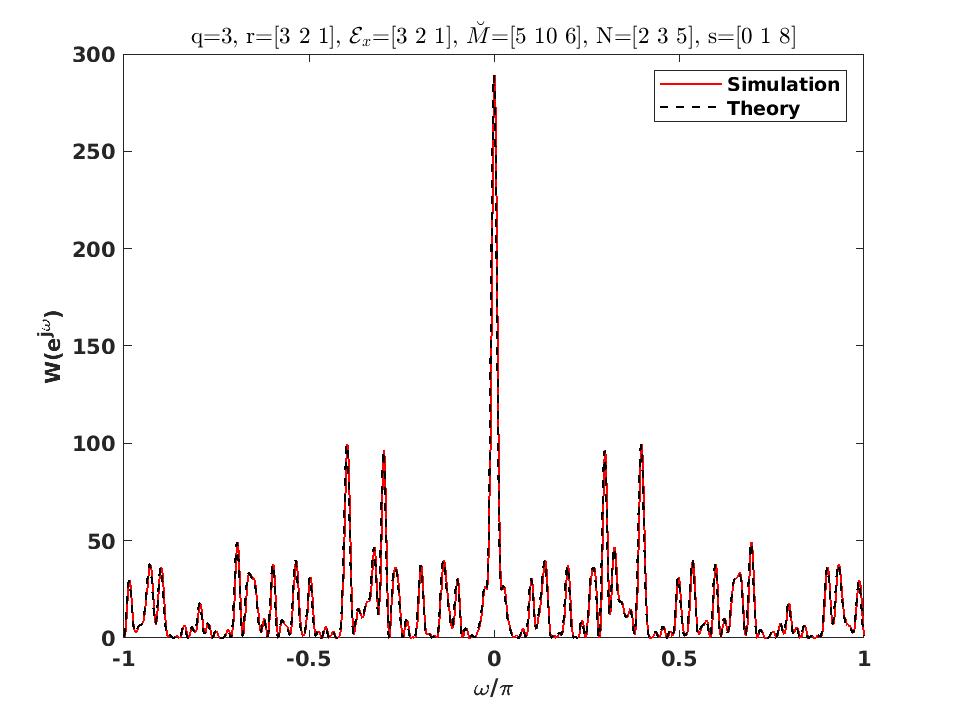}
	\includegraphics[width=0.49\textwidth]{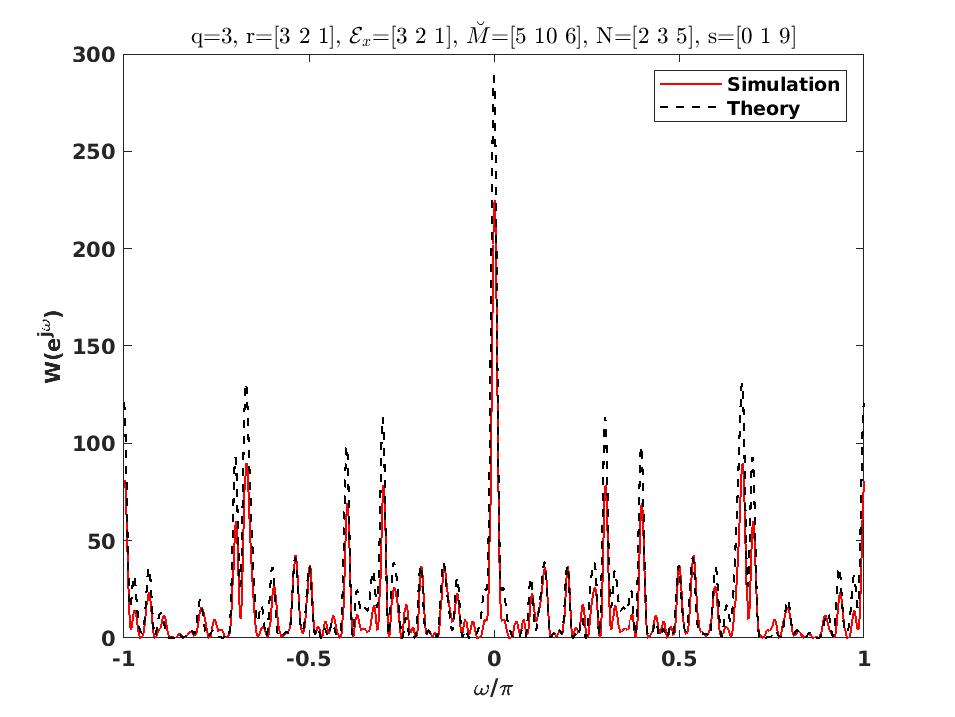}\\
	\includegraphics[width=0.5\textwidth]{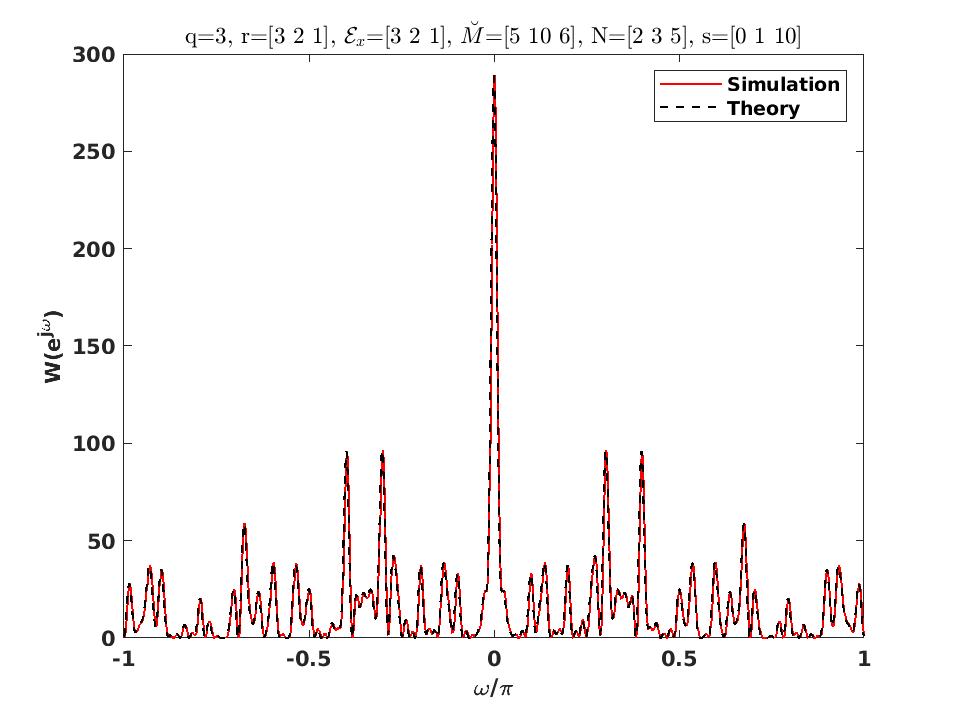}%
	\includegraphics[width=0.5\textwidth]{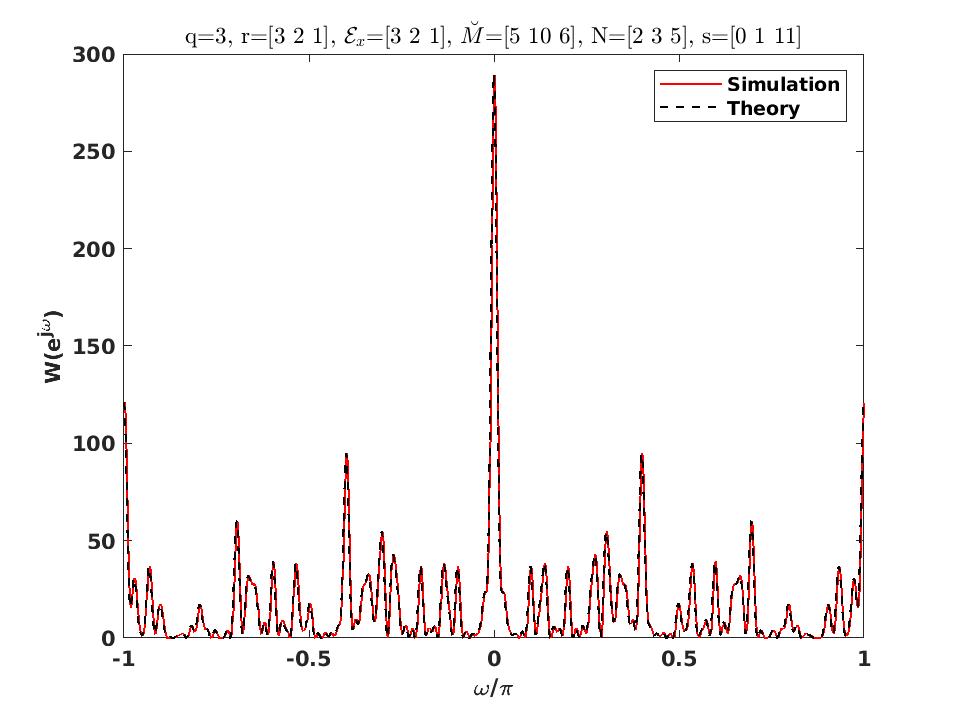}\\
	\includegraphics[width=0.5\textwidth]{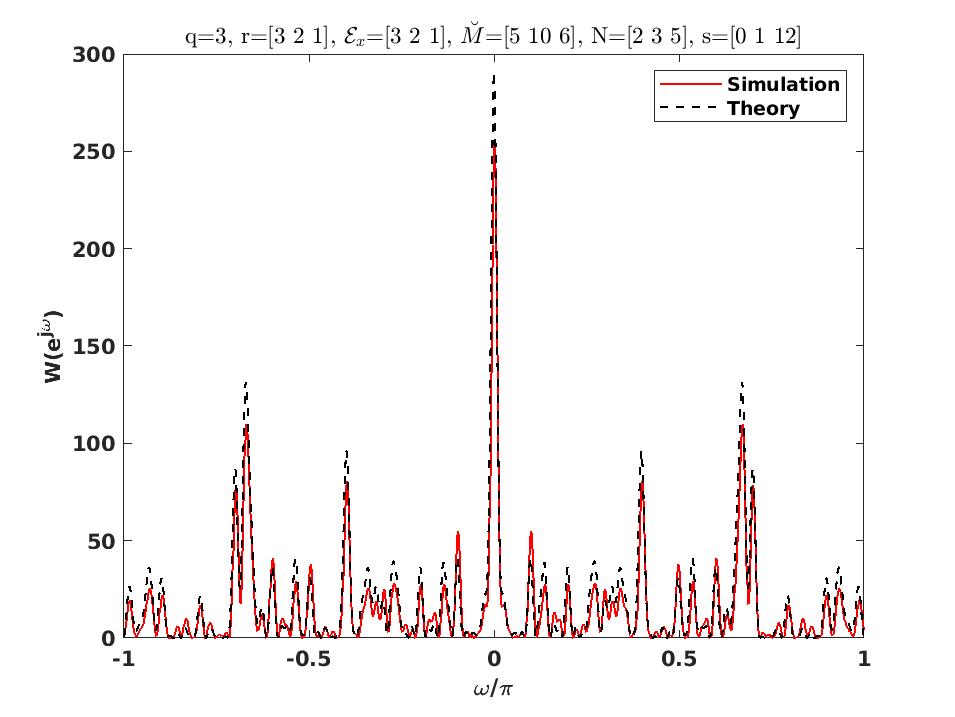}%
	\includegraphics[width=0.5\textwidth]{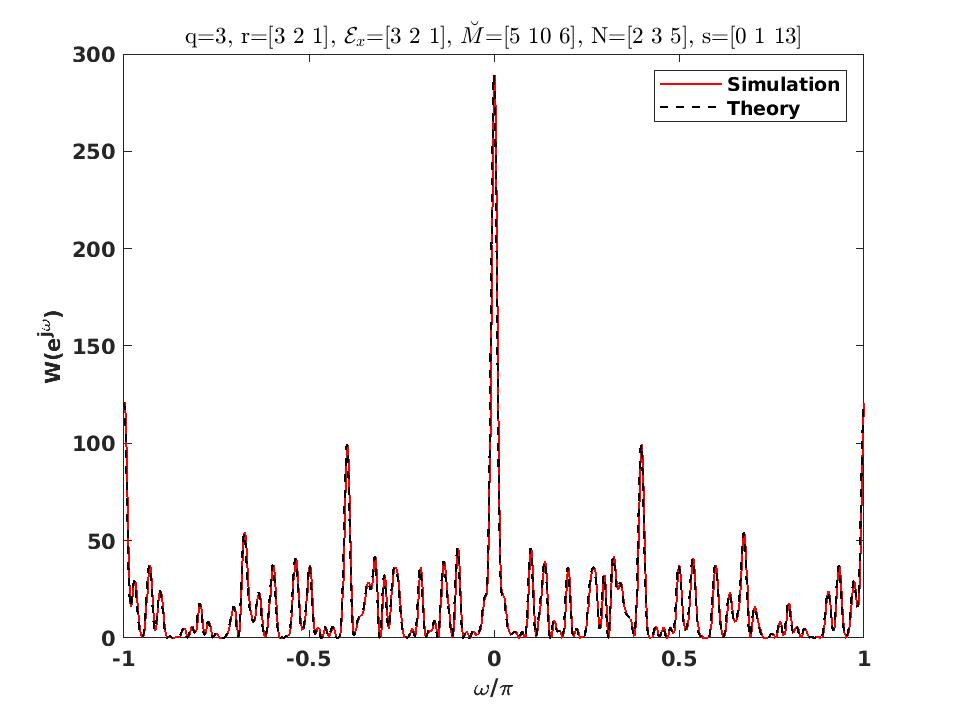}
	\caption{Bias window for generalized ExSCA as in  Fig.~\ref{fig:ExSCADiS_E1E2} for $8\leq s_3\leq 13$.}
	\label{fig:gen_ExSCA_set5_s8_13}
\end{figure*}
\section{Conclusion}
\label{sec:conclusion}
An adjustable pivot co-prime array is presented with an additional pivot selection design parameter $s$. A novel extremely sparse co-prime structure is proposed with APCA as a special case. This structure has better resolution. However, it does not provide continuous difference values/lags. This may seem to be a disadvantage for certain spectrum estimation methods which require matrix inversion. But prior works in this area have investigated interpolation of missing values. This paper uses the correlogram spectral estimation method which works with missing lag. Here, interpolation is not necessary but can be considered in the future for possible improvement. Futhermore, a generalized extremely sparse co-prime array is proposed with several existing co-prime schemes as its special case. The paper also describes the multidimensional and hybrid extremely sparse co-prime array as an extension of the 1-D theory. Simulation results show the effect of pivot selection, sparsity factor, and several other design parameters for temporal power spectrum estimation. Low latency estimation was also demonstrated. 
\ifCLASSOPTIONcaptionsoff
  \newpage
\fi
\ifCLASSOPTIONcaptionsoff
  \newpage
\fi
\bibliographystyle{IEEEtran}
\bibliography{refs}

\begin{thebibliography}{100}
\providecommand{\url}[1]{#1}
\csname url@samestyle\endcsname
\providecommand{\newblock}{\relax}
\providecommand{\bibinfo}[2]{#2}
\providecommand{\BIBentrySTDinterwordspacing}{\spaceskip=0pt\relax}
\providecommand{\BIBentryALTinterwordstretchfactor}{4}
\providecommand{\BIBentryALTinterwordspacing}{\spaceskip=\fontdimen2\font plus
\BIBentryALTinterwordstretchfactor\fontdimen3\font minus
  \fontdimen4\font\relax}
\providecommand{\BIBforeignlanguage}[2]{{%
\expandafter\ifx\csname l@#1\endcsname\relax
\typeout{** WARNING: IEEEtran.bst: No hyphenation pattern has been}%
\typeout{** loaded for the language `#1'. Using the pattern for}%
\typeout{** the default language instead.}%
\else
\language=\csname l@#1\endcsname
\fi
#2}}
\providecommand{\BIBdecl}{\relax}
\BIBdecl

\bibitem{1.10}
C.~E. Shannon, ``Communication in the presence of noise,'' \emph{Proceedings of
  the IEEE}, vol.~86, no.~2, pp. 447--457, Feb 1998.

\bibitem{1.11}
H.~Nyquist, ``Certain topics in telegraph transmission theory,''
  \emph{Transactions of the American Institute of Electrical Engineers},
  vol.~47, no.~2, pp. 617--644, April 1928.

\bibitem{1.13}
E.~T. Whittaker, ``Xviii.—on the functions which are represented by the
  expansions of the interpolation-theory,'' \emph{Proceedings of the Royal
  Society of Edinburgh}, vol.~35, p. 181–194, 1915.

\bibitem{1.14}
V.~A. Kotel'nikov, ``On the transmission capacity of “ether” and wire in
  electrocommunications,'' \emph{Material for the First All-Union Conference of
  Questions of Communication, Izd. Red. Upr. Svyazi RKKA, Moscow}, 1933.

\bibitem{1.9}
M.~Mishali and Y.~C. Eldar, ``Sub-\uppercase{N}yquist sampling,'' \emph{IEEE
  Signal Processing Magazine}, vol.~28, no.~6, pp. 98--124, Nov 2011.

\bibitem{1.5}
D.~Donoho, ``Compressed sensing,'' \emph{IEEE Trans. Inf. Theory}, vol.~52,
  no.~4, pp. 1289--1306, Apr. 2006.

\bibitem{1.6}
E.~J. Candes and T.~Tao, ``Near-optimal signal recovery from random
  projections: Universal encoding strategies?'' \emph{IEEE Trans. Inf. Theory},
  vol.~52, no.~12, pp. 5406--5425, Dec. 2006.

\bibitem{1.7}
R.~G. Baraniuk, ``Compressive sensing,'' \emph{IEEE Signal Process. Mag.},
  vol.~24, no.~4, pp. 118--121, Jul. 2007.

\bibitem{survey_CSS1}
A.~Mousavi, M.~Rezaee, and R.~Ayanzadeh, ``A survey on compressive sensing:
  Classical results and recent advancements,'' 2019, arXiv: 1908.01014
  [math.OC].

\bibitem{survey_CSS2}
S.~{Qaisar}, R.~M. {Bilal}, W.~{Iqbal}, M.~{Naureen}, and S.~{Lee},
  ``Compressive sensing: From theory to applications, a survey,'' \emph{Journal
  of Communications and Networks}, vol.~15, no.~5, pp. 443--456, 2013.

\bibitem{survey_CSS3}
\BIBentryALTinterwordspacing
F.~Salahdine, N.~Kaabouch, and H.~{El Ghazi}, ``A survey on compressive sensing
  techniques for cognitive radio networks,'' \emph{Physical Communication},
  vol.~20, pp. 61 -- 73, 2016. [Online]. Available:
  \url{http://www.sciencedirect.com/science/article/pii/S1874490716300386}
\BIBentrySTDinterwordspacing

\bibitem{survey_CSS4}
K.~V. {Siddamal}, S.~P. {Bhat}, and V.~S. {Saroja}, ``A survey on compressive
  sensing,'' in \emph{2015 2nd International Conference on Electronics and
  Communication Systems (ICECS)}, 2015, pp. 639--643.

\bibitem{survey_CSS5}
M.~{Rani}, S.~B. {Dhok}, and R.~B. {Deshmukh}, ``A systematic review of
  compressive sensing: Concepts, implementations and applications,'' \emph{IEEE
  Access}, vol.~6, pp. 4875--4894, 2018.

\bibitem{survey_CSS6}
S.~K. {Sharma}, E.~{Lagunas}, S.~{Chatzinotas}, and B.~{Ottersten},
  ``Application of compressive sensing in cognitive radio communications: A
  survey,'' \emph{IEEE Communications Surveys Tutorials}, vol.~18, no.~3, pp.
  1838--1860, 2016.

\bibitem{survey_CSS7}
Z.~Li, W.~Xu, X.~Zhang, and J.~Lin, ``A survey on one-bit compressed sensing:
  theory and applications,'' \emph{Frontiers of Computer Science}, vol.~12, pp.
  217--230, 2018.

\bibitem{survey_CSS8}
J.~C. Ye, ``Compressed sensing mri: a review from signal processing
  perspective,'' \emph{BMC Biomedical Engineering}, vol.~1, 2019.

\bibitem{survey_CSS9}
M.~Yousufi, M.~Amir, U.~Javed, M.~Tayyib, S.~Abdullah, H.~Ullah, I.~Qureshi,
  K.~S. Alimgeer, M.~W. Akram, and K.~Khan, ``Application of compressive
  sensing to ultrasound images: A review,'' \emph{BioMed Research
  International}, vol. 2019, 2019.

\bibitem{survey_CSS10}
I.~Orovi{\'c}, V.~Papic, C.~Ioana, X.~Li, and S.~Stankovic, ``Compressive
  sensing in signal processing: Algorithms and transform domain formulations,''
  \emph{Mathematical Problems in Engineering}, vol. 2016, pp. 1--16, 2016.

\bibitem{survey_CSS11}
R.~Carrillo, A.~Ramirez, G.~Arce, K.~Barner, and B.~Sadler, ``Robust
  compressive sensing of sparse signals: A review,'' \emph{EURASIP Journal on
  Advances in Signal Processing}, vol. 2016, p. 108, 10 2016.

\bibitem{survey_CSS12}
\BIBentryALTinterwordspacing
S.~Salari, F.~Chan, and Y.-T. Chan, ``Applications of compressive sampling
  technique to radar and localization,'' in \emph{Advanced Electronic
  Circuits}, M.~Niu, Ed.\hskip 1em plus 0.5em minus 0.4em\relax Rijeka:
  IntechOpen, 2018, ch.~5. [Online]. Available:
  \url{https://doi.org/10.5772/intechopen.75072}
\BIBentrySTDinterwordspacing

\bibitem{survey_CSS13}
T.~Wimalajeewa and P.~K. Varshney, ``Application of compressive sensing
  techniques in distributed sensor networks: A survey,'' 2017, arXiv:
  1709.10401 [eess.SP].

\bibitem{survey_CSS14}
\BIBentryALTinterwordspacing
D.~Gurve, D.~Delisle-Rodriguez, T.~Bastos-Filho, and S.~Krishnan, ``Trends in
  compressive sensing for eeg signal processing applications,'' \emph{Sensors
  (Basel, Switzerland)}, vol.~20, no.~13, July 2020. [Online]. Available:
  \url{https://europepmc.org/articles/PMC7374282}
\BIBentrySTDinterwordspacing

\bibitem{survey_CSS15}
M.~Don, ``Can compressive sensing solve your sensor and measurement problems?''
  vol.~7, p.~19, 05 2020.

\bibitem{10.3}
A.~Goldsmith and I.~Maric, ``Capacity of cognitive radio networks,''
  \emph{Principles of Cognitive Radio:. Cambridge: Cambridge University Press},
  vol.~11, pp. 41--101, 2012.

\bibitem{1.1}
H.~Sun, A.~Nallanathan, C.~X. Wang, and Y.~Chen, ``Wideband spectrum sensing
  for cognitive radio networks: A survey,'' \emph{IEEE Wireless Commun.},
  vol.~20, no.~2, pp. 74--81, Apr. 2013.

\bibitem{1.21}
M.~Mishali and Y.~C. Eldar, ``From theory to practice: Sub-\uppercase{N}yquist
  sampling of sparse wideband analog signals,'' \emph{IEEE Journal of Selected
  Topics in Signal Processing}, vol.~4, no.~2, pp. 375--391, April 2010.

\bibitem{1.24}
M.~{Mishali}, Y.~C. {Eldar}, O.~{Dounaevsky}, and E.~{Shoshan}, ``Xampling:
  Analog to digital at sub-\uppercase{N}yquist rates,'' \emph{IET Circuits,
  Devices Systems}, vol.~5, no.~1, pp. 8--20, January 2011.

\bibitem{1.25}
M.~{Mishali}, Y.~C. {Eldar}, and A.~J. {Elron}, ``Xampling: Signal acquisition
  and processing in union of subspaces,'' \emph{IEEE Transactions on Signal
  Processing}, vol.~59, no.~10, pp. 4719--4734, Oct 2011.

\bibitem{1.22}
A.~Moffet, ``Minimum-redundancy linear arrays,'' \emph{IEEE Transactions on
  Antennas and Propagation}, vol.~16, no.~2, pp. 172--175, Mar 1968.

\bibitem{MHA}
G.~S. {Bloom} and S.~W. {Golomb}, ``Applications of numbered undirected
  graphs,'' \emph{Proceedings of the IEEE}, vol.~65, no.~4, pp. 562--570, 1977.

\bibitem{4.52}
P.~Pal and P.~P. Vaidyanathan, ``Nested arrays: a novel approach to array
  processing with enhanced degrees of freedom,'' \emph{IEEE Trans. Signal
  Process.}, vol.~58, no.~8, pp. 4167--4181, Aug. 2010.

\bibitem{4.7}
P.~P. Vaidyanathan and P.~Pal, ``Sparse sensing with co-prime samplers and
  arrays,'' \emph{IEEE Trans. Signal Process.}, vol.~59, no.~2, pp. 573--586,
  Feb. 2011.

\bibitem{Nested1}
C.~{Liu} and P.~P. {Vaidyanathan}, ``Super nested arrays: Linear sparse arrays
  with reduced mutual coupling—part i: Fundamentals,'' \emph{IEEE
  Transactions on Signal Processing}, vol.~64, no.~15, pp. 3997--4012, 2016.

\bibitem{Nested2}
S.~Li and D.~Xie, ``Compressed symmetric nested arrays and their application
  for direction-of-arrival estimation of near-field sources,'' \emph{Sensors},
  vol.~16, 11 2016.

\bibitem{Nested3}
H.~{Huang}, B.~{Liao}, X.~{Wang}, X.~{Guo}, and J.~{Huang}, ``A new nested
  array configuration with increased degrees of freedom,'' \emph{IEEE Access},
  vol.~6, pp. 1490--1497, 2018.

\bibitem{Nested5}
J.~{Shi}, G.~{Hu}, X.~{Zhang}, and H.~{Zhou}, ``Generalized nested array:
  Optimization for degrees of freedom and mutual coupling,'' \emph{IEEE
  Communications Letters}, vol.~22, no.~6, pp. 1208--1211, 2018.

\bibitem{Nested6}
P.~{Gupta} and M.~{Agrawal}, ``Design and analysis of the sparse array for
  \uppercase{D}o\uppercase{A} estimation of noncircular signals,'' \emph{IEEE
  Transactions on Signal Processing}, vol.~67, no.~2, pp. 460--473, 2019.

\bibitem{Nested4}
X.~{Lin}, P.~{Gong}, L.~{He}, and J.~{Li}, ``Widened nested array:
  configuration design, optimal array and \uppercase{D}o\uppercase{A}
  estimation algorithm,'' \emph{IET Microwaves, Antennas Propagation}, vol.~14,
  no.~5, pp. 440--447, 2020.

\bibitem{4.62}
P.~Pal and P.~P. Vaidyanathan, ``Coprime sampling and the music algorithm,'' in
  \emph{2011 Digital Signal Processing and Signal Processing Education Meeting
  (DSP/SPE)}, Jan 2011, pp. 289--294.

\bibitem{20.1}
A.~Raza, W.~Liu, and Q.~Shen, ``Thinned coprime arrays for
  \uppercase{D}o\uppercase{A} estimation,'' in \emph{2017 25th European Signal
  Processing Conference (EUSIPCO)}, Aug 2017, pp. 395--399.

\bibitem{4.44}
Q.~Si, Y.~D. Zhang, and M.~G. Amin, ``Generalized coprime array configurations
  for direction-of-arrival estimation,'' \emph{IEEE Trans. Signal Process.},
  vol.~63, no.~6, pp. 1377--1390, Mar. 2015.

\bibitem{20.4}
S.~A. {Alawsh} and A.~H. {Muqaibel}, ``Three-level prime arrays for sparse
  sampling in direction of arrival estimation,'' in \emph{2016 IEEE
  Asia-Pacific Conference on Applied Electromagnetics (APACE)}, 2016, pp.
  277--281.

\bibitem{20.2}
D.~Bush and N.~Xiang, ``n-tuple coprime sensor arrays,'' \emph{The Journal of
  the Acoustical Society of America}, vol. 142, no.~6, Dec 2017.

\bibitem{20.3}
\BIBentryALTinterwordspacing
S.~A. Alawsh and A.~H. Muqaibel, ``Multi-level prime array for sparse
  sampling,'' \emph{IET Signal Processing}, vol.~12, pp. 688--699(11), August
  2018. [Online]. Available:
  \url{https://digital-library.theiet.org/content/journals/10.1049/iet-spr.2017.0252}
\BIBentrySTDinterwordspacing

\bibitem{20.5}
S.~A. {Alawsh} and A.~H. {Muqaibel}, ``Sparse \uppercase{D}o\uppercase{A}
  estimation based on multi-level prime array with compression,'' \emph{IEEE
  Access}, vol.~7, pp. 70\,828--70\,841, 2019.

\bibitem{20.6}
------, ``Three-level prime arrays with compressed subarray for
  \uppercase{D}o\uppercase{A} estimation using compressive sensing,'' in
  \emph{2019 2nd IEEE Middle East and North Africa COMMunications Conference
  (MENACOMM)}, 2019, pp. 1--5.

\bibitem{20.7}
D.~Bush and N.~Xiang, ``Investigations on n-tuple coprime arrays,'' \emph{The
  Journal of the Acoustical Society of America}, vol. 143, pp. 1851--1851, 03
  2018.

\bibitem{semiCA}
\BIBentryALTinterwordspacing
K.~Adhikari, ``Beamforming with semi-coprime arrays,'' \emph{The Journal of the
  Acoustical Society of America}, vol. 145, no.~5, pp. 2841--2850, 2019.
  [Online]. Available: \url{https://doi.org/10.1121/1.5100281}
\BIBentrySTDinterwordspacing

\bibitem{CAMPs}
\BIBentryALTinterwordspacing
W.~Wang, S.~Ren, and Z.~Chen, ``Unified coprime array with multi-period
  subarrays for direction-of-arrival estimation,'' \emph{Digital Signal
  Processing}, vol.~74, pp. 30 -- 42, 2018. [Online]. Available:
  \url{http://www.sciencedirect.com/science/article/pii/S1051200417302786}
\BIBentrySTDinterwordspacing

\bibitem{multidim1}
P.~P. {Vaidyanathan} and P.~{Pal}, ``Theory of sparse coprime sensing in
  multiple dimensions,'' \emph{IEEE Transactions on Signal Processing},
  vol.~59, no.~8, pp. 3592--3608, 2011.

\bibitem{multidim2}
\BIBentryALTinterwordspacing
X.~Zhang, W.~Zheng, W.~Chen, and Z.~Shi, ``Two-dimensional
  \uppercase{D}o\uppercase{A} estimation for generalized coprime planar arrays:
  A fast-convergence trilinear decomposition approach,'' \emph{Multidimensional
  Syst. Signal Process.}, vol.~30, no.~1, p. 239–256, Jan. 2019. [Online].
  Available: \url{https://doi.org/10.1007/s11045-018-0553-9}
\BIBentrySTDinterwordspacing

\bibitem{multidim3}
P.~Gong, T.~Ahmed, and J.~Li, ``Three-dimensional coprime array for massive
  \uppercase{MIMO}: Array configuration design and \uppercase{2D}
  \uppercase{D}o\uppercase{A} estimation,'' \emph{Wireless Communications and
  Mobile Computing}, vol. 2020, pp. 1--14, 01 2020.

\bibitem{multidim4}
G.~Wang, Z.~Fei, S.~Ren, and X.~Li, ``Improved \uppercase{2D} coprime array
  structure with the difference and sum coarray concept,'' \emph{Electronics},
  vol.~9, p. 273, 02 2020.

\bibitem{rect1}
K.~{Adhikari} and B.~{Drozdenko}, ``Symmetry-imposed rectangular coprime and
  nested arrays for direction of arrival estimation with multiple signal
  classification,'' \emph{IEEE Access}, vol.~7, pp. 153\,217--153\,229, 2019.

\bibitem{ShapeL1}
\BIBentryALTinterwordspacing
P.~Gong, X.~Zhang, and W.~Zheng, ``Unfolded coprime \uppercase{L}-shaped arrays
  for two-dimensional direction of arrival estimation,'' \emph{International
  Journal of Electronics}, vol. 105, no.~9, pp. 1501--1519, 2018. [Online].
  Available: \url{https://doi.org/10.1080/00207217.2018.1460874}
\BIBentrySTDinterwordspacing

\bibitem{ShapeL2}
A.~Elbir, ``L-shaped coprime array structures for \uppercase{D}o\uppercase{A}
  estimation,'' \emph{Multidimensional Systems and Signal Processing}, 05 2019.

\bibitem{ShapeL3}
M.~{Yang}, J.~{Ding}, B.~{Chen}, and X.~{Yuan}, ``Coprime \uppercase{L}-shaped
  array connected by a triangular spatially-spread
  electromagnetic-vector-sensor for two-dimensional direction of arrival
  estimation,'' \emph{IET Radar, Sonar Navigation}, vol.~13, no.~10, pp.
  1609--1615, 2019.

\bibitem{ShapeL4}
Q.~{Liu}, X.~{Yi}, L.~{Jin}, and W.~{Chen}, ``Two dimensional direction of
  arrival estimation for co-prime \uppercase{L}-shaped array using sparse
  reconstruction,'' in \emph{2015 8th International Congress on Image and
  Signal Processing (CISP)}, 2015, pp. 1499--1503.

\bibitem{ShapeL5}
Z.~{Zhang}, Y.~{Guo}, Y.~{Huang}, and P.~{Zhang}, ``A 2-\uppercase{D}
  \uppercase{D}o\uppercase{A} estimation method with reduced complexity in
  unfolded coprime \uppercase{L}-shaped array,'' \emph{IEEE Systems Journal},
  pp. 1--4, 2019.

\bibitem{ShapeL6}
P.~{Gong}, X.~{Zhang}, and W.~{Zheng}, ``{Unfolded coprime \uppercase{L}-shaped
  arrays for two-dimensional direction of arrival estimation},''
  \emph{International Journal of Electronics}, vol. 105, no.~9, pp. 1501--1519,
  Sep. 2018.

\bibitem{ShapeL7}
\BIBentryALTinterwordspacing
B.~Hu, W.~Lv, and X.~Zhang, ``\uppercase{2D}-\uppercase{D}o\uppercase{A}
  estimation for co-prime \uppercase{L}-shaped arrays with propagator method,''
  in \emph{Proceedings of the 2015 4th National Conference on Electrical,
  Electronics and Computer Engineering}.\hskip 1em plus 0.5em minus 0.4em\relax
  Atlantis Press, 2015/12, pp. 1551--1556. [Online]. Available:
  \url{https://doi.org/10.2991/nceece-15.2016.279}
\BIBentrySTDinterwordspacing

\bibitem{ShapeL8}
X.~{Li}, X.~{Wang}, W.~{Wang}, and S.~{Ren}, ``Generalized \uppercase{L}-shaped
  array based on the difference and sum coarray concept,'' \emph{IEEE Access},
  vol.~8, pp. 140\,456--140\,466, 2020.

\bibitem{ShapeL9}
L.~Zhang, S.~Ren, X.~Li, G.~Ren, and X.~Wang, ``Generalized
  \uppercase{L}-shaped nested array concept based on the fourth-order
  difference co-array,'' \emph{Sensors}, vol.~18, p. 2482, 08 2018.

\bibitem{ShapeL10}
D.-l. YANG, W.-t. LIU, Q.-l. CHENG, Z.-x. XIA, and X.-f. ZHANG,
  ``\uppercase{2D}-\uppercase{D}o\uppercase{A} estimation for coprime
  \uppercase{L}-shaped arrays with music algorithm,'' \emph{DEStech
  Transactions on Computer Science and Engineering}, 10 2017.

\bibitem{ShapeV1}
A.~Elbir, ``V-shaped sparse arrays for 2-\uppercase{D}
  \uppercase{D}o\uppercase{A} estimation,'' \emph{Circuits, Systems, and Signal
  Processing}, vol.~38, 11 2018.

\bibitem{ShapeCirc1}
J.~{Zhao} and C.~{Ritz}, ``Co-prime circular microphone arrays and their
  application to direction of arrival estimation of speech sources,'' in
  \emph{ICASSP 2019 - 2019 IEEE International Conference on Acoustics, Speech
  and Signal Processing (ICASSP)}, 2019, pp. 800--804.

\bibitem{ShapeCirc2}
\BIBentryALTinterwordspacing
G.~jun Jiang, X.~peng Mao, and Y.~tan Liu, ``Coprime sparse circular array with
  little angular dependence and reduced mutual coupling,'' \emph{AEU -
  International Journal of Electronics and Communications}, vol. 117, p.
  153051, 2020. [Online]. Available:
  \url{http://www.sciencedirect.com/science/article/pii/S1434841119325476}
\BIBentrySTDinterwordspacing

\bibitem{ShapeCirc3}
T.~Basikolo, K.~ICHIGE, and H.~Arai, ``Nested circular array and its concentric
  extension for underdetermined direction of arrival estimation,'' \emph{IEICE
  Transactions on Communications}, vol. E101.B, 10 2017.

\bibitem{ShapeCirc4}
S.~{Wandale}, T.~{Basikolo}, and K.~{Ichige}, ``Super nested sparse circular
  array for high resolution \uppercase{D}o\uppercase{A} estimation,'' in
  \emph{2019 IEEE International Symposium on Circuits and Systems (ISCAS)},
  2019, pp. 1--5.

\bibitem{DFT_filter}
C.~{Liu} and P.~P. {Vaidyanathan}, ``Coprime \uppercase{DFT} filter bank
  design: Theoretical bounds and guarantees,'' in \emph{2015 IEEE International
  Conference on Acoustics, Speech and Signal Processing (ICASSP)}, April 2015,
  pp. 3861--3865.

\bibitem{MIN_defense}
Y.~Liu, J.~Buck, and R.~Bautista, ``Spatial power spectral estimation using
  coprime sensor array with the min processor,'' \emph{Journal of the
  Acoustical Society of America}, vol. 139, no.~04, pp. 2109--2110, 2016.

\bibitem{I2}
V.~Chavali, ``Coprime and nested array processing of the elba island sonar data
  set,'' M. S. thesis, George Mason University, 2017.

\bibitem{4.45}
Y.~D. Zhang, M.~G. Amin, and B.~Himed, ``Sparsity-based
  \uppercase{D}o\uppercase{A} estimation using co-prime arrays,'' in \emph{2013
  IEEE International Conference on Acoustics, Speech and Signal Processing},
  May 2013, pp. 3967--3971.

\bibitem{F2}
C.~Zhou, Y.~Gu, Y.~D. Zhang, Z.~Shi, T.~Jin, and X.~Wu, ``Compressive
  sensing-based coprime array direction-of-arrival estimation,'' \emph{IET
  Communications}, vol.~11, pp. 1719--1724(5), August 2017.

\bibitem{F3}
Z.~Cheng, Y.~Zhao, H.~Li, and P.~Shui, ``Two-dimensional
  \uppercase{D}o\uppercase{A} estimation algorithm with co-prime array via
  sparse representation,'' \emph{Electronics Letters}, vol.~51, pp.
  2084--2086(2), December 2015.

\bibitem{F5}
F.-G. Yan, S.~Liu, J.~Wang, M.~Jin, and Y.~Shen,
  ``\BIBforeignlanguage{English}{Fast \uppercase{D}o\uppercase{A} estimation
  using co-prime array},'' \emph{\BIBforeignlanguage{English}{Electronics
  Letters}}, vol.~54, pp. 409--410(1), April 2018.

\bibitem{F1}
G.~D. Martino and A.~Iodice, ``Passive beamforming with coprime arrays,''
  \emph{IET Radar, Sonar and Navigation}, vol.~11, pp. 964--971(7), June 2017.

\bibitem{F4}
K.~Liu and Y.~D. Zhang, ``\BIBforeignlanguage{English}{Coprime array-based
  robust beamforming using covariance matrix reconstruction technique},''
  \emph{\BIBforeignlanguage{English}{IET Communications}}, vol.~12, pp.
  2206--2212(6), October 2018.

\bibitem{16.1}
Y.~Gu, C.~Zhou, N.~A. Goodman, W.~Z. Song, and Z.~Shi, ``Coprime array adaptive
  beamforming based on compressive sensing virtual array signal,'' in
  \emph{2016 IEEE International Conference on Acoustics, Speech and Signal
  Processing (ICASSP)}, March 2016, pp. 2981--2985.

\bibitem{4.40}
P.~P. Vaidyanathan and P.~Pal, ``System identification with sparse coprime
  sensing,'' \emph{IEEE Signal Processing Letters}, vol.~17, no.~10, pp.
  823--826, Oct 2010.

\bibitem{4.28}
S.~Ren, Z.~Zeng, C.~Guo, and X.~Sun, ``Wideband spectrum sensing based on
  coprime sampling,'' in \emph{22nd Int. Conf. Telecommunications (ICT)}, 2015,
  pp. 348--352.

\bibitem{4.32}
P.~Pal and P.~P. Vaidyanathan, ``Soft-thresholding for spectrum sensing with
  coprime samplers,'' in \emph{IEEE Sensor Array and Multichannel Signal
  Processing Workshop (SAM)}.\hskip 1em plus 0.5em minus 0.4em\relax IEEE,
  2014, pp. 517--520.

\bibitem{Speec}
J.~{Zhao} and C.~{Ritz}, ``Investigating co-prime microphone arrays for speech
  direction of arrival estimation,'' in \emph{2018 Asia-Pacific Signal and
  Information Processing Association Annual Summit and Conference (APSIPA
  ASC)}, 2018, pp. 1658--1664.

\bibitem{mimo1}
M.~{Priyadarsini} and C.~{Srinivasarao}, ``Beamforming in \uppercase{MIMO}
  radar using coprime array,'' in \emph{2018 2nd International Conference on
  Electronics, Materials Engineering Nano-Technology (IEMENTech)}, 2018, pp.
  1--4.

\bibitem{mimo2}
C.~{Li}, L.~{Gan}, and C.~{Ling}, ``2\uppercase{D} \uppercase{MIMO} radar with
  coprime arrays,'' in \emph{2018 IEEE 10th Sensor Array and Multichannel
  Signal Processing Workshop (SAM)}, 2018, pp. 612--616.

\bibitem{mimo3}
\BIBentryALTinterwordspacing
J.~Li, L.~He, Y.~He, and X.~Zhang, ``Joint direction of arrival estimation and
  array calibration for coprime \uppercase{MIMO} radar,'' \emph{Digital Signal
  Processing}, vol.~94, pp. 67 -- 74, 2019, special Issue on Source
  Localization in Massive MIMO. [Online]. Available:
  \url{http://www.sciencedirect.com/science/article/pii/S1051200419300946}
\BIBentrySTDinterwordspacing

\bibitem{mimo4}
J.~Shi, G.~Hu, X.~Zhang, F.~Sun, W.~Zheng, and Y.~Xiao, ``Generalized co-prime
  \uppercase{MIMO} radar for \uppercase{D}o\uppercase{A} estimation with
  enhanced degrees of freedom,'' \emph{IEEE Sensors Journal}, vol.~PP, pp.
  1--1, 12 2017.

\bibitem{doppler1}
J.~{Pan}, C.~{Zhou}, B.~{Liu}, and K.~{Jiang}, ``Joint
  \uppercase{D}o\uppercase{A} and doppler frequency estimation for coprime
  arrays and samplers based on continuous compressed sensing,'' in \emph{2016
  CIE International Conference on Radar (RADAR)}, 2016, pp. 1--5.

\bibitem{spacetime1}
C.~{Liu} and P.~P. {Vaidyanathan}, ``Coprime arrays and samplers for space-time
  adaptive processing,'' in \emph{2015 IEEE International Conference on
  Acoustics, Speech and Signal Processing (ICASSP)}, 2015, pp. 2364--2368.

\bibitem{CR1}
Y.~Zhao and S.~Xiao, ``Sparse multiband signal spectrum sensing with
  asynchronous coprime sampling,'' \emph{Cluster Computing}, vol.~22, 03 2019.

\bibitem{U_S_NCC2018}
U.~V. Dias and S.~Srirangarajan, ``Co-prime sampling and cross-correlation
  estimation,'' in \emph{24th National Conference on Communications (NCC)}, Feb
  2018.

\bibitem{UVD_PHD}
U.~V. Dias, ``Sub-\uppercase{N}yquist \uppercase{C}o-\uppercase{P}rime
  \uppercase{S}ensing: \uppercase{T}oo \uppercase{L}ittle cannot
  \uppercase{B}elittle \uppercase{Y}ou,'' Doctoral Thesis, Department of
  Electrical Engineering, Indian Institute of Technology Delhi, New Delhi, July
  2020.

\bibitem{I1}
P.~{Pakrooh}, L.~L. {Scharf}, and A.~{Pezeshki}, ``Modal analysis using
  co-prime arrays,'' \emph{IEEE Transactions on Signal Processing}, vol.~64,
  no.~9, pp. 2429--2442, May 2016.

\bibitem{S1}
X.~Huang, Z.~Yan, S.~Jing, H.~Fang, and L.~Xiao, ``Co-prime sensing-based
  frequency estimation using reduced single-tone snapshots,'' \emph{Circuits,
  Systems, and Signal Processing}, vol.~35, no.~9, pp. 3355--3366, 2016.

\bibitem{U_S_1}
U.~V. Dias and S.~Srirangarajan, ``Co-prime arrays and difference set
  analysis,'' in \emph{25th European Signal Processing Conference (EUSIPCO)},
  2017, pp. 961--965.

\bibitem{UVD_extended}
U.~V. Dias, ``Extended (conventional) co-prime arrays and difference set
  analysis: Low latency approach,'' 2020, arXiv: 2003.05474 [eess.SP].

\bibitem{UVD_CACIS}
------, ``Sub-\uppercase{N}yquist coprime sensing with compressed inter-element
  spacing - low latency approach,'' \emph{ICTACT Journal on Communication
  Technology}, vol.~11, no.~1, pp. 2126--2137, 2020.

\bibitem{UVD_supernyquist}
------, ``Super-\uppercase{N}yquist co-prime sensing,'' 2020, arXiv: 2010.00858
  [eess.SP].

\bibitem{7.1}
P.~Stoica and R.~L. Moses, \emph{Spectral Analysis of Signals}.\hskip 1em plus
  0.5em minus 0.4em\relax Upper Saddle River, New Jersey: Pearson Prentice
  Hall, 2005, vol. 452.

\bibitem{7.17}
\BIBentryALTinterwordspacing
E.~Axell. (2011, June 29) Lecture notes on nonparametric spectral estimation.
  [Online]. Available: \url{http://www.commsys.isy.liu.se/ADE/axell-notes.pdf}
\BIBentrySTDinterwordspacing

\bibitem{U_S_2}
U.~V. Dias and S.~Srirangarajan, ``Co-prime sampling jitter analysis,'' in
  \emph{25th European Signal Processing Conference (EUSIPCO)}, 2017, pp.
  1180--1184.

\bibitem{UVD_jitter_CAMP}
U.~V. Dias, ``Co-prime sensing with multiple periods and difference set
  analysis in the presence of sampling jitter,'' 2020, arXiv: 2003.07248
  [eess.SP].

\bibitem{Dilation1}
G.~{Qin}, Y.~D. {Zhang}, and M.~G. {Amin}, ``\uppercase{D}o\uppercase{A}
  estimation exploiting moving dilated nested arrays,'' \emph{IEEE Signal
  Processing Letters}, vol.~26, no.~3, pp. 490--494, 2019.

\bibitem{Dilation2}
Y.~{Zhou}, Y.~{Li}, and C.~{Wen}, ``The multi-level dilated nested array for
  direction of arrival estimation,'' \emph{IEEE Access}, vol.~8, pp.
  43\,134--43\,144, 2020.

\bibitem{Dilation3}
S.~{Li} and X.~{Zhang}, ``A novel moving sparse array geometry with increased
  degrees of freedom,'' in \emph{ICASSP 2020 - 2020 IEEE International
  Conference on Acoustics, Speech and Signal Processing (ICASSP)}, 2020, pp.
  4767--4771.

\end{thebibliography}

\end{document}